\documentclass[12pt]{article}
\usepackage{amsmath,amsthm,amssymb,mathtools,bm,hyperref,xcolor,multirow}
\usepackage{graphicx,psfrag,epsf}
\usepackage{enumerate}
\usepackage{natbib}
\usepackage{url} % not crucial - just used below for the URL 
\usepackage[nodisplayskipstretch]{setspace}
\usepackage[ruled,vlined]{algorithm2e}
\SetAlgoSkip{} 
\newcommand{\blind}{0}

\theoremstyle{plain}
\newtheorem{theorem}{Theorem}
\newtheorem{lemma}{Lemma}
\newtheorem{corollary}{Corollary}

\theoremstyle{definition}
\newtheorem{example}{Example}

    \usepackage{paralist}
    \usepackage[compact]{titlesec}
    \titlespacing{\section}{0pt}{0ex}{0ex}
    \titlespacing{\subsection}{0pt}{0ex}{0ex}
    \titlespacing{\subsubsection}{0pt}{0ex}{0ex}

\begin{document}

\def\spacingset#1{\renewcommand{\baselinestretch}%
{#1}\small\normalsize} \spacingset{1}

%%%%%%%%%%%%%%%%%%%%%%%%%%%%%%%%%%%%%%%%%%%%%%%%%%%%%%%%%%%%%%%%%%%%%%%%%%%%%%

  \title{\bf 
    Priors and parameterizations for Bayesian regression with categorical variables and interactions 
    % The interplay of priors and parameterizations for Bayesian regression with categorical covariates
    %parameterizations and priors matter for Bayesian regression with categorical covariates and interactions 
    %The Overlooked Role of parameterization in Bayesian Regression with Categorical Covariates and Interactions
    %Shrinkage Meets parameterization: Efficient Bayesian Regression with Categorical Covariates
  }
  \if0\blind
    {   
  \author{Daniel R. Kowal\thanks{Corresponding author. Associate Professor, Department of Statistics and Data Science, Cornell University (\href{mailto:dan.kowal@cornell.edu}{dan.kowal@cornell.edu}). 
    Research  was sponsored by the National Institute of Environmental Health Sciences (R01ES028819) and the National Science Foundation (SES-2435310). The findings and conclusions in this publication are those of the author(s) and do not necessarily represent the views of the NIH, the U.S. government, or the North Carolina Department of Health and Human Services, Division of Public Health.} \ and 
    Inkoo Lee\thanks{Assistant Professor, Department of Epidemiology and Biostatistics, University of Georgia}
}} \fi

  \maketitle

\bigskip
\begin{abstract}
Regression analysis commonly features categorical variables and their interactions to model heterogeneous, group-specific effects. Such interactions are notorious for weak signals and cumbersome parameterizations, and thus urgently require shrinkage. However, the overwhelming default for linear models---using dummy coding, removing a reference group, and applying shrinkage---leads to unwelcome biases, asymmetries, and parameter incongruities across models. We resolve these limitations with a new parameterization-prior combination that couples abundance-based constraints for identifiability with customized shrinkage priors for categorical effects and interactions (BayesABCs). BayesABCs offers a surprising and unique invariance property: the joint posterior of all main effects is (nearly) unchanged by the addition of categorical interactions. We show how this, in conjunction with our shrinkage priors, effectively removes the principal drawbacks of including categorical interactions, thereby making heterogeneous effect estimation broadly feasible. We enforce the identifiability constraints via conditioning and provide a model representation that facilitates efficient posterior computing and direct analysis. Across simulations with a variety of categorical effects and interactions, BayesABCs consistently produces more accurate estimates and predictions, sharper inferences, and unmatched robustness to extraneous interactions relative to competing parameterizations and priors. We apply BayesABCs to study disparities in the error rates of automatic speech recognition systems.
%Thus, ABCs facilitate heterogeneous effect estimation without any notable drawbacks for estimating or interpreting the main effects. 
%Consequently, we anticipate an excess of interaction terms; we design structured shrinkage priors to effectively remove the unimportant ones. 
%How about computing? prior construction? 
% educational outcomes for children in North Carolina. \if0\blind An \texttt{R} package \texttt{lmabc} is available. \fi
\end{abstract}

\noindent%
\vfill

\newpage
%\spacingset{1.45} % DON'T change the spacing!
\spacingset{1.75} % DON'T change the spacing!

\section{Introduction}
\label{sec-intro}
Despite advancements in statistical and machine learning, linear models remain exceptionally useful. Among many benefits, they are conducive to careful model construction, with interpretable parameters, engineered features, and priors adapted to challenging (e.g., high-dimensional) settings. 
%They are amenable to careful model construction with a variety of covariate data types and engineered features; admit customized priors for structured or high-dimensional settings; produce interpretable model output and predictions; are computationally favorable; and are both general and generalizable, including for nonlinear effects  and data dependencies. 
Of particular interest are \emph{categorical modifiers}, where a categorical variable (or factor) is interacted with a continuous or categorical variable to estimate heterogeneous, group-specific effects. In these often noisy, low signal settings, Bayesian models are appealing because of the shrinkage and uncertainty quantification they provide. However, Bayesian inference %---especially for  categorical effects and their interactions--- 
depends critically on both the model \emph{parameterization} and the \emph{prior}. It is at this junction where existing approaches have significant flaws.

Consider an outcome $y \in \mathbb{R}$,  continuous covariates $\bm x \in \mathbb{R}^p$, and  categorical covariates $\bm c = (c_1,\ldots,c_K)'$ where each $c_k$ indexes the $L_k$ levels of the $k$th categorical variable. We study the class of linear models that include categorical-continuous (CX) and categorical-categorical (CC) interactions (or modifiers), referred to as \emph{cat-modified} models:
%Among the most popular models for estimating heterogeneous, group-specific effects are linear models with categorical-continuous (CX) and categorical-categorical (CC) interactions, referred to as \emph{cat-modified} models: 
\begin{equation}\label{reg-cm}
\mu(\bm x, \bm c) = \alpha_0 + \bm x' \bm \alpha + \sum_{k=1}^K \beta_{k, c_k} +  \sum_{k=1}^K  \bm x' \bm \gamma_{k, c_k} + \sum_{k=1}^{K-1} \sum_{k'=k+1}^K \gamma_{k,k',c_k, c_{k'}}
\end{equation}
where $\mu(\bm x, \bm c)$ models the  expectation of $y$, $\bm \alpha = (\alpha_1,\ldots,\alpha_p)'$ are  continuous main effects, $\{\beta_{k,c_k}\}_{c_k=1}^{L_k}$ are  main effects for the $k$th categorical variable, $\bm \gamma_{k,c_k} = (\gamma_{1,k, c_k},\ldots, \gamma_{p,k, c_k})'$ are all CX interactions with the $k$th categorical variable, and $\{\gamma_{k,k',c_k, c_{k'}}\}_{c_k, c_{k'}}$ are CC interactions between the $k$th and $k'$th categorical variables.  Special cases include   \emph{main-only} models with continuous and categorical variables but no %\texttt{y $\sim$ x$_1$ + \ldots + x$_p$ + c$_1$ + \ldots + c$_K$} 
 interactions (all $\gamma_{j,k, c_k} =0$ and $\gamma_{k,k',c_k, c_{k'}}= 0$); \emph{$K$-way ANOVA models}  with only categorical variables (all $\alpha_j = 0$ and $\gamma_{j,k,c_k} = 0$), with or without CC interactions; and
%, e.g., \texttt{y $\sim$ c$_1$ + \ldots + c$_K$},  that may also include CC interactions; 
\emph{ANCOVA models} with $K=1$ categorical variable and $p=1$ continuous variable, with or without the CX interaction.  %In  Wilkinson notation, the cat-modified model is \texttt{y $\sim$ (x$_1$ + \ldots + x$_p$)*(c$_1$ + \ldots + c$_K$) + c$_1$*c$_2$ + \ldots + c$_{K-1}$*c$_K$}, with similar expressions for these special cases. 
% that exclude (all $\gamma_{1,1,c_1} = 0$; \texttt{y $\sim$ x$_1$ + c$_1$}) or include (\texttt{y $\sim$ x$_1$ + c$_1$ + x$_1$:c$_1$}) the CX interaction. 
We focus on  pairwise CX and CC (not continuous-continuous) interactions only for simplicity.

We highlight three defining features of a cat-modified model. First, its purpose is to estimate heterogeneous  effects: 
%\begin{equation} \label{joint-slopes}
$\dot\mu_{x_j}(\bm c) \coloneqq \mu(x_j+1, \bm x_{-j}, \bm c) - \mu(x_j, \bm x_{-j}, \bm c) = \alpha_j + \sum_{k=1}^K \gamma_{j, k, c_k}, 
$ %\end{equation}
so the effect  of $x_j$ varies across mutual groups  $\bm c$ due to the CX interaction parameters $\gamma_{j, k, c_k}$. 
%which shows the central role of the CX interaction parameters $\gamma_{j, k, c_k}$. %For instance, ANCOVA models have group-specific intercepts and may have a common slope or group-specific slopes depending on whether the CX interaction is omitted (\texttt{y $\sim$ x$_1$ + c$_1$}) or included (\texttt{y $\sim$ x$_1$ + c$_1$ + x$_1$:c$_1$}). 
Similarly, the CC interactions $\{\gamma_{k,k',c_k, c_{k'}}\}_{c_k, c_{k'}}$ determine the group-specific intercepts and allow for intersections of mutual groups that impact the outcome beyond the individual main effects $\{\beta_{k,c_k}\}$.

Second, model \eqref{reg-cm} is overparameterized. Each CX interaction $\gamma_{j,k,c_k}$ is not identifiable from the main effect $\alpha_j$, and similar for  $\beta_{k,c_k}$, $\alpha_0$, and   $\gamma_{k,k',c_k, c_{k'}}$. Implicitly, model \eqref{reg-cm} uses dummy coding of categorical variables and falls into the ``dummy variable trap".  Although it is possible to fit \eqref{reg-cm} unconstrained with the help of informative priors, this has its drawbacks: parameters lose interpretability, the posterior may be improper, algorithms behave erratically, and performance suffers %---even for identified quantities like $\mu$ 
(see Sections~\ref{sec-sims} and \ref{sec-asr}). 
Instead, the predominant identification strategy is \emph{reference group encoding} (RGE): a reference group is chosen for each categorical variable, say $c_k = 1$, and removed ($\beta_{k,1} = 0,\gamma_{j,k,1} = 0, \gamma_{k,k',1,1} =0$ for all $j,k,k'$). This is the default in all major statistical software (\texttt{R}, Python, SAS, MATLAB, Stata, etc.), nearly all textbook treatments of linear models, and nearly all current statistical methods (see below). However, RGE is problematic: it creates unwelcome biases, asymmetries, and parameter incongruities.  We illustrate with an example.
%We illustrate some of these issues with a simple example. 
%\vspace{-1mm} 
\begin{example}\label{ex-anova}
A one-way ANOVA model is $\mu(c) = \alpha_0 + \beta_c$, where $c \in \{1,\ldots,L\}$ denotes the categorical variable's levels (omitting $k$ subscripts). This is   typically paired with Gaussian errors for the observed outcome
%Consider a categorical variable with $L$ levels. Given data $\{r_i, y_i\}_{i=1}^n$ that records the group and the outcome, the one-way ANOVA model is
%\begin{equation}\label{main-only-anova}
    %y_i = \alpha_0 + \beta_{c_i} + \epsilon_i, \quad [\epsilon_i \mid \tau] \stackrel{iid}{\sim} %N(0,\tau^{-1}).%, \quad i=1,\ldots,n.
%\end{equation}
%Suppose we pair \eqref{main-only-anova} with 
and priors such as  $\alpha_0\sim N(0, \lambda_\alpha^2)$ and $ \beta_c   \sim N(0, \lambda_\beta^2)$, all independent.  
Besides the parameters $\alpha_0$ and $\bm \beta = (\beta_1,\ldots,\beta_L)'$, we are interested in group means,  $\mu_c \coloneqq \mu(c)$, and their contrasts, $\delta_{c,c'} \coloneqq \mu_c - \mu_{c'}$. %= \beta_c - \beta_{r'}$. %This collection of parameterization, priors, and inferential targets remains relevant more generally, for instance with CX interactions ($\alpha_0$ would play the role of a continuous main effect  and $\beta_{r}$ the CX interaction coefficient). 

For identifiability, RGE fixes $\beta_1 = 0$. The implied priors for the group means are $\mu_1 \sim N(0, \lambda_\alpha^2)$ and $\mu_c \sim N(0, \lambda_\alpha^2 + \lambda_\beta^2)$ for $c > 1$, and for the contrasts  $\delta_{c,1} \sim N(0, \lambda_\beta^2)$ and $\delta_{c,c'} \sim N(0, 2 \lambda_\beta^2)$ for $c \ne c' > 1$. However, applying shrinkage to the parameters, say $\beta_c  \to 0$, creates bias: the group means are pulled toward that of the reference group, $\mu_c \to \mu_1$. This is especially concerning for variables like race, sex, or religion, where shrinkage under RGE induces \emph{biased} (by race, sex, religion) estimates. It also suppresses deviations from the reference group ($\delta_{c,1} \to 0$), which is commonly a dominant group  (White, male, Protestant, etc.). 
Shrinkage to a reference group can be reasonable, such as when $c=1$ is a control and $c>1$ treatments, but it is not a favorable target in general. Besides this bias, RGE leads to asymmetry: posterior inference depends on the choice of reference group, with prior variances for group means and contrasts that are larger for non-reference groups. 

These concerns about RGE persist more generally with cat-modified models \eqref{reg-cm}:  for CX interactions, replace $\alpha_0$ with $\alpha_j$, $\beta_c$ with $\gamma_{j,k,c_k}$, and $\mu_c$ with $\dot\mu_{x_j}(\bm c)$. 
%While this behavior may be appropriate in some settings, such as when $r=1$ is a control and $r>1$ are treatments, it is problematic in other settings, such as when $R$ is a protected group like race, sex, or religion, and the reference group is a dominant group like White, male, Protestant, etc. 
%For $L=2$, the contrast behavior is unremarkable, since there are no contrasts that exclude the reference group, but the variance for the group-specific effect is inflated for the non-reference level.  
%Thus, results depend on the choice of reference group, and shrinkage of $\{\beta_c\}$ pulls the group-specific effects towards that of the reference group. 

%It is also possible to fit the model without any constraints on $\{\beta_c\}$, which we call an  \emph{overparameterized} model (Over). This approach has significant drawbacks: the nonidentified parameters are difficult to interpret, the posterior may be improper for certain priors, and traditional (MCMC) algorithms may be unreliable (see Section~\ref{sec-asr}). 
\end{example} %\vspace{-1mm}

The third feature of \eqref{reg-cm} is that it includes an enormous collection of potential models with varying parameter meanings. Rarely would an analyst begin with \eqref{reg-cm} in its fullest form. Instead, one typically would begin with a main-only model and incorporate interactions carefully and incrementally. Yet critically, with RGE, the parameter interpretations change drastically across version of \eqref{reg-cm}. For instance, under a main-only model, each $x_j$-effect is common across groups, $\alpha_j = \dot\mu_{x_j}(\bm c)$. Including CX interactions and using RGE,   $\alpha_j = \dot\mu_{x_j}(\bm 1)$ is the $x_j$-effect \emph{for the reference groups}. These quantities share the same names (``main effects") and notation ($\alpha_j$), yet the inclusion of CX interactions changes their meanings, their estimates, and usually inflates their uncertainties. This impedes clear regression analysis of heterogeneous effects and can produce misleading conclusions \citep{kowal2024regression}.

The central point is that Bayesian inference with cat-modified models must consider the interplay of \emph{parameterization} and \emph{prior}. It is natural to prioritize shrinkage priors that pull (CX and CC) interactions toward zero to eliminate unwarranted heterogeneity. Yet the prior alone is insufficient: it is the parameterization that determines the shrinkage targets and thus model behaviors. Here, we construct a Bayesian model parameterization based on abundance-based constraints (ABCs\footnote{No relation to ``approximate Bayesian computation".}), which identify parameters using group-weighted averages  (Section~\ref{sec-methods}). To enforce parameter identification for Bayesian modeling, we specify a prior in the overparameterized space  of \eqref{reg-cm} and then condition on the (linear) constraints. This is accompanied by a tractable, lower-dimensional representation that enables posterior analysis and efficient MCMC sampling. We refer to this approach % that combines Bayesian modeling for  \eqref{reg-cm} with ABCs for identification
as \emph{BayesABCs}.

A main argument for BayesABCs is a unique \emph{posterior invariance} property for \eqref{reg-cm}: the joint posterior for main effects is (nearly) unchanged by the addition of categorical interactions (Section~\ref{sec-theory}). Crucially, we may include CC and CX interactions---and learn potential heterogeneities---without incurring any costs for main effect posterior inference. %This cannot occur for RGE. 
This also ensures favorable shrinkage targets  for heterogeneous effects: %With RGE, group-specific effects such as $\dot\mu_{x_j}(\bm c)$ are pulled toward the reference group effect. 
%With BayesABCs, 
the group-specific effects are pulled toward group-averaged effects, which are equivalently main effects from the \emph{main-only} model. Accordingly, we design shrinkage priors that aggressively shrink null (interaction) effects to zero and  leverage the grouping structures among categorical effects and their interactions (Section~\ref{subsec-priors}). Whereas invariance opens the door for a zoo of categorical interactions, BayesABCs provides multiple protections against extraneous ones.

These advantages lead to clear practical gains over other parameterization-prior combinations.  Simulating data for several versions of \eqref{reg-cm}---one-way ANOVAs, two-way ANOVAs with and without CC interactions, and general main-only and cat-modified models---we show how BayesABCs produces more accurate estimates, sharper inferences, and unmatched robustness to extraneous categorical interactions (Section~\ref{sec-sims}). We apply BayesABCs to analyze error rates for automatic speech recognition systems and highlight its clear, interpretable, and MCMC-efficient posterior inferences for both main and heterogeneous effects (Section~\ref{sec-asr}).  This example also includes a subject identifier for repeated measurements, and shows that BayesABCs remains effective with high-dimensional categorical variables (124 levels) and for linear mixed models  with random intercepts and slopes.

%Uniquely, BayesABCs leads to clear and robust conclusions about main effects, , and demonstrates excellent MCMC efficiency. 

%Second, we develop efficient computing strategies for posterior inference under ABCs, which apply for other linear constraints. Third, we design and evaluate multiple competing shrinkage priors for cat-modified models, from simple ridge-like priors to structured global-local shrinkage priors, with or without CC heredity whereby CC interactions inherit shrinkage parameters from the parent main effects. A
%In fact, posterior invariance with BayesABCs holds under noninformative and informative priors

%The conditions are mild and consider both noninformative and informative priors, but ultimately the theory explains what we observe in practice for structured shrinkage priors. 

% Why is this good? 
%   can include interactions for free
%   main effects are good shrinkage targets

%Using CX interactions as an example, the main effect $\alpha_j$ is the shrinkage target of the $x_j$-effect $\dot\mu_{x_j}(\bm c)$  in  \eqref{joint-slopes} (i.e., when $\gamma_{j,k,c_k} \to 0$). With RGE, that target is determined by the reference groups---which introduces biases (by race, sex, religion, etc.), undesirable asymmetries for posterior inference, and incongruence across versions of \eqref{reg-cm}. 

ABCs have been used intermittently in various forms, although not (to our knowledge) for Bayesian analysis. Classically, ANOVA models used  linear constraints for identifiability \citep{Scheffe1999}. %\citep{Scheffe1999,Fujikoshi1993}, 
%such as RGE, ABCs, and sum-to-zero (STZ) constraints.   
\cite{Sweeney1972} suggested ABCs for  ANCOVA models without interactions. Related parameterizations appeared in \cite{Park2021} %and \cite{Park2023} 
to estimate interactions but not main effects and in \cite{stokell2021modelling} for penalized estimation with $K$-way ANOVAs without interactions. From a contrast coding perspective, ABCs are linked to weighted effects coding \citep{TeGrotenhuis2017a}. In the context of \eqref{reg-cm}, \cite{kowal2026facilitating} advocated for ABCs with ordinary least squares (OLS) estimation, including OLS invariance results for main effects. Those results used disparate conditions for different versions of \eqref{reg-cm} and only considered point estimates and standard errors. By comparison, we provide overarching, unifying conditions for invariance and study nontrivial posterior distributions that are joint across main effects, marginal over other parameters, induced by different priors, and conditional on identifying constraints. The importance of shrinkage also cannot be overstated. While invariance protects \emph{main} effects  against extraneous categorical interactions, it does not protect predictions or distinguish  signal from noise among interactions. Our shrinkage priors are designed for these tasks, leading to massive empirical improvements in estimation, prediction, and inference for BayesABCs over OLS. 

Bayesian models for categorical effects and interactions have focused on priors but not parameterizations.  \cite{dixon1991bayesian} used a ridge prior for  two-way ANOVA, where all CC interactions shared a common shrinkage parameter. %\cite{jones2011bayesian} allowed separate variances for each interaction order. 
MAIHDA \citep{evans2018multilevel}  is widely popular in the social sciences to model heterogeneity across  intersections of user-specified ``social strata" (e.g., Black and female, high SES and Hispanic, etc.), and models them with a single, exchangeable random effect.  For one-way ANOVA models, \cite{pauger2019bayesian} introduced sparsity priors on all pairwise differences to fuse levels, although \cite{stokell2021modelling} cautioned that such priors favor groups of unequal sizes. \cite{alt2025hierarchical} used grouped horseshoe priors for heterogeneous treatment effects. %  in randomized clinical trials. 
However, \emph{all} of these approaches used RGE. Notable exceptions include \cite{gelman2005analysis}, who focused on batched priors for categorical effects but did not impose identifiability constraints, and \cite{garcia2021variable}, who deliberately avoided RGE for Bayesian variable selection of categorical effects but did not estimate coefficients or include interactions. 
% and inherit its limitations.

%First, the \emph{main-only} model 
%\begin{equation}
%    \label{reg-main}
%    \mu^M(\bm x, \bm c) = \alpha_0^M + \bm x' \bm \alpha^M + \sum_{k=1}^K \beta_{k, c_k}^M
%\end{equation}
%includes continuous and categorical main effects but no interactions.   Second, the \emph{cat-modified} model is

Frequentist methods have prioritized penalized estimation for selecting and fusing levels of a categorical covariate in ANOVAs  \citep{bondell2009simultaneous,gertheiss2010,stokell2021modelling}, including interactions \citep{post2013factor}. \cite{Lim2015} selected  interaction effects using a variant of the group lasso penalty and %enforced strong hierarchies 
that 
only permits an interaction if both parent main effects are active. These methods are primarily split between RGE and sum-to-zero constraints. %; only \cite{stokell2021modelling} used ABCs. 
\cite{dondelinger2020joint} used group lasso-type estimation for a cat-modified ANCOVA model, but did not use any constraints and thus did not identify or estimate any main effects. In general, these approaches do not natively provide uncertainty quantification and require selection of one or more tuning parameters.

\section{Model parameterization}\label{sec-methods}

\subsection{Abundance-based constraints (ABCs)}\label{subsec-abcs}
Suppose we observe data   $\{y_i, \bm x_i, \bm c_i\}_{i=1}^n$. 
We identify the categorical effects, CX interactions, and CC interactions for the cat-modified model \eqref{reg-cm} using ABCs, which are linear constraints based on joint proportions $\bm{\hat \pi}$ across all categorical variables $\bm C = (C_1,\ldots, C_K)'$. % in the data $\{\bm c_i\}_{i=1}^n$. 
We prefer sample proportions: they are always available, lead to exact invariance properties  (Section~\ref{sec-theory}), and behave well with missing levels or CC interaction cells (Appendix~\ref{app-theory}).  ABCs for categorical main effects and CX interactions are  
\begin{subequations}\label{abcs-gen}
    \begin{align}
        \label{abcs-gen-cat}
        \mathbb{E}_{\bm{\hat \pi}} (\beta_{k, C_k}) &= 0, \quad k=1,\ldots, K \\
        \label{abcs-gen-cts}
        \mathbb{E}_{\bm{\hat \pi}} (\gamma_{j,k, C_k}) &= 0, \quad k=1,\ldots, K, \quad j=1,\ldots,p.
    \end{align}
\end{subequations}
We clarify that $\mathbb{E}_{\bm{\hat \pi}}$ is taken over the categorical variables  $\bm C$ and used for notational convenience and interpretability, not as part of a model or prior for the regression coefficients. For instance, \eqref{abcs-gen-cat} simply expresses a linear constraint $\sum_{\ell = 1}^{L_k} \hat \pi_{k, \ell} \beta_{k, \ell} = 0$, where $\{\hat \pi_{k,\ell}\}_{\ell=1}^{L_k}$ are the proportions for each categorical variable $C_k$, and similarly for \eqref{abcs-gen-cts}. 

The use of $\mathbb{E}_{\bm{\hat \pi}}$ in \eqref{abcs-gen-cat}--\eqref{abcs-gen-cts} also communicates the \emph{group average} interpretation of the main effects in model \eqref{reg-cm} under ABCs. The purpose of a cat-modified model is to estimate heterogeneous effects. ABCs identify main effects as the average of these effects:
 \begin{equation}\label{joint-slopes-avg}
\alpha_j = \mathbb{E}_{\bm{\hat \pi}} \{\dot\mu_{x_j}(\bm C)\}, \quad j=1,\ldots,p.
\end{equation}
%i.e., the continuous main effects are group-averages of the group-specific slopes.
Now,  CX interaction parameters are group-specific deviations from the  average effect: $\gamma_{j,k,c_k} = \mathbb{E}_{\bm{\hat \pi}_{-k}} \{\dot\mu_{x_j}(\bm C_{-k}, c_k)\} - \mathbb{E}_{\bm{\hat \pi}} \{\dot\mu_{x_j}(\bm C)\}$, where $\bm{\hat \pi}_{-k}$ are joint proportions over the remaining categorical variables $\bm C_{-k}$. This partial expectation only matters for other categorical variables interacted with $x_j$; if there are no others, then $\gamma_{j,k,c_k} = \dot\mu_{x_j}(c_k) -\mathbb{E}_{\bm{\hat \pi}} \{\dot\mu_{x_j}(\bm C)\}$. 
Unlike RGE, neither main effects nor interaction effects are bound to a reference group. 

  ABCs for CC interactions use  \emph{conditional} proportions for all $(C_k,C_{k'})$:
\begin{equation}\label{abcs-cat-cat-gen}
    \begin{split}
    \mathbb{E}_{\bm{\hat \pi}_{C_k \mid C_{k'}=\ell}}(\gamma_{k,k', C_k, \ell}) &= 0, \quad \ell=1,\ldots,L_{k'}\\
    \mathbb{E}_{\bm{\hat \pi}_{C_{k'} \mid C_{k}=\ell}}(\gamma_{k,k', \ell, C_{k'}}) &= 0, \quad \ell=1,\ldots,L_{k}
    \end{split}
\end{equation}
Like \eqref{abcs-gen-cat}--\eqref{abcs-gen-cts}, ABCs for CC interactions can be expressed as a linear constraint with $\bm{\hat\pi}$. Taking $K=2$ for illustration, \eqref{abcs-cat-cat-gen} is equivalent to $\sum_{r=1}^{L_1} \hat\pi_{rs}\gamma_{1,2,r,s} = 0$ for all $s=1,\ldots,L_2$ and  $\sum_{s=1}^{L_2} \hat\pi_{rs}\gamma_{1,2,r,s} = 0$ for all $r=1,\ldots,L_1$. We only must enforce \eqref{abcs-cat-cat-gen} for all but one interaction term (e.g., $\ell = 1,\ldots, L_k - 1$) and the constraint is guaranteed to hold for the remaining term ($\ell = L_k$) without loss of generality (\citealp{kowal2026facilitating}, Lemma~2). 

This particular CC interaction constraint has several useful properties. First,    $\alpha_0 = \mathbb{E}_{\bm{\hat \pi}}\{\mu(\bm 0, \bm C)\}$ is the \emph{group-averaged} intercept. All categorical main effects ($\beta_{k, c_k}$) and CC interactions ($\gamma_{k,k', c_k, c_{k'}}$) are identified in relation to this group average. Second, \eqref{abcs-cat-cat-gen} facilitates unique \emph{invariance} properties for the categorical main effects  (Corollary~\ref{cor-anova}), whereby the joint posterior for main effects is unchanged by the addition of CC interactions. And third, \eqref{abcs-cat-cat-gen} provides internal coherence: if we were to combine the interacted CC variables into a single categorical variable (e.g., race-sex) with $L_kL_{k'}$ levels and include it as a main effect, then \eqref{abcs-gen-cat} would hold for this new categorical variable. %Thus, it is unquestionably preferable to include both categorical main effects and the CC interaction.  

\subsection{Priors, posteriors, and computing with linear constraints}\label{subsec-enforce}
%{\color{blue} Mechanism: write a prior then \emph{condition} on constraints. This gives certain properties (under conditionally Gaussian priors) [include a lemma]. We can explore these using a QR decomposition [describe both prior and posterior distributions]. That also makes for convenient computing [algorithm].}
We provide a general mechanism to construct shrinkage priors that 1) enforce ABCs and 2) %provides favorable properties for shrinkage, and 3) 
admit efficient algorithms for posterior inference. Suppose the model contains a linear predictor $\bm X \bm \theta$ with $n\times p$ covariate matrix $\bm X$ and parameter $\bm \theta \in \mathbb{R}^p$ subject to 
%\begin{equation}\label{gen-con}
$\bm A \bm \theta = \bm 0$, 
%\end{equation}
where $\bm A$ is $m \times p$ with rank $m$, so $\bm \theta$ contains $p_\mathcal{N} = p - m$  free parameters. The cat-modified model \eqref{reg-cm} with ABCs \eqref{abcs-gen} and \eqref{abcs-cat-cat-gen} can be expressed in this way. 
%The central questions are 1) how to specify a prior jointly with the linear constraints, 2) what properties follow for the prior and posterior, and 3) how to compute the posterior distribution efficiently? 

With BayesABCs, we pursue a natural Bayesian approach: specify an unconstrained prior on $\bm \theta$ and then \emph{condition} on the constraints. Consider the class of (conditionally) Gaussian priors $\bm \theta \sim N_p(\bm 0, \bm \Sigma_\theta)$, which includes many popular shrinkage priors; extensions are available for nonzero  mean.  The task is then to construct the conditional version $[\bm \theta \mid \bm A \bm \theta = \bm 0]$. This is not difficult: $(\bm\theta, \bm A \bm \theta)$ are jointly Gaussian so the required conditional is available. However, there are implications for posterior computing. One approach is to sample from the unconstrained posterior  $[\bm \theta \mid \bm y]$ and then apply an appropriate shift to acquire a draw from the constrained posterior $[\bm \theta \mid \bm y, \bm A \bm \theta = \bm 0]$ (\citealp{gelfand2010handbook}; see Appendix~\ref{app-proofs-repar}). This requires sampling and computing in $p$ dimensions, which is wasteful in the presence of many constraints so that  $p_\mathcal{N} \ll p$.

Instead, we apply a reparameterization that facilitates both analysis and computing. First, take the QR decomposition of the transposed constraint matrix:
\begin{equation}\label{qr-con}
\bm A' = \bm Q \bm R = (\bm Q_{\mathcal{R}} ,   \bm Q_{\mathcal{N}})( \bm R_{\mathcal{R}}', \bm 0')'
%\bm A' = \bm Q \bm R = \begin{pmatrix} \bm Q_{\mathcal{R}} &  \bm Q_{\mathcal{N}} \end{pmatrix} \begin{pmatrix} \bm R_{\mathcal{R}} \\ \bm 0 \end{pmatrix}
\end{equation}
%Let $\bm A' = (\bm Q_{\mathcal{R}}: \bm Q_{\mathcal{N}}) (\bm R_{A}' : \bm 0)' = \bm Q_{\mathcal{R}} \bm R_{\mathcal{R}}$ be the QR decomposition of the transposed constraint matrix, 
where $\bm Q'\bm Q = \bm Q \bm Q' = \bm I_{p}$, $\bm Q_{\mathcal{R}} \in \mathbb{R}^{p \times m}$, $\bm Q_{\mathcal{N}} \in \mathbb{R}^{p \times p_{\mathcal{N}}}$, and the columns of $\bm Q_{\mathcal{R}}$ and  $\bm Q_{\mathcal{N}}$ span the row space of $\bm A$ and the null space of $\bm A$, respectively.    %By design, $\bm A \bm Q_{\mathcal{N}} = \bm 0$. 
Define the rotation $\bm Q' \bm \theta = (\bm \theta_{\mathcal{R}}', \bm \theta_{\mathcal{N}}')'$ so that  $\bm \theta =  \bm Q_{\mathcal{R}} \bm \theta_{\mathcal{R}} + \bm Q_{\mathcal{N}} \bm \theta_{\mathcal{N}}$. Now, the linear constraint  is equivalent to  $\bm \theta_{\mathcal{R}} = \bm 0$, leading to the reparameterization $\bm \theta = \bm Q_{\mathcal{N}}  \bm \theta_{\mathcal{N}}$. Any such $\bm\theta$ satisfies $\bm A \bm \theta = \bm 0$. 

This indicates the general strategy: find the posterior of the lower-dimensional, unconstrained parameter $\bm \theta_{\mathcal{N}} \in \mathbb{R}^{p_{\mathcal{N}}}$ and set $\bm \theta = \bm Q_{\mathcal{N}}  \bm \theta_{\mathcal{N}}$. First, it is convenient to derive an unconstrained prior for $\bm \theta_{\mathcal{N}}$ that induces the  constrained prior on $\bm \theta$. 
\begin{lemma} \label{lem-prior-equiv}
    For a (conditionally) Gaussian prior $\bm \theta \sim N_p (\bm 0, \bm \Sigma_\theta)$, the constrained prior satisfies $[\bm \theta \mid \bm A \bm \theta = \bm 0] \stackrel{d}{=} \bm Q_{\mathcal{N}} \bm \theta_{\mathcal{N}}$ where $\bm \theta_{\mathcal{N}} \sim N_{p_\mathcal{N}}\{\bm 0,  (\bm Q_{\mathcal{N}}'\bm \Sigma_\theta^{-1}\bm Q_{\mathcal{N}})^{-1}\}.$
\end{lemma}
%The proof is similar to the posterior version (see below) and omitted. 
The constrained prior is then $[\bm \theta \mid \bm A \bm \theta = \bm 0] \sim N_p(\bm 0, \bm Q_{\mathcal{N}} (\bm Q_{\mathcal{N}}'\bm \Sigma_\theta^{-1}\bm Q_{\mathcal{N}})^{-1} \bm Q_{\mathcal{N}}')$. A useful special case occurs for a ridge prior, $\bm \Sigma_\theta =  \lambda_\theta^2 \bm I_p$, where  orthonormality simplifies $\bm \theta_\mathcal{N} \sim N_{p_\mathcal{N}}(\bm 0, \lambda_\theta^2 \bm I_{p_\mathcal{N}})$ and $[\bm \theta \mid \bm A \bm \theta = \bm 0] \sim N_p(\bm 0,  \lambda_\theta^2 \bm Q_{\mathcal{N}}  \bm Q_{\mathcal{N}}')$. Now, observe that $\bm Q_{\mathcal{N}}\bm Q_{\mathcal{N}}'$ is an orthogonal projector for the null space of $\bm A$. As a consequence, the covariance may be written explicitly in terms of the constraints: $\lambda_\theta^2 \bm Q_{\mathcal{N}}\bm Q_{\mathcal{N}}' = \lambda_\theta^2\{\bm I_{p} - \bm A' (\bm A \bm A')^{-1} \bm A\}$. This is insightful for prior and posterior analysis (see Section~\ref{subsec-anova}).
%We revisit this simplification for a one-way ANOVA model in Section~\ref{subsec-anova}. 
%For instance, with a single constraint $\bm A = \bm \pi'$, we have $\bm Q_{\mathcal{N}}\bm Q_{\mathcal{N}}' = \bm I_{p} - \bm \pi \bm \pi'/\Vert \bm \pi \Vert^2$

%Then, by reparameterizing $\bm \gamma = \bm Q_{\mathcal{N}} \bm \gamma_{\mathcal{N}}$  the constraint $\bm A \bm \gamma = \bm 0$ is satisfied for \emph{any} $\bm \gamma_{\mathcal{N}} \in \mathbb{R}^{p_{\mathcal{N}}}$. 

Focusing on the Gaussian linear model $[\bm y \mid \bm\theta] \sim N_n(\bm X \bm \theta, \tau^{-1} \bm I_n)$, 
we provide the constrained posterior distribution and proposed sampling algorithm in Algorithm~\ref{alg:constr}   and establish its validity in Appendix~\ref{app-proofs-repar} (Lemma~\ref{lem-prior-equiv} follows immediately). Using Lemma~\ref{lem-prior-equiv} and rewriting the linear predictor $\bm X \bm \theta = \bm X_{\mathcal{N}} \bm \theta_{\mathcal{N}}$ with $\bm X_{\mathcal{N}} = \bm X \bm Q_{\mathcal{N}}$, it is a standard exercise to derive and sample the posterior of $\bm \theta_{\mathcal{N}} $ and thus $\bm \theta$.  The sampling step involves only $p_{\mathcal{N}}$-dimensional or $p_{\mathcal{N}} \times p_{\mathcal{N}}$ quantities, so there is no wasted computing effort on the $m$ constrained components. 
Computing $\bm Q_{\mathcal{N}}$ and $\bm X_{\mathcal{N}}$ is a one-time cost prior to any sampling. 

\begin{algorithm}[h]
\SetAlgoLined  
\begin{enumerate}
    \item Sample the unconstrained parameter, $[\bm \theta_{\mathcal{N}} \mid \bm y] \sim N_{p_\mathcal{N}}(\bm \mu_{\mathcal{N}}, \bm \Sigma_{\mathcal{N}})$ where $\bm\Sigma_{\mathcal{N}} = (\tau \bm X_{\mathcal{N}}'\bm X_{\mathcal{N}} + \bm Q_{\mathcal{N}}'\bm\Sigma_\theta^{-1} \bm Q_{\mathcal{N}})^{-1}$ and $\bm \mu_{\mathcal{N}} = \tau \bm \Sigma_{\mathcal{N}} \bm X_{\mathcal{N}}' \bm y$;  \vspace{-2mm}
    %from \eqref{unconstr-post};
    \item Compute $\bm \theta =\bm Q_{\mathcal{N}}  \bm \theta_{\mathcal{N}}$.  \vspace{-2mm}
\end{enumerate}  
    %Return  $\bm \theta =\bm Q_{\mathcal{N}}  \bm \theta_{\mathcal{N}}$
 \caption{Sampling from $[\bm \theta \mid \bm y, \bm A \bm\theta = \bm 0]$ under a Gaussian linear model.} \label{alg:constr}
\end{algorithm}   \vspace{-2mm}

With Algorithm~\ref{alg:constr} in hand, posterior inference for BayesABCs with cat-modified models follows familiar Gibbs sampling steps for Bayesian linear models. Appendix~\ref{app-computing} provides the details of these sampling blocks:  all regression coefficients, the error precision $\tau$, the prior variance parameters for our shrinkage priors  (Section~\ref{subsec-priors}), and posterior predictives. 

Notice that conditioning is not the only path to enforce ABCs  in a Bayesian model. Appendix~\ref{app-centering} discusses an alternative approach based on parameter centering and which remains compatible with our algorithms  and theory (Section~\ref{sec-theory}). Because the model parameterization remains the same, different mechanisms to enforce $\bm A \bm \theta = \bm 0$ only change the prior. 
%However, it is arguably less favorable in combination with shrinkage priors, so  we do not pursue it further.%; see Appendix~\ref{app-centering} for details. 

%{\bf Remark:} For the general case with continuous and categorical main effects and cat-cts and cat-cat interactions, we will be able to apply this strategy blockwise. All coefficients will be sampled jointly, but the reparameterization occurs separately for each block of constrained coefficients (due to prior independence). Thus, the joint posterior precision will be block diagonal. 

%Suppose we have the model
%\[
%[\bm y \mid \bm\theta] \sim N_n(\bm X \bm \theta, \tau^{-1} \bm I_n), \quad \bm\theta \sim N_p(\bm 0, \tau^{-1}\bm \Sigma_\theta)
%\]
%\[\bm y = \bm X \bm \theta + \bm \epsilon, \quad \bm \epsilon \sim N_n(0, \tau^{-1} \bm I_n)\]
%and a prior $\bm\theta \sim N(\bm 0, \tau^{-1}\bm \Sigma_\theta)$. Fix $\tau = 1$ for simplicity. 

\subsection{Example: one-way ANOVA}\label{subsec-anova}
 %Notice also that under the ANOVA model,
%\[
%\bar y_c \stackrel{indep}{\sim} N(\mu, (\tau n \hat\pi_c)^{-1}), \quad c=1,\ldots,L
%\]
%where $\bar y_c$ is the sample mean in group $c$ and $n \hat\pi_c$ is the sample size in group $c$.  

To understand the behavior of this BayesABCs prior construction, we revisit the one-way ANOVA (Example~\ref{ex-anova}). Beginning with the same (unconstrained) priors, we now condition on $\bm{\hat \pi}'\bm\beta = 0$  to enforce ABCs \eqref{abcs-gen-cat}. % ; the initial results below remain valid for other choices of $\bm{\hat\pi}$, including RGE and STZ. %For now, we only require $\pi_r \ge 0$ with $\sum_r \pi_r = 1$, which includes RGE, STZ, and ABCs as special cases; subsequent consideration of ABCs will use the notation $\bm \pi = \bm{\hat \pi}$. 
Applying Lemma~\ref{lem-prior-equiv},  $[\bm\beta \mid \bm{\hat \pi}'\bm\beta = 0]$ is Gaussian with mean zero and covariance $\bm \Sigma_\beta = \lambda_\beta^2 (\bm I_L - \bm{\hat \pi}\bm{\hat \pi}'/\Vert\bm{\hat \pi}\Vert^2).$ Notice that the prior depends on $\bm X$ akin to Zellner's $g$-prior, here 
via the categorical proportions $\bm{\hat\pi}$. 
The prior variances $\mbox{Var}(\beta_c \mid \bm{\hat \pi}'\bm\beta = 0) = \lambda_\beta^2(1 - \hat \pi_c^2/\Vert \bm{\hat \pi} \Vert^2)$ are largest when $\hat \pi_c$ is small. At first glance, this seems unappealing: we generally prefer more shrinkage for smaller groups, and that usually occurs with smaller prior variance. We show that this concern does not materialize: greater shrinkage \emph{does} occur for smaller groups. %A contributing factor is that the constraint induces negative dependencies,   $\mbox{Cov}(\beta_c, \beta_{c'}\mid \bm{\hat \pi}'\bm\beta = 0) = -\lambda_\beta^2\hat\pi_c\hat\pi_{c'}/\Vert\bm{\hat\pi}\Vert^2$ for $c\ne c'$. Both effects are natural: $\bm{\hat  \pi}' \bm \beta = 0$ keeps $\beta_c$ close to zero when $\hat\pi_c$ is large and forces cancellations among  $\bm\beta$. 
Similar results can be derived for the group means $\bm \mu = (\mu_1,\ldots,\mu_L)'$, % with $\mu_c = \mu(c)$, 
which are mean zero Gaussian with covariance $\bm \Sigma_\mu = \lambda_\alpha^2 \bm 1\bm 1' + \bm \Sigma_\beta$. %The contrasts $\delta_{r,r'} = \mu_r - \mu_{r'}$ are also mean zero Gaussian with variance $\mbox{Var}(\delta_{r,r'}  \mid \bm{\hat\pi}'\bm\beta = 0) = \lambda_\beta^2\{ 2 - (\hat\pi_r - \hat\pi_{r'})^2/\Vert \bm{\hat\pi} \Vert^2\}$, which is smallest for imbalanced groups (large $\vert \hat \pi_r - \hat \pi_{r'} \vert$), and readily computable covariances. 

Consider $L=2$ groups with $\bm{\hat \pi} = (\hat \pi_1, \hat \pi_2)'$. Under ABCs, $\beta_2 = -(\hat\pi_1/\hat\pi_2)\beta_1$ and the contrast is $\delta  \coloneqq \mu_1 - \mu_2 =  \beta_1 + (\hat \pi_1/\hat \pi_2) \beta_1 = (1-\hat \pi_1)^{-1} \beta_1$. %, writing $\mu_c = \mu(c)$. 
The group means may be written exclusively in terms of $\alpha_0$ and $\delta$,
\begin{equation}\label{mu-two-groups}
    \mu_1 = \alpha_0 + \hat \pi_2 \delta, \quad \mu_2 = \alpha_0  -\hat \pi_1 \delta
\end{equation}
and similarly, $\beta_1 = \hat \pi_2 \delta$ and $\beta_2 = - \hat \pi_1 \delta$. With BayesABCs, it is especially convenient to reparameterize to  $(\alpha_0, \delta)$ for posterior analysis. Clearly, $[\delta\mid \bm{\hat \pi}'\bm\beta = 0]$ is Gaussian with mean zero and variance $\lambda_\beta^2/(\hat\pi_1^2 + \hat \pi_2^2)$, which is maximized for balanced groups  ($\hat\pi_1 = 1/2$) and minimized for imbalanced groups, i.e., when one group is rare  ($\hat \pi_1 \approx 0$ or $\hat \pi_1 \approx 1$). %Further, $\delta$ and $\alpha_0$ are \emph{a priori} independent due to  prior independence of $\alpha_0$ and $\bm \beta$. 

Unique advantages of BayesABCs appear upon considering the likelihood. Specifically,  assume a Gaussian model $y_i = \mu(c_i) +\epsilon_i$, with $\epsilon_i\sim N(0,\tau^{-1})$ independently for $i=1,\ldots,n$. The likelihood contributions for $(\alpha_0, \delta)$ are fully characterized by 
\begin{equation}\label{two-like}
    \bar y \sim N(\alpha_0, (\tau n)^{-1}) \quad \perp \quad  \bar y_1 - \bar y_2 \sim N\left\{\delta, ( \tau n \hat\pi_1\hat\pi_2)^{-1}\right\}.
\end{equation}
The result for $\bar y$ is a unique consequence of ABCs. %  and previews our more general (posterior) invariance results in Section~\ref{sec-theory}.  
The other results are generic: independence follows from $\mbox{Cov}(\bar y, \bar y_1 - \bar y_2) = 0$, while for $\bar y_1 - \bar y_2$, notice that the usual variance contribution from the group sizes is  $
(1/n_1 + 1/n_2) = %(n_1 + n_2)/(n_1 n_2) = 
(n \hat \pi_1 \hat \pi_2)^{-1}$ with $n_c = \hat \pi_c n$ for $c=1,2$. 

Because ABCs decompose \eqref{two-like} into distinct terms for $\alpha_0$ and $\delta$, and since $\alpha_0$ and $\delta$ are \emph{a priori} independent,    $\alpha_0$ and $\delta$ are also \emph{a posteriori} independent and we may derive their posteriors separately. The posterior for $\alpha_0$ is simple and familiar: $[\alpha_0 \mid \bm y, \bm{\hat\pi}'\bm\beta = 0] \sim N(m_\alpha, v_\alpha)$ where $m_\alpha = (\tau n)(\tau n + \lambda_\alpha^{-2})^{-1} \bar y$ and  $v_\alpha = (\tau n + \lambda_\alpha^{-2})^{-1}.$ This is exactly the posterior distribution for $\alpha_0$ under a model that omits  $\bm \beta$ (see Corollary~\ref{cor-anova-k}). This \emph{posterior invariance} (here, to inclusion of $\bm \beta$) is unique to BayesABCs and is substantially generalized in Section~\ref{sec-theory}. 
The posterior for the contrast is $[\delta \mid \bm y, \bm{\hat\pi}'\bm\beta = 0] \sim N(m_\delta, v_\delta)$, where
\begin{equation}\label{delta-moments}
m_\delta = \frac{\tau n}{\tau n + \rho(\lambda_\beta)} (\bar y_1 - \bar y_2), \quad v_\delta = \frac{1}{\hat \pi_1 \hat \pi_2 \{\tau n + \rho(\lambda_\beta)\}}, \quad \rho(\lambda_\beta) = \frac{1}{\lambda_\beta^2}\left(\frac{\hat \pi_1}{\hat \pi_2} + \frac{\hat \pi_2}{\hat \pi_1}\right).
    %m_\delta = \frac{\tau n \hat \pi_1 \hat \pi_2}{\tau n \hat \pi_1 \hat \pi_2 + v_0^{-1}} (\bar y_1 - \bar y_2), \quad
    %v_\delta = \frac{1}{\tau n \hat \pi_1 \hat \pi_2 + v_0^{-1}}.
    %\rho(\lambda_\beta) = \frac{1}{\lambda_\beta^2}\frac{\pi_1^2 + \pi_2^2}{\hat \pi_1 \hat \pi_2}.
\end{equation}
We express the posterior moments  in terms of the data model precision $\tau$, the sample size $n$, and a quantity $\rho(\lambda_\beta)$ that depends on $\lambda_\beta$ and $\bm{\hat \pi}$ and combines with $\tau n$ as if it were a prior precision. It ranges from $\rho(\lambda_\beta) = 2/\lambda_\beta^2$ for balanced groups to  larger, unbounded values for  imbalanced groups. 
%and combines with $\tau n$ like a prior precision.  For balanced groups, $\rho(\lambda_\beta) = 2/\lambda_\beta^2$; for  imbalanced groups, %, say $\hat \pi_2 \approx 0$, then $\rho(\lambda_\beta) \approx \lambda_\beta^{-2} (\hat \pi_1/\hat \pi_2)$ 
 %$\rho(\lambda_\beta)$ is larger. 
When one group is rare, the shrinkage induced by ABCs increases and the estimated difference between groups is more conservative (smaller $\vert m_\delta \vert$). 
% For general constraints $\bm \pi$,  the contrast posterior $[\delta \mid \bm y, \bm \pi'\bm\beta = 0]$ instead uses the prior term $\rho(\lambda_\beta) = \Vert \bm \pi \Vert ^2/(\lambda_\beta^2\hat \pi_1 \hat \pi_2)$. 

Applying \eqref{mu-two-groups},  the joint posterior for $\mu_1$ and $\mu_2$ 
is Gaussian and can be derived by combining the independent contributions from $\alpha_0$ and $\delta$ and computing the covariance of $\mu_1$ and $\mu_2$. Specifically,  $[\mu_1 \mid \bm y, \bm{\hat\pi}'\bm\beta = 0] \sim N(m_{\mu_1}, v_{\mu_1})$ with
\begin{equation}\label{group-effect-moments}
\begin{split}
    m_{\mu_1} &= m_\alpha + \hat \pi_2 m_\delta% \\ 
    %\mathbb{E}(\mu_1 \mid \bm y) &= \mathbb{E}(\alpha_0 \mid \bm y) + (1 - \hat \pi_1) \mathbb{E}(\delta \mid \bm y) \\
    %&
    = \frac{\tau n}{\tau n + \lambda_\alpha^{-2}} \bar y + \hat \pi_2 \frac{\tau n}{\tau n + \rho(\lambda_\beta)} (\bar y_1 - \bar y_2)\\
    %\mbox{Var}(\mu_1 \mid \bm y) &= \mbox{Var}(\alpha_0 \mid \bm y) + (1-\hat \pi_1)^2 \mbox{Var}(\delta \mid \bm y) \\
    v_{\mu_1} &= v_\alpha + \hat \pi_2^2 v_\delta %\\
    %&
    = \frac{1}{\tau n + \lambda_\alpha^{-2}} +  \frac{\hat \pi_2/\hat \pi_1}{\tau n   + \rho(\lambda_\beta)}. 
\end{split}
\end{equation}
and similarly for $\mu_2$. 
%noticing that $(1-\hat \pi_1)^2/(\hat \pi_1 \hat \pi_2) = \hat \pi_2/\hat \pi_1$. 
%The moments for $\mu_2$ are obtained  by reversing the group indexes $\{1,2\}$. 
%and similarly, 
%\begin{align*}
%    \mathbb{E}(\mu_2 \mid \bm y) &= %\mathbb{E}(\alpha_0 \mid \bm y) - (1 - \hat \pi_2) \mathbb{E}(\delta \mid \bm y) = 
 %   \frac{\tau n}{\tau n + \lambda_\alpha^{-2}} \bar y + (1 - \hat \pi_2) \frac{\tau n}{\tau n + \rho(\lambda_\beta)} (\bar y_2 - \bar y_1)\\
 %   \mbox{Var}(\mu_2 \mid \bm y) &= %\mbox{Var}(\alpha_0 \mid \bm y) + (1-\hat \pi_2)^2 \mbox{Var}(\delta \mid \bm y) \\
 %   \frac{1}{\tau n + \lambda_\alpha^{-2}} +  \frac{\hat \pi_1/\hat \pi_2}{\tau n   + \rho(\lambda_\beta)}.
%\end{align*}
The posterior covariance is  $\mbox{Cov}(\mu_1, \mu_2 \mid \bm y, \bm{\hat\pi}'\bm\beta = 0) = v_\alpha - \hat \pi_1 \hat \pi_2 v_\delta = (\tau n + \lambda_\alpha^{-2})^{-1} -  \{ \tau n + \rho(\lambda_\beta)\}^{-1}$.

These results offer insights about the group means, which are inferential targets under any identification (see Section~\ref{sec-sims-anova}). First, the BayesABCs posteriors neatly combine contributions from the intercept and  the contrast: 1) they are independent (\emph{a priori} and \emph{a posteriori}), 2) the contribution from $\alpha_0$ is exactly as in the simpler model that omits categorical effects, and 3) the contribution from $\delta$ is weighted by the other group's proportion. The latter point is intuitive: a larger group contributes more to the overall average and  requires less information from the contrast. No such decomposition occurs for  RGE.

Second, notice the posterior variances. The role of group sizes is quite clear:  after subtracting out the shared variance due to $\alpha_0$, the posterior variances of the group means differ only by the groups' squared odds, $(v_{\mu_1} - v_\alpha)/(v_{\mu_2} - v_\alpha) =  \hat \pi_2^2/\hat \pi_1^2$. 
%\[
%\frac{v_{\mu_1} - v_\alpha}{v_{\mu_2} - v_\alpha} =  \hat \pi_2^2/\hat \pi_1^2.
%\]
The larger group  is guaranteed to have a smaller posterior variance. Now compare this posterior variance to that of the no-shrinkage estimator $\bar y_c$ with variance $(\tau n \hat \pi_c)^{-1}$. %, i.e., $\lambda_\alpha \to \infty$ and $ \lambda_\beta \to \infty$. 
Both terms that contribute to $v_{\mu_c}$ in \eqref{group-effect-moments} are bounded above by the diffuse cases, so    $
v_{\mu_c} \le (\tau n \hat \pi_c)^{-1}$ with equality only when both $\lambda_\alpha \to \infty$ and $ \lambda_\beta \to \infty$. Thus, any shrinkage ($\lambda_\alpha < \infty$ or $\lambda_\beta < \infty$) guarantees a variance reduction. Further, the posterior covariance shows
 that $\mu_1$ and $\mu_2$ are more tightly coupled when $\alpha_0$ is \emph{a posteriori} uncertain (large $v_\alpha$),  $\bm \beta$ is \emph{a priori} certain (small $\lambda_\beta$), or the groups are imbalanced. % (one group is rare). %The latter case corresponds to larger $\mbox{Var}(\bar y_1 - \bar y_2)$ (see \eqref{two-like}) 

Finally, we resolve the concerns about relative shrinkage for smaller groups. We examine $(\bar y_1 - m_{\mu_1})^2$, so larger values imply greater shrinkage.
%Compare each posterior mean $m_{\mu_c}$ to the group sample mean $\bar y_c$; larger discrepancies correspond to greater shrinkage. 
Notice that $\bar y_1 = \bar y + \hat \pi_2 (\bar y_1 - \bar y_2)$, %(e.g., consider $m_{\mu_r}$ in \eqref{group-effect-moments} with $\lambda_\alpha \to \infty$ and $ \lambda_\beta \to \infty$) 
so  
$\bar y_1 - m_{\mu_1} = \lambda_\alpha^{-2}(\tau n + \lambda_\alpha^{-2})^{-1} \bar y + \hat \pi_2 \rho(\lambda_\beta)\{\tau n + \rho(\lambda_\beta)\}^{-1} (\bar y_1 - \bar y_2)$ 
%\bar y_1 - m_{\mu_1} = \frac{\lambda_\alpha^{-2}}{\tau n + \lambda_\alpha^{-2}} \bar y + \hat \pi_2 \frac{\rho(\lambda_\beta)}{\tau n + \rho(\lambda_\beta)} (\bar y_1 - \bar y_2)$ 
%\[
%\bar y_1 - m_{\mu_1} = \frac{\lambda_\alpha^{-2}}{\tau n + \lambda_\alpha^{-2}} \bar y + \hat \pi_2 \frac{\rho(\lambda_\beta)}{\tau n + \rho(\lambda_\beta)} (\bar y_1 - \bar y_2)
%\]
and similarly for $c=2$. The first term will be identical for both groups, so we focus on the latter term (or fix $\bar y =0$). Then $(\bar y_1 - m_{\mu_1})^2 = \hat \pi_2^2 [\rho(\lambda_\beta)\{\tau n + \rho(\lambda_\beta)\}^{-1}]^2 (\bar y_1 - \bar y_2)^2$, 
%\[\Vert \bar y_1 - m_{\mu_1} \Vert^2 = \hat \pi_2^2 \left\{\frac{\rho(\lambda_\beta)}{\tau n + \rho(\lambda_\beta)}\right\}^2 (\bar y_1 - \bar y_2)^2,\]
while $c=2$  swaps $\hat \pi_1^2$ for $\hat \pi_2^2$. The shrinkage applied to $\mu_c$ is \emph{largest} for smaller groups, as appropriate. Inspecting only the prior variances of $\beta_c$ or $\mu_c$---and ignoring the correlations induced by the constraints---
%---which are both larger for small groups with BayesABCs---
does not provide a complete picture of the posterior shrinkage behavior. 
For both groups, greater shrinkage of the group means occurs when data are small or noisy (small $\tau n$), groups are imbalanced (large $\rho(\lambda_\beta)$), or estimated group differences are large (large $\Vert \bar y_1 - \bar y_2 \Vert^2)$.  The  posterior for $\bm \beta$ can be derived similarly, % to \eqref{group-effect-moments}, 
instead using \eqref{delta-moments} and $\beta_1 = \hat \pi_2 \delta$ and $\beta_2 = - \hat \pi_1 \delta$.

%For completeness, the remaining model parameters have posteriors $[\beta_1 \mid \bm y, \bm{\hat\pi}'\bm\beta = 0] \sim N(\hat \pi_2 m_\delta, \hat\pi_2^2 v_\delta)$ and  $[\beta_2 \mid \bm y, \bm{\hat\pi}'\bm\beta = 0] \sim N(-\hat \pi_1 m_\delta, \hat \pi_1^2 v_\delta)$ with  $\mbox{Cov}(\beta_1, \beta_2 \mid \bm y, \bm{\hat\pi}'\bm\beta = 0) = -\hat\pi_1 \hat\pi_2 v_\delta = \{\tau n + \rho(\lambda_\beta)\}^{-1}$. 

%, say $\hat \pi_2 < \hat \pi_1$. 

%By comparison, consider RGE: $\beta_1 = 0$ implies that $\alpha_0 = \mu_1$ and $\delta = -\beta_2$. Now, $\alpha$ and $\delta$ are \emph{a priori} independent, but not \emph{a posteriori} independent since the likelihoods will not factorize. The contributions from $\alpha_0$ and $\delta$ are not orthogonal, and posterior invariance of $\alpha_0$ will not hold.

%\cite{stokell2021modelling} use ABCs for penalized estimation of the main-only, one-way ANOVA model, claiming that this constraint ``treats the levels more symmetrically, but also takes into account the prevalence of levels, so the fitted values corresponding to more prevalent levels effectively undergo less regularisation."

\section{Theory}\label{sec-theory}

\subsection{Main effect posteriors are invariant to categorical modifiers}\label{subsec-invar}
The central tension of categorical interactions is that they are essential for estimating heterogeneous effects, but they completely change main effect estimates, inferences, and interpretations. With RGE, these impacts are uniformly negative: main effects anchor at a reference group, uncertainties usually inflate, and interpretations are more convoluted. BayesABCs completely resolves this tension. The reason is a unique and general \emph{posterior invariance} property: the joint posterior of all main effects is (nearly) unchanged by the addition of categorical interactions. We identify a sufficient condition that holds exactly for several versions of \eqref{reg-cm} and otherwise provides accurate empirical approximations. 
%We establish sufficient conditions for both non-informative and potentially informative priors, and show that it holds empirically for the more sophisticated structured shrinkage priors from Section~\ref{subsec-priors}. 

To proceed generally, consider a {base model} $\mathcal{M}_{01}$ that includes covariates ($n\times p_0$) $\bm X_0$  and ($n \times p_1$) $\bm X_1$, along with  an {augmented} model $\mathcal{M}_{012}$ that also includes ($n\times p_2)$ $\bm X_2$:
\begin{align*}
&(\mathcal{M}_{01}) \quad  && \bm y = \bm X_0 \bm\beta +  \bm X_1 \bm \alpha  +\bm \epsilon, \quad && [\bm \epsilon \mid \tau] \sim N_n(\bm 0, \tau^{-1} \bm I_n) \\
&(\mathcal{M}_{012}) \quad &&  \bm y = \bm X_0 \bm \beta + \bm X_1 \bm \alpha    + \bm X_2 \bm \gamma +  \bm \epsilon, \quad && [\bm \epsilon \mid \tau] \sim N_n(\bm 0, \tau^{-1} \bm I_n).
\end{align*}
The augmented parameters are linearly constrained,  $\bm A \bm \gamma = \bm 0$. 
In our context, $\mathcal{M}_{01}$ is a main-only model and $\mathcal{M}_{012}$ is a cat-modified model: $\bm X_1$ includes main effects, $\bm X_2$  includes interactions between columns of $\bm X_1$ and categorical variables, and $\bm A$ encodes ABCs.  $\bm X_0$ is included in both models but is not of primary interest. We do not separately annotate the parameters in $\mathcal{M}_{01}$ in $\mathcal{M}_{012}$, only their  posterior distributions $p_{\mathcal{M}_{01}}(\cdot \mid \bm y)$ and $p_{\mathcal{M}_{012}}(\cdot \mid \bm y)$.

%\begin{equation}\label{mo-gen}
%    \bm y = \bm X_0 \bm\beta +  \bm X_1 \bm \alpha  +\bm \epsilon, \quad [\bm \epsilon \mid \tau] \sim N_n(\bm 0, \tau^{-1} \bm I_n)
%\end{equation}
%\begin{equation}\label{cm-gen}
%    \bm y = \bm X_0 \bm \beta + \bm X_1 \bm \alpha    + \bm X_2 \bm \gamma +  \bm \epsilon, \quad [\bm \epsilon \mid \tau] \sim N_n(\bm 0, \tau^{-1} \bm I_n). 
%\end{equation}

The goal is to establish conditions under which $p_{\mathcal{M}_{01}}(\bm \alpha \mid \bm y) = p_{\mathcal{M}_{012}}(\bm \alpha \mid \bm y)$, so that the posterior of $\bm \alpha$ is \emph{invariant} to the addition of $\bm X_2$. This requires marginalization over unknowns: %under $\mathcal{M}_{01}$ with joint posterior $ p_{\mathcal{M}_{01}}(\bm \alpha, \bm \beta, \tau \mid \bm y)$, 
%we target 
$ p_{\mathcal{M}_{01}}(\bm \alpha \mid \bm y) = \int \int  p_{\mathcal{M}_{01}}(\bm \alpha, \bm \beta, \tau \mid \bm y) \ d \bm\beta \ d \tau %\quad p_{\mathcal{M}_{012}}(\bm \alpha \mid \bm y) = \int \int \int  p_{\mathcal{M}_{012}}(\bm \alpha, \bm \beta,\bm\gamma, \tau \mid \bm y) \ d \bm\beta \ d \bm\gamma \ d \tau,
$   and 
%while $\mathcal{M}_{012}$ also requires marginalization over $\bm \gamma$: 
$%\[
p_{\mathcal{M}_{012}}(\bm \alpha \mid \bm y) = \int \int \int  p_{\mathcal{M}_{012}}(\bm \alpha, \bm \beta,\bm\gamma, \tau \mid \bm y) \ d \bm\beta \ d \bm\gamma \ d \tau.
$ %\]
% where  $p_{\mathcal{M}_{012}}(\bm \alpha, \bm \beta,\bm\gamma, \tau \mid \bm y)$ is the joint posterior under $\mathcal{M}_{012}$. 
%and the latter must  account for the constraint on $\bm\gamma$. 
One might suspect that such invariance occurs if the augmented covariates are \emph{orthogonal} to existing covariates, invoking   Frisch-Waugh-Lovell \citep{FW1933,Lovell1963}. However, categorical interactions cannot be orthogonal to main effects (except in trivial cases). %Even if the true augmented coefficients are null, adding $\bm X_2$ will alter the posterior distribution of $\bm \alpha$. 
Instead, we pursue an alternative route that leverages the linear constraint $\bm A \bm \gamma = \bm 0$ and is empirically viable for ABCs.

We analyze joint priors of the form
\begin{equation}\label{gen-prior}
p(\bm \alpha, \bm \beta, \bm \gamma, \tau) \propto \tau^{-1} p(\bm \alpha \mid \tau)\ p(\bm \gamma \mid \tau)
\end{equation}
where $p(\bm \gamma \mid \tau)$ is omitted for $\mathcal{M}_{01}$. Implicitly, \eqref{gen-prior} features a flat prior on $\bm\beta$  and a Jeffreys prior on the error variance.  For simplicity, we take $p(\bm \alpha \mid \tau) \propto 1$ here and reproduce the analysis under  a  Gaussian prior %$[\bm \alpha \mid \tau] \sim N(\bm \alpha_0, \tau^{-1}\bm \Sigma_{\alpha_0})$ 
in  Appendix~\ref{app-theory}. The critical term is $\bm \gamma$, which must enforce the linear constraint: we consider a flat prior $p(\bm \gamma \mid \tau) \propto  \mathbb{I}\{\bm A \bm \gamma = \bm 0\}$ and a Gaussian prior $[\bm \gamma \mid \tau] \sim N(\bm \gamma_0, \tau^{-1}\bm \Sigma_{\gamma_0})\mathbb{I}\{\bm A \bm \gamma = \bm 0\}$. Analysis with these constrained priors is facilitated by the approach in Section~\ref{subsec-enforce}. Accordingly, let $p_\mathcal{N} = p_2 - m$ count the free parameters in $\bm\gamma$.

The main condition for posterior invariance depends on the covariates and the constraints. Let $\bm E_{10} 
=  \bm X_1 - \bm H_0 \bm X_1$ and $\bm E_{20} =  \bm X_2 - \bm H_0 \bm X_2$ be the residuals from regressing each column of $\bm X_1$ (respectively, $\bm X_2$) on $\bm X_0$, with $\bm H_0 = \bm X_0(\bm X_0'\bm X_0)^{-1}\bm X_0'$. These quantities arise from marginalizing over $\bm\beta$. Let $\bm e_{10,j}$ be the $j$th column of $\bm E_{10}$. The main condition requires that the rows of $\bm E_{10}'\bm E_{20}$ are linear combinations of the rows of $\bm A$: 
\begin{equation}\label{cond}
    \bm e_{10,j}'\bm E_{20} \in \mbox{row}(\bm A), \quad j=1,\ldots,p_1.
\end{equation}
More directly, we arrive at \eqref{cond} via $\bm E_{10}'\bm E_{20} \bm Q_{\mathcal{N}} = \bm 0$, where $\bm Q_{\mathcal{N}}$ comes from \eqref{qr-con}; this  is also equivalent to  $\bm X_1'\bm E_{20} \bm Q_{\mathcal{N}} = \bm 0$ or $\bm E_{10}'\bm X_2 \bm Q_{\mathcal{N}} = \bm 0$. If not identically zero, $\Vert \bm E_{10}'\bm E_{20} \bm Q_{\mathcal{N}} \Vert$ can be established as a bound for the discrepancies between the $\mathcal{M}_{01}$ and $\mathcal{M}_{012}$ posteriors, leading to the approximate posterior invariance we observe empirically (see Section~\ref{sec-asr}).

We provide some preliminary intuition for \eqref{cond}, followed by specific examples in Section~\ref{subsec-demyst}. Orthogonality of the residualized covariates  $\bm E_{10}'\bm E_{20} = \bm 0$ is sufficient, but typically cannot occur for cat-modified models. Instead, consider $\bm A$: with ABCs, the rows of $\bm A$ are populated by the group proportions  $\bm{\hat\pi}'$, with a specific form that depends on \eqref{abcs-gen} or \eqref{abcs-cat-cat-gen}. For cat-modified models, $\bm X_{2}$ includes categorical interactions, so the entries of $\bm e_{10,j}'\bm E_{20}$ correspond to those categorical interactions. Informally, \eqref{cond} requires that each entry accumulates proportional to that categorical (or CC) group's abundance. 
% so \eqref{cond} is satisfied when $\bm e_{10,j}'\bm E_{20} \propto \bm{\hat\pi}'$. 
%For cat-modified models, . 
For some models (e.g., ANOVAs), this holds exactly; in other models, it is an approximation that can be expressed precisely in terms of $\bm X_0$, $\bm X_1$, or $\bm X_2$. Neither requires any assumptions about $\bm y$. %Instead, \eqref{cond} elicits a more general and useful condition that delivers the same kinds of benefits.  %No assumptions are needed for $\bm y$ or the true data-generating process. Examples are provided below. 

To state the main result, we first define the ingredients of the $\mathcal{M}_{01}$ posterior. Under $p(\bm\alpha)\propto 1$, the marginal posterior of $\bm \alpha$ requires $\bm{\hat \alpha} = \bm{\hat\Sigma}_\alpha\bm E_{10}'(\bm y - \bm H_0 \bm y)$ and $\bm{\hat\Sigma}_\alpha = (\bm E_{10}'\bm E_{10})^{-1}$, which are the OLS estimator and the unscaled covariance matrix, respectively, under $\mathcal{M}_{01}$. Similarly, let $\hat S_{\mathcal{M}_{01}}^2$  and $\hat S_{\mathcal{M}_{012}}^2$ be  the estimated OLS residual variances under $\mathcal{M}_{01}$ and $\mathcal{M}_{012}$, respectively. 
%Similarly, let  $\bm{\hat\gamma}  = \bm{\hat\Sigma}_{\gamma} (\bm E_{20}'\bm e_0 +    \bm{\Sigma}_{\gamma_0}^{-1} \bm \gamma_{0})$ and $\bm{\hat\Sigma}_{\gamma} = \bm Q_{\mathcal{N}}\{\bm Q_{\mathcal{N}}'(\bm E_{20}'\bm E_{20} + \bm{\Sigma}_{\gamma_0}^{-1})\bm Q_{\mathcal{N}}\}^{-1}\bm Q_{\mathcal{N}}' $ be the posterior mean and unscaled posterior covariance, respectively, of $\bm\gamma$ under $\mathcal{M}_{012}$ (see Section~\ref{subsec-enforce} and Appendix~\ref{app-proofs-repar}). 
%Since $\bm{\hat\Sigma}_{\gamma}$ is not invertible due to the linear constraint, we require a generalized inverse with $\bm{\hat\Sigma}_{\gamma} \bm{\hat\Sigma}_{\gamma}^{-} \bm{\hat\Sigma}_{\gamma}= \bm{\hat\Sigma}_{\gamma}$. 
Let $t_\nu(\bm\mu, \bm \Sigma)$ denote a multivariate $t$-distribution with expectation $\bm \mu$,  covariance $\nu(\nu-2)^{-1}\bm\Sigma$, % (defined when $\nu > 2$), 
and density 
$p (\bm x)  \propto \{ \nu + \Vert \bm x - \bm \mu \Vert_{\bm\Sigma^{-1}}^2 %(\bm x - \bm \mu)' \bm{\Sigma}^{-1} (\bm x - \bm \mu)
\}^{-(\nu + p)/2}$ for $\bm x \in \mathbb{R}^p$ using the notation $\Vert \bm v \Vert_{\bm M}^2 = \bm v'\bm M\bm v$, and let $\bm{\Sigma}^{-} $ be a generalized inverse such that $\bm{\Sigma} \bm{\Sigma}^{-} \bm{\Sigma}= \bm{\Sigma}$. 
%The main result is the following:
\begin{theorem}\label{thm-flat}
    Under the joint  prior \eqref{gen-prior} with  $p(\bm \alpha \mid \tau) \propto 1$, we have the following:
    \begin{compactenum}
        \item Under model $\mathcal{M}_{01}$, the marginal posterior is $
            [\bm \alpha \mid \bm y] \sim t_{n- p_0 - p_1}(\bm{\hat\alpha}, \hat S_{\mathcal{M}_{01}}^2 \bm{\hat\Sigma}_\alpha)$. %, where $\bm{\hat \alpha} = \bm{\hat\Sigma}_\alpha\bm E_{10}'(\bm y - \bm H_0 \bm y)$, $\bm{\hat\Sigma}_\alpha = (\bm E_{10}'\bm E_{10})^{-1}$, 
        %with density  $p_{\mathcal{M}_{01}}(\bm \alpha \mid \bm y)  \propto \big\{ (n - p_0 - p_1)\hat S_{\mathcal{M}_{01}}^2 + (\bm \alpha - \bm{\hat \alpha})' \bm{\hat \Sigma}_\alpha^{-1} (\bm \alpha - \bm{\hat \alpha})\big\}^{-(n - p_0)/2}$;
        %where and $\hat S_{\mathcal{M}_{01}}^2$ are the OLS estimate, (unscaled) OLS covariance matrix, and estimated OLS residual variance under $\mathcal{M}_{01}$. 
        
        \item Under model $\mathcal{M}_{012}$, the flat (constrained) prior $p(\bm \gamma \mid \tau) \propto \mathbb{I}\{\bm A \bm \gamma = \bm 0\}$, and condition \eqref{cond}, the marginal posterior is  $[\bm \alpha \mid \bm y] \sim t_{n- p_0 - p_1-p_{\mathcal{N}}}(\bm{\hat\alpha}, \hat S_{\mathcal{M}_{012}}^2 \bm{\hat\Sigma}_\alpha)$. %  
        %with density $p_{\mathcal{M}_{012}}(\bm \alpha \mid \bm y) \propto \big\{ (n - p_0 - p_1 - p_{\mathcal{N}})\hat S_{\mathcal{M}_{012}}^2 + (\bm \alpha - \bm{\hat \alpha})' \bm{\hat \Sigma}_\alpha^{-1} (\bm \alpha - \bm{\hat \alpha})\big\}^{-(n - p_0-p_{\mathcal{N}})/2}$; 
        %where $\hat S_{\mathcal{M}_{012}}^2$ is  the estimated OLS residual variance under $\mathcal{M}_{012}$. 
        \item Under model $\mathcal{M}_{012}$, the Gaussian (constrained) prior $[\bm \gamma \mid \tau] \sim N(\bm \gamma_0, \tau^{-1}\bm \Sigma_{\gamma_0})\mathbb{I}\{\bm A \bm \gamma = \bm 0\}$, and condition \eqref{cond}, the marginal posterior is $
        [\bm \alpha \mid \bm y] \sim t_{n- p_0 - p_1}(\bm{\hat\alpha}, \hat S_{\mathcal{M}_{012}(G)}^2 \bm{\hat\Sigma}_\alpha)$, 
        %with density $p_{\mathcal{M}_{012}}(\bm \alpha \mid \bm y) \propto \big\{ (n - p_0 - p_1)\hat S_{\mathcal{N},1, \mathcal{M}_{012}}^2 + (\bm \alpha - \bm{\hat \alpha})' \bm{\hat \Sigma}_\alpha^{-1} (\bm \alpha - \bm{\hat \alpha}) \big\}^{-(n - p_0)/2}$, 
        where $\hat S_{\mathcal{M}_{012}(G)}^2 = \hat S_{\mathcal{M}_{01}}^2 + (n - p_0 - p_1)^{-1} ( \Vert \bm \gamma_{0}\Vert_{\bm{\Sigma}_{\gamma_0}^{-1}}^2  - \Vert \bm{\hat\gamma} \Vert_{\bm{\hat\Sigma}_\gamma^{-}}^2)$, $\bm{\hat\gamma}  = \bm{\hat\Sigma}_{\gamma} (\bm E_{20}'\bm e_0 +    \bm{\Sigma}_{\gamma_0}^{-1} \bm \gamma_{0})$,    and $\bm{\hat\Sigma}_{\gamma} = \bm Q_{\mathcal{N}}\{\bm Q_{\mathcal{N}}'(\bm E_{20}'\bm E_{20} + \bm{\Sigma}_{\gamma_0}^{-1})\bm Q_{\mathcal{N}}\}^{-1}\bm Q_{\mathcal{N}}' $. % are the posterior mean and (unscaled) posterior covariance, respectively, of $\bm\gamma$ under $\mathcal{M}_{012}$.
    \end{compactenum}
    \end{theorem}
    %Despite substantial differences among these models, %---whether or not to include (cat-modifiers) $\bm X_2$ and the prior on $\bm \gamma$---
    The marginal posteriors $[\bm \alpha \mid \bm y]$ are remarkably similar: they share the same  expectations, unscaled covariances, and family of distributions.  There are only two potential sources of differences: the scalings and the degrees of freedom, which both contribute to the posterior covariances. A practical concern is that including categorical interactions unnecessarily will inflate the posterior uncertainty about $\bm\alpha$. %, especially when $\bm X_1$ and $\bm X_2$ are correlated. 
    Crucially, BayesABCs avoids this. %With a flat prior on $\bm\gamma$, $\mathcal{M}_{012}$ yields a smaller marginal posterior variance whenever $\hat S_{\mathcal{M}_{012}}^2  \lesssim S_{\mathcal{M}_{01}}^2$, omitting the degrees of freedom discrepancy by $p_{\mathcal{N}}$ that becomes negligible for large $n$. 
    %otherwise a term for the $p_{\mathcal{N}}$ reduction in degrees of freedom  from $\mathcal{M}_{01}$ to $\mathcal{M}_{012}$. 
    %$ (n- p_0 - p_1-p_{\mathcal{N}})(n- p_0 - p_1-p_{\mathcal{N}}-2)^{-1} \hat S_{\mathcal{M}_{012}}^2 <  (n- p_0 - p_1)(n- p_0 - p_1-2)^{-1}\hat S_{\mathcal{M}_{01}}^2$, which also shows the $p_{\mathcal{N}}$ reduction in degrees of freedom  from $\mathcal{M}_{01}$ to $\mathcal{M}_{012}$.  
    %For  large $n$, this difference in degrees of freedom becomes negligible, and the inequality holds when 
    %so  has smaller posterior variance whenever its 
    %the $R^2$-adjusted is larger for $\mathcal{M}_{012}$ than $\mathcal{M}_{01}$.   
    %than $\mathcal{M}_{01}$ whenever the variability explained is sufficiently larger on a per-parameter basis: $ \hat S_{\mathcal{M}_{012}}^2 <  \hat S_{\mathcal{M}_{01}}^2$  occurs when $(\Vert \bm e_{\mathcal{M}_{01}}\Vert^2 -  \Vert \bm e_{\mathcal{M}_{012}}\Vert^2)/\Vert \bm e_{\mathcal{M}_{01}}\Vert^2 > p_{\mathcal{N}}/(n - p_0 - p_1 - 2)$, where $\bm e_{\mathcal{M}_{01}}$ and $\bm e_{\mathcal{M}_{012}}$ are the OLS residuals under $\mathcal{M}_{01}$ and $\mathcal{M}_{012}$, respectively. 
   %This also holds for OLS standard errors under ABCs \citep{kowal2026facilitating}. %, for which \eqref{cond} also provides a unifying condition. 
   Focusing on the (constrained) Gaussian prior for $\bm\gamma$, 
   %With a Gaussian prior on $\bm\gamma$, 
   the degrees of freedom are identical under $\mathcal{M}_{01}$ and $\mathcal{M}_{012}$ and the scaling terms differ by %clear:
    %\begin{equation}\label{scale-diff}
       $ \hat S_{\mathcal{M}_{012}(G)}^2 - \hat S_{\mathcal{M}_{01}}^2 = (n - p_0 - p_1)^{-1} ( \Vert \bm \gamma_{0}\Vert_{\bm{\Sigma}_{\gamma_0}^{-1}}^2  - \Vert \bm{\hat\gamma} \Vert_{\bm{\hat\Sigma}_\gamma^{-}}^2),$ 
    %\end{equation}
    where $\bm{\hat\gamma}$ is the posterior mean  and $\bm{\hat\Sigma}_\gamma$ the (unscaled) posterior covariance  of $\bm\gamma$ under $\mathcal{M}_{012}$ (see Section~\ref{subsec-enforce} and Appendix~\ref{app-proofs-repar}). 
    Except in unusual cases, this difference is negative: the posterior variance of each $\alpha_j$ is \emph{smaller} under $\mathcal{M}_{012}$  than $\mathcal{M}_{01}$.    This increase in power under $\mathcal{M}_{012}$ grows with $\Vert \bm{\hat\gamma} \Vert_{\bm{\hat\Sigma}_{\gamma}^{-}}^2$, i.e., when the estimated coefficients on $\bm X_2$ (categorical interactions) are large in magnitude, precision, or both. Counterexamples only arise when the prior mean is nonzero  \emph{and} the posterior for $\bm\gamma$ is less informative than the prior (e.g., an informative prior that conflicts with the data).%, but even so $ \hat S_{\mathcal{M}_{012}(G)}^2 - \hat S_{\mathcal{M}_{01}}^2$ vanishes as $n\to\infty$. %We take $\bm\gamma_0 = \bm0$ so $\hat S_{\mathcal{M}_{012}(G)}^2 < \hat S_{\mathcal{M}_{01}}^2$ without exception. 
 %Crucially, ABCs do \emph{not} require $\bm{\hat\alpha} \to \bm 0$
    
    %The other, less important source of differences among the marginal posteriors is the degrees of freedom.  Under $\mathcal{M}_{012}$ and a flat prior on $\bm \gamma$,  the degrees of freedom are reduced from $\mathcal{M}_{01}$ by $p_{\mathcal{N}}$; this discrepancy matters less when $n$ is large. Instead using a Gaussian prior on $\bm\gamma$, there are no differences in the degrees of freedom %for $[\bm \alpha \mid \bm y]$ 
    %under $\mathcal{M}_{01}$ or $\mathcal{M}_{012}$.%, regardless of how informative the prior is. 
    
    %The relationship in \eqref{scale-diff} also motivates our use of structured shrinkage priors. Suppose $\bm X_2$ is extraneous (i.e., with true $\bm \gamma = \bm 0$). The ideal prior will shrink $\bm{\hat\gamma} \to \bm 0$, which zeros out the difference in \eqref{scale-diff}. Perhaps surprisingly, extraneously including $\bm X_2$---which in our case contains cat-modifiers and thus is correlated with $\bm X_0$ and $\bm X_1$---has no adverse effects on the marginal posterior of $\bm \alpha$. Namely, the posterior expectation, covariance, and joint distribution of $[\bm \alpha \mid \bm y]$ are all invariant. This encourages our use of shrinkage priors but also relieves the burden of variable selection and sparse estimation: we do not require $\bm{\hat\gamma} = \bm 0$ to achieve posterior invariance. 

Together, the constraints $\bm A \bm \gamma = \bm 0$ and condition \eqref{cond} provide remarkable stability for $\bm\alpha$. Adding $\bm X_2$ to move from $\mathcal{M}_{01}$ to $\mathcal{M}_{012}$---regardless of whether these variables are important, whether they are correlated with $\bm X_0$ or $\bm X_1$, and without assumptions on $\bm y$---has no adverse impact on posterior inference for $\bm \alpha$. %Under a   Gaussian prior on the augmented coefficients $\bm \gamma$, all discrepancies between the marginal posteriors under $\mathcal{M}_{01}$ and $\mathcal{M}_{012}$ vanish with the difference between the scaling terms \eqref{scale-diff}. This difference goes to zero as $n\to\infty$, but also for estimators that shrink to zero,  $ \bm{\hat\gamma} \to \bm 0$. This is particularly important when $\bm X_2$ is extraneous---and our shrinkage priors are designed precisely for this setting.
These results apply more generally, including under a Gaussian prior $[\bm \alpha \mid \tau] \sim N(\bm \alpha_0, \tau^{-1}\bm \Sigma_{\alpha_0})$:  the marginal posterior $[\bm \alpha \mid \bm y]$ under $\mathcal{M}_{01}$ differs from Theorem~\ref{thm-flat}, but the discrepancies between $\mathcal{M}_{01}$ and $\mathcal{M}_{012}$ (with flat or Gaussian priors on $\bm \gamma$) are identical to those in Theorem~\ref{thm-flat} (see Appendix~\ref{app-theory}).

%{\color{blue} how about mentioniong that \eqref{cond} unifies the frequentist OLS results from \cite{kowal2026facilitating}?}
%While the posterior moments and scaling terms change from Theorem~\ref{thm-flat}, the themes are the same. Despite a new prior for $\bm\alpha$, posterior invariance persists under the same conditions. In fact, the discrepancies between $\mathcal{M}_{01}$ and $\mathcal{M}_{012}$ (for flat or Gaussian priors) for the scaling terms and the degrees of freedom exactly match those from Theorem~\ref{thm-flat}. 

\subsection{Demystifying the main condition}\label{subsec-demyst}
The outstanding question from Section~\ref{subsec-invar} is, when does   condition \eqref{cond} apply? %We provide several examples to illustrate both 1) the breadth and utility of this condition and 2) how it is essentially distinct from orthogonality. 
There are two notable cases where condition \eqref{cond} holds exactly. These apply only for BayesABCs and have no restrictions on the data-generating process (DGP). 
The first is the $K$-way ANOVA model as an augmentation to the intercept-only model: $\mathcal{M}_{01} = $ \texttt{y $\sim$ 1} and $\mathcal{M}_{012} = $ \texttt{y $\sim$ 1 + c$_1$ + $\ldots$ + c$_K$}, using Wilkinson notation. For any categorical covariate $c_k$, let $\bm Z_k$ denote its $n \times L_k$ matrix of indicator variables, so each column is a dummy variable for one of the $L_k$ groups. 
\begin{corollary}\label{cor-anova-k}
    Suppose $\bm X_0 = \emptyset$ is omitted, $\bm X_1 = \bm 1$, and $\bm X_2 = (\bm Z_1, \ldots, \bm Z_K)$. Using ABCs, condition \eqref{cond} holds exactly. 
\end{corollary}
Posterior inference with the base model is unimpeded by the addition of arbitrary many categorical covariates---significant or not---as long as ABCs are enforced. 
Although Corollary~\ref{cor-anova-k} does not involve categorical interactions, it immediately validates condition \eqref{cond} for a widely-used class of models and showcases the central connection with ABCs. 
% Corollary~\ref{cor-anova-k} also remains valid with arbitrary many continuous covariates in $\bm X_2$, as long as they are centered (and unconstrained). 

The second is a two-way ANOVA model with or without interactions: $\mathcal{M}_{01} = $ \texttt{y $\sim$ 1 + c$_1$ + c$_2$} and  $\mathcal{M}_{012}= $ \texttt{y $\sim$ 1 + c$_1$ + c$_2$ + c$_1$:c$_2$}. For categorical covariates $c_k$ and $c_{k'}$, let $\bm Z_{k:k'}$ denote the $n \times L_kL_{k'}$ interaction matrix, so each column indicates mutual belonging. 
%For two-way ANOVA models, condition \eqref{cond} holds exactly---but only with ABCs.  For clarity, let \texttt{race} and \texttt{sex} be two categorical variables with $L_R$ and $L_S$ levels, respectively. We aim to apply Theorems~\ref{thm-flat} and \ref{thm-normal} to $\mathcal{M}_{01} = $ \texttt{y $\sim$ race + sex} and $\mathcal{M}_{012}= $ \texttt{y $\sim$ race + sex + race:sex}. Specifically, let $\bm Z_1$ be an $n \times L_R$ matrix of indicator variables for \texttt{race}, i.e., the first column of $\bm Z_1$ equals one only if each observation $i$ belongs to the first \texttt{race} group, $i=1,\ldots,n$, and so forth. Similarly define $\bm Z_2$  ($n \times L_R$) for \texttt{sex}. Let $\bm Z_{1:2}$ denote the $n \times L_RL_S$ interaction matrix, so each column is an indicator of mutual belonging to a \texttt{race}, \texttt{sex} pairing. 
\begin{corollary}\label{cor-anova}
    Suppose %$\mathcal{M}_{01}$ is \texttt{y $\sim$ race + sex} and $\mathcal{M}_{012}$ is \texttt{y $\sim$ race + sex + race:sex}, so 
    $\bm X_0 = \emptyset$ is omitted, $\bm X_1 = (\bm 1, \bm Z_1, \bm Z_2)$, and $\bm X_2 = \bm Z_{1:2}$. Using ABCs, condition \eqref{cond} holds exactly. 
\end{corollary}
Contrast this  with orthogonality:  $\bm X_1'\bm X_2 = \bm 0$ cannot occur except in trivial cases. % (for which the interaction is not estimable). 
Yet  \eqref{cond} holds exactly. % , with the only requirement that ABCs are used for identification.  
For the two-way ANOVA model with BayesABCs, joint posterior inference for \emph{all} $(1 + L_1 + L_2)$ main effects is invariant to  the CC interaction \texttt{c$_1$:c$_2$}.% in the sense of Theorem~\ref{thm-flat}. % (and Theorem~\ref{thm-normal}). 
%There is no cost to including the interaction. 

Next, we consider models for which condition \eqref{cond} does not hold automatically, but may be expressed clearly as a requirement on the covariates. %For these cases, %condition \eqref{cond} will not hold exactly in practice; rather, 
We view this as an approximation that explains the near-invariance we observe empirically (see Section~\ref{sec-asr}). 
A simple example is an ANCOVA model:  $\mathcal{M}_{01} = $ \texttt{y $\sim$ 1 + c$_1$ + x} and  $\mathcal{M}_{012}= $ \texttt{y $\sim$ 1 + c$_1$ + x + c$_1$:x}, where the CX interaction introduces group-specific slopes. 
%The main-only model has group-specific intercepts with a common slope while the cat-modified model has group-specific intercepts and slopes. 
With ABCs, posterior invariance for the coefficient on $x$ is assured when the variance of $x$ is the same within each group.  %Precisely, let $\hat \sigma_{x[r]}^2 = n_r^{-1} s_{x[r]}^2 - \bar x_r^2$ denote the (scaled) sample variance for $x$ by each \texttt{race} $r=1,\ldots,L_R$, where $s_{x[r]}^2 = \sum_{r_i = r} x_i^2$,  $\bar x_r = n_r^{-1} \sum_{r_i=r} x_i$, and $n_r = n \hat \pi_r$. 
\begin{corollary}\label{cor-x}
    Suppose 
    $\bm X_0 = \bm Z_1$, $\bm X_1 = \bm x = (x_1,\ldots,x_n)'$, and $\bm X_2 = \mbox{diag}(\bm x)\bm Z_1$. Using ABCs, condition \eqref{cond} holds whenever $\hat \sigma_{x[1]}^2 = \cdots = \hat \sigma_{x[L_1]}^2$, where $\hat \sigma_{x[c]}^2 = n_c^{-1} \sum_{c_i = c} (x_i - \bar x_c)^2$. %,
    %n_r^{-1} s_{x[r]}^2 - \bar x_r^2$, $s_{x[r]}^2 = \sum_{r_i = r} x_i^2$,  
    %$\bar x_r = n_r^{-1} \sum_{r_i=r} x_i$, and $n_r = n \hat \pi_r$. 
    % for all $r=1,\ldots,L_R$. %, where  $\hat \sigma_{x[r]}^2 = n_r^{-1} s_{x[r]}^2 - \bar x_r^2$, $s_{x[r]}^2 = \sum_{r_i = r} x_i^2$, and $\bar x_r = n_r^{-1} \sum_{r_i=r} x_i$. 
\end{corollary}
%If this does not occur, we may contest that the main-only model $\mathcal{M}_{01}$ lacks interpretability because it disallows group-specific $x$-effects.  
Neither independence nor uncorrelatedness of $x$ and $\texttt{c}_1$ is needed: the distribution $[x \mid \texttt{c}_1]$ is free to vary  across groups  so long as the variances are similar. Even when they are unequal, the BayesABCs cat-modified model remains applicable. It simply means that the cat-modified model slope $\alpha_1 = \mathbb{E}_{\bm{\hat\pi}}\{\mu_x'(C_1)\}$ is distinct from the main-only model slope. 
%Yet the main $x$-effect under the cat-modified ANCOVA remains a reasonable target even when the scales are unequal. In that case, a one-unit change in $x$ holds different meanings for different groups. With ABCs,  $\alpha_1 = \mathbb{E}_{\bm{\hat\pi}}\{\mu_x(C_1)\}$ is the average of group-specific slopes. 

%This matters because often we are most interested in heterogeneous effects due to the complexities of the associations among \texttt{x}, $\texttt{c}_1$, and \texttt{y}. In practice, we view the equal-scale assumption as an approximation that delivers {approximate} posterior invariance. Empirically, we find that Theorems~\ref{thm-flat} and \ref{thm-normal} apply even under moderate violations of this assumption. 

If we expand Corollary~\ref{cor-x} to include the main categorical effects, 
%The benefits afforded by Corollary~\ref{cor-x} not only are unique for ABCs (which ensure \eqref{cond} and thus Theorems~\ref{thm-flat} and \ref{thm-normal}), but also highlight distinct behavior between continuous and categorical effects. To see this, consider expanding Corollary~\ref{cor-x} to include the coefficients on \texttt{race}: 
$\bm X_0 = \emptyset$, $\bm X_1 = (\bm x,\bm Z_1)$, and $\bm X_2 = \mbox{diag}(\bm x)\bm Z_1$, then a sufficient condition for \eqref{cond} is  equal means, $\bar x_1 = \cdots = \bar x_{L_1}$, along with  equal variances. These show two distinct paths to \eqref{cond}: equal-means implies $\bm Z_1'\bm X_2 = \bm Z_1'\mbox{diag}(\bm x) \bm Z_1 =\bm 0$ and equal-scales implies $\bm x'\bm X_2 = \bm x' \mbox{diag}(\bm x) \bm Z_1 \propto  \bm{\hat \pi}'$. Equal-means is more restrictive and requires that  $x$ and $\texttt{c}_1$ are uncorrelated. Even so, orthogonality $\bm X_1'\bm X_2 = \bm0$ is not needed---and in fact cannot occur for this model. 

Lastly, we expand to multiple continuous covariates, $x_1,\ldots,x_p$, seeking  posterior invariance for all $p$ coefficients. The first case has two categorical variables and we add the CC interaction: $\mathcal{M}_{01} = $ \texttt{y $\sim$ 1 + c$_1$ + c$_2$ + x$_1$ + $\cdots$ + x$_p$} and $\mathcal{M}_{012}= $ \texttt{y $\sim$ 1 + c$_1$ + c$_2$ + x$_1$ + $\cdots$ + x$_p$ + c$_1$:c$_2$}. For each continuous covariate $j=1,\ldots,p$, let $\bar e_{j[rs]}$ be the average of the residuals $\bm e_{10,j} = \bm x_j - \bm H_0 \bm x_j$ %with $\bm H_0 = \bm X_0(\bm X_0'\bm X_0)^{-1}\bm X_0'$  
    over observations with $\texttt{c$_1$} =r$ and $\texttt{c$_2$} =s$. 
%holds under general dependencies among the continuous covariates $\bm x_1,\ldots, \bm x_p$ and the categorical covariates \texttt{c$_2$} and \texttt{c$_2$}, as long as the associations between each $\bm x_j$ and  \texttt{c$_1$} and \texttt{c$_2$} are additive rather than interactive. 
\begin{corollary}\label{cor-cat-cat}
    Suppose   $\bm X_0 = (\bm 1, \bm Z_1, \bm Z_2)$, $\bm X_1 = (\bm x_1,\ldots, \bm x_p)$, and $\bm X_2 = \bm Z_{1:2}$. 
    Using ABCs, condition \eqref{cond} holds when $\bar e_{j[rs]} = \bar e_{j[11]}$ for  $r=1,\ldots, L_1$, $s=1,\ldots, L_2$, and  $j=1,\ldots,p$. 
\end{corollary}
Unlike with two-way ANOVAs, posterior invariance  is not automatic (Corollary~\ref{cor-anova}). Instead, Corollary~\ref{cor-cat-cat} requires that associations between each $x_j$ and  \texttt{c$_1$} and \texttt{c$_2$} are additive rather than interactive, but otherwise allows general dependencies.
%Theorems~\ref{thm-flat} and \ref{thm-normal} deliver the benefits of orthogonality (namely, posterior invariance) with fewer assumptions. By comparison, orthogonality would prohibit linear associations between $\bm x_j$ and \texttt{race} or \texttt{sex}: $\bm X_1'\bm X_2 = \bm 0$ 
%Again, compare  \eqref{cond} with orthogonality. Condition \eqref{cond} allows for dependencies between each $\bm x_j$ and \texttt{race} and \texttt{sex}, but only additive ones. % : regressing \texttt{x$_j \sim $ race + sex + race:sex} must have zero OLS coefficients for \texttt{race:sex}. 
%By comparison,  orthogonality   
%requires  $\bar x_{j[rs]} = 0$ for all $r,s,j$, where $\bar x_{j[rs]}$ is the average of $\bm x_j$ over observations with $\texttt{race} =r$ and $\texttt{sex} =s$. Instead, Corollary~\ref{cor-cat-cat} permits significantly more general dependencies among the covariates. For practical use, replace the main condition of  Corollary~\ref{cor-cat-cat} with an approximation, $\bar e_{j[rs]} \approx \bar e_{j[11]}$, for approximate posterior invariance. Adding more covariates to $\bm X_0$ will improve this approximation in general.%, which broadens the utility of  Corollary~\ref{cor-cat-cat}. 

The second case is more general: $\mathcal{M}_{01} = $ \texttt{y $\sim$ 1 + c$_1$ + x$_1$ + $\cdots$ + x$_p$ + $\cdots$} includes categorical $\texttt{c}_1$, continuous $x_1,\ldots,x_p$, and arbitrary covariates, while  $\mathcal{M}_{01} = $ \texttt{y $\sim$ 1 + c$_1$ + x$_1$ + $\cdots$ + x$_p$ + c$_1$:x$_1$ + $\cdots$ + c$_1$:x$_p$ + $\cdots$} adds CX interactions between $\texttt{c}_1$ and each $x_j$. Let $\bm e_{10,j} = \bm x_j - \bm H_0 \bm x_j$ be the residuals from regressing $\bm x_j$ on $\bm X_0$, where $\bm X_0$ include $\texttt{c}_1$ and the arbitrary covariates,  
and define
$\widehat{\mbox{Cov}}_{[c]}(\bm v_1, \bm v_2) = n_c^{-1} \sum_{c_i = c} (v_{i1} - \bar v_{1[c]})(v_{i2} - \bar v_{2[c]})$ to be the (scaled) sample covariance for observations with $\texttt{c}_1 = c$. 
\begin{corollary}\label{cor-cat-cts}
    Suppose   $\bm X_0 = (\bm Z_1, \ldots)$,   $\bm X_1 = (\bm x_1,\ldots, \bm x_p)$, and $\bm X_2 = (\mbox{diag}(\bm x_1)\bm Z_{1}, \ldots, \mbox{diag}(\bm x_p)\bm Z_{1})$.
    %$\bm X_2 = (\bm Z_{x_1}, \ldots, \bm Z_{x_p})$ with $\bm Z_{x_k} = \mbox{diag}(\bm x_k)\bm Z_{1}$. 
    %Let $\bm e_{10,j} = \bm x_j - \bm H_0 \bm x_j$ be the residuals from regressing $\bm x_j$ on $\bm X_0$. 
    Using ABCs, condition \eqref{cond} holds when $\widehat{\mbox{Cov}}_{[1]}(\bm e_{10,j}, \bm x_k) = \ldots =\widehat{\mbox{Cov}}_{[L_1]}(\bm e_{10,j}, \bm x_k)$  for each $j,k=1,\ldots,p$.%, where $\bm e_{10,j} = \bm x_j - \bm H_0 \bm x_j$. 
\end{corollary}
With additional  main effects and interactions, Corollary~\ref{cor-cat-cts} generalizes the equal-scales from Corollary~\ref{cor-x} for  residuals due to $\bm X_0$ and equal-covariances due to $p > 1$. Notably, these conditions are separate for each $(x_j,x_k)$ pair---equality holds across groups, not necessarily $(x_j,x_k)$ pairs---yet still ensure posterior invariance for all $p$ coefficients. 

\section{Structured shrinkage priors}\label{subsec-priors}
To this point, we have prioritized the \emph{parameterization} of cat-modified models, using comparatively simple priors that nonetheless deliver compelling properties for posterior inference (Sections~\ref{subsec-anova} and \ref{sec-theory}). 
%Indeed, BayesABCs has excellent empirical performance even with simple ridge-like priors (Section~\ref{sec-sims-lm}). 
We now design \emph{priors} for   \eqref{reg-cm}. Shrinkage is critical: 
although main effect posteriors are robust to extraneous categorical interactions (Section~\ref{sec-theory}), we still must 1) distinguish signal from noise among \emph{heterogeneous} (interaction) effects and 2) regularize unnecessary coefficients to maintain accurate predictions. Our priors aim to aggressively shrink noise, minimally shrink signals, and leverage the grouping structures among  categorical effects and their interactions. 
Structured shrinkage priors have been used successfully in other contexts, including continuous-continuous interactions \citep{griffin2017hierarchical} and groups of continuous covariates \citep{boss2024group}. Here, we must contend with categorical effects, interactions, and identifying constraints.

With BayesABCs, we express the priors in the unconstrained, overparameterized space and then condition on ABCs.  
We offer two versions: \emph{grouped} shrinkage priors   use a shared shrinkage parameter for all coefficients within a categorical covariate (or interaction), while \emph{local} shrinkage priors   include local shrinkage parameters for each coefficient, grouped hierarchically within a categorical covariate (or interaction). Both  use a flat prior on the intercept, $p(\alpha_0) \propto 1$, and horseshoe priors \citep{carvalho2010horseshoe} for the continuous main effects, $ [\alpha_j \mid \lambda_{\alpha_j}] \sim N(0, \lambda_{\alpha_j}^2)$ and $[\lambda_{\alpha_j} \mid \tau_0] \sim C^+(0, \tau_0)$ for $j=1,\ldots,p$, 
%\begin{equation}\label{prior-main-x}
%    [\alpha_j \mid \lambda_{\alpha_j}] \stackrel{indep}{\sim} N(0, \lambda_{\alpha_j}^2), \quad [\lambda_{\alpha_j} \mid \tau_0] \stackrel{iid}{\sim} C^+(0, \tau_0)%, \quad \tau_0 \sim C^+(0,1)
%\end{equation}
where 
$C^+$ denotes a half-Cauchy distribution.  %In each case, we also include the option to omit the CC heredity. 
Both also include a \emph{CC heredity} variation where the CC interactions inherit shrinkage parameters from their parent main effects, which pulls CC interactions toward zero whenever either main effect is small. With ABCs, we do \emph{not} consider heredity for CX interactions: it is plausible for a group-averaged effect  to be near zero, $\alpha_j = \mathbb{E}_{\bm{\hat\pi}}\{\dot\mu_{x_j}(\bm C)\} \approx 0 $,  while group-specific effects are not. Whether $\beta_{k,c_k} \approx 0$ has no bearing on sparsity of $\gamma_{j,k, c_k}$. 

\subsection{Grouped shrinkage priors}\label{subsub-grp}
For the grouped prior, categorical main effects  follow a group-structured horseshoe prior: 
\begin{equation}\label{prior-main-cat}
       [\beta_{k,c_k} \mid \lambda_{\beta_k}] \sim N(0, \lambda_{\beta_k}^2), %\phi_{\beta_{k,c_k}}^2/w_{k,c_k}^2), 
        \quad [\lambda_{\beta_k} \mid\tau_0] \sim C^+(0, \tau_0)
\end{equation}
for levels  $c_k = 1,\ldots,L_k$ for each categorical variable $ k=1,\ldots,K$. Besides the subsequent conditioning on ABCs \eqref{abcs-gen-cat}, this   prior differs from the horseshoe prior through the shared shrinkage parameter $\lambda_{\beta_k}$, and resembles the batched priors of \cite{gelman2005analysis}. 
Here, $\lambda_{\beta_k}$ is capable of pulling $\{\beta_{k,c_k}\}_{c_k=1}^{L_k}$ to zero simultaneously, thus removing the $k$th categorical variable.  The global shrinkage parameter  $\tau_0 \sim C^+(0,1)$ is shared across all continuous and categorical main effects and serves as a proxy for their overall sparsity. 
%The global shrinkage parameter $\tau_0$ is a proxy for the overall sparsity among all (continuous and categorical) main effects. 

%This includes the option to (inversely) weight by group abundances, which ensures equitable shrinkage behavior and is discussed below.

Priors for CX and CC interactions are structured similarly, with interaction-specific shrinkage parameters (denoted by $\phi$) and a global shrinkage parameter ($\tau_*$) shared only among the interactions. The CX  and CC interaction terms   follow
\begin{equation}\label{prior-cat-int}
    \begin{split}
       &[\gamma_{j:k,c_k} \mid \phi_{\gamma_{j:k}}] 
       \sim N(0, \phi_{\gamma_{j:k}}^2), 
       \quad [\phi_{\gamma_{j:k}} \mid  \tau_*] \sim C^+(0, \tau_{*}),\\
       &[\gamma_{k:k',c_k,c_{k'}} \mid \phi_{\gamma_{k:k'}}] 
       \sim N(0, \phi_{\gamma_{k:k'}}^2), 
       \quad [\phi_{\gamma_{k:k'}} \mid   \tau_*] \sim C^+(0,  \tau_{*}) %[\phi_{\gamma_{k:k'}} \mid  \lambda_{\beta_k}, \lambda_{\beta_{k'}}, \tau_*] \sim C^+(0, \lambda_{\beta_k} \lambda_{\beta_{k'}}\tau_{*}) 
       \end{split}
\end{equation}
%\begin{equation}\label{prior-cat-int}
%    \begin{split}
%       &[\gamma_{j:k,c_k} \mid \phi_{\gamma_{j:k}}] 
%       \sim N(0, \phi_{\gamma_{j:k}}^2), 
%       \quad [\phi_{\gamma_{j:k}} \mid \lambda_{\alpha_j}, \lambda_{\beta_k}, \tau_*] \sim C^+(0, \lambda_{\alpha_j}\lambda_{\beta_k}\tau_{*}),\\
%       &[\gamma_{k:k',c_k,c_{k'}} \mid \phi_{\gamma_{k:k'}}] 
%       \sim N(0, \phi_{\gamma_{k:k'}}^2), 
%       \quad [\phi_{\gamma_{k:k'}} \mid  \lambda_{\beta_k}, \lambda_{\beta_{k'}}, \tau_*] \sim C^+(0, \lambda_{\beta_k} \lambda_{\beta_{k'}}\tau_{*}) 
%       \end{split}
%\end{equation}
% the categorical-categorical interaction terms follow 
%\begin{equation}\label{prior-cat-cat}
%       [\gamma_{k:k',c_k,c_{k'}} \mid \phi_{\gamma_{k:k'}}]        \sim N(0, \phi_{\gamma_{k:k'}}^2), 
       %\sim N\Big(0, \frac{\phi_{\gamma_{k:k',c_k,c_{k'}}}^2}{w_{k:k',c_k, c_{k'}}^2}\Big), 
%       \quad [\phi_{\gamma_{k:k'}} \mid  \lambda_{\beta_k}, \lambda_{\beta_{k'}}, \tau_*] \sim C^+(0, \lambda_{\beta_k} \lambda_{\beta_{k'}}\tau_{*}) 
%\end{equation}
and  $\tau_* \sim C^+(0,1)$. %To omit CC heredity, we modify  $[\phi_{\gamma_{k:k'}} \mid   \tau_*] \sim C^+(0,  \tau_{*})$. 
To include CC heredity, we modify $\phi_{\gamma_{k:k'}}  \sim C^+(0, \lambda_{\beta_k} \lambda_{\beta_{k'}}\tau_{*})$: 
%$[\phi_{\gamma_{k:k'}} \mid  \lambda_{\beta_k}, \lambda_{\beta_{k'}}, \tau_*] \sim C^+(0, \lambda_{\beta_k} \lambda_{\beta_{k'}}\tau_{*})$:  
when a parent categorical effect is unimportant, the corresponding $\lambda_{\beta_k}$ or $\lambda_{\beta_{k'}}$ will be near zero, thus pulling the categorical interaction coefficients toward zero. 
%Either way, the  shrinkage terms ($\phi$) provide flexibility to accommodate a mix of strong, moderate, and near-zero effects. %Finally, the global shrinkage parameter $\tau_*$ is shared only among the cat-modifier effects, and thus offers a way to shrink all interaction terms to zero. 

Incorporating ABCs by conditioning on \eqref{abcs-gen} and \eqref{abcs-cat-cat-gen} is straightforward and follows directly from Section~\ref{subsec-enforce}. Notably, the grouped shrinkage prior is highly favorable for computing: the categorical- and interaction-specific variances factorize as in ridge prior example following Lemma~\ref{lem-prior-equiv}, leading to Gibbs sampling steps for regression coefficients and variance components that resemble those in the standard horseshoe (see Appendix~\ref{app-px-group}).

\subsection{Local shrinkage priors}\label{subsub-loc}
The local prior includes level-specific shrinkage ($c_k = 1,\ldots,L_k$) arranged hierarchically for each categorical variable ($k=1,\ldots,K$):
\begin{equation}\label{prior-main-cat-local}
       [\beta_{k,c_k} \mid \phi_{\beta_{k,c_k}}] \sim N(0, \phi_{\beta_{k,c_k}}^2), %\phi_{\beta_{k,c_k}}^2/w_{k,c_k}^2), 
       \quad [\phi_{\beta_{k,c_k}} \mid \lambda_{\beta_k}] \sim C^+(0, \lambda_{\beta_k})
\end{equation}
and $\lambda_{\beta_k}$ has the same prior as in \eqref{prior-main-cat}. Again, $\lambda_{\beta_k}$ controls the total shrinkage for the $k$th categorical covariate. Yet with  \eqref{prior-main-cat-local},  
%This hierarchical structure incorporates level-specific parameters $\phi_{\beta_{k,c_k}}$ that allow for some (or many) of the level-specific effects to be strong, even if the remainder are near zero. Now, $\lambda_{\beta_k}$ shrinks all coefficients for the categorical variable (e.g., race), but 
$\phi_{\beta_{k,c_k}}$ allows for differential shrinkage among the levels. This is most useful when many levels are present (large $L_k$) but few  are important. 
%Specifically, even with the convenient Gamma parameter-expansions for the half-Cauchy distribution, the full conditional distributions for the local shrinkage parameters are no longer available in closed form, nor are they independent. 

 %Further, the local shrinkage coefficients are completely separate for the continuous and categorical terms, which helps mitigate the scaling issues across data types (Gelman 2008; check to make sure this is important enough to mention!). 

CX and CC interactions are constructed similarly to \eqref{prior-cat-int}, now with level-specific terms:
\begin{equation}\label{prior-cat-int-local}
    \begin{split}
       &[\gamma_{j:k,c_k} \mid \phi_{\gamma_{j:k,c_k}}] 
       \sim N(0, \phi_{\gamma_{j:k,c_k}}^2), 
       \quad [\phi_{\gamma_{j:k,c_k}}  \mid \tau_*] \sim C^+(0, \tau_{*}), \\
       &       [\gamma_{k:k',c_k,c_{k'}} \mid \phi_{\gamma_{k:k',c_k,c_{k'}}}] 
       \sim N(0, \phi_{\gamma_{k:k',c_k,c_{k'}}}^2), 
       \quad [\phi_{\gamma_{k:k',c_k,c_{k'}}} \mid \tau_*] \sim C^+(0, \tau_{*})%\phi_{\gamma_{k:k',c_k,c_{k'}}} \sim C^+(0, \lambda_{\beta_k} \lambda_{\beta_{k'}}\tau_{*}) 
    \end{split}
\end{equation}
%\begin{equation}\label{prior-cat-int-local}
%    \begin{split}
%       &[\gamma_{j:k,c_k} \mid \phi_{\gamma_{j:k,c_k}}] 
%       \sim N(0, \phi_{\gamma_{j:k,c_k}}^2), 
%       \quad [\phi_{\gamma_{j:k,c_k}} \mid \lambda_{\alpha_j}, \lambda_{\beta_k}, \tau_*] \sim C^+(0, \lambda_{\alpha_j}\lambda_{\beta_k}\tau_{*}), \\
%       &       [\gamma_{k:k',c_k,c_{k'}} \mid \phi_{\gamma_{k:k',c_k,c_{k'}}}] 
%       \sim N(0, \phi_{\gamma_{k:k',c_k,c_{k'}}}^2), 
%       \quad [\phi_{\gamma_{k:k',c_k,c_{k'}}} \mid  \lambda_{\beta_k}, \lambda_{\beta_{k'}}, \tau_*] \sim C^+(0, \lambda_{\beta_k} \lambda_{\beta_{k'}}\tau_{*}) 
%    \end{split}
%\end{equation}
and  $\tau_* \sim C^+(0,1)$ as before. For 
CC heredity, substitute
$\phi_{\gamma_{k:k',c_k,c_{k'}}} \sim C^+(0, \lambda_{\beta_k} \lambda_{\beta_{k'}}\tau_{*})$. %$[\phi_{\gamma_{k:k',c_k,c_{k'}}} \mid   \lambda_{\beta_k}, \lambda_{\beta_{k'}}, \tau_*] \sim C^+(0, \lambda_{\beta_k} \lambda_{\beta_{k'}}\tau_{*})$.  
Although it possible to include additional layers in the hierarchy (e.g., grouped within each CX or CC interaction term), we caution that additional half-Cauchy layers induce increasingly heavy tails and can degrade MCMC efficiency. 
 
%CC heredity may be omitted by substituting $\phi_{\gamma_{k:k',c_k,c_{k'}}} \sim C^+(0, \tau_{*})$. The remaining parameters have the same priors as in the grouped case. 

Although these local shrinkage priors \eqref{prior-main-cat-local} and \eqref{prior-cat-int-local} appear to be relatively straightforward modifications of the horseshoe prior, conditioning on ABCs alters certain sampling steps. Sampling the regression coefficients again follows from Section~\ref{subsec-enforce}.  However, due to the linear constraint and its effect on the conditional priors, 
the standard Gamma parameter expansions for half-Cauchy priors  do not yield the usual full conditional distributions for local shrinkage parameters. Fortunately, they do provide excellent proposal distributions, leading to an efficient Metropolis-within-Gibbs sampling algorithm (see Appendix~\ref{app-px-local}).

\section{Simulations}\label{sec-sims}
We conduct three simulation studies. Section~\ref{sec-sims-anova} evaluates the interplay of priors and parameterizations in a one-way ANOVA model. % , which only includes categorical main effects. 
Section~\ref{sec-sims-lm} features  multiple, dependent categorical and continuous covariates  and compares how main-only and cat-modified model performance differs under competing parameterizations and priors.   Appendix~\ref{app-sims-twoway} considers two-way ANOVA models and evaluates shrinkage with or without CC heredity. 

\subsection{Simulations for one-way ANOVA}\label{sec-sims-anova}
We evaluate estimation and inference for an imbalanced one-way ANOVA. Our target is the group means,   $ \mu_c = \alpha_0 + \beta_c$ for $c=1,\ldots,L$, which is not tied to any parameterization, and we generate data without enforcing any identification constraints. % (trivially, Over is satisfied). 
We vary number of groups $L \in \{5, 15\}$, the sample size $n \in \{200, 1000\}$, and the sparsity $L_{*} \in \{0, 3, L\}$, which is the number of distinct group means. Along with an intercept $\alpha_0 = 1$, we set $\beta_c = c$ for $c=1,\ldots, L_{*}$ and $\beta_c = 0$ otherwise. To create an  imbalanced categorical variable, we allocate  groups 
$c=1,\ldots,\lfloor L/2 \rfloor$  with $1/4$ the probability of the remaining groups (i.e., $\pi_1 = 0.071$ and $\pi_L = 0.286$ for $L=5$ while $\pi_1 = 0.026$ and $\pi_L = 0.103$ for $L=15$). The outcomes $y_i$ are generated from a one-way ANOVA with Gaussian errors and signal-to-noise ratio SNR~$=0.25$. We simulate 100 datasets from this DGP for each design.

We include two versions of BayesABCs: \texttt{ABCs(grp)} uses grouped shrinkage priors (Section~\ref{subsub-grp}) and \texttt{ABCs(loc)} uses local shrinkage priors (Section~\ref{subsub-loc}). %Since no interactions are present, \texttt{ABCs(ridge)} is nearly identical to \texttt{ABCs(grp)}  and thus excluded. 
For Bayesian competitors, \texttt{RGE} sets $\beta_1=0$ as the reference group and \texttt{Over} (for ``overparameterized") does not enforce any constraints; both use horseshoe priors. All Bayesian models report posterior means and 95\% highest posterior density (HPD) intervals based on 5000 MCMC draws after discarding a burn-in of 5000. For frequentist alternatives, we include \texttt{SCOPE} \citep{stokell2021modelling} using the default specifications in the \texttt{R} package \texttt{CatReg}, and \texttt{OLS} computes OLS estimates and confidence intervals (for $\mu_c$, these  will be identical across RGE, ABCs, etc.).

Point estimation is evaluated by root mean squared errors (RMSEs) for $\{\mu_c\}_{c=1}^L$ (Table~\ref{tab:rmse_results_n200}). BayesABCs methods are highly competitive (in the top two) across all group sizes and sparsity levels. Other methods are less consistent: \texttt{SCOPE} excels under complete sparsity but is not competitive under partial sparsity, while \texttt{OLS} performs best with no sparsity but poorly otherwise.  In each case, \texttt{RGE} is  less accurate than both \texttt{ABCs(loc)} and \texttt{ABCs(grp)}.

\begin{table}[h] \small
    \centering
    \begin{tabular}{l cc cc cc}
        % --- Top-level header: # signals ---
        & \multicolumn{2}{c}{$\text{\# signals} = 0$}
        & \multicolumn{2}{c}{$\text{\# signals} = 3$}
        & \multicolumn{2}{c}{$\text{\# signals} = L$} \\
        % --- Second-level header: L values ---
        Method
            & $L=5$ & $L=15$
            & $L=5$ & $L=15$
            & $L=5$ & $L=15$ \\
        \hline
        \texttt{ABCs(loc)}   & {\bf 0.159} & 0.165 & {\bf 0.433} & {\bf 0.342} & 0.513 & 2.847 \\
        \texttt{ABCs(grp)}   & {\bf 0.157} & {\bf 0.150} & {\bf 0.433} & 0.427 & {\bf 0.471} & {\bf 2.502} \\
        \texttt{Over}        & 0.179 & 0.183 & 0.467 & {\bf 0.324} & 0.498 & 2.776 \\
        \texttt{RGE}         & 0.172 & 0.179 & 0.456 & 0.380 & 0.571 & 3.138 \\
        \texttt{SCOPE}       & 0.185 & {\bf 0.157} & 0.574 & 0.415 & 0.504 & 2.807 \\
        \texttt{OLS}         & 0.369 & 0.704 & 0.492 & 0.403 & {\bf 0.431} & {\bf 2.576} \\
        \hline
    \end{tabular}
    \caption{\small RMSEs for $\mu_c = \alpha_0 + \beta_c$ in a one-way ANOVA with $n=200$ (see Appendix~\ref{app-sims-anova} for $n=1000$). %($n=200$, $\mbox{SNR} = 0.25$, imbalanced groups). 
    Smallest two RMSEs in each case are bold. BayesABCs produces one or both of the most accurate estimates in every case.}
    \label{tab:rmse_results_n200}
\end{table}

Next, we compute the average widths of 95\% HPD credible or confidence intervals and  empirical coverage for $\{\mu_c\}_{c=1}^L$ (Table~\ref{tab:mciw_ecp_results_n200}). Smaller intervals that achieve the nominal coverage are preferred. 
\texttt{SCOPE} does not natively provide interval estimates and is excluded. Both \texttt{ABCs(loc)} and \texttt{ABCs(grp)} typically provide much sharper inference than competing methods, and either achieve the nominal coverage or come closer than Bayesian competitors (\texttt{Over} or \texttt{RGE}). Grouped shrinkage   is generally preferable to local shrinkage, with the only exception occurring for partial sparsity ($L=15, L_* = 3$).

\begin{table}[h] %\small
    \centering
    \resizebox{\linewidth}{!}{%
    \begin{tabular}{l cc cc cc}
        % --- Top-level header: # signals ---
        & \multicolumn{2}{c}{$\text{\# signals} = 0$}
        & \multicolumn{2}{c}{$\text{\# signals} = 3$}
        & \multicolumn{2}{c}{$\text{\# signals} = L$} \\
        % --- Second-level header: L values ---
        Method
            & $L=5$ & $L=15$
            & $L=5$ & $L=15$
            & $L=5$ & $L=15$ \\
        \hline
        \texttt{ABCs(loc)}   & {\bf 0.940} (98\%) & {\bf 1.101} (100\%) & 1.794 (92\%) & {\bf 1.194} (95\%) & 1.730 (91\%) & 8.352 (89\%) \\
        \texttt{ABCs(grp)}   & {\bf 0.915} (98\%) & {\bf 1.035} (100\%) & \emph{1.852} (93\%) & 1.369 (93\%) & \emph{1.695} (93\%) & \emph{8.290} (92\%) \\
        \texttt{Over}        & 1.022 (98\%) & 1.215 (100\%) & 1.872 (92\%) & {\bf 1.169} (96\%) & 1.724 (91\%) & 8.176 (88\%) \\
        \texttt{RGE}         & 0.943 (98\%) & 1.165 (100\%) & \emph{1.821} (93\%) & 1.085 (90\%) & 1.666 (88\%) & 7.592 (83\%) \\
        \texttt{SCOPE}       & ---          & ---           & ---          & ---          & ---          & ---          \\
        \texttt{OLS}         & 1.483 (93\%) & 2.700 (95\%)  & 1.976 (93\%) & 1.546 (95\%) & \emph{1.738} (93\%) & {\bf 9.878} (95\%) \\
        \hline
    \end{tabular}}
    \caption{\small Mean credible/confidence interval widths (and empirical coverage) for $\mu_c  = \alpha_0 + \beta_c$ in a one-way ANOVA with $n=200$ (see Appendix~\ref{app-sims-anova} for $n=1000$). Narrowest two intervals subject to nominal coverage (or closest to it) in each case are bold (or italicized, respectively). BayesABCs produces one or both of the most precise and well-calibrated intervals in every case.}
    \label{tab:mciw_ecp_results_n200}
\end{table}

In summary, BayesABCs provides accurate estimates and sharp inference for group means in a one-way ANOVA. This occurs even though the DGP has imbalanced categorical variables and does not satisfy ABCs (or RGE). It suggests that the BayesABCs parameterization and priors combine favorably, even without categorical interactions. 

\subsection{Simulations for sparse and cat-modified linear models}\label{sec-sims-lm}
%\subsubsection{Design and competing methods}
%We study how estimation and inference in cat-modified models depends on 1) the model parameterization, 2) the priors, and 3) whether the cat-modifiers are truly necessary or extraneous. %To do so, 
We now define a DGP with multiple, dependent categorical and continuous variables, multiple identifications satisfied concurrently (RGE, ABCs, Over), %---thus ensuring meaningful comparisons for parameter estimation and inference---
and an interaction parameter  to toggle between necessary ($\gamma \ne 0$)  and extraneous ($\gamma = 0$) categorical interactions. For convenience, denote the categorical covariates  \texttt{race} and \texttt{sex} with arbitrary levels $\{\texttt{a}, \texttt{b}, \texttt{c}, \texttt{d}\}$ and $\{\texttt{uu}, \texttt{vv}\}$, respectively. \texttt{race} is generated with imbalanced groups, $(\pi_a, \pi_b,\pi_c,\pi_d) = (0.4, 0.3, 0.2, 0.1)$, followed by \texttt{sex}  conditional on \texttt{race} with 
$(\pi_{uu}, \pi_{vv})_{\cdot \mid \texttt{a}} = (0.4, 0.6)$,
$(\pi_{uu}, \pi_{vv})_{\cdot \mid \texttt{b}} = (0.6, 0.4)$,
$(\pi_{uu}, \pi_{vv})_{\cdot \mid \texttt{c}} = (0.7, 0.3)$, and
$(\pi_{uu}, \pi_{vv})_{\cdot \mid \texttt{d}} = (0.2, 0.8)$. For $p=10$ continuous covariates, $j=1,3,5,7,9$ are dependent with \texttt{race} (and thus \texttt{sex}) using the conditional distributions $[x_j \mid \texttt{race} = \texttt{a}] \sim 5 + N(0, 1)$,  % x[race == 'A'] = 5 + 1*(rnorm(n = sum(race == 'A')))
$[x_j \mid \texttt{race} = \texttt{b}] \sim \sqrt{12}\ \mbox{Uniform}(0, 1)$,   % x[race == 'B'] = 0 + (sqrt(12)*runif(n = sum(race == 'B')))
$[x+j \mid \texttt{race} = \texttt{c}] \sim  -5 + t_8(0, 1)$, and % x[race == 'C'] = -5 + 1*(sqrt((8-2)/8)*rt(n = sum(race == 'C'), df = 8))
$[x_j \mid \texttt{race} = \texttt{d}] \sim  \mbox{Gamma}(1,1)$. % x[race == 'D'] = 0 + (rgamma(n = sum(race == 'D'), shape = 1, rate = 1))
Both the mean and the distributional families are group-specific, while all groups share the same scale. The remaining covariates are independent $x_j \sim N(0,1)$, $j=2,4,6,8,10$.
 
The outcomes $y_i$ are generated from a Gaussian linear model with SNR~$=1$. The regression coefficients are sparse and otherwise  satisfy RGE and ABCs (Over is satisfied trivially). The nonzero coefficients are the intercept  $\alpha_0=1$,   main $x$-effects $\alpha_j = 1$ for $j=1,\ldots,5$,   \texttt{race} main effects $\beta_b = 1$ and $\beta_c = \beta_d = -1$, and categorical modifiers $\gamma_{b,j} = \gamma$ and $\gamma_{c,j} = \gamma_{d,j} = -\gamma$ for $j=1,\ldots,5$. All reference group coefficients are zero ($\beta_a  = 0$, $\beta_{uu} = 0$, $\gamma_{a,uu} = 0$, and $\gamma_{a, j} = 0 $ for all $j=1,\ldots,p$) and ABCs are satisfied under the population (but not sample) proportions. %Because \texttt{sex} has two groups, its main effects and interactions must be zero to satisfy both RGE and ABCs. 
Finally,  $\gamma$ determines the \texttt{x:race} interaction: we consider $\gamma = 0$ for \emph{extraneous} interactions or $\gamma = 0.5$ for \emph{necessary} but weak interactions. We vary the sample size $n \in \{200, 1000\}$ and generate 500 datasets for each design.

%For each synthetic dataset, we fit the main-only  model \texttt{y $\sim$ x$_1$ + \ldots + x$_p$ + sex + race} and the cat-modified model \texttt{y $\sim$ (x$_1$ + \ldots + x$_p$ + sex)*race} that includes \texttt{race} interactions with all continuous covariates and \texttt{sex}, both under ABCs and RGE. The main-only model is favored when $\gamma = 0$, while the opposite is true for $\gamma > 0$. Either way, the true data-generating process is sparse, so both models include many extraneous parameters. The main-only model includes 15 identifiable parameters (9 true signals)  while the cat-modified model includes 48 identifiable parameters. When $\gamma = 0$, the cat-modified model estimates 33 extraneous (identified) parameters; even when $\gamma > 0$, only 15 of those cat-modifier effects are nonzero.  

%\subsubsection{Competing methods}
We primarily compare Bayesian models under different parameterizations or priors. We include BayesABCs methods \texttt{ABCs(grp)} and \texttt{ABCs(loc)}, Bayesian competitors \texttt{RGE} and \texttt{Over}, and the frequentist estimator \texttt{OLS}, now using ABCs \citep{kowal2026facilitating}, from Section~\ref{sec-sims-anova}. Here, \texttt{RGE} uses the true reference groups defined in the DGP. We also include a simplified BayesABCs model \texttt{ABCs(ridge)}, which uses normal-inverse-gamma priors with distinct variances for each block of continuous main effects, categorical main effects, CX interactions, and CC interactions (see Appendix~\ref{app-ridge}). This represents a middle ground between ABCs without shrinkage (\texttt{OLS}) and ABCs with aggressive, structured shrinkage (\texttt{ABCs(grp)} and \texttt{ABCs(loc)}). 
For each model specification, we fit two versions: a main-only model, \texttt{y $\sim$ x$_1$ + \ldots + x$_p$ + sex + race}, and a  cat-modified model, \texttt{y $\sim$ (x$_1$ + \ldots + x$_p$ + sex)*race}, where \texttt{race} interacts with all other covariates. The latter is more suitable when $\gamma \ne 0$, but because of the sparsity in the DGP, neither corresponds to the ground truth in any case. Each method receives a \texttt{main} or \texttt{mod} prefix accordingly, e.g., \texttt{main:ABCs(grp)}.

%{\bf Key takeaways: extraneous case} (\texttt{coef(x:r) = 0}):
    % Compared to OLS (w/ ABCs), shrinkage reduces interval width without sacrificing coverage (see ridge and OLS); also notice how the gaps are rather small between main-only and cat-modified versions, despite the extraneous parameters
    % over and rge undercover, while loc gains precision while maintaining calibration
    % parameterization and shrinkage both matter! the ridge gets most of the way there 

% {\bf Key takeaways: necessary case} (\texttt{coef(x:r) = 0.5}):
    %shrinkage still matters
    %still a failure to cover for rge, over

%Note: for the conditional mean (not just parameters, as above), rge and over are more competitive (but not best). there it really matters having the "right" model (main vs mod)

%This is not true of all model parameters. 
%In fact, main-only models agree on continuous main effects in the sense that OLS estimates of $\bm \alpha$ are identical for any minimally sufficient identification (RGE, ABCs, etc.) of the categorical main effects.  That is not the case for cat-modified models. Similarly, OLS estimates of $\mu(\bm x)$ are  invariant to the identification, both for main-only and cat-modified models. Thus, both targets address the interplay of model parameterization and shrinkage, while $\bm\alpha$ especially isolates differences between main-only and cat-modified models.  

Here, we assess estimation and inference for the continuous main effects $\bm \alpha$ with $n=1000$. The DGP always satisfies ABCs, RGE, and Over regardless of $\gamma$, so all methods target the same $\bm \alpha$. In Appendix~\ref{app-sims-lm}, we present results for $n=200$ (the conclusions are similar) and for the regression function $\mu(\bm x) = \mathbb{E}(y \mid \bm x)$ (summarized below).  %In conjunction, these targets assess how estimation, inference, and prediction with cat-modified models depend on the model parameterization, the priors, and whether the cat-modifiers are necessary or extraneous. 

\begin{figure}[h]
\centering
\includegraphics[width=.49\textwidth]{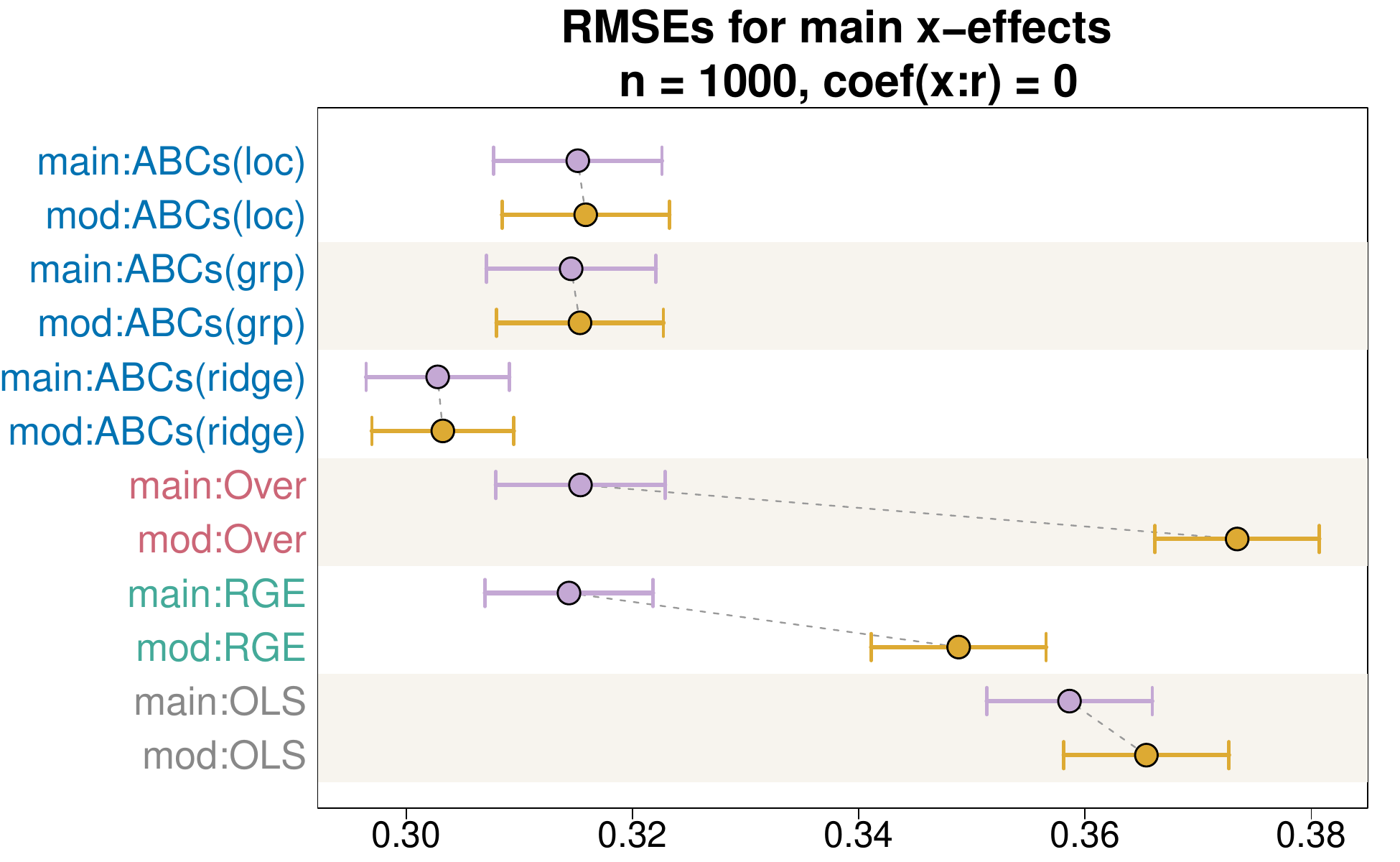}
\includegraphics[width=.49\textwidth]{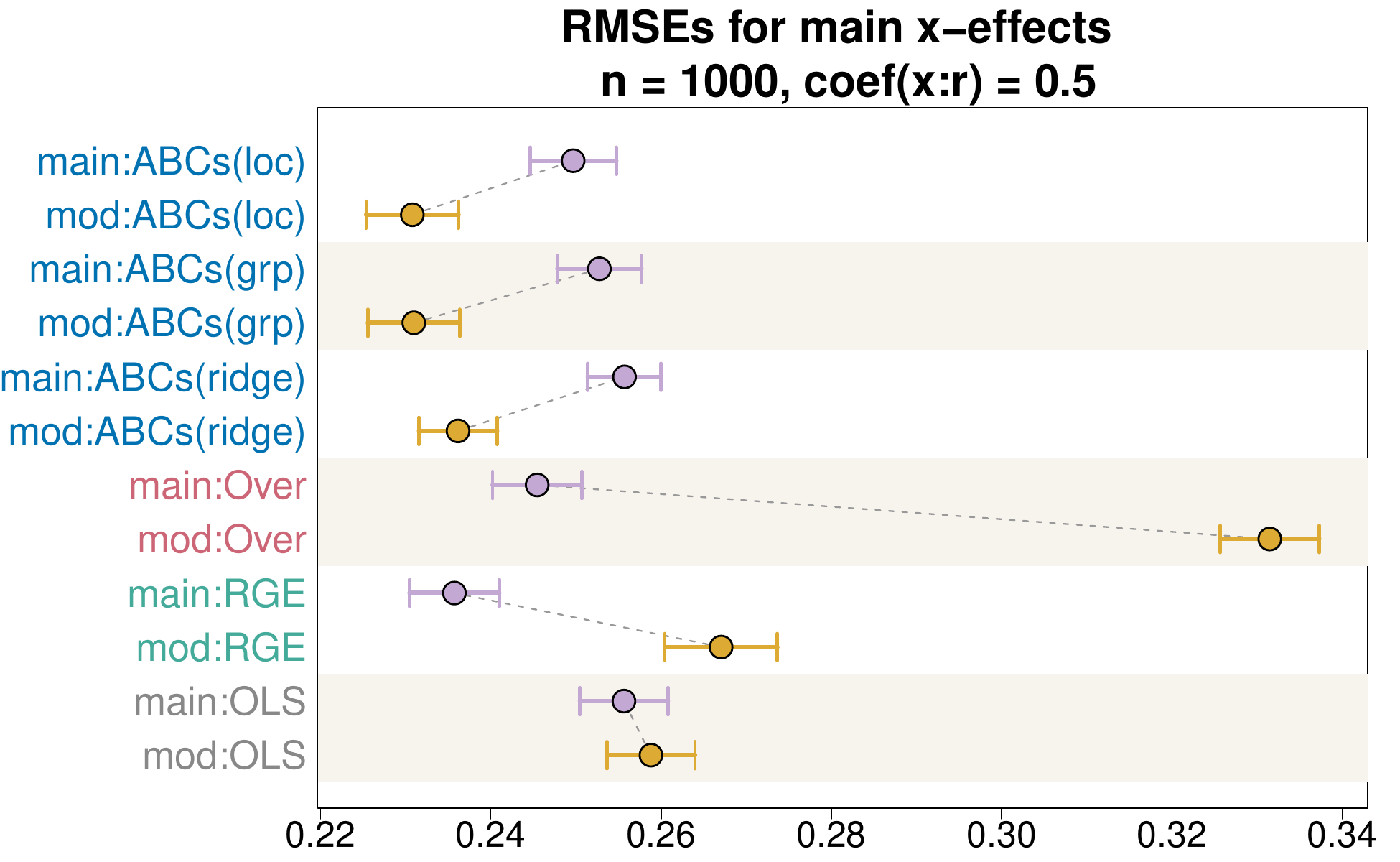}
\caption{\small RMSEs (mean $\pm 2$ SEs) for the continuous main  effects $\bm \alpha$. Dashed lines connect the main-only and cat-modified versions of each method. When categorical modifiers are extraneous (left), BayesABCs is uniquely accurate under \emph{both} the main-only and cat-modified versions. When some categorical modifiers are necessary (right), BayesABCs with structured shrinkage priors (\texttt{ABCs(loc)} or \texttt{ABCs(grp)})  elevates the cat-modified model to be most accurate. 
\label{fig:sims-rmse-x}}
\end{figure}

Figure~\ref{fig:sims-rmse-x} evaluates point estimation of $\bm \alpha$. BayesABCs estimates are the most accurate in each case. More interesting is to compare the main-only and cat-modified versions of each method. When all categorical interactions are null (Figure~\ref{fig:sims-rmse-x}, left), the Bayesian main-only models with aggressive shrinkage priors (\texttt{ABCs(loc)}, \texttt{ABCs(grp)}, \texttt{Over}, and \texttt{RGE}) perform similarly, as expected. However, critical differences emerge for the cat-modified versions: for BayesABCs, the accuracy is nearly identical to each main-only counterpart; yet for \texttt{Over} and \texttt{RGE} there is a steep decline. This is posterior invariance of BayesABCs in action. % :  with ABCs, the cat-modified models do not pay a price for including null interactions. 
Alternatively, when some categorical interactions are necessary (Figure~\ref{fig:sims-rmse-x}, right),  models with ABCs \emph{and} structured shrinkage priors (\texttt{ABCs(loc)} or \texttt{ABCs(grp)}) elevate the cat-modified model to be most accurate. 
Neither \texttt{Over} nor \texttt{RGE} produce accurate estimates under the cat-modified model. Uniquely, BayesABCs enables reliable estimation of main effects under cat-modified models---regardless of whether the interactions are null or active. 
%---across a variety of shrinkage priors. %   (\texttt{ABCs(loc)}, \texttt{ABCs(grp)}, \texttt{ABCs(ridge)}). 

\begin{figure}[h]
\centering
\includegraphics[width=.49\textwidth]{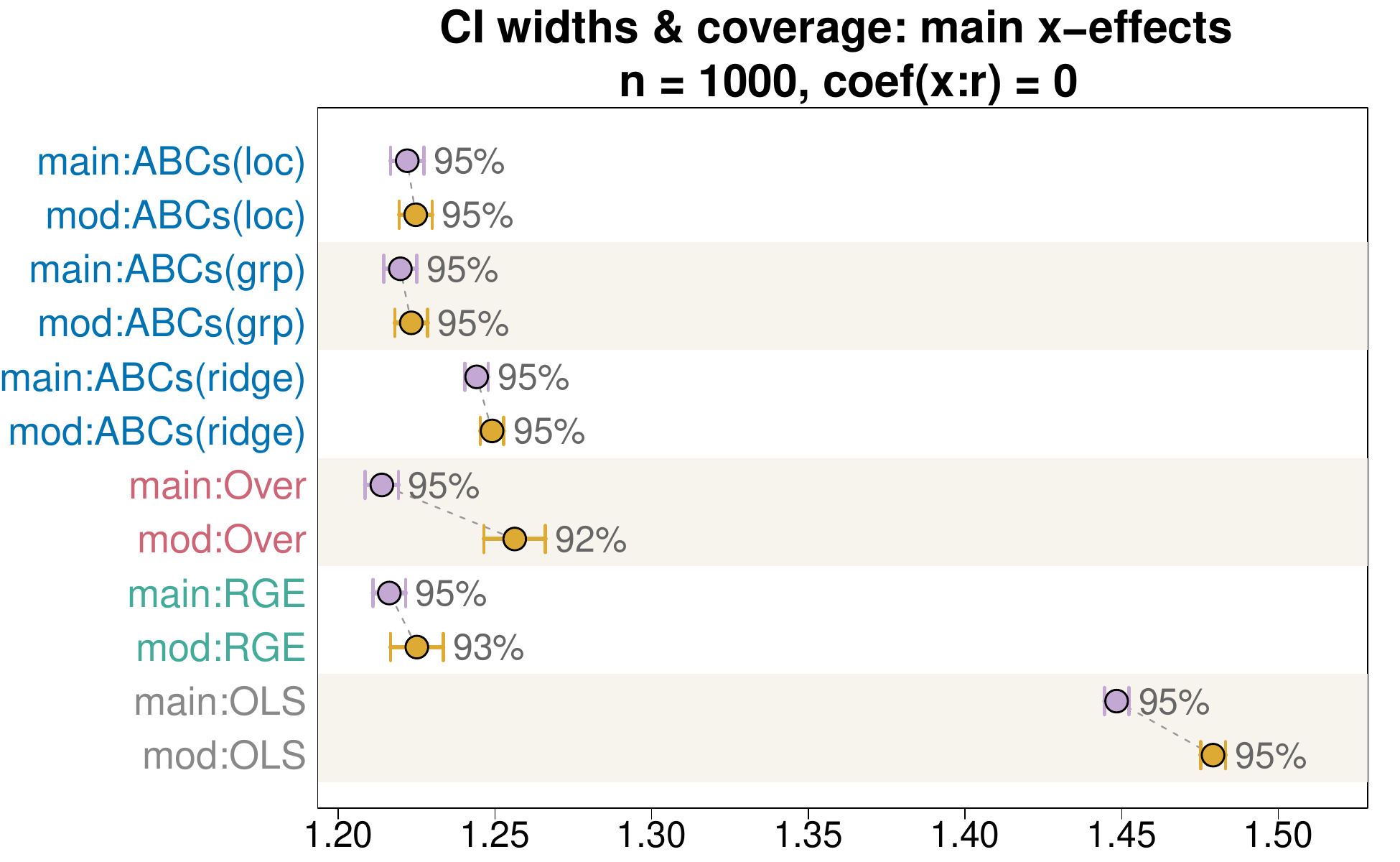}
\includegraphics[width=.49\textwidth]{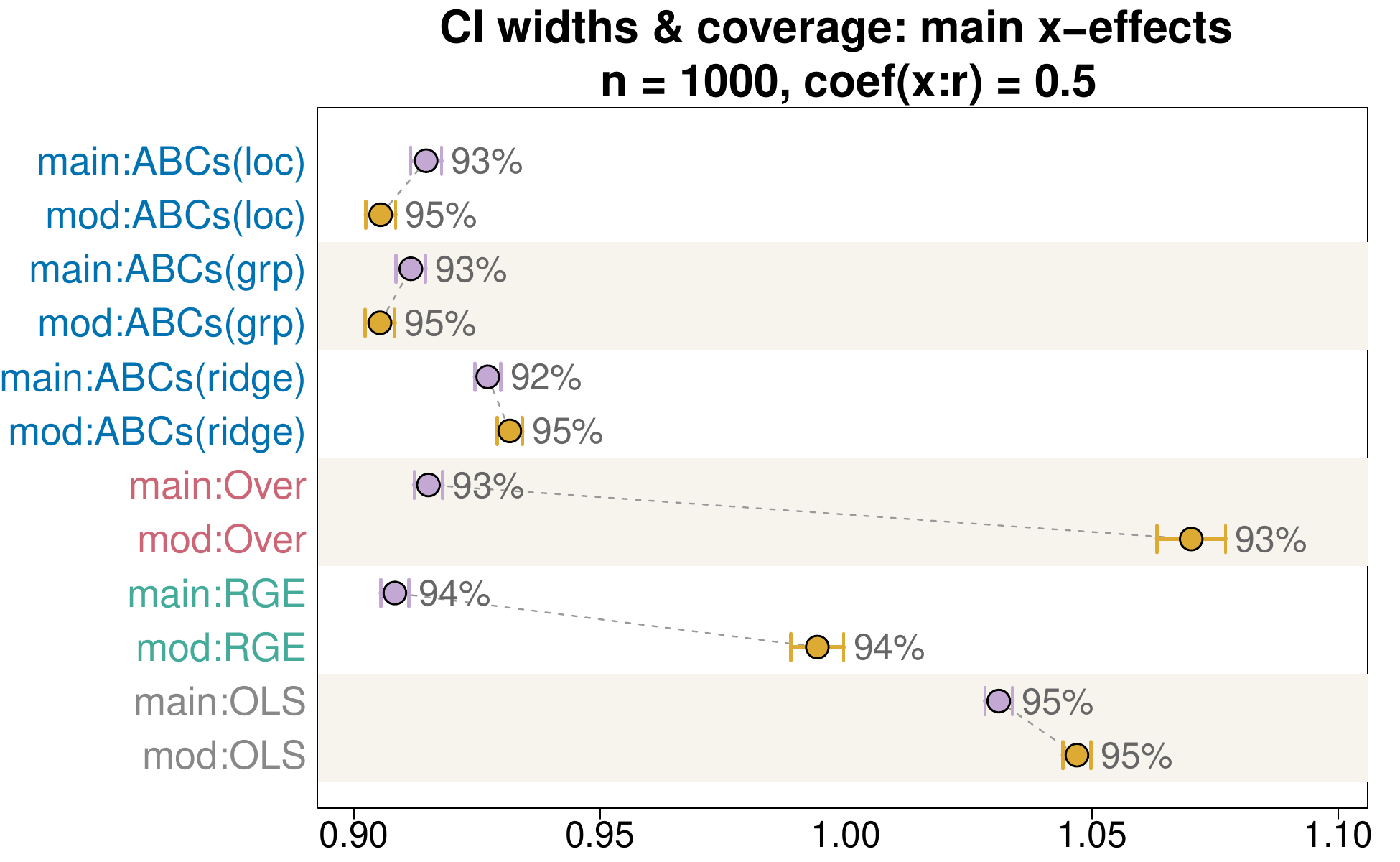}
\caption{\small Mean interval widths (mean $\pm 2$ SEs) and empirical coverage (annotations) for the continuous main  effects $\bm \alpha$. Dashed lines connect the main-only and cat-modified versions of each method. Only Bayesian models with ABCs are favorable to categorical interactions: there is little cost in precision when extraneous (left) and notable gains  when some are necessary (right).
\label{fig:sims-ci-x}}
\end{figure}

Mean credible/confidence interval widths and empirical coverages for $\bm\alpha$ are in Figure~\ref{fig:sims-ci-x}. The story is familiar: BayesABCs methods do not incur a cost (i.e., wider intervals) for including null interactions, but can provide sharper inference when some interactions are necessary. Here, we clearly see the advantages of aggressive shrinkage, where \texttt{ABCs(loc)} and \texttt{ABCs(grp)} have narrower intervals than \texttt{ABCs(ridge)} and \texttt{OLS}.  As expected, \texttt{main:RGE} and \texttt{main:Over} both perform well: for main-only models, the parameterization only refers to categorical main effects, which has little impact on the continuous main effects. %, and the horseshoe priors are well-suited to the sparse DGP. %, and he horseshoe priors employed by these methods is well-suited for the sparse DGP. 
However, their cat-modified versions are noncompetitive, either due to undercoverage ($\gamma = 0$) or imprecision ($\gamma = 0.5$).

%models that use ABCs dominate \texttt{RGE} and \texttt{Over}. Posterior invariance (Section~\ref{subsec-invar}) is apparent and impactful: with ABCs, the cat-modified models do not pay a price for including null interactions (top left). This is not the case for \texttt{RGE} or \texttt{Over}. When correctly including interactions (bottom left), $\bm \alpha$ is estimated more accurately for the cat-modified models that use both ABCs and a shrinkage prior (not \texttt{OLS}). Perhaps surprisingly, the best strategy combines ABCs with \emph{some} shrinkage: \texttt{ABCs(ridge)} outperforms \texttt{OLS}, which uses ABCs  without shrinkage, and  \texttt{RGE} and  \texttt{Over}, which use aggressive shrinkage without ABCs, and often matches or beats ABCs with more sophisticated shrinkage priors. 

Finally, we summarize estimation and inference for $\mu(\bm x)$ (see Appendix~\ref{app-sims-lm}). For prediction, the decision to include or exclude categorical interactions is  vital: when some are active, the main-only models are much less accurate than the cat-modified versions (Figures~\ref{fig:sims-rmse-mu} and \ref{fig:sims-ci-mu}, right). However, when all are extraneous, only the BayesABCs models with ABCs \emph{and} shrinkage priors  (\texttt{ABCs(loc)}, \texttt{ABCs(grp)}, \texttt{ABCs(ridge)}) keep the cat-modified version competitive with the main-only version (Figures~\ref{fig:sims-rmse-mu} and \ref{fig:sims-ci-mu}, left).

%RMSEs (mean $\pm 2$ SEs) for the regression function $\mu(\bm x)$ with $n=1000$ (top) and $n=200$ (bottom). Dashed lines connect the main-only and cat-modified versions of each method. When all cat-modifiers are extraneous (left), Bayesian models with ABCs keep the cat-modified version competitive with the main-only version. When some   are necessary (right), the cat-modified Bayesian models offer massive gains in accuracy over their main-only counterparts. 

%Mean interval widths (mean $\pm 2$ SEs) and empirical coverage (annotations) for the regression function $\mu(\bm x)$ with $n=1000$ (top) and $n=200$ (bottom). Dashed lines connect the main-only and cat-modified versions of each method. When cat-modifiers are extraneous (left), the cat-modified models are less precise than their main-only counterparts, although ABCs slightly narrow that gap. When some cat-modifiers are necessary (right), the main-only models substantially undercover while \texttt{ABCs(grp)} and \texttt{ABCs(loc)} are most precise. 

\section{Application}\label{sec-asr}
We analyze errors in automatic speech recognition (ASR) systems. Heterogeneous effects are pivotal to determine if ASR systems are less accurate for  certain subgroups of speakers, such as by race/ethnicity, gender, or age. The dataset  is curated by  \cite{koenecke2020racial} and combines \href{https://oraal.github.io/coraal}{Corpus of Regional African American Language} 
and \href{https://web.stanford.edu/dept/linguistics/VoCal/}{Voices of California} data. 
For each of $i=1,\ldots, n = 8817$ recordings, let $y_i$ be the average word error rate (WER) across ASR systems by Google, IBM, Amazon, Microsoft, and Apple, which are tightly coupled. %(correlations $\ge 0.70$). 
Continuous variables are the speaker's age  and the recording's duration, both standardized. Categorical variables include the speaker's race/ethnicity (Black or White) and gender (female or male). The speakers are 35\% Black female, 23\% Black male, 21\% White male, and 20\% White female, although this is not the same for recordings (e.g., only 17\% of recordings are by Black males).  
%The recordings are 34\% black female, 25\% white female, 24\% white male, and 17\% black male.

We begin with two models: a main-only model   \texttt{y $\sim$ race + gender + age + duration} and a cat-modified model that adds CX and CC interactions, \texttt{y $\sim$ race + gender + age + duration + race:(gender + age + duration) + gender:(age + duration)}, to obtain race- and gender-specific slopes and a race-gender interaction. We fit both models using Bayesian competitors from Section~\ref{sec-sims-lm}: \texttt{ABCs(grp)}, \texttt{RGE} (reference groups White and male, as commonly chosen), and \texttt{Over}   (see Figure~\ref{fig:panel-stack-ridge} for a version with all ridge priors).

\begin{figure}[!h]
\centering
\includegraphics[width=.75\textwidth]{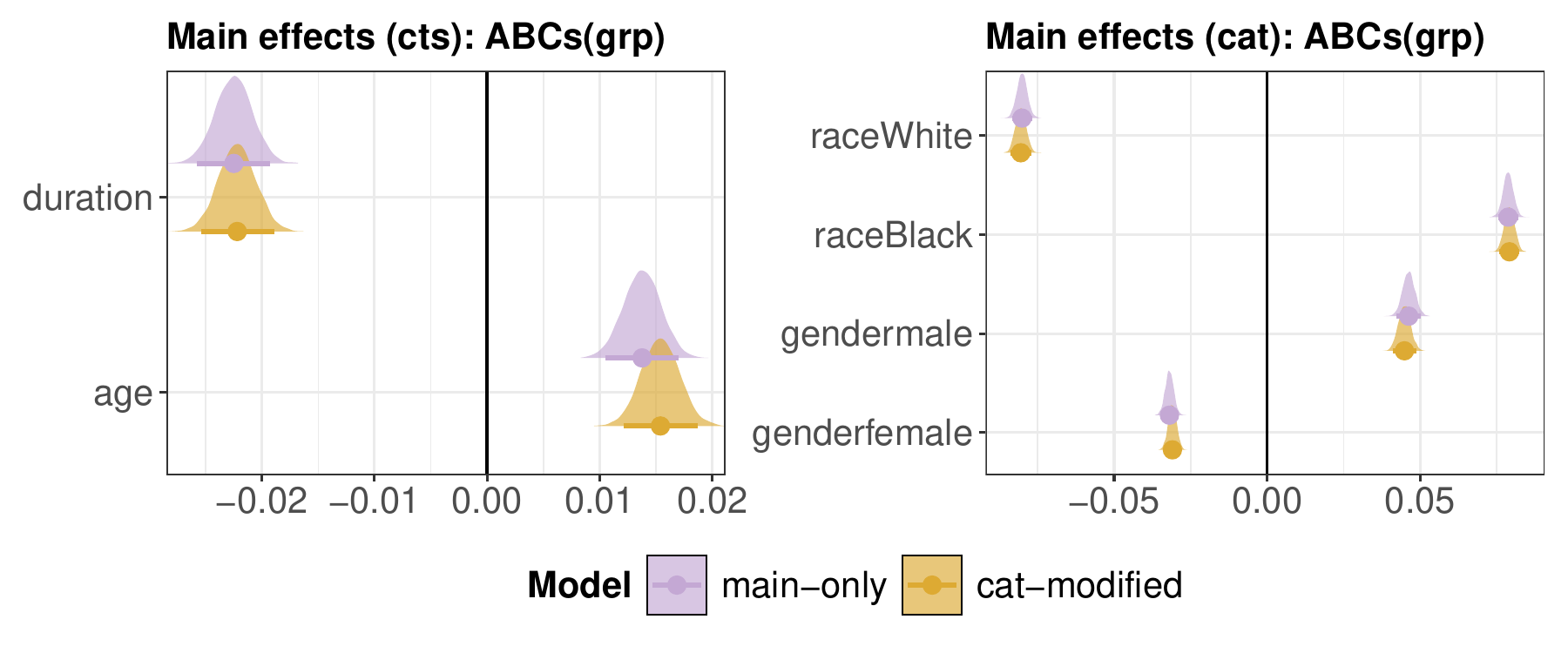}
\includegraphics[width=.75\textwidth]{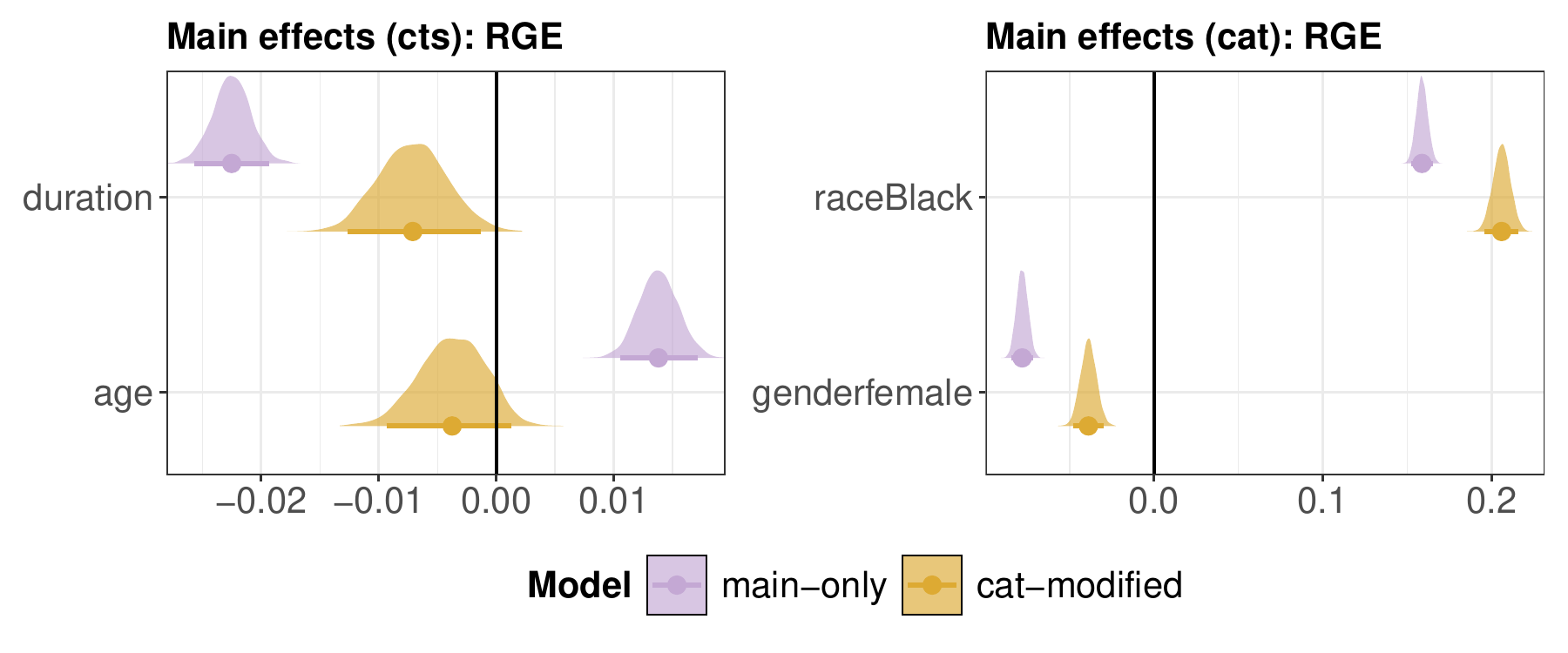}
\includegraphics[width=.75\textwidth]{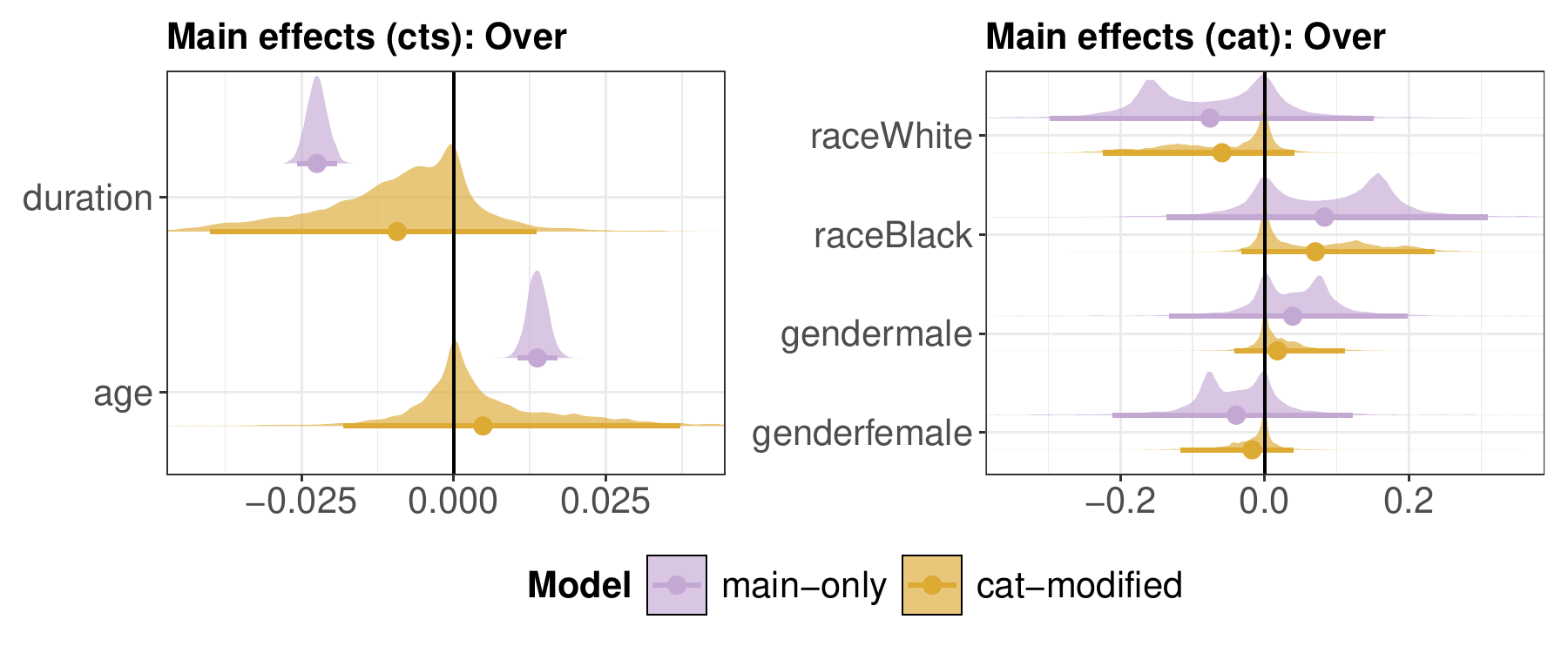}
\caption{\small Posteriors for continuous (left) and categorical (right) main effects under main-only (blue) and cat-modified (gold) models with BayesABCs (top), RGE (middle), and Over (bottom). Points indicate posterior means and bars are 95\% equal-tailed intervals.
Uniquely for BayesABCs, the main effect posteriors are nearly identical for the main-only and cat-modified models, leading to clear and unanimous conclusions. 
\label{fig:panel-stack}}
\end{figure}

We present the main effect posteriors for all models and parameterizations in Figure~\ref{fig:panel-stack}.  BayesABCs demonstrates the posterior invariance anticipated by our theory: main effect inference is nearly unchanged by the addition of multiple categorical interactions. In fact, all of the parameterizations agree on the continuous main effects from the \emph{main-only} model; yet only BayesABCs recoups this for the cat-modified model. With BayesABCs, the additional heterogeneity does not complicate interpretations or conclusions:  it is clear that higher WERs occur for older, Black, and male speakers and for shorter recordings.

There are no such benefits for RGE and Over. With RGE, the added CX and CC interactions alter and attenuate both continuous effects, which now refer to age and duration \emph{for White male} speakers. RGE shrinkage pulls each race- and gender-specific effect toward that for White  males, thus creating biases and suppressing potential deviations from these groups. With Over, the  nonidentifiability of categorical effects and interactions is problematic: the categorical main effect posteriors are highly dispersed and multimodal, and the continuous main effect posteriors are erratic under the cat-modified model.

% MCMC efficiency:
%   ABCs:
%   summary(effectiveSize(as.mcmc(post_main_abc)))
%    Min. 1st Qu.  Median    Mean 3rd Qu.    Max. 
%    4647    4821    4994    4898    5000    5000 
%   summary(effectiveSize(as.mcmc(post_mod_abc)))
%    Min. 1st Qu.  Median    Mean 3rd Qu.    Max. 
%    2165    4063    4632    4193    5000    5000 
%
%   Over:
%   summary(effectiveSize(as.mcmc(cbind(fit_main_over$post_alpha0, fit_main_over$post_beta))))
%    Min. 1st Qu.  Median    Mean 3rd Qu.    Max. 
%    1647    1732    2544    3030    3773    6012 
%   summary(effectiveSize(as.mcmc(cbind(fit_mod_over$post_alpha0, fit_mod_over$post_beta))))
%    Min. 1st Qu.  Median    Mean 3rd Qu.    Max. 
%   315.9   611.5  1095.6  1087.2  1269.3  2266.3 
%
%   RGE
%   summary(effectiveSize(as.mcmc(fit_main_rge$post_beta)))
%   Min. 1st Qu.  Median    Mean 3rd Qu.    Max. 
%   4445    4699    4892    4920    5113    5452 
%   summary(effectiveSize(as.mcmc(fit_mod_rge$post_beta)))
%   Min. 1st Qu.  Median    Mean 3rd Qu.    Max. 
%   1858    2224    4140    3436    4615    4684 

\begin{figure}[h]
\centering
\includegraphics[width=.8\textwidth]{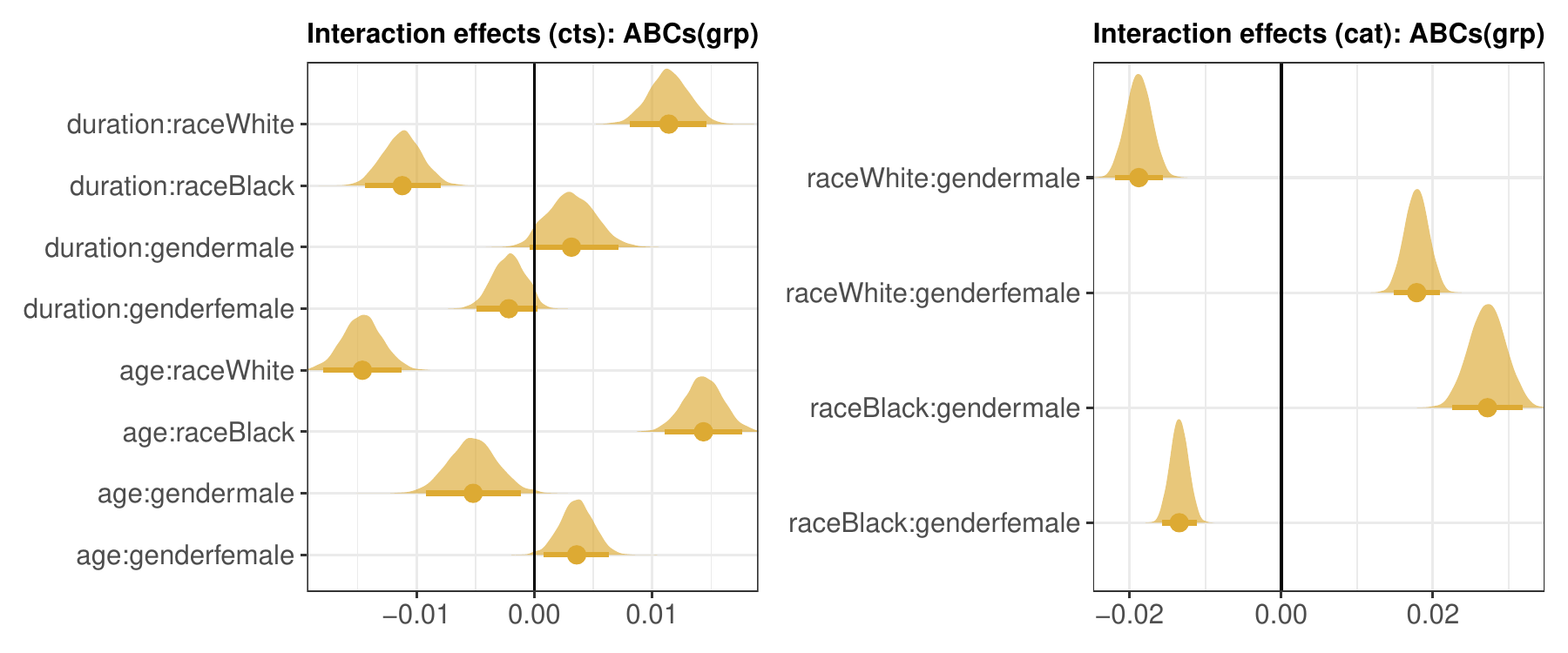}
\caption{\small BayesABCs posteriors for CX (left) and CC (right) interactions for ASR data. 
\label{fig:interact-abcs}}
\end{figure}

With BayesABCs, the cat-modified model also provides interpretable and informative heterogeneous effect inference.  The CX and CC interaction  posteriors are in Figure~\ref{fig:interact-abcs}. Although higher WERs were identified separately for Black speakers and male speakers, the intersection (Black male) corresponds to even higher WERs. The other CC interactions offer a counterweight to their main effects, as the signs are opposite. 
%The other CC interactions appear as a counterweight to the main effects, e.g., lower WERs were identified for both White speakers and female speakers, but WERs for White female speakers are not as small as predicted by the individual categories. 
The (directionally) positive age effect nearly doubles for Black speakers but vanishes for White speakers. Similar effects---amplification for Black speakers and attenuation for White speakers---occur for duration. Gender interactions are more modest, with minor amplifications of age and duration effects for female speakers.

%{\bf Remark:} Large dataset $(n = 8817)$ with potentially many cat levels ($L_1 = L_2 = 2$ for race and sex, but $L_3 = 124$ for ID). 

% For the group-specific terms $\mu(\bm 0, r ,s)$, we see that Black:male is pulled the most toward the others (smallest group, largest $\bar y_{rs}$). 

Now we consider an additional categorical variable: an identifier for the speaker (\texttt{ID}) in each recording. Classically, this is a linear mixed model with random effect \texttt{ID} to account for within-speaker dependence among recordings. Our Bayesian models require no distinction between fixed and random effects, so it is yet another categorical covariate---albeit a high-dimensional one ($L_3 = 124$ speakers). We add \texttt{ID} to the main-only model, %as a ``random intercept", 
often coded \texttt{(1 | ID)} for random intercepts, and add \texttt{ID} interactions with age and duration to the cat-modified model,   \texttt{(1 + age + duration | ID)}, for random slopes. %Now, the main-only and cat-modified models have 128 and 379 identified coefficients, respectively.
%(``random") slopes. 

The detailed results are in Appendix~\ref{app-app}.   Despite the additional high-dimensional categorical covariate and interactions, BayesABCs again maintains posterior invariance of main effects. Some conclusions are unchanged: higher WERs occur for Black and male speakers and for shorter recordings (Figures~\ref{fig:panel-stack-id}--\ref{fig:panel-stack-id-ridge}), with amplified duration effects for Black speakers (Figure~\ref{fig:interact-abcs-id}). Now, however, the age effect and its interactions vanish along with the  race-gender interaction. The limitations of RGE and Over reappear, including for both horseshoe (Figure~\ref{fig:panel-stack-id}) and ridge (Figure~\ref{fig:panel-stack-id-ridge}) priors.

Finally, we note that the parameterization and prior also matter for MCMC efficiency \citep{gelfand1995efficient}. We compute  effective sample sizes from 5000 MCMC samples of the main effects in each cat-modified model and summarize using the 1st quartile across $\{\alpha_0, \bm \alpha, \bm\beta\}$. The gaps across parameterizations are enormous: without \texttt{ID}, we  obtain   $4063$ for BayesABCs, but only $2224$ for RGE and $612$  for Over; with \texttt{ID} and using ridge priors for simplicity, we obtain  $5000$ for BayesABCs, but only $148$ for RGE and $159$  for Over. %\cite{zanella2021multilevel}
% ABCs
%   Min. 1st Qu.  Median    Mean 3rd Qu.    Max. 
%   3863    5000    5000    4965    5000    5894 
% Over
%   Min. 1st Qu.  Median    Mean 3rd Qu.    Max. 
%  75.85  158.52  183.92  422.82  571.87 4651.96
% RGE
%   Min. 1st Qu.  Median    Mean 3rd Qu.    Max. 
%  55.73  148.18  167.11  468.92  622.89 5000.00
  
%The addition of IDs removes the significant interaction effects. We still see the posterior invariance comparing models that includes ID \emph{interactions}, but that is different from comparing models that do or do not contain main effects (ID). 

% Using GBPVs to select blocks of covariates...

\section{Conclusion}\label{sec-disc}
Linear models effectively use categorical variables and their interactions to estimate heterogeneous, group-specific effects. Shrinkage is crucial to handle potentially many low-signal interactions. However, shrinkage depends on the model parameterization, and default parameterizations (RGE) have significant flaws. We introduced BayesABCs to overcome these limitations. This new parameterization-prior combination ensures favorable shrinkage targets, interpretable parameter inference across distinct models, and eliminates the usual downsides with heterogeneous models. Specifically, BayesABCs guarantees (nearly) identical posterior inference for all main effects regardless of whether categorical interactions are included. Simultaneously, grouped and local shrinkage priors adapt to the structures among categorical effects and their interactions and aggressively shrink extraneous effects to zero.  
The practical advantages of BayesABCs are abundant: across multiple simulation studies and a real data analysis, BayesABCs provides accurate estimates and predictions, sharp inferences, unmatched robustness to extraneous interactions, and clear, robust, and interpretable conclusions about both main and heterogeneous effects. BayesABCs also admits a convenient and general representation of the (constrained) prior, which facilitates analysis and efficient posterior computing.

% What did we do?
% posterior invariance
% study shrinkage
% computing
% shrinkage priors
% emprical stuff

%and computing 

One promising avenue for future work is to replace the continuous shrinkage priors with sparsity %(e.g., spike-and-slab) 
priors. This is uniquely appealing for BayesABCs: the implicit Bayesian model averaging  may benefit from posterior invariance, and parameter inference requires shared parameter (e.g., main effect) meanings across submodels. A related and compelling case was made by \cite{semken2022specification} for average treatment effects across multiple, heterogeneous models. Further, while we focused on linear models, the issues of parameterization, shrinkage priors, and computing with categorical effects and interactions remain relevant for nonlinear models \citep{deshpande2025flexbart}. Finally, the problematic behavior of RGE is not unique to Bayesian models and also occurs for penalized estimation.

\subsection*{Supplementary materials}
Supplementary material includes a PDF with computing details (Appendix~\ref{app-computing}), additional theory (Appendix~\ref{app-theory}), additional simulations (Appendix~\ref{app-sims}), additional data analysis (Appendix~\ref{app-app}), proofs of all results (Appendix~\ref{app-proofs}), and alternative approaches (Appendix~\ref{app-centering}); and \texttt{R} code for reproducibility. Data are publicly available at \url{https://github.com/stanford-policylab/asr-disparities}. 

%\subsection*{Disclosure}
%Research was supported in part by the National Science Foundation (SES-2435310). 
{\bf Disclosure:} 
The authors have no competing interests to declare. AI use was strictly limited to 1) checking all proofs for correctness and code for errors and 2)  constructing \texttt{R} functions for plotting results. 
%The findings and conclusions in this publication are those of the author(s) and do not necessarily represent the views of the U.S. government.

\bibliographystyle{apalike}
\bibliography{BayesianABC}

\clearpage
\appendix

\counterwithin{figure}{section}
\counterwithin{table}{section}
\counterwithin{equation}{section}
\counterwithin{theorem}{section}
\counterwithin{lemma}{section}
\counterwithin{corollary}{section}

\begin{center} %\onehalfspacing
    \LARGE {\bf 
    Supplement to \\
    ``Priors and parameterizations for Bayesian regression with categorical covariates and interactions"}
      \if0\blind
    \large{\\ \vspace{5mm} Daniel R. Kowal and Inkoo Lee}
    \fi
  %\author{Daniel R. Kowal\thanks{Associate Professor, Department of Statistics and Data Science, Cornell University and Department of Statistics, Rice University (\href{mailto:dan.kowal@cornell.edu}{dan.kowal@cornell.edu}). 
    %\fi
\end{center}

     This supplementary file includes  computing details (Appendix~\ref{app-computing}), additional theory (Appendix~\ref{app-theory}), additional simulations (Appendix~\ref{app-sims}), additional data analysis (Appendix~\ref{app-app}), proofs of all results (Appendix~\ref{app-proofs}), and alternative approaches (Appendix~\ref{app-centering})

\section{Computing}\label{app-computing}
We implement a Gibbs sampler for BayesABCs for cat-modified models under various shrinkage priors. Proceeding generally, consider Gaussian linear models of the form
\[
\bm y = \bm X \bm \theta + \bm \epsilon, \quad  \bm \epsilon \sim N_n(\bm 0, \tau^{-1} \bm I_n)
\]
with (conditionally) Gaussian priors $[\bm \theta \mid \bm \Sigma_\theta] \sim N_p(\bm 0, \bm \Sigma_\theta)$ subject to linear constraints $\bm A \bm \theta = \bm 0$ (as in Section~\ref{subsec-enforce}), a Jeffreys prior $p(\tau) \propto \tau^{-1} $ (as in Section~\ref{subsec-invar}), and shrinkage priors implemented via $\bm \Sigma_\theta$ (as in Section~\ref{subsec-priors}). The algorithm cycles through the following terms, including a posterior predictive sampling step at $n_{test}$ testing points $\bm X_{test}$:
\begin{algorithm}[h]
\SetAlgoLined  
\begin{enumerate}
    \item Sample $[\tau \mid \bm y, -] \sim \mbox{Gamma}(n/2, \Vert \bm y - \bm X \bm \theta\Vert^2/2)$;
    \item Sample $[\bm \theta \mid \bm y, \bm A\bm\theta = \bm 0, -]$ using Algorithm~\ref{alg:constr};
    \item Sample $[\bm \Sigma_\theta \mid \bm y, -]$ using algorithms for shrinkage prior parameters (Appendix~\ref{app-px-group}, Appendix~\ref{app-px-local}, Appendix~\ref{app-ridge});
    \item Sample $[\bm{\tilde y}_{test} \mid \bm y, -] \sim N_{n_{test}}(\bm X_{test} \bm \theta, \tau^{-1} \bm I_{n_{test}})$
\end{enumerate} 
 \caption{Gibbs sampler for posterior (predictive) inference with BayesABCs.} \label{alg:gibbs}
\end{algorithm}

Steps~2 and 3 both must account for the linear constraint $\bm A \bm \theta = \bm 0$.  Algorithm~\ref{alg:constr} provides a generic solution for Step~2. For cat-modified models \eqref{reg-cm}, we can further improve efficiency by applying the reparameterizations blockwise. The continuous main effects require no constraints; each categorical main effect and each CX interaction is constrained by a separate row of $\bm A$; and CC interactions have blocks of constraints in $\bm A$. Thus, %$\bm A$ is block diagonal under \eqref{abcs-gen} and \eqref{abcs-cat-cat-gen}. 
in practice, we do not form $\bm A$ and $\bm Q_\mathcal{N}$ in their entirety across all coefficients $\bm \theta$ for Algorithm~\ref{alg:constr}, but rather construct separate  $\bm A$ and  $\bm Q_\mathcal{N}$ for each of these instances.  Specifically, we form each column block of the reparameterized design $\bm X_\mathcal{N}$ and each block of the posterior precision $\bm Q_{\mathcal{N}}'\bm\Sigma_\theta^{-1} \bm Q_{\mathcal{N}}$ separately, and then sample all coefficients jointly as in Algorithm~\ref{alg:constr}. 

Step~3 requires consideration of the particular shrinkage prior. Because ours involve half-Cauchy distributions,  we adopt the standard Gamma-Gamma parameter expansion. Specifically, we can equivalently express $[\lambda \mid \tau] \sim C^+(0,\tau)$ via $\eta_\lambda=\lambda^{-2}$ with 
%$\lambda = \eta_\lambda^{-1/2}$ and 
\begin{equation}\label{gg-px}
    [\eta_\lambda \mid \xi_\lambda] \sim \mbox{Gamma}(1/2, \xi_\lambda), \quad [\xi_\lambda \mid \tau] \sim \mbox{Gamma}(1/2, 1/\tau^2).
\end{equation}
%is $$ and $$ 
For standard horseshoe priors on the main continuous effects---and for Bayesian competitors \texttt{RGE} and \texttt{Over}---we apply \eqref{gg-px} for each half-Cauchy standard deviation parameter and sample from the Gamma full conditionals (see Appendix~\ref{app-px-group} for examples). In fact, this is sufficient for Gibbs sampling with \texttt{RGE} and \texttt{Over}: we simply replace Step 2 with the usual Gaussian full conditional distribution for (unconstrained) regression coefficients, where \texttt{RGE} omits the columns of $\bm X$ corresponds to reference groups.

To illustrate the careful accounting necessary for categorical coefficients  (main effects and CX/CC interactions) under ABCs, consider a categorical main effect $\bm \beta$ with $L$ levels, dropping additional subscripts for simplicity. The analog of the local shrinkage prior (Section~\ref{subsub-loc}) is $\beta_c \mid \lambda_c \sim N(0, \lambda_c^2)$, or equivalently $[\bm \beta\mid \bm \Sigma_\beta] \sim N_L(\bm 0, \bm\Sigma_\beta)$  with 
$\bm \Sigma_\beta = \mbox{diag}(\lambda_1^2,\ldots,\lambda_L^2)$. The prior on $\{\lambda_c\}_{c=1}^L$ will dictate the shrinkage behavior for $\bm \beta$. However, with ABCs, the constraint  transforms the prior into   $
[\bm \beta \mid  \bm \Sigma_\beta, \bm A \bm \beta = \bm 0] \stackrel{d}{=} \bm Q_{\mathcal{N}} \bm \beta_{\mathcal{N}}$, where $[\bm \beta_{\mathcal{N}}\mid \bm \Sigma_\beta] \sim N_{L_\mathcal{N}}\{\bm 0,  (\bm Q_{\mathcal{N}}'\bm \Sigma_\beta^{-1}\bm Q_{\mathcal{N}})^{-1}\}$  from Lemma~\ref{lem-prior-equiv}, $L_\mathcal{N} = L-1$ for categorical main effects, and $\bm Q_{\mathcal{N}}$ is defined in \eqref{qr-con} based on $\bm A$, which here is  $\bm A = \bm{\hat \pi}'$ under \eqref{abcs-gen-cat}. Specifically, this prior on the lower-dimensional, unconstrained parameters is 
 %The subsequent discussions focus primarily on $\bm \beta$ and $\bm \beta_{\mathcal{N}}$. However, we must also consider the implications for the components of $\bm\Sigma_\beta$, which will receive their own priors to encourage appropriate shrinkage behaviors. Namely, 
\begin{equation}\label{constr-cond-prior}
p( \bm \beta_{\mathcal{N}} \mid \bm \Sigma_\beta ) = \vert (2 \pi)^{-1} \bm Q_{\mathcal{N}}'\bm \Sigma_\beta^{-1}\bm Q_{\mathcal{N}} \vert^{1/2} \exp\left( -\frac{1}{2} \bm \beta_{\mathcal{N}}'\bm Q_{\mathcal{N}} ' \bm \Sigma_\beta^{-1} \bm Q_{\mathcal{N}} \bm \beta_{\mathcal{N}}\right).
\end{equation}
For \emph{grouped} shrinkage priors (Section~\ref{subsub-grp}), the prior simplifies: $\lambda_c = \lambda$ and $\bm \Sigma_\beta = \lambda^2 \bm I_L$. The contribution from the joint prior is then
%\begin{equation}\label{constr-joint-prior}
\[
p( \bm \beta_{\mathcal{N}}, \lambda) \propto \lambda^{-L_\mathcal{N}}\exp\left( -\frac{1}{2\lambda^2} \Vert  \bm \beta_{\mathcal{N}}\Vert^2\right) \ p(\lambda)
\]%\end{equation}
since $\Vert  \bm \beta_{\mathcal{N}}\Vert^2 = \Vert \bm Q_{\mathcal{N}} \bm \beta_{\mathcal{N}}\Vert^2$. Notably, this is effectively the same as  assigning the ridge-like prior $[\bm \beta_{\mathcal{N}} \mid \lambda] \sim N(\bm 0, \lambda^2 \bm I_{L_\mathcal{N}})$ directly on the reparameterized coefficients. Thus, posterior derivations may proceed directly using $\bm \beta_{\mathcal{N}}$. Completing this example with $[\lambda \mid \tau] \sim C^+(0,\tau)$ and the parameter-expansion \eqref{gg-px}, the full conditionals are simply
\[
    [\eta \mid -]\sim \mbox{Gamma}\{1/2 + (L-1)/2, \xi + \Vert \bm \beta_{\mathcal{N}} \Vert^2/2\}, \quad 
    [\xi \mid -] \sim \mbox{Gamma}(1, \eta + 1/\tau^2),    
\]
similar to the usual unconstrained case. We adapt this approach for the grouped shrinkage priors \eqref{prior-main-cat}--\eqref{prior-cat-int} in  Appendix~\ref{app-px-group}. 

%Priors on $\lambda^2$ often take the form of a (conditional) inverse-Gamma prior: $\lambda \sim C^+(0,A)$ can be written $[\lambda^{-2} \mid \xi_{\lambda}] \sim \mbox{Gamma}(1/2, \xi_{\lambda})$ and $\xi_{\lambda} \sim \mbox{Gamma}(1/2, 1/A^2)$. The resulting full conditional distributions are simply $[\lambda^{-2} \mid -] \sim \mbox{Gamma}(1/2 + p_\mathcal{N}/2, \xi_\lambda + \Vert   \bm \beta_{\mathcal{N}}\Vert^2/2)$ and $[\xi_\lambda \mid -] \sim \mbox{Gamma}(1, \lambda^{-2} + A^{-2})$, 

For local shrinkage priors, the same simplifications do not occur. Consider the initial (unconstrained) prior, $\beta_c \mid \lambda_c \sim N(0, \lambda_c^2)$, now accompanied by $[\lambda_c \mid \tau] \sim C^+(0, \tau)$ for $c=1,\ldots,L$. Adopting the parameter expansion strategy \eqref{gg-px} for each $\eta_c = \lambda_c^{-2}$, the usual full conditionals---i.e., without constraints $\bm A \bm \theta = \bm 0$---would be 
\begin{equation}\label{gg-fcd}
    [\eta_c \mid -]\sim \mbox{Gamma}(1, \xi_c + \beta_c^2/2), \quad 
    [\xi_c \mid -] \sim \mbox{Gamma}(1, \eta_c + 1/\tau^2),    
\end{equation}
independently across $c=1,\ldots,L$. However, the full conditionals for $\{\eta_c\}$ are incorrect: the prior described for $\beta_c$ does not yet incorporate the ABCs. The distributions in \eqref{gg-fcd} for $\bm \eta = (\eta_1,\ldots, \eta_L)'$ derive from the unnormalized posterior
\begin{equation}\label{local-prop}
    J(\bm \eta) = p(\bm \eta) \prod_{c=1}^L \{\eta_c^{1/2} \exp(-\eta_c \beta_c^2/2)\}
\end{equation}
%where $p(\bm \eta)$ is the product of  $\mbox{Gamma}(1/2, \xi_c)$ densities, 
but in fact the model requires 
\begin{equation}\label{local-target}
    q(\bm \eta) = p(\bm \eta) \ \vert  \bm Q_{\mathcal{N}}'\bm \Sigma_\eta^{-1}\bm Q_{\mathcal{N}} \vert^{1/2} \prod_{c=1}^L  \exp(-\eta_c \beta_c^2/2)
%\exp\left( -\sum_{c=1}^L \eta_c \beta_c^2/2\right)
\end{equation}
where $\bm \Sigma_\eta^{-1} = \mbox{diag}(\bm \eta)$ and $\bm \beta = \bm Q_{\mathcal{N}} \bm \beta_{\mathcal{N}}$. Comparing these terms, we see that 
\begin{equation}\label{local-logratio}
\log \frac{q(\bm \eta)}{J(\bm \eta)} = \frac{1}{2} \log \vert  \bm Q_{\mathcal{N}}'\bm \Sigma_\eta^{-1}\bm Q_{\mathcal{N}} \vert - \frac{1}{2} \sum_{c=1}^L \log \eta_c
\end{equation}
so the exponential term cancels and the (log) ratio does not depend on $\bm \beta$. 

Our strategy is to use the familiar full conditionals \eqref{local-prop} as a proposal distribution jointly for $\bm \eta$, and then take a Metropolis-Hastings adjustment toward the target \eqref{local-target}. The log-acceptance probability is computed by applying \eqref{local-logratio} to the proposed draw of $\bm \eta$ and then subtracting off \eqref{local-logratio}  for the previous value of $\bm\eta$. The local scale parameters $\{\eta_c\}_{c=1}^L$ are accepted or rejected jointly, but in our experience the acceptance rates remain high (around 60\% or larger) even when $L$ is large. The sampling steps for $[\xi_c \mid -]$ are unchanged. We detail this approach for the local shrinkage priors \eqref{prior-main-cat-local}--\eqref{prior-cat-int-local} in Appendix~\ref{app-px-local}. 

\subsection{Gibbs sampling for grouped shrinkage priors}\label{app-px-group}
The basic strategy is to apply the parameter expansion \eqref{gg-px} to each half-Cauchy term in \eqref{prior-main-cat}--\eqref{prior-cat-int}. For clarity, we first rewrite all priors in terms of the precisions: 
\begin{align*}
        &[\alpha_j \mid \eta_{\alpha_j}] \sim N(0, \eta_{\alpha_j}^{-1}), \quad  [\beta_{k,c_k} \mid \lambda_{\beta_k}] \sim N(0, \eta_{\beta_k}^{-1})\\
        &[\gamma_{j:k,c_k} \mid \eta_{\gamma_{j:k}}] \sim N(0, \eta_{\gamma_{j:k}}^{-1}) , \quad [\gamma_{k:k',c_k,c_{k'}} \mid \eta_{\gamma_{k:k'}}] 
       \sim N(0, \eta_{\gamma_{k:k'}}^{-1})
\end{align*}
independently, ranging over indices $j=1,\ldots,p$, $c_k = 1,\ldots,L_k$, and $k=1,\ldots K$. The precisions are structured hierarchically with parameter-expanded Gamma accompaniments:
\begin{align*}
    & [\eta_{\alpha_j} \mid \xi_{\alpha_j}] \sim \mbox{Gamma}(1/2, \xi_{\alpha_j}), \quad     && [\xi_{\alpha_j} \mid \eta_{\tau_0}] \sim \mbox{Gamma}(1/2, \eta_{\tau_0})\\
    & [\eta_{\beta_k} \mid \xi_{\beta_k}] \sim \mbox{Gamma}(1/2, \xi_{\beta_k}), \quad     && [\xi_{\beta_k} \mid \eta_{\tau_0}]  \sim \mbox{Gamma}(1/2, \eta_{\tau_0})\\
    & [\eta_{\tau_0} \mid \xi_{\tau_0}] \sim \mbox{Gamma}(1/2, \xi_{\tau_0}), \quad     && [\xi_{\tau_0}] \sim \mbox{Gamma}(1/2, 1)\\
    & [\eta_{\gamma_{j:k}} \mid \xi_{\gamma_{j:k}}] \sim \mbox{Gamma}(1/2, \xi_{\gamma_{j:k}}), \quad     && [\xi_{\gamma_{j:k}} \mid \eta_{\tau_*}] \sim \mbox{Gamma}(1/2, \eta_{\tau_*})\\
    %    & [\eta_{\gamma_{j:k}} \mid \xi_{\gamma_{j:k}}] \sim \mbox{Gamma}(1/2, \xi_{\gamma_{j:k}}), \quad     && [\xi_{\gamma_{j:k}} \mid \eta_{\tau_*}, \eta_{\alpha_j}, \eta_{\beta_k}] \sim \mbox{Gamma}(1/2, \eta_{\tau_*}\eta_{\alpha_j} \eta_{\beta_k})\\
        & [\eta_{\gamma_{k:k'}} \mid \xi_{\gamma_{k:k'}}] \sim\mbox{Gamma}(1/2, \xi_{\gamma_{k:k'}}), \quad     && [\xi_{\gamma_{k:k'}} \mid \eta_{\tau_*},  \eta_{\beta_k}, \eta_{\beta_{k'}}] \sim\mbox{Gamma}(1/2, \eta_{\tau_*}\eta_{\beta_k} \eta_{\beta_{k'}})\\
        & [\eta_{\tau_*} \mid \xi_{\tau_*}]  \sim  \mbox{Gamma}(1/2, \xi_{\tau_*}), \quad     && [\xi_{\tau_*}]  \sim \mbox{Gamma}(1/2, 1).
\end{align*}
We first present the case with CC heredity and discuss the alternative below. %Similar strategies are available for CX heredity, but it is not necessary for ABCs so we do not pursue it further. 

Due to the shared parameter structure, careful bookkeeping of the interaction terms is required. To index interactions, let $(j, k) \in \mathcal{J}$ denote (unordered) pairs of interactions between continuous variable $j$ and categorical variable $k$ and  %For a particular continuous (respectively, categorical) variable $j$, let $\mathcal{J}_j = \{k: (j,k) \in \mathcal{J}\}$ be the set of categorical (respectively, continuous) variables interacted with it. 
similarly define $(k,k') \in \mathcal{I}$ for the CC interactions. For a particular categorical variable $k$, let  $\mathcal{I}_k = \{k':(k,k') \in \mathcal{I}\}$  be the set of categorical variables interacted with it. 
Then we arrive at the following full conditional distributions first for the precision parameters,
\begin{align*}  
        [\eta_{{\alpha_j}} \mid -] &\sim \mbox{Gamma}(1,\ \xi_{{\alpha_j}} + \alpha_j^2/2)\\
    %[\eta_{{\alpha_j}} \mid -] &\sim \mbox{Gamma}\Big(1 + \vert \mathcal{J}_j \vert/2,\ \xi_{{\alpha_j}} + \alpha_j^2/2 +  \sum_{k \in \mathcal{J}_j}  \eta_{\tau_*}\eta_{{\beta_k}}\xi_{{\gamma_{j:k}}}\Big)\\
    %[\eta_{{\beta_k}} \mid -] &\sim \mbox{Gamma}\Big\{ 1/2 + (L_k-1)/2  + \vert \mathcal{J}_k \vert /2 + \vert \mathcal{I}_k \vert/2, \\
    %&\quad \quad\quad \quad \quad 
    %\xi_{{\beta_k}} + \Vert \bm \beta_k \Vert^2/2 +   \sum_{j\in \mathcal{J}_k} \eta_{\tau_*}\eta_{{\alpha_j}}\xi_{{\gamma_{j:k}}} + \sum_{k'\in \mathcal{I}_k}  \eta_{\tau_*} \eta_{{\beta_{k'}}} \xi_{{\gamma_{k:k'}}}  \Big\}\\
    [\eta_{{\beta_k}} \mid -] &\sim \mbox{Gamma}\Big\{ 1/2 + (L_k-1)/2  + \vert \mathcal{I}_k \vert/2, \ \xi_{{\beta_k}} + \Vert \bm \beta_k \Vert^2/2  + \sum_{k'\in \mathcal{I}_k}  \eta_{\tau_*} \eta_{{\beta_{k'}}} \xi_{{\gamma_{k:k'}}}  \Big\}\\
    [\eta_{{\gamma_{j:k}}}\mid -] &\sim \mbox{Gamma}(1/2 + (L_k-1)/2,\ \xi_
    {{\gamma_{j:k}}} + \Vert \bm \gamma_{j:k} \Vert^2/2 )\\
    [\eta_{{\gamma_{k:k'}}} \mid -] & \sim \mbox{Gamma}(1/2 + (L_k-1)(L_{k'}-1)/2,\ \xi_{{\gamma_{k:k'}}} + \Vert \bm \gamma_{k:k'} \Vert^2/2)\\
    [\eta_{\tau_0} \mid -] &\sim \mbox{Gamma}\Big\{(1  + p  + K)/2,\ \xi_{\tau_0} + \sum_{j=1}^p\xi_{{\alpha_j}} + \sum_{k=1}^K\xi_{{\beta_k}} \Big\} \\ 
    [\eta_{\tau_*} \mid -] &\sim \mbox{Gamma}\Big(1/2 + \vert \mathcal{J} \vert/2 + \vert \mathcal{I} \vert /2 ,  \  \xi_{\tau_*} + \sum_{(j,k) \in \mathcal{J}} \xi_{{\gamma_{j:k}}} + \sum_{(k,k') \in \mathcal{I}}  \eta_{{\beta_k}}\eta_{{\beta_{k'}}}\xi_{{\gamma_{k:k'}}}
    \Big)
%    \xi_{\tau_*} + \sum_{(j,k) \in \mathcal{J}} \eta_{{\alpha_j}} \eta_{{\beta_k}}\xi_{{\gamma_{j:k}}} + \sum_{(k,k') \in \mathcal{I}}  \eta_{{\beta_k}}\eta_{{\beta_{k'}}}\xi_{{\gamma_{k:k'}}}
%    \Big\}
\end{align*}
where $\bm \beta_k = (\beta_{1, k},\ldots, \beta_{L_k, k})'$ and similarly for $\bm \gamma_{j:k}$ and $\bm \gamma_{k:k'}$, 
followed by the full conditionals for the parameter-expanded terms,
\begin{align*}
        [\xi_{{\alpha_j}}\mid -] &\sim \mbox{Gamma}(1,\ \eta_{{\alpha_j}} + \eta_{\tau_0})
    \\
    [\xi_{{\beta_k}}\mid -] &\sim \mbox{Gamma}(1,\ \eta_{{\beta_k}} + \eta_{\tau_0})
    \\
    [\xi_{{\gamma_{j:k}}}\mid -] &\sim \mbox{Gamma}(1,\ \eta_{{\gamma_{j:k}}} + \eta_{\tau_*} )
        %[\xi_{{\gamma_{j:k}}}\mid -] &\sim \mbox{Gamma}(1,\ \eta_{{\gamma_{j:k}}} + \eta_{{\alpha_j}}\eta_{{\beta_k}}\eta_{\tau_*} )
    \\
    [\xi_{{\gamma_{k:k'}}}\mid -] &\sim \mbox{Gamma}(1,\ \eta_{{\gamma_{k:k'}}} + \eta_{{\beta_k}}\eta_{{\beta_{k'}}}\eta_{\tau_*} )\\
    [\xi_{\tau_0}\mid -] &\sim \mbox{Gamma}(1,\ \eta_{\tau_0} + 1)
    \\
    [\xi_{\tau_*}\mid -] &\sim \mbox{Gamma}(1,\ \eta_{\tau_*} + 1)
\end{align*}
with each (conditionally) independent across indices. Note that the squared norms of $\bm \beta_k$, $\bm \gamma_{j:k}$, and $\bm \gamma_{k:k'}$ can be computed for the lower-dimensional, unconstrained parameters due to orthogonality of $\bm Q_{\mathcal{N}}$. These squared norms contribute to the Gamma rate parameter for the respective precision $\eta$, while the number of identified parameters  contributes to the Gamma shape.  Each of $\eta$ and $\xi$ contribute to the other's Gamma rate parameter and tally $1/2$ for the shape, which then accumulates for parameters that are shared across indices. % (e.g., $\eta_{\beta_k}$ appears in every $\xi_{\gamma_{k:k'}}$ density for which $k \in \mathcal{J}_j$, etc.). 

 Without CC heredity, one parameter expansion is modified to omit the parent shrinkage terms, $[\xi_{\gamma_{k:k'}} \mid \eta_{\tau_*}] \sim\mbox{Gamma}(1/2, \eta_{\tau_*})$, leading to the modified full conditionals 
 \begin{align*}
     [\eta_{{\beta_k}} \mid -] &\sim \mbox{Gamma}\big\{ 1/2 + (L_k-1)/2, \ \xi_{{\beta_k}} + \Vert \bm \beta_k \Vert^2/2\Big\} \\
     [\eta_{\tau_*} \mid -] &\sim \mbox{Gamma}\Big(1/2 + \vert \mathcal{J} \vert/2 + \vert \mathcal{I} \vert /2 ,  \  \xi_{\tau_*} + \sum_{(j,k) \in \mathcal{J}} \xi_{{\gamma_{j:k}}} + \sum_{(k,k') \in \mathcal{I}}  \xi_{{\gamma_{k:k'}}}
    \Big)\\
    [\xi_{{\gamma_{k:k'}}}\mid -] &\sim \mbox{Gamma}(1,\ \eta_{{\gamma_{k:k'}}} + \eta_{\tau_*} ).
 \end{align*}
This is our default specification and completes Step~3 of Algorithm~\ref{alg:gibbs}.

\subsection{Metropolis-within-Gibbs sampling for local shrinkage priors}\label{app-px-local}
Applying the strategy from \eqref{local-logratio} to the priors \eqref{prior-main-cat-local}--\eqref{prior-cat-int-local}, we now have the \emph{proposals}
\begin{align*}  
    [\eta_{{\beta_{k,c_k}}}^* \mid -] & \sim \mbox{Gamma}(1,\ \xi_{{\beta_{k,c_k}}} + \beta_{k,c_k}^2 /2) \\
    %\sim \mbox{Gamma}(1,\ \xi_{{\beta_{k,c_k}}} + \beta_{k,c_k}^2w_{k,c_k}^2/2) \\
    [\eta_{{\gamma_{j:k,c_k}}}^* \mid -] & \sim \mbox{Gamma}(1,\ \xi_{{\gamma_{j:k,c_k}}} + \gamma_{j:k,c_k}^2/2)    \\
    [\eta_{{\gamma_{k:k',c_k,c_{k'}}}}^* \mid -] & \sim \mbox{Gamma}(1,\ \xi_{{\gamma_{k:k',c_k,c_{k'}}}} + \gamma_{k:k',c_k,c_{k'}}^2/2 )
    %\sim \mbox{Gamma}(1,\ \xi_{{\gamma_{k:k',c_k,c_{k'}}}} + \gamma_{k:k',c_k,c_{k'}}^2w_{k:k',c_k, c_{k'}}^2/2 )
\end{align*}
which are proposed independently across variables ($k=1,\ldots,K$) but accepted or rejected jointly across each set of levels ($c_k = 1,\ldots, L_k$), and similar for interactions. We annotate those terms with an asterisk to keep them distinct from the direct Gibbs sampling steps.

The remaining precision terms are more standard, including CC heredity for now:
\begin{align*}  
        [\eta_{{\alpha_j}} \mid -] \sim \mbox{Gamma}&(1,\ \xi_{{\alpha_j}} + \alpha_j^2/2)\\
    %[\eta_{{\alpha_j}} \mid -] &\sim \mbox{Gamma}\Big(1 + \sum_{k \in \mathcal{J}_j} L_k/2,\ \xi_{{\alpha_j}} + \alpha_j^2/2 +  \sum_{k \in \mathcal{J}_j} \sum_{c_k=1}^{L_k} \eta_{\tau_*}\eta_{{\beta_k}}\xi_{{\gamma_{j:k,c_k}}}\Big)\\
    [\eta_{{\beta_k}} \mid -] \sim \mbox{Gamma}&\Big\{ (1 + L_k + \sum_{(k,k') \in \mathcal{I}} L_k  L_{k'})/2,\\
     & \quad  \xi_{{\beta_k}} + \sum_{c_k = 1}^{L_k} \xi_{{\beta_{k,c_k}}} + \sum_{k' \in \mathcal{I}_k} \sum_{c_k=1}^{L_k} \sum_{c_{k'} = 1}^{L_{k'}} \eta_{\tau_*} \eta_{{\beta_{k*}}} \xi_{{\gamma_{k:k',c_k,c_{k'}}}}  \Big\}\\
    %[\eta_{{\beta_k}} \mid -] &\sim \mbox{Gamma}\Big\{ 1/2 + L_k/2  + \sum_{j \in\mathcal{J}_k}  L_k/2 + \sum_{(k,k') \in \mathcal{I}} L_k  L_{k'}/2,\\
    %& \quad \quad \quad \quad \xi_{{\beta_k}} + \sum_{c_k = 1}^{L_k} \xi_{{\beta_{k,c_k}}} +  \sum_{j \in \mathcal{J}_k}  \sum_{c_k = 1}^{L_k} \eta_{\tau_*}\eta_{{\alpha_j}}\xi_{{\gamma_{j:k,c_k}}} + \sum_{k' \in \mathcal{I}_k} \sum_{c_k=1}^{L_k} \sum_{c_{k'} = 1}^{L_{k'}} \eta_{\tau_*} \eta_{{\beta_{k*}}} \xi_{{\gamma_{k:k',c_k,c_{k'}}}}  \Big\}\\
    [\eta_{\tau_0} \mid -] \sim \mbox{Gamma}&\Big\{(1  + p  + K)/2,\ \xi_{\tau_0} + \sum_{j=1}^p\xi_{{\alpha_j}} + \sum_{k=1}^K\xi_{{\beta_k}} \Big\} \\ 
    [\eta_{\tau_*} \mid -]  \sim \mbox{Gamma}&\Big\{(1  + \sum_{(j,k) \in \mathcal{J}} L_k + \sum_{(k,k') \in \mathcal{I}} L_kL_{k'})/2 ,\\  
        & \quad \xi_{\tau_*} + \sum_{(j,k) \in \mathcal{J}} \sum_{c_k=1}^{L_k}  \xi_{{\gamma_{j:k,c_k}}} + \sum_{(k,k') \in \mathcal{I}} \sum_{c_k=1}^{L_k} \sum_{c_{k'}=1}^{L_{k'}} \eta_{{\beta_k}}\eta_{{\beta_{k'}}}\xi_{{\gamma_{k:k',c_k,c_{k'}}}}
    %& \xi_{\tau_*} + \sum_{(j,k) \in \mathcal{J}} \sum_{c_k=1}^{L_k} \eta_{{\alpha_j}} \eta_{{\beta_k}}\xi_{{\gamma_{j:k,c_k}}} + \sum_{(k,k') \in \mathcal{I}} \sum_{c_k=1}^{L_k} \sum_{c_{k'}=1}^{L_{k'}} \eta_{{\beta_k}}\eta_{{\beta_{k'}}}\xi_{{\gamma_{k:k',c_k,c_{k'}}}}
    \Big\}
    \end{align*}
Finally, the parameter expansion terms follow:
\begin{align*}  
    [\xi_{{\alpha_j}}\mid -] &\sim \mbox{Gamma}(1,\ \eta_{{\alpha_j}} + \eta_{\tau_0})
    \\
    [\xi_{{\beta_k}}\mid -] &\sim \mbox{Gamma}(1,\ \eta_{{\beta_k}} + \eta_{\tau_0})
    \\
    [\xi_{\tau_0}\mid -] &\sim \mbox{Gamma}(1,\ \eta_{\tau_0} + 1)
    \\
    [\xi_{\tau_*}\mid -] &\sim \mbox{Gamma}(1,\ \eta_{\tau_*} + 1)\\
    [\xi_{{\beta_{k,c_k}}}\mid -] &\sim \mbox{Gamma}(1,\ \eta_{{\beta_{k,c_k}}} + \eta_{{\beta_k}})
    \\
    [\xi_{{\gamma_{j:k,c_k}}}\mid -] &\sim \mbox{Gamma}(1,\ \eta_{{\gamma_{j:k,c_k}}} + \eta_{\tau_*} )
%[\xi_{{\gamma_{j:k,c_k}}}\mid -] &\sim \mbox{Gamma}(1,\ \eta_{{\gamma_{j:k,c_k}}} + \eta_{{\alpha_j}}\eta_{{\beta_k}}\eta_{\tau_*} )
    \\
    [\xi_{{\gamma_{k:k',c_k,c_{k'}}}}\mid -] &\sim \mbox{Gamma}(1,\ \eta_{{\gamma_{k:k',c_k,c_{k'}}}} + \eta_{{\beta_k}}\eta_{{\beta_{k'}}}\eta_{\tau_*} )
\end{align*}
all mutually independent. Note that only the local interaction terms are different from the grouped shrinkage case. Without CC heredity, we modify:
\begin{align*}
    [\eta_{{\beta_k}} \mid -] \sim \mbox{Gamma}&\Big\{ (1 + L_k)/2,\ \xi_{{\beta_k}} + \sum_{c_k = 1}^{L_k} \xi_{{\beta_{k,c_k}}}\Big\}\\
        [\eta_{\tau_*} \mid -]  \sim \mbox{Gamma}&\Big\{(1  + \sum_{(j,k) \in \mathcal{J}} L_k + \sum_{(k,k') \in \mathcal{I}} L_kL_{k'})/2 ,\\  
        & \quad \xi_{\tau_*} + \sum_{(j,k) \in \mathcal{J}} \sum_{c_k=1}^{L_k}  \xi_{{\gamma_{j:k,c_k}}} + \sum_{(k,k') \in \mathcal{I}} \sum_{c_k=1}^{L_k} \sum_{c_{k'}=1}^{L_{k'}} \xi_{{\gamma_{k:k',c_k,c_{k'}}}}
    \Big\} \\
        [\xi_{{\gamma_{k:k',c_k,c_{k'}}}}\mid -] \sim \mbox{Gamma}&(1,\ \eta_{{\gamma_{k:k',c_k,c_{k'}}}} + \eta_{\tau_*} ).
\end{align*}

\subsection{Ridge prior specification}\label{app-ridge}

The ridge prior (\texttt{ABCs(ridge)}) assumes normal-inverse-gamma priors for all coefficients, with variances specific to each block of continuous main effects, $[\alpha_j \mid \lambda_{\alpha}] \sim N(0, \lambda_{\alpha}^2)$, categorical main effects $[\beta_{k, c_k} \mid \lambda_{\beta}] \sim N(0, \lambda_{\beta}^2)$, CX interactions $[\gamma_{j:k, c_k} \mid \lambda_{\gamma_1}] \sim N(0, \lambda_{CX}^2)$ and CC interactions $[\gamma_{k:k',c_k,c_{k'}} \mid \lambda_{\gamma_2}] \sim N(0, \lambda_{CC}^2)$, and $\lambda_{\alpha}^{-2}, \lambda_\beta^{-2}, \lambda_{CX}^{-2}, \lambda_{CC}^{-2} \sim \mbox{Gamma}(a_\lambda, b_\lambda)$ with $a_\lambda = b_\lambda = 0.01$ by default. Each precision parameter is updated in Step~3 of Algorithm~\ref{alg:gibbs} from its Gamma full conditional. 
This prior lacks the aggressive shrinkage of horseshoe-like priors, but still provides regularization specific to each parameter block.   

\iffalse %%%%%%%%%%%%%%%%%%%%%%%%%%
\subsection{Grouped heredity prior}
As an alternative to the grouped shrinkage prior, we considered a grouped heredity variation that retained the priors on  $\bm \alpha$ and $\bm \beta_k$ but modified the interaction priors. Specifically, this prior takes a single, global scale for all CX interactions and then brings in the parent scales directly for CC interaction variances:
\begin{align*}
    [\gamma_{j:k, c_k} \mid \lambda_{CX}] & \sim N(0, \lambda_{CX}^2), \quad \lambda_{CX}  \sim C^+(0,1) \\
    [\gamma_{k,k', c_k, c_{k'}} \mid \lambda_{\beta_k}, \lambda_{\beta_{k'}}, \lambda_{CC}] &\sim N(0, \lambda_{\beta_k}^2 \lambda_{\beta_{k'}}^2 \lambda_{CC}^2), \quad   \lambda_{CC} \sim C^+(0,1).
\end{align*}
Effectively, this prior collapses \eqref{prior-cat-int} to remove the (partially) local scale layer. 

% also tried a version with $\eta_{\alpha_j} \sim \mbox{Gamma}(a_0, \eta_\tau_0)$ and similar for $\eta_{\beta_k}$ with $a_0 = 2$ to get $t$-distributed marginals for the mains.
\fi %%%%%%%%%%%%%%%%%%%%%%%%%%

\section{Additional theory}\label{app-theory}

\subsection{Posterior invariance under Gaussian priors}
In Theorem~\ref{thm-flat}, we established posterior invariance under a flat prior for $\bm \alpha$. Here, we provide an analogous result under a Gaussian prior for $\bm \alpha$. The central terms for the marginal posterior are now
$\bm{\hat \alpha}_{G} = \bm{\hat \Sigma}_{\alpha_{G}} (\bm E_{10}'\bm e_0 + \bm \Sigma_{\alpha_0}^{-1} \bm \alpha_0)$ and $\bm{\hat \Sigma}_{\alpha_{G}} = (\bm E_{10}'\bm E_{10} + \bm\Sigma_{\alpha_0}^{-1})^{-1}$. Then we have the following:     
\begin{theorem}\label{thm-normal}
    Under the joint  prior \eqref{gen-prior} with   $[\bm \alpha \mid \tau] \sim N(\bm \alpha_0, \tau^{-1}\bm \Sigma_{\alpha_0})$, we have the following:
    \begin{enumerate}
        \item Under model $\mathcal{M}_{01}$, the marginal posterior is
        \[
        [\bm \alpha \mid \bm y] \sim t_{n-p_0}(\bm{\hat \alpha}_{G}, \hat S_{\mathcal{M}_{01}(G)}^2 \bm{\hat \Sigma}_{\alpha_G} ) 
        \]
        %with density  $p_{\mathcal{M}_{01}}(\bm \alpha \mid \bm y) \propto \big\{ (n -  p_0)\hat S_{\mathcal{N},\mathcal{M}_{01}}^2 + (\bm \alpha - \bm{\hat \alpha}_{\mathcal{N}})' \bm{\hat \Sigma}_{\alpha_{\mathcal{N}}}^{-1}(\bm \alpha - \bm{\hat \alpha}_{\mathcal{N}})\big\}^{-(n - p_0 + p_1)/2}$;
        where  $ \hat S_{\mathcal{M}_{01}(G)}^2 
        = \hat S_{0}^2 + (n-p_0)^{-1}(\Vert \bm\alpha_0 \Vert_{\bm\Sigma_{\alpha_0}^{-1}}^2  - \Vert \bm{\hat \alpha}_{G} \Vert_{\bm{\hat \Sigma}_{\alpha_{G}}^{-1}}^2)$ and $\hat S_{0}^2 = (n-p_0)^{-1} \Vert \bm e_0 \Vert^2$ is the estimated OLS residual variance using only $\bm X_0$. 

        \item  Under model $\mathcal{M}_{012}$, the flat (constrained) prior $p(\bm \gamma \mid \tau) \propto \mathbb{I}\{\bm A \bm \gamma = \bm 0\}$, and condition \eqref{cond}, the marginal posterior is
        \[
            [\bm \alpha \mid \bm y] \sim t_{n-p_0-p_{\mathcal{N}}}(\bm{\hat \alpha}_{G}, \hat S_{\mathcal{M}_{012}(G)}^2 \bm{\hat \Sigma}_{\alpha_{G}} )
        \]
        %with density $p_{\mathcal{M}_{012}}(\bm \alpha \mid \bm y) \propto \big\{ (n -  p_0 - p_{\mathcal{N}})\hat S_{\mathcal{N},\mathcal{M}_{012}}^2 + (\bm \alpha - \bm{\hat \alpha}_{\mathcal{N}})' \bm{\hat \Sigma}_{\alpha_{\mathcal{N}}}^{-1}(\bm \alpha - \bm{\hat \alpha}_{\mathcal{N}}) \big\}^{-(n - p_0 - p_{\mathcal{N}} + p_1)/2}$; 
        where  $\hat S_{\mathcal{M}_{012}(G)}^2 = \hat S_{\mathcal{M}_{02}}^2 + (n-p_0 - p_{\mathcal{N}})^{-1}( \Vert \bm\alpha_0 \Vert_{\bm\Sigma_{\alpha_0}^{-1}}^2 - \Vert \bm{\hat \alpha}_{G} \Vert_{\bm{\hat \Sigma}_{\alpha_{G}}^{-1}}^2)
        $ and $\hat S_{\mathcal{M}_{02}}^2$ is the estimated OLS residual variance using only $\bm X_0$ and $\bm X_{2}$ (subject to $\bm A \bm \gamma=\bm 0$).

        \item Under model $\mathcal{M}_{012}$, the Gaussian (constrained) prior $[\bm \gamma \mid \tau] \sim N(\bm \gamma_0, \tau^{-1}\bm \Sigma_{\gamma_0})\mathbb{I}\{\bm A \bm \gamma = \bm 0\}$, and condition \eqref{cond}, the marginal posterior is 
        \[
        [\bm \alpha \mid \bm y] \sim t_{n-p_0}(\bm{\hat \alpha}_{G}, \hat S_{\mathcal{M}_{012}(GG)}^2 \bm{\hat \Sigma}_{\alpha_{G}} )
        \]
        %with density $p_{\mathcal{M}_{012}}(\bm \alpha \mid \bm y) \propto \big\{ (n -  p_0 )\hat S_{\mathcal{N},\mathcal{N},\mathcal{M}_{012}}^2 + (\bm \alpha - \bm{\hat \alpha}_{\mathcal{N}})' \bm{\hat \Sigma}_{\alpha_{\mathcal{N}}}^{-1}(\bm \alpha - \bm{\hat \alpha}_{\mathcal{N}}) \big\}^{-(n - p_0 + p_1)/2}$, 
        where $\hat S_{\mathcal{M}_{012}(GG)}^2 = \hat S_{\mathcal{M}_{01}(G)}^2 + (n-p_0)^{-1}   ( \Vert \bm \gamma_{0}\Vert_{\bm{\Sigma}_{\gamma_0}^{-1}}^2  - \Vert \bm{\hat\gamma} \Vert_{\bm{\hat\Sigma}_{\gamma}^{-}}^2)$. 
    \end{enumerate}
    \end{theorem}
    Compared to the flat prior for $\bm \alpha$, the Gaussian prior leads to marginal posteriors $[\bm \alpha \mid \bm y]$ with different moments, different scalings, and degrees of freedom that no longer subtract off $p_1$. However, comparing across versions in Theorem~\ref{thm-normal}---thus investigating posterior invariance---the differences between them are effectively the same as in Theorem~\ref{thm-flat}. Specifically, the posterior means and unscaled covariances are identical to one another, while the degrees of freedom and scalings differ in the same way as Theorem~\ref{thm-flat}.

%Note: $\hat S_{\mathcal{M}_{02}}^2$ is the estimated OLS residual variance using only $\bm X_0$ and $\bm X_{2}$, where the coefficients on $\bm X_2$ satisfy the same linear constraints as for $\mathcal{M}_{012}$. 

\subsection{BayesABCs with missing levels and interaction cells}
We show that BayesABCs maintains reasonable prior and posterior behavior when either 1) one (or more) levels of a categorical variable is not observed or 2) one (or more) cells of a CC interaction is not observed. %The main result is that BayesABCs automatically maintains reasonable prior and posterior behavior. This further supports our use of the \emph{sample} proportions $\bm{\hat\pi}$ in the ABCs. 

First, consider a one-way ANOVA (Example~\ref{ex-anova}) with $\hat \pi_1 = 0$ so that group $c=1$ is unobserved; the same arguments apply for $K$-way ANOVAs, ANCOVAs, and multiple missing levels. The ABCs \eqref{abcs-gen-cat} become $\sum_{c=2}^L \hat\pi_c\beta_c = 0$. If $\beta_1$ is \emph{a priori} independent of $\bm \beta_{-1} = (\beta_2,\ldots,\beta_L)'$, as in our priors, then the constrained prior factorizes: $p(\bm \beta \mid \bm{\hat \pi}'\bm\beta = 0)= p(\beta_1) \ p(\bm\beta_{-1} \mid  \bm{\hat \pi}_{-1}'\bm\beta_{-1} = 0)$ with $\bm{\hat\pi}_{-1} = (\hat\pi_2,\ldots,\hat\pi_L)'$. Because $\beta_1$ does not appear in the likelihood, it is clear that $p(\beta_1 \mid \bm y) = p(\beta_1)$, as anticipated (the same holds conditional on ABCs). For the remaining coefficients $\bm\beta_{-1}$, it is clear that the constraint,  the prior, and the posterior all act as though the missing level $c=1$ had been deleted from the categorical variable. Even so, the proposed algorithms (Appendix~\ref{app-computing}) remain valid without modification. %, although it will be more efficient to remove $\beta_1$ and use $L-1$ levels. 
This missingness does not preclude posterior inference for $\mu_1 = \alpha_0 + \beta_1$, where the data and priors from groups $c>1$ inform $\alpha_0$ and the prior  informs $\beta_1$. 

Next, we consider CC interactions and the implications of missing cells on the ABCs \eqref{abcs-cat-cat-gen}. The subsequent conclusions about (conditional) priors, likelihoods, and posteriors follow as in the one-way ANOVA case. The simplest version includes two categorical variables with two groups each, $L_1 =L_2 = 2$, so that any missing level (say $\hat\pi_{11} = 0$) renders the CC interaction inestimable (to see this, notice that there are $2\times 2 = 4$ groups, only $3$ of which are observed, but the model has $3$ identified parameters from the main-only model already). In this degenerate case, $\hat\pi_{11} = 0$ and ABCs \eqref{abcs-cat-cat-gen} together imply that $\gamma_{12} = \gamma_{21} = \gamma_{22} = 0$. Thus, the only free CC interaction parameter is $\gamma_{11}$, which satisfies $p(\gamma_{11} \mid \bm y) = p(\gamma_{11})$ since it does not 
appear in the likelihood for the observed data. Again, the proposed algorithms (Appendix~\ref{app-computing}) remain valid without modification.%: the draws of $\gamma_{11}$ will be from the prior and the remaining draws of $\bm \gamma_{-(11)}$ will be zero. 

Now consider $L_1 = 3$ and $L_2 =2$, so that missing interaction cells do not automatically render $\bm \gamma$ inestimable. Suppose  $\hat \pi_{11}=0$ but all other cells are observed.  The CC interaction $\bm \gamma$ reduces from two to one identified parameters. Since 
ABCs \eqref{abcs-cat-cat-gen} include $\hat\pi_{11}\gamma_{11} + \hat\pi_{12}\gamma_{12} = 0$, it follows that $\gamma_{12} = 0$. The remaining ABCs simplify to $\hat \pi_{c_1 1}\gamma_{c_1 1} + \hat \pi_{c_1 2}\gamma_{c_12} = 0$ for $c_1 = 2,3$ and $\hat \pi_{2 c_2} \gamma_{2 c_2} + \hat \pi_{3 c_2} \gamma_{3 c_2} = 0$ for $c_2=1,2$. These effectively collapse  to  ABCs from the $L_1 = L_2 =2$ case. Specifically, they do not involve $c_1=1$:  $\gamma_{11}$ is free but will not appear in the likelihood and  $\gamma_{12} = 0$. The $L_1=L_2=2$ case has one identified CC interaction coefficient, and thus the rank is correct here.

If there are multiple missing cells, the consequences depend on the arrangement. First, suppose both $\hat \pi_{11} = 0$ and $\hat \pi_{12} =0$, so the $c_1=1$ case is never observed. This is only a minor modification of the previous case: $\gamma_{12}$ is no longer set to zero, but rather is free and does not appear in the likelihood, like $\gamma_{11}$. The remaining CC interaction coefficients are unmodified, while the main effect for $c_1 = 1$ is impacted as in the one-way ANOVA example above. Alternatively, if 
both $\hat \pi_{11} = 0$ and $\hat \pi_{22} =0$, then the CC interaction is not estimable. The pattern is similar to the $L_1=L_2=2$ case: $\gamma_{11}$ and $\gamma_{22}$ are free but do not appear in the likelihood, while all other $\gamma_{rs} = 0$. 

In summary, the BayesABCs response to missing levels or CC interaction cells depends on the dimensions $L_k$ and the missingness pattern. Regardless of the identifying constraints, the accompanying parameters do not appear in the likelihood. Because ABCs involve $\bm{\hat\pi}$, the unobserved levels and cells also will not appear in the constraints. What is appealing is that 1) the ensuing constraints remain reasonable and 2) the general computing approach  for BayesABCs does not require any modifications for missing levels or CC interaction cells.

\section{Additional simulation results}\label{app-sims}

\subsection{Simulations for two-way ANOVA}\label{app-sims-twoway}
We evaluate estimation and inference for a two-way ANOVA model. Our DGP ensures that 1) neither ABCs nor RGE is favored, 2) the categorical variables are dependent, and 3) we may have null or active CC interactions. The categorical variables, nominally \texttt{race} and \texttt{sex}, are generated  as in Section~\ref{sec-sims-lm}. The outcome variable is generated from a sparse Gaussian linear model with intercept $\alpha_0 =1$, \texttt{race} main effects $\beta_{\texttt{a}} = 1$ and $\beta_{\texttt{c}} = -1$,  interactions $\gamma_{\texttt{b}:\texttt{vv}} = \gamma$ and $\gamma_{\texttt{d}:\texttt{uu}} = -\gamma$, and all other coefficients zero. Among the active coefficients, neither ABCs nor RGE is satisfied (Over is satisfied trivially). We vary the sample size $n \in \{200, 1000\}$ and the interaction effect $\gamma \in \{0, 0.5\}$, fix the SNR = $0.25$,  and generate 500 synthetic datasets in each case. 

We use the same competing methods as in Section~\ref{sec-sims-lm}, again with main-only and cat-modified versions. 
Here, the main-only model is \texttt{y $\sim$ 1 + race + sex} and the cat-modified model is \texttt{y $\sim$ race + sex + race:sex}. Additionally, we include \texttt{ABCs(grp-her)} and \texttt{ABCs(loc-her)} which use grouped and local shrinkage priors, respectively, that incorporate CC heredity shrinkage for interactions (see Section~\ref{subsec-priors}). The CC heredity priors only apply in the presence of interactions, so we do not fit a main-only version of these models. Our metrics are RMSEs and mean credible/confidence interval widths and empirical coverage for $\mathbb{E}(Y \mid \texttt{race} = r, \texttt{sex} = s) = \mu(r,s)$ across all \texttt{race} and \texttt{sex} levels. This target is not anchored to any model parameterization. 

%  mu_true = 1 +  I(race=='A') - I(race == 'C') + gamma*I(race == 'B' &  sex == 'vv') -  gamma*I(race == 'D' &  sex == 'uu')

\begin{figure}[h]
\centering
\includegraphics[width=.49\textwidth]{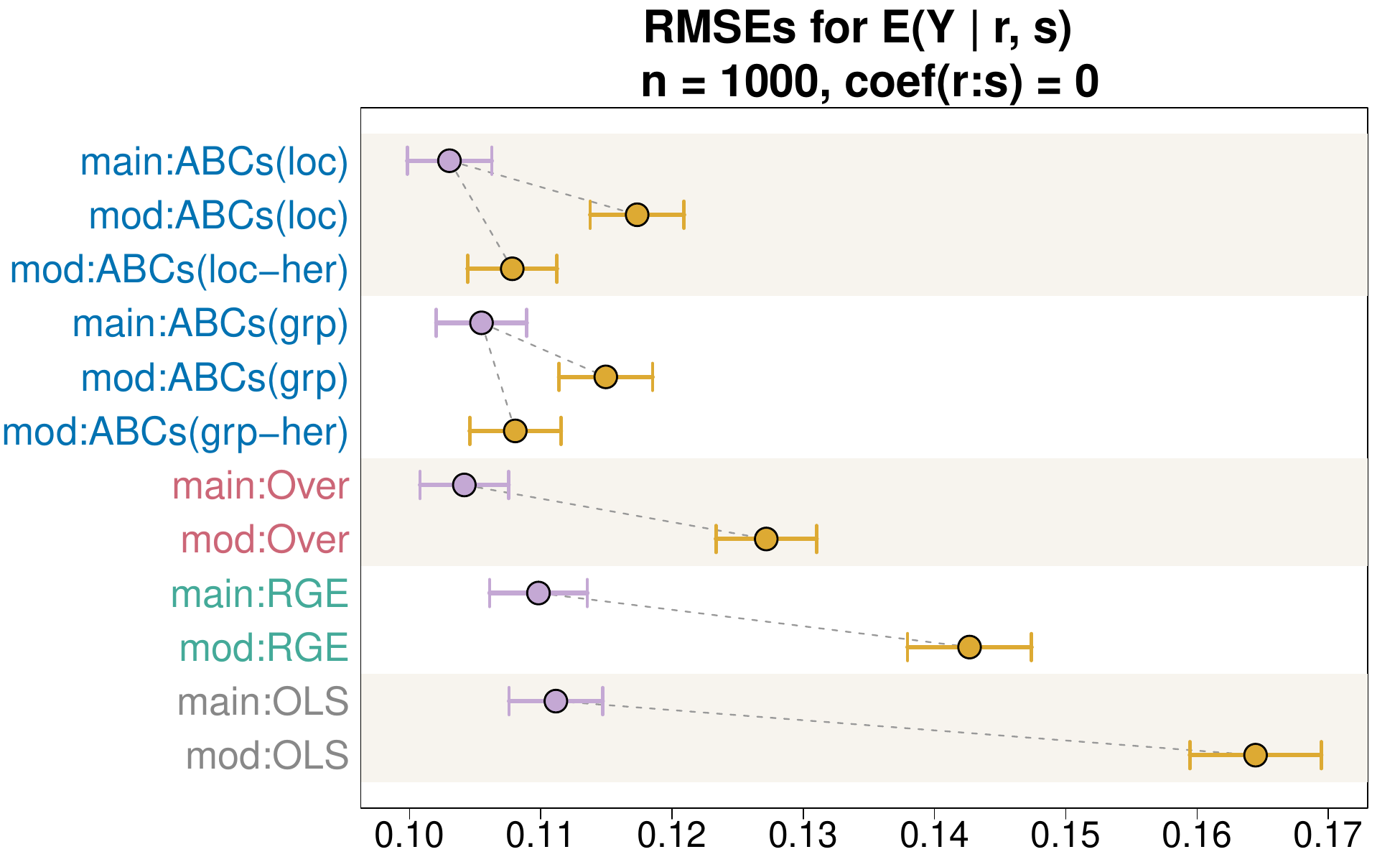}
\includegraphics[width=.49\textwidth]{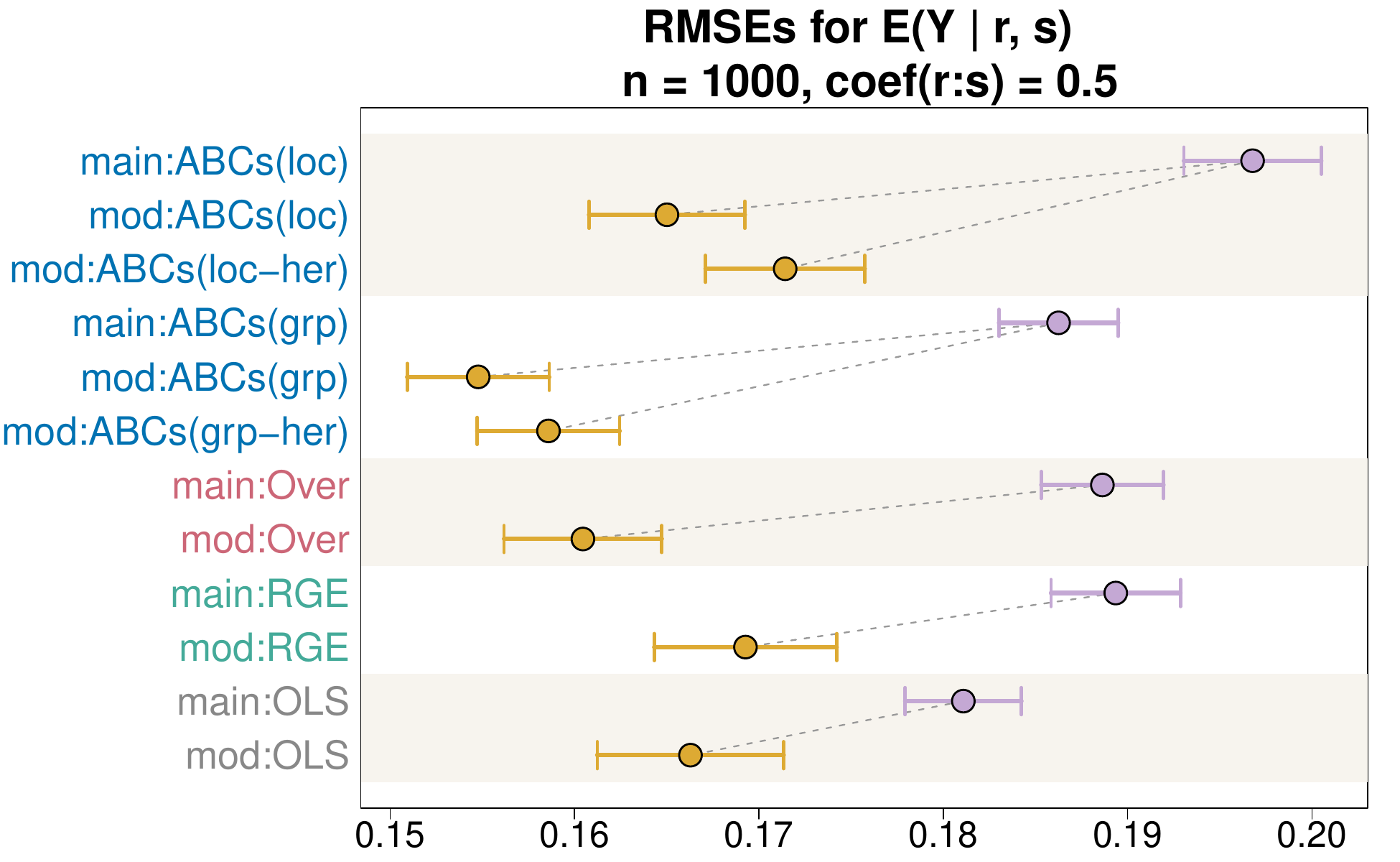}
\includegraphics[width=.49\textwidth]{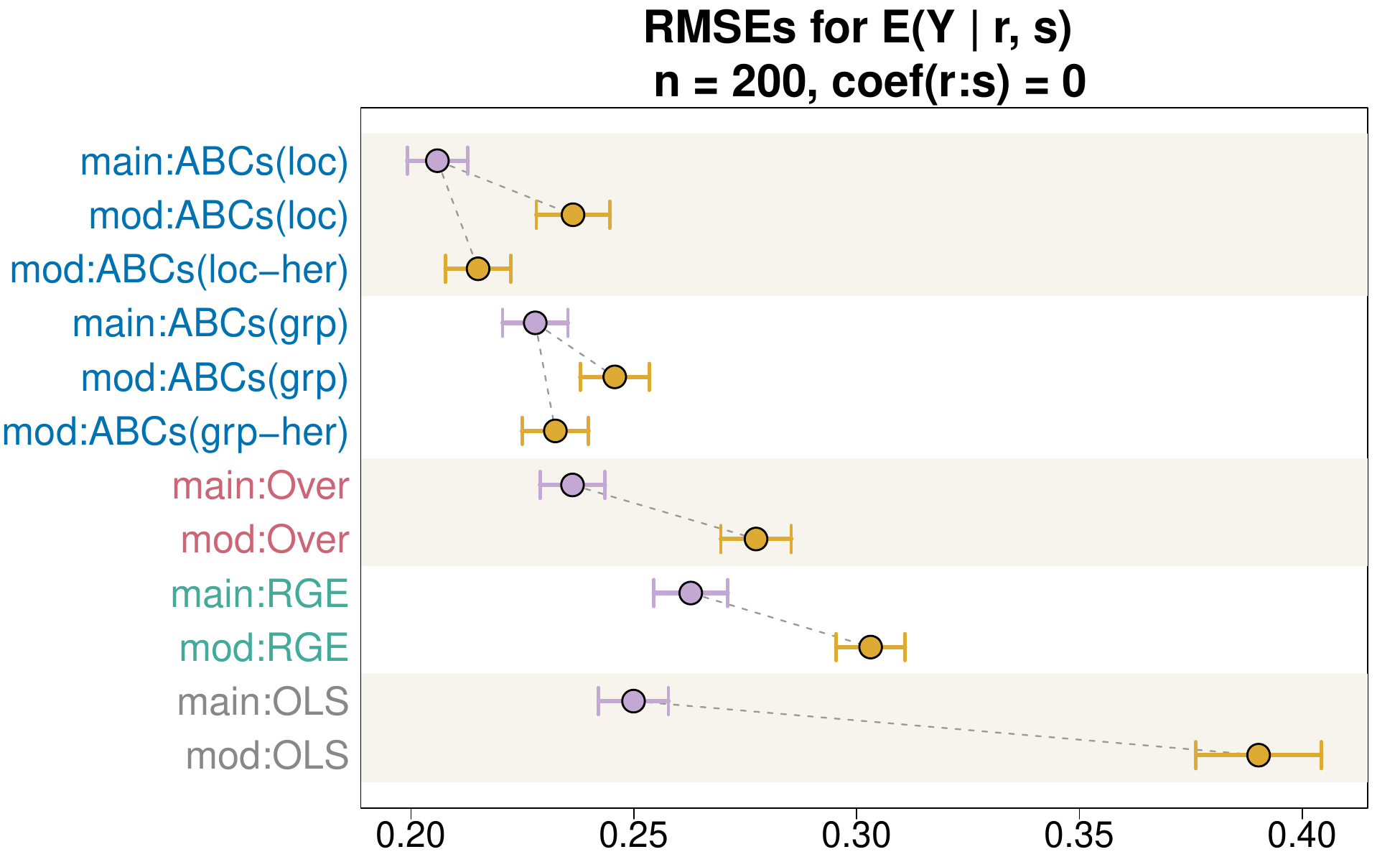}
\includegraphics[width=.49\textwidth]{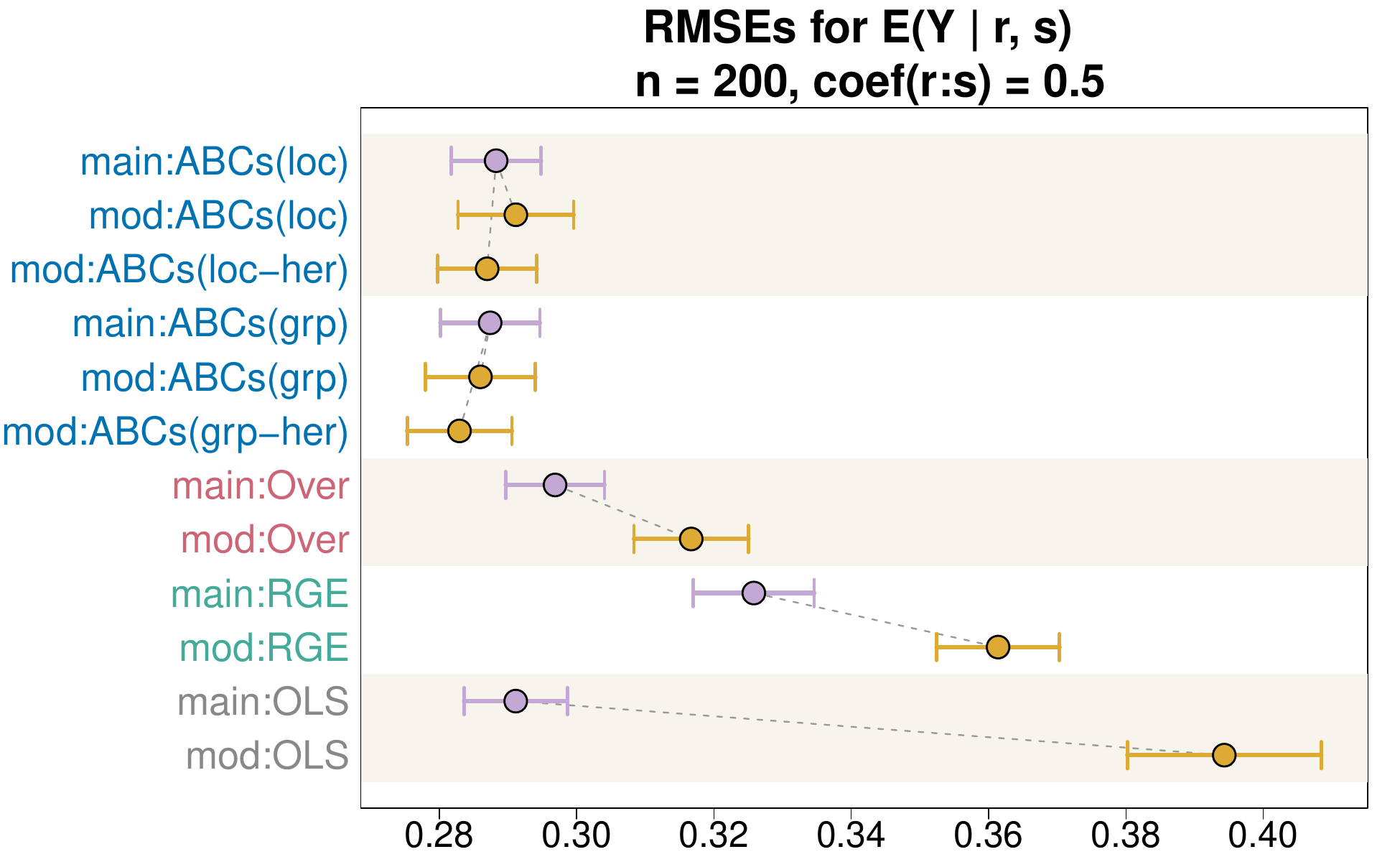}
\caption{\small RMSEs (mean $\pm 2$ SEs) for a two-way ANOVA expectations $\mu(r,s)$ with $n=1000$ (top) and $n=200$ (bottom). Dashed lines connect the main-only and cat-modified versions of each method. When the interaction is null (left), BayesABCs with CC heredity shrinkage keep the cat-modified version most competitive with the main-only version. When the interaction is active (right), cat-modified BayesABCs models perform best, with preference for the non-heredity prior when $n$ is large. 
\label{fig:sims-rmse-twoway}}
\end{figure}

Point estimation is evaluated in Figure~\ref{fig:sims-rmse-twoway}. When the interaction is null (Figure~\ref{fig:sims-rmse-twoway}, left), main-only models are naturally more accurate than their cat-modified counterparts. Among main-only models, \texttt{ABCs(loc)} is most accurate, followed by \texttt{Over} and \texttt{ABCs(grp)}. Because this case excludes interactions, it represents an extension of the one-way ANOVA model simulations (Section~\ref{sec-sims-anova}) to include two categorical variables. Here, we also compare the main-only and cat-modified versions of each model. BayesABCs produce the most competitive cat-modified models, especially with CC heredity priors. In fact, both cat-modified BayesABCs models with CC heredity match the performance of the \emph{main-only} \texttt{RGE} model. By comparison, \texttt{Over}, \texttt{RGE}, and \texttt{OLS} significantly deteriorate when null interactions are included. 

Alternatively, when the interaction is (partly) active (Figure~\ref{fig:sims-rmse-twoway}, right), BayesABCs cat-modified models show substantial improvements. This is made clear by comparing the results for different sample sizes. When $n=200$, the cat-modified versions of \texttt{Over}, \texttt{RGE}, and \texttt{OLS} do not actually improve upon their main-only counterparts, whereas BayesABCs methods offer slight gains. When $n=1000$, all methods favor their cat-modified versions, with \texttt{ABCs(grp)}---omitting CC heredity---providing the highest accuracy.

\begin{figure}[h]
\centering
\includegraphics[width=.49\textwidth]{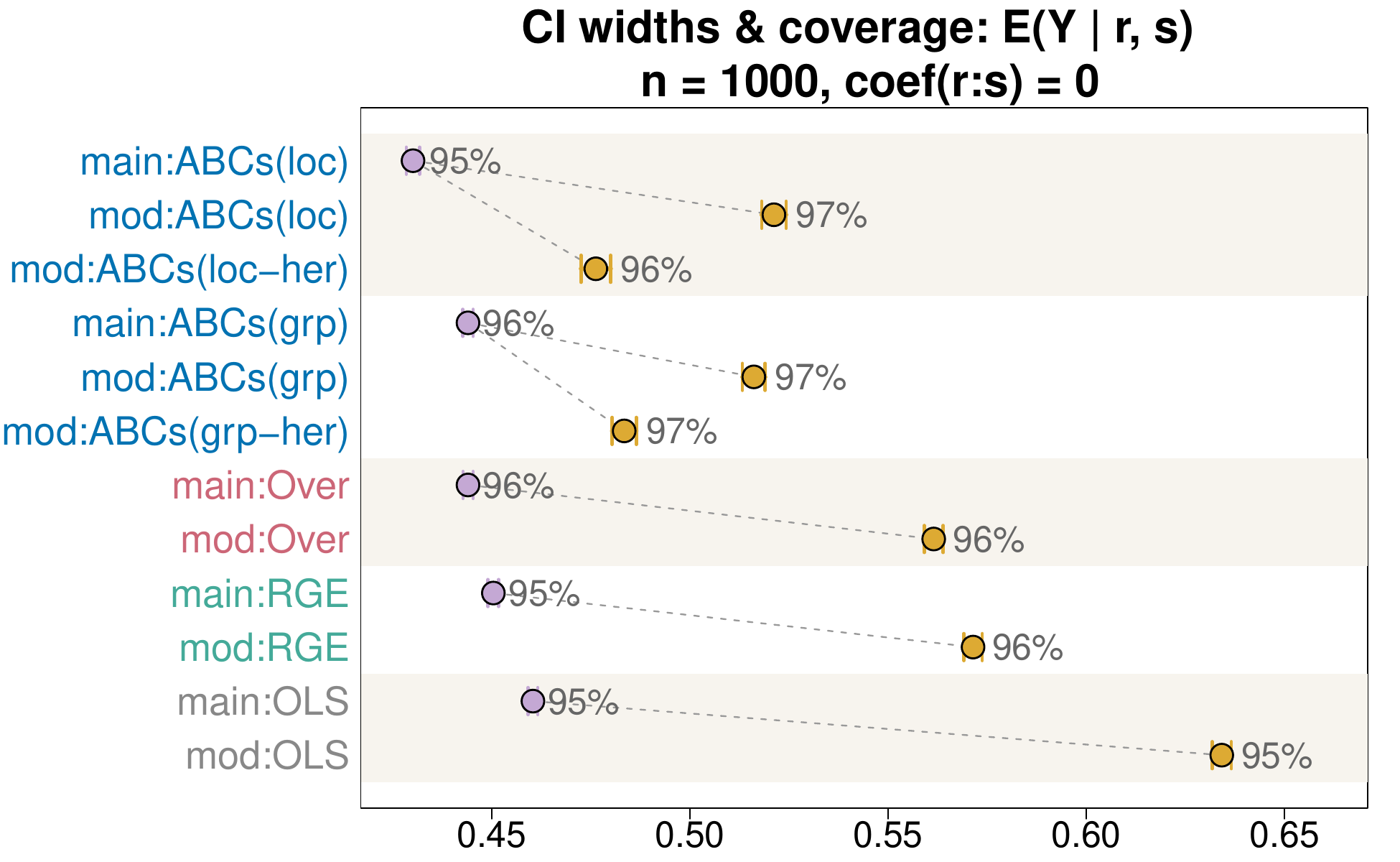}
\includegraphics[width=.49\textwidth]{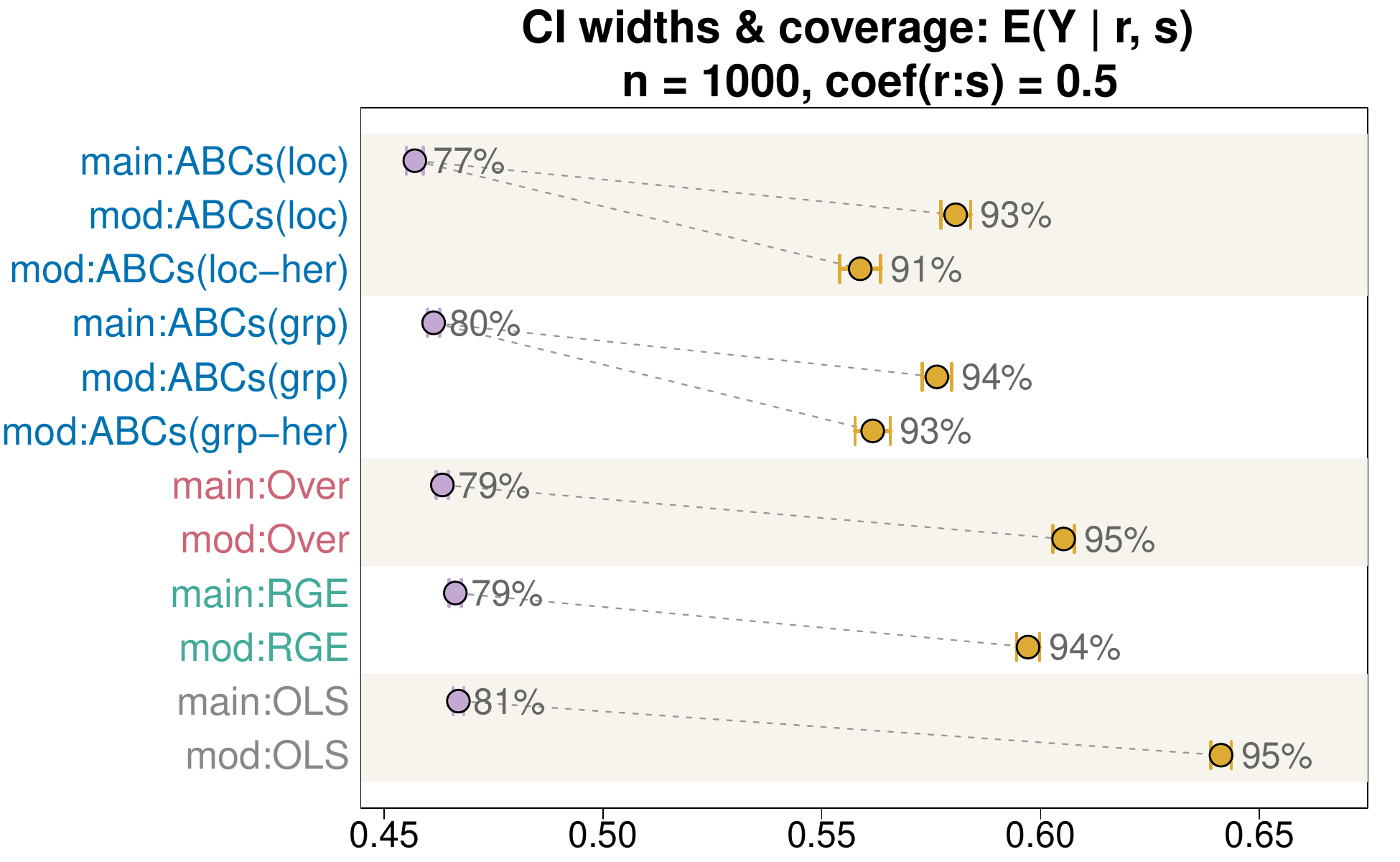}
\includegraphics[width=.49\textwidth]{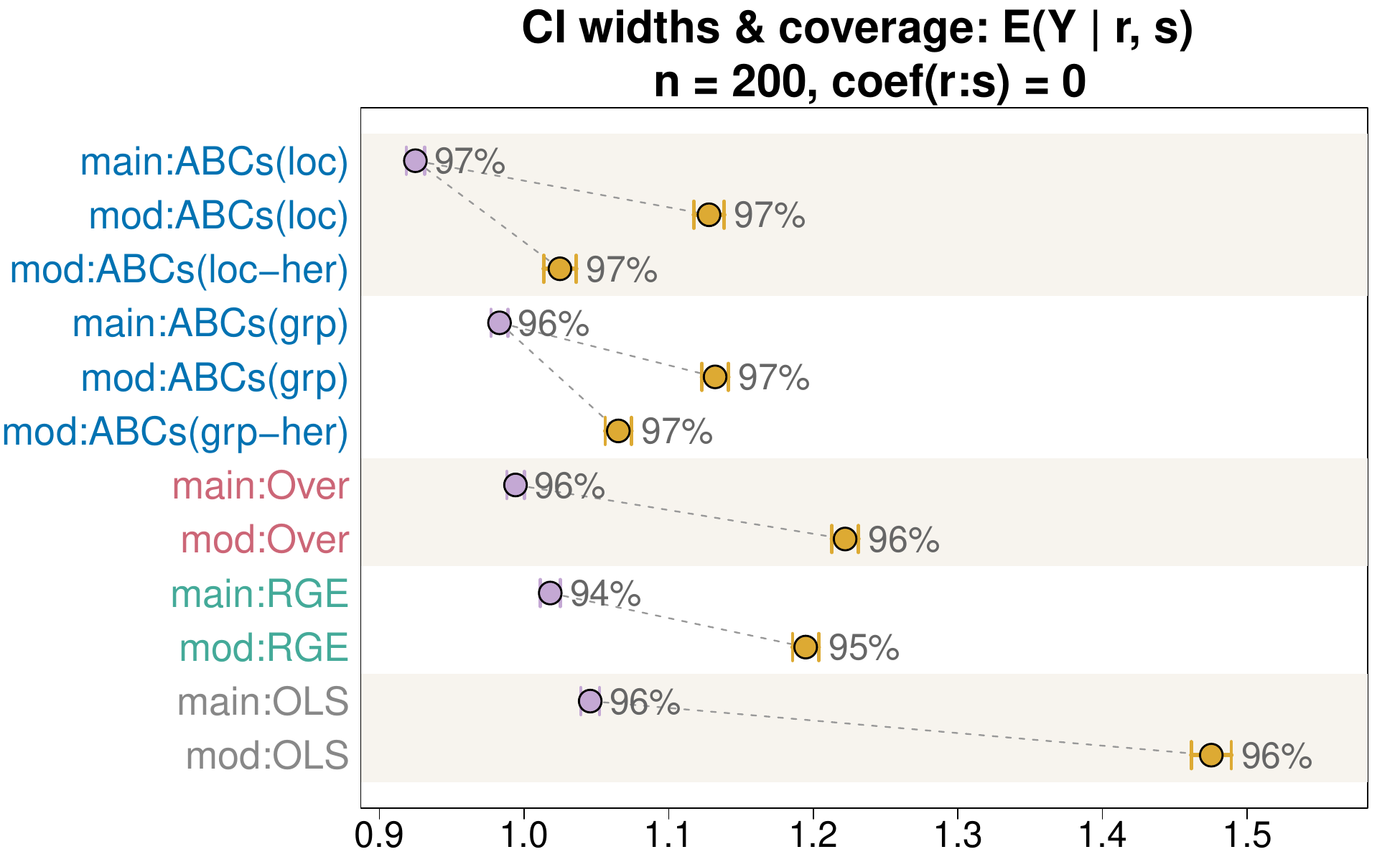}
\includegraphics[width=.49\textwidth]{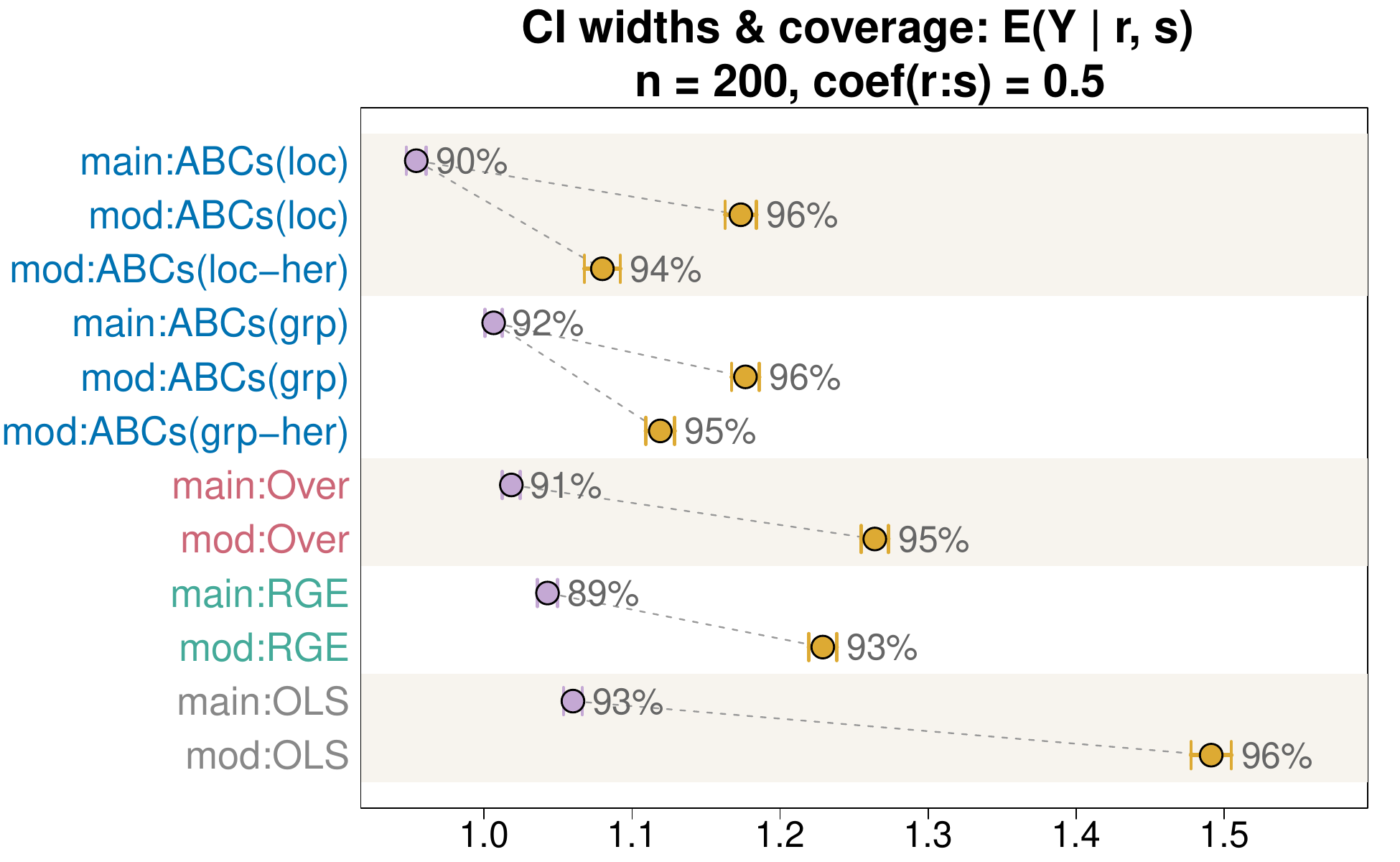}
\caption{\small Mean interval widths (mean $\pm 2$ SEs) and empirical coverage (annotations) for a two-way ANOVA expectations $\mu(r,s)$ with $n=1000$ (top) and $n=200$ (bottom). Dashed lines connect the main-only and cat-modified versions of each method. When the interaction is null (left), main-only models provide the most precise and well-calibrated inference; among cat-modified models, BayesABCs with CC heredity shrinkage is most competitive. When the interaction is active (right), the BayesABCs cat-modified models  provide the narrowest and (nearly) calibrated intervals. 
\label{fig:sims-ci-twoway}}
\end{figure}

Inference is evaluated in Figure~\ref{fig:sims-ci-twoway}. The behavior of main-only models is predictable: their intervals are narrow and well-calibrated when interactions are null (Figure~\ref{fig:sims-ci-twoway}, left) but miscalibrated when interactions are (partly) active (Figure~\ref{fig:sims-ci-twoway}, right). In the former case, BayesABCs provides the most precise intervals among main-only models, which extends the one-way ANOVA simulation results (Section~\ref{sec-sims-anova}) to multiple categorical main effects. Cat-modified models are less competitive in this setting, but BayesABCs methods perform the best, especially with CC heredity. When interactions are (partly) active (Figure~\ref{fig:sims-ci-twoway}, right), the story is again different across $n\in\{200,1000\}$. When $n=200$, CC heredity is favorable and provides narrow and (nearly) calibrated intervals. Yet when $n=1000$, CC heredity leads to undercoverage, which is partly mitigated by omitting the CC heredity (e.g., \texttt{ABCs(grp)}).

These results demonstrate that BayesABCs has a most favorable parameterization-prior combination. For main-only models with truly null interactions, BayesABCs outperforms Over and RGE with horseshoe priors in both estimation and inference. This confirms and extends the results one-way ANOVA models (Section~\ref{sec-sims-anova}) to multiple categorical covariates. 
For cat-modified models---regardless of whether the interactions are truly active or null---BayesABCs again outperforms these competing parameterization-prior combinations. We note that the difference in performance between BayesABCs main-only and cat-modified models does not contradict Corollary~\ref{cor-anova}, which applies only for the categorical main effects, not $\mu(r,s) = \alpha_0 + \beta_r + \beta_s + \gamma_{r:s}$. 
Finally, the performance of CC heredity shrinkage is mixed: it appears beneficial when $n$ is small or interactions are null. This is precisely the setting where we expect additional shrinkage to help. However, CC heredity shrinkage also tends to decrease MCMC efficiency (in terms of effective sample size of regression coefficients). Thus, we do not include it by default, but recommend it for small sample sizes.

\subsection{Supporting results: one-way ANOVA}\label{app-sims-anova}
Section~\ref{sec-sims-anova} reported results for the one-way ANOVA with $n=200$. In Tables~\ref{tab:rmse_results_n1000} and \ref{tab:mciw_ecp_results_n1000}, we provide estimation and inference results, respectively, for $n=1000$. 

\begin{table}[ht] \small
    \centering
    \begin{tabular}{l cc cc cc}
        % --- Top-level header: # signals ---
        & \multicolumn{2}{c}{$\text{\# signals} = 0$}
        & \multicolumn{2}{c}{$\text{\# signals} = 3$}
        & \multicolumn{2}{c}{$\text{\# signals} = L$} \\
        %\cline{2-3} \cline{4-5} \cline{6-7}
        % --- Second-level header: L values ---
        Method
            & $L=5$ & $L=15$
            & $L=5$ & $L=15$
            & $L=5$ & $L=15$ \\
        \hline
        \texttt{ABCs(loc)}   & {\bf 0.080} & 0.077 & {\bf 0.219} & {\bf 0.166} & 0.213 & 1.207 \\
        \texttt{ABCs(grp)}   & {\bf 0.078} & {\bf 0.069} & {\bf 0.227} & 0.175 & {\bf 0.208} & {\bf 1.110} \\
        \texttt{Over}        & 0.090 & 0.085 & 0.242 & {\bf 0.141} & 0.214 & 1.201 \\
        \texttt{RGE}         & 0.083 & 0.083 & 0.257 & 0.270 & 0.225 & 1.420 \\
        \texttt{SCOPE}       & 0.098 & {\bf 0.067} & 0.264 & 0.188 & 0.220 & 1.176 \\
        \texttt{OLS}         & 0.176 & 0.303 & 0.234 & 0.176 & {\bf 0.206} & {\bf 1.104} \\
        \hline
    \end{tabular}
        \caption{\small RMSEs for $\mu_c = \alpha_0 + \beta_c$ in a one-way ANOVA with $n=1000$. Smallest two RMSEs in each case are bold. BayesABCs produces one or both of the most accurate estimates in every case, with the same top methods as for $n=200$.}
            \label{tab:rmse_results_n1000}
\end{table}

 \begin{table}[ht] \small
    \centering
    \resizebox{\linewidth}{!}{%
    \begin{tabular}{l cc cc cc}
        % --- Top-level header: # signals ---
        & \multicolumn{2}{c}{$\text{\# signals} = 0$}
        & \multicolumn{2}{c}{$\text{\# signals} = 3$}
        & \multicolumn{2}{c}{$\text{\# signals} = L$} \\
        % --- Second-level header: L values ---
        Method
            & $L=5$ & $L=15$
            & $L=5$ & $L=15$
            & $L=5$ & $L=15$ \\
        \hline
        \texttt{ABCs(loc)}   & 0.442 (99\%) & {\bf 0.506} (100\%) & {\bf 0.842} (94\%) & {\bf 0.600} (95\%) & 0.768 (93\%) & 4.187 (93\%) \\
        \texttt{ABCs(grp)}   & {\bf 0.434} (99\%) & {\bf 0.482} (100\%) & {\bf 0.853} (94\%) & 0.650 (96\%) & {\bf 0.759} (95\%) & {\bf 4.056} (94\%) \\
        \texttt{Over}        & 0.476 (99\%) & 0.551 (100\%) & 0.868 (94\%) & {\bf 0.545} (98\%) & 0.767 (94\%) & 4.134 (92\%) \\
        \texttt{RGE}         & {\bf 0.436} (99\%) & 0.527 (100\%) & 0.875 (93\%) & 0.501 (92\%) & 0.767 (92\%) & 3.904 (89\%) \\
        \texttt{SCOPE}       & ---          & ---           & ---          & ---          & ---          & ---          \\
        \texttt{OLS}         & 0.652 (95\%) & 1.147 (96\%)  & 0.868 (95\%) & 0.664 (96\%) & {\bf 0.764} (95\%) & {\bf 4.185} (96\%) \\
        \hline
    \end{tabular}}
    \caption{\small Mean credible/confidence interval widths (and empirical coverage) for $\mu_c = \alpha_0 + \beta_c$ in a one-way ANOVA with $n=1000$. Narrowest two intervals subject to near-nominal coverage ($\pm 1$\%) in each case are bold. BayesABCs produces one or both of the most precise and well-calibrated intervals in every case.}
    \label{tab:mciw_ecp_results_n1000}
\end{table}

\subsection{Supporting results: sparse and cat-modified linear models}\label{app-sims-lm}
\begin{figure}[!h]
\centering
\includegraphics[width=.49\textwidth]{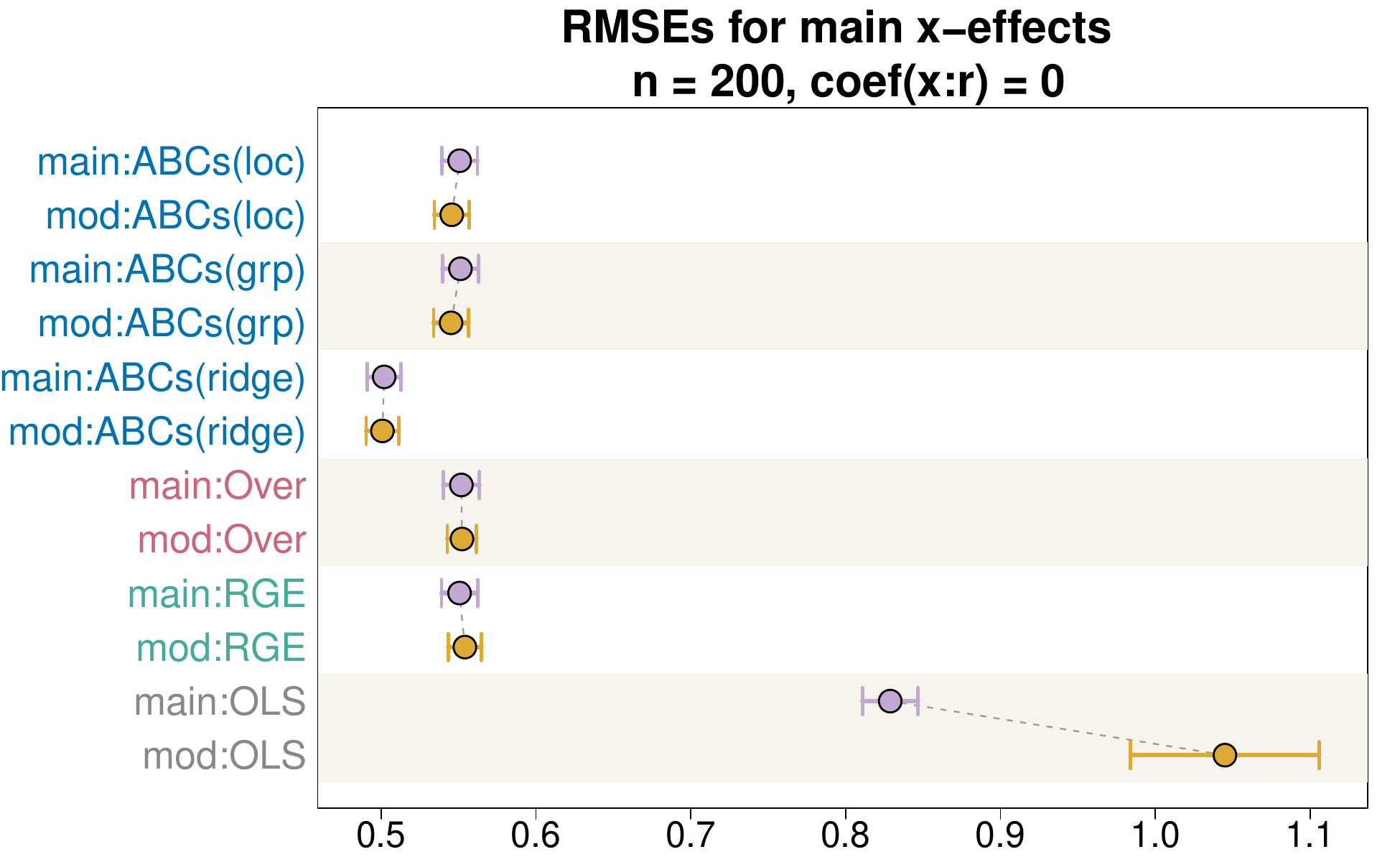}
\includegraphics[width=.49\textwidth]{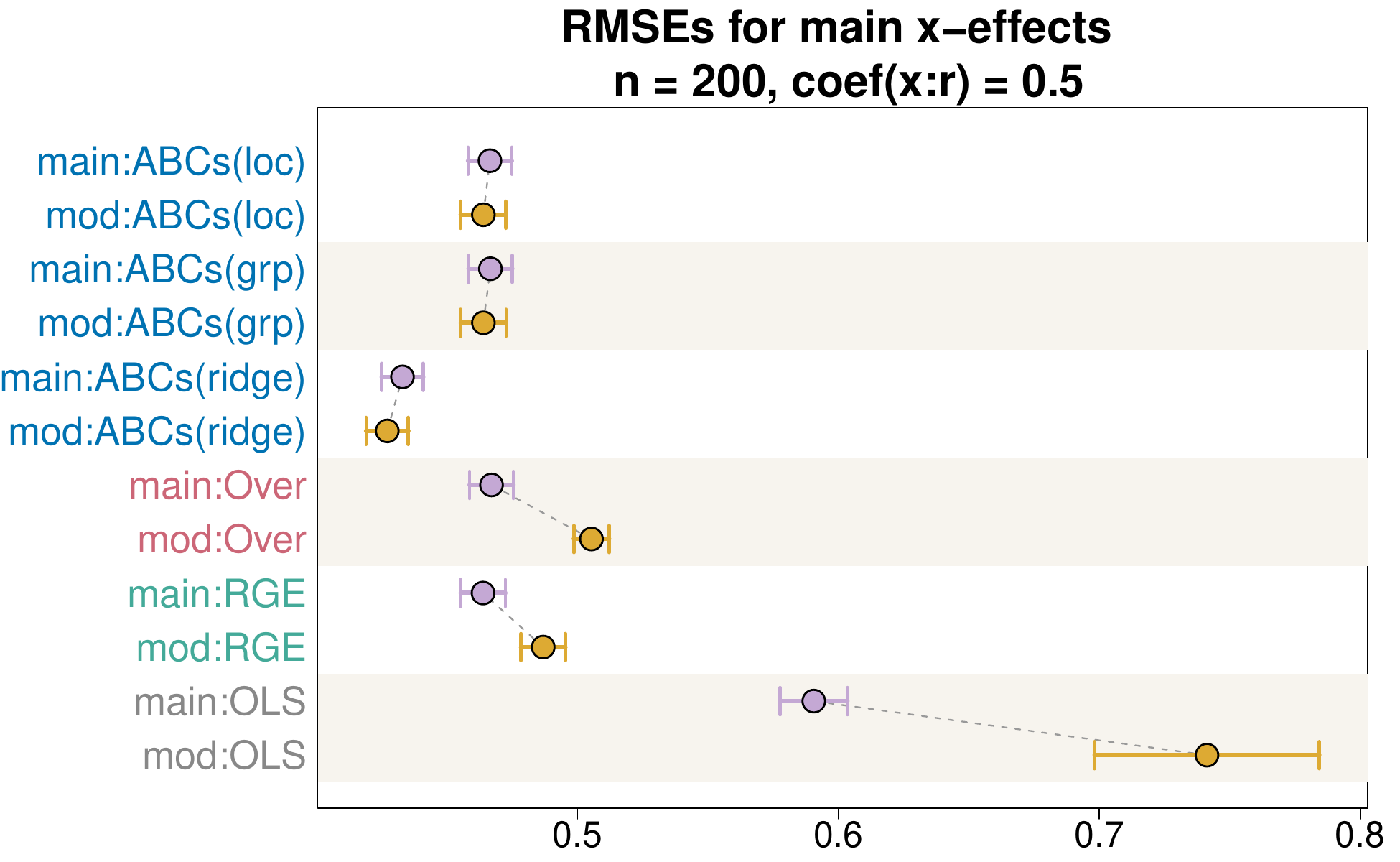}
\caption{\small RMSEs (mean $\pm 2$ SEs) for the continuous main  effects $\bm \alpha$ when categorical interactions are extraneous (left) or when some are necessary (right). Dashed lines connect the main-only and cat-modified versions of each method. Owing to the small sample size ($n=200$), BayesABCs with a simple ridge prior performs best in both cases. 
\label{fig:sims-rmse-x-200}}
\end{figure}

\begin{figure}[!h]
\centering
\centering
\includegraphics[width=.49\textwidth]{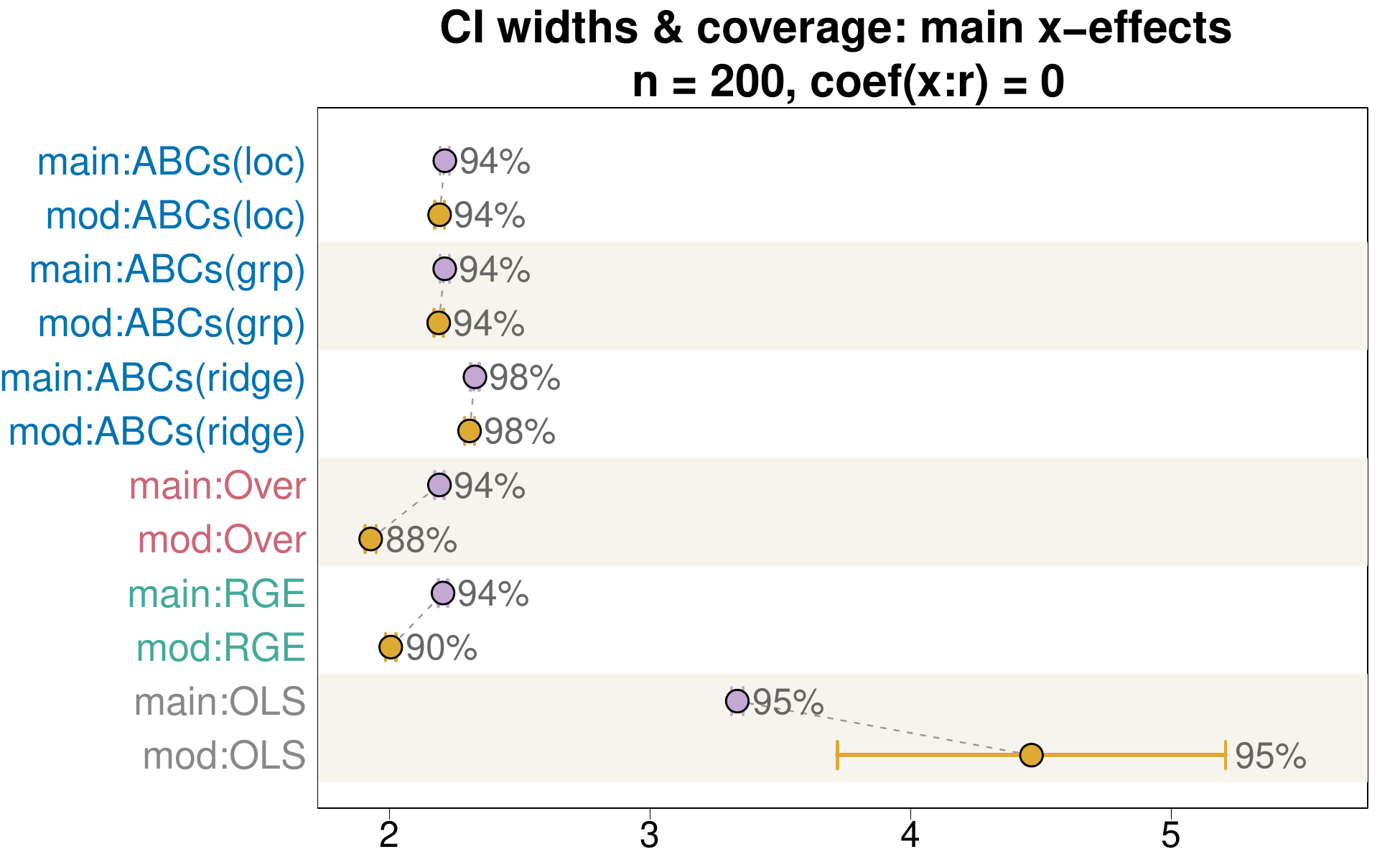}
\includegraphics[width=.49\textwidth]{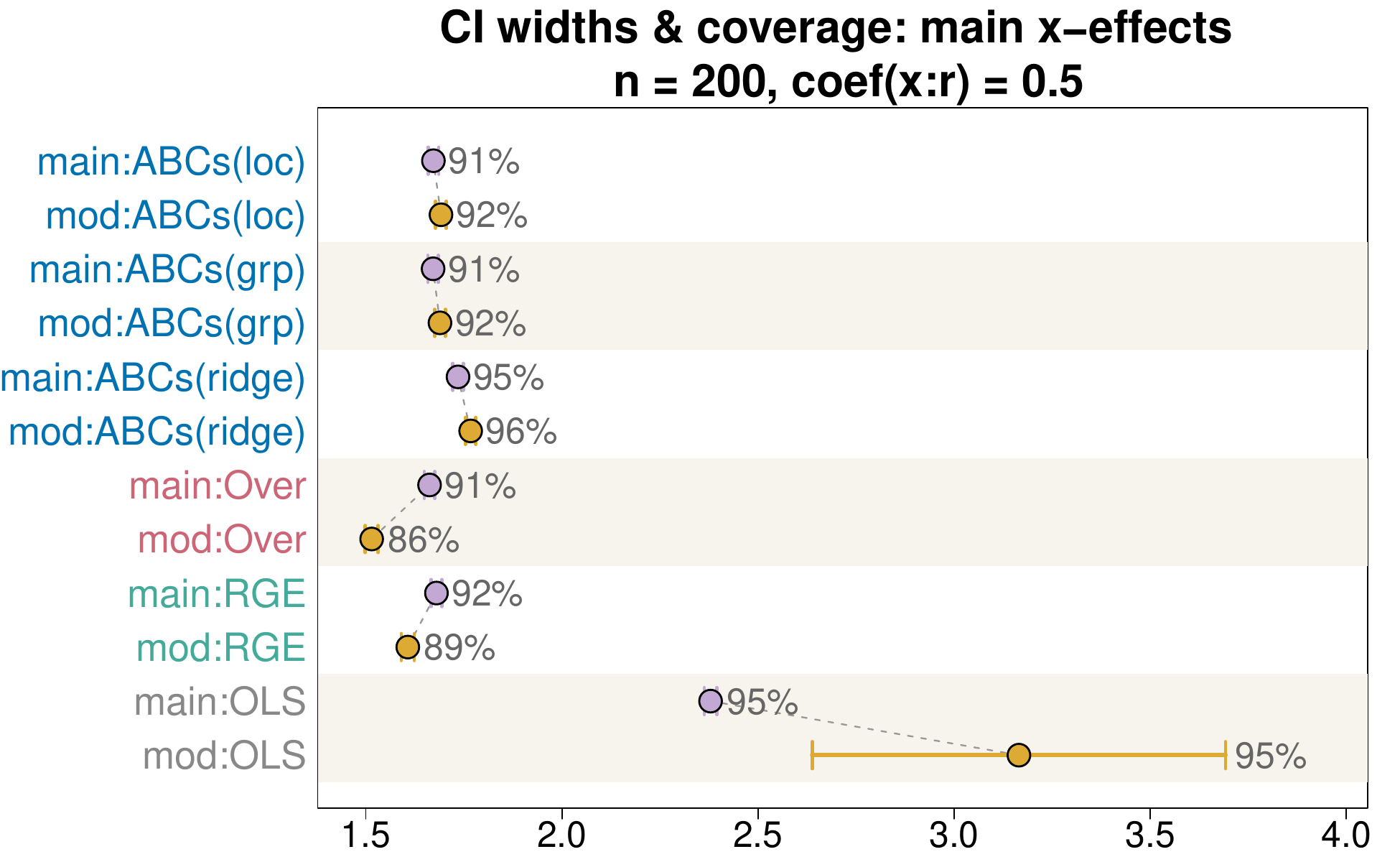}
\caption{\small Mean interval widths (mean $\pm 2$ SEs) and empirical coverage (annotations) for the continuous main  effects $\bm \alpha$ with $n=200$.  Dashed lines connect the main-only and cat-modified versions of each method. BayesABCs with grouped or local shrinkage priors are sharpest when  categorical interactions are extraneous (left), while cat-modified \texttt{RGE} and \texttt{Over} models substantially undercover. When categorical interactions are necessary (right), all aggressive shrinkage priors suffer from undercoverage, so \texttt{ABCs(ridge)} is best. 
\label{fig:sims-ci-200}}
\end{figure}

\begin{figure}[!h]
\centering
\includegraphics[width=.49\textwidth]{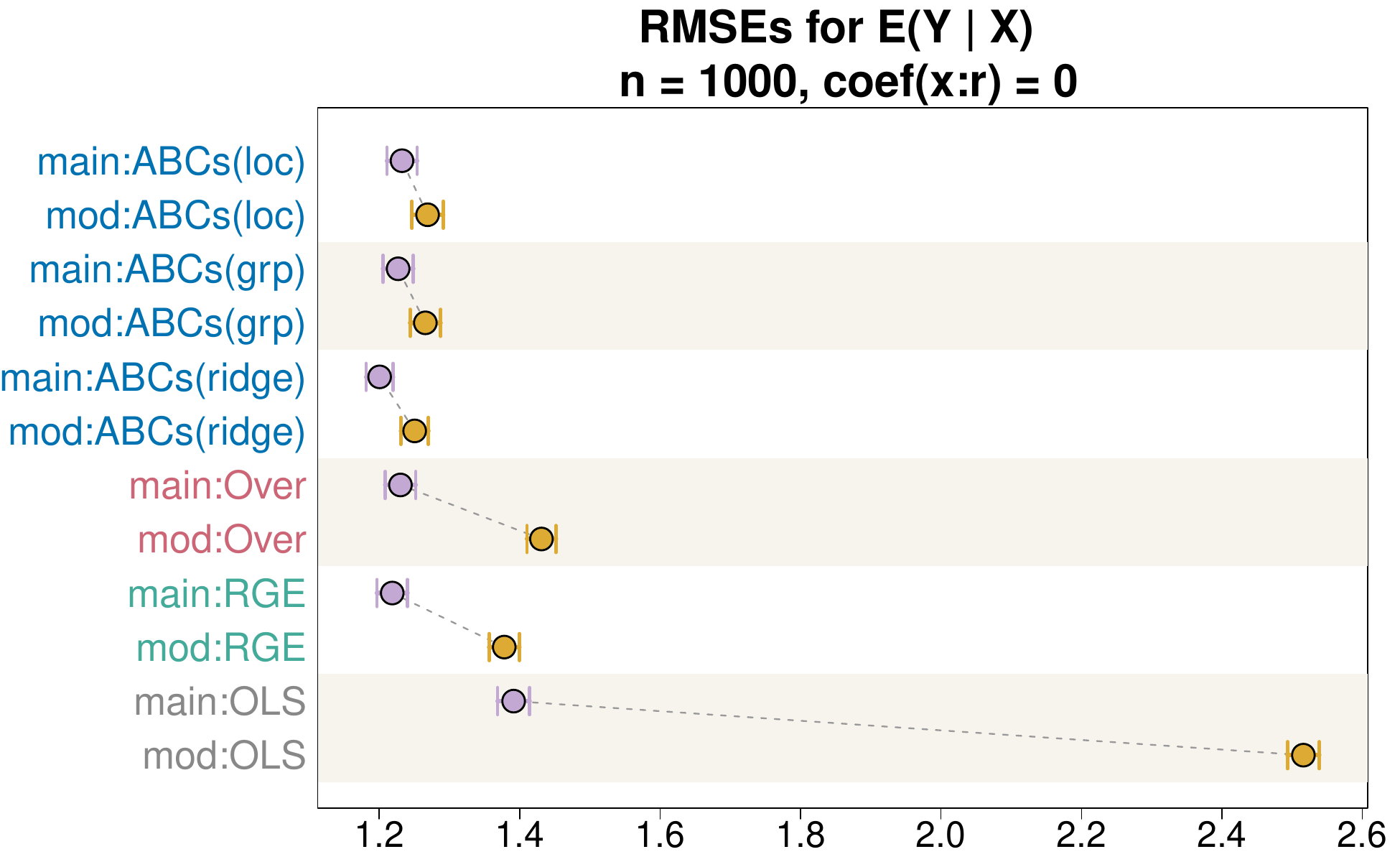}
\includegraphics[width=.49\textwidth]{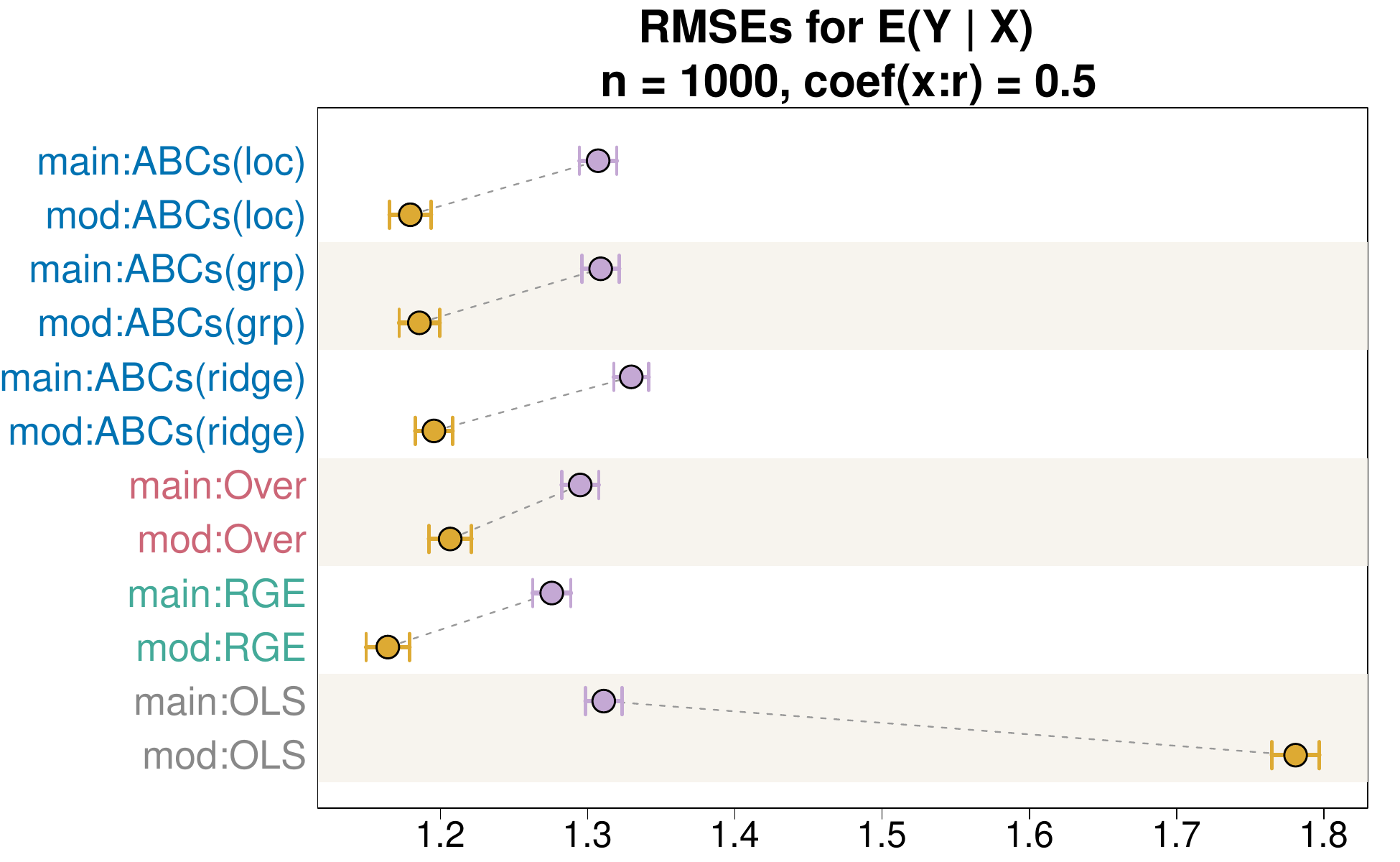}
\includegraphics[width=.49\textwidth]{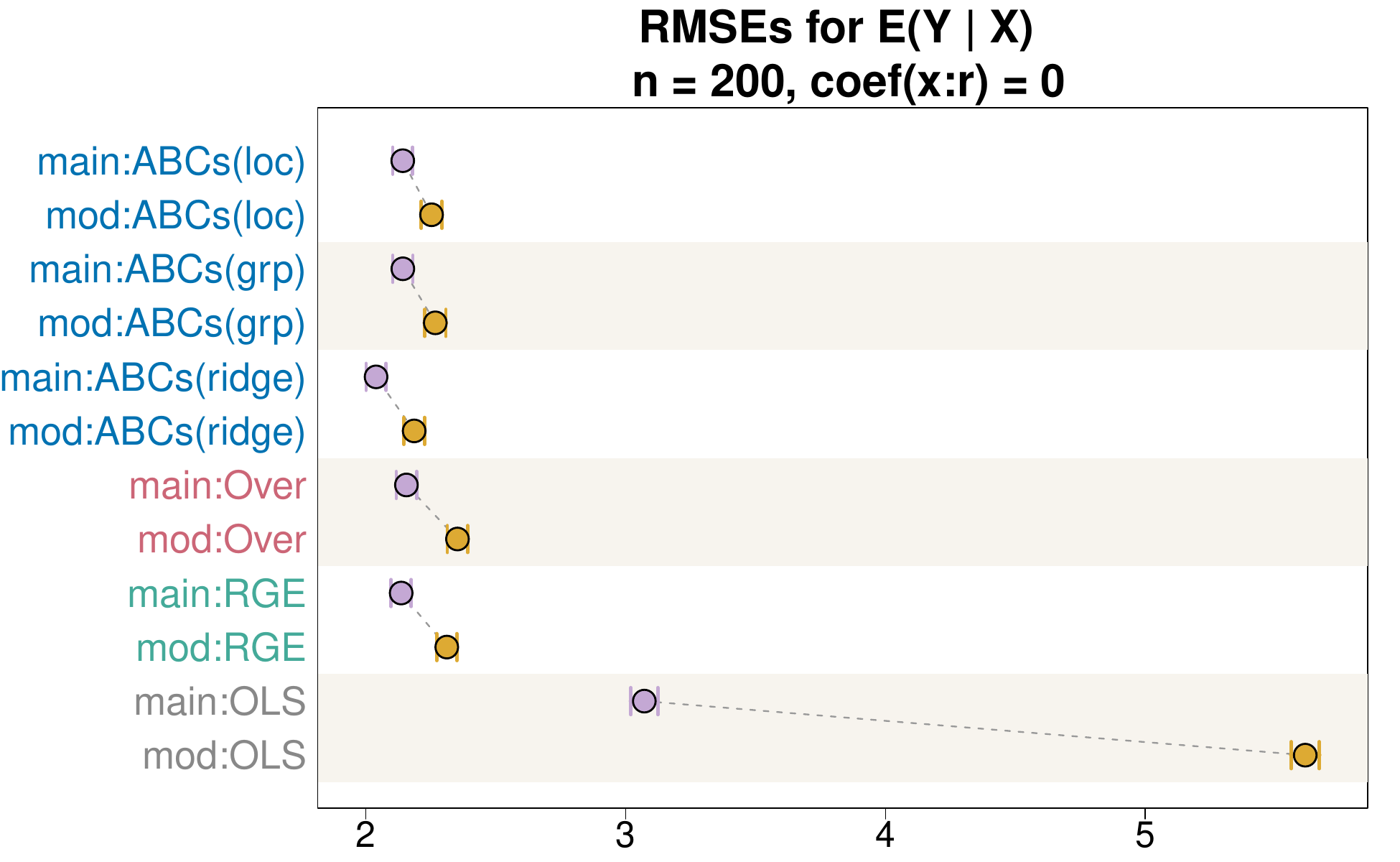}
\includegraphics[width=.49\textwidth]{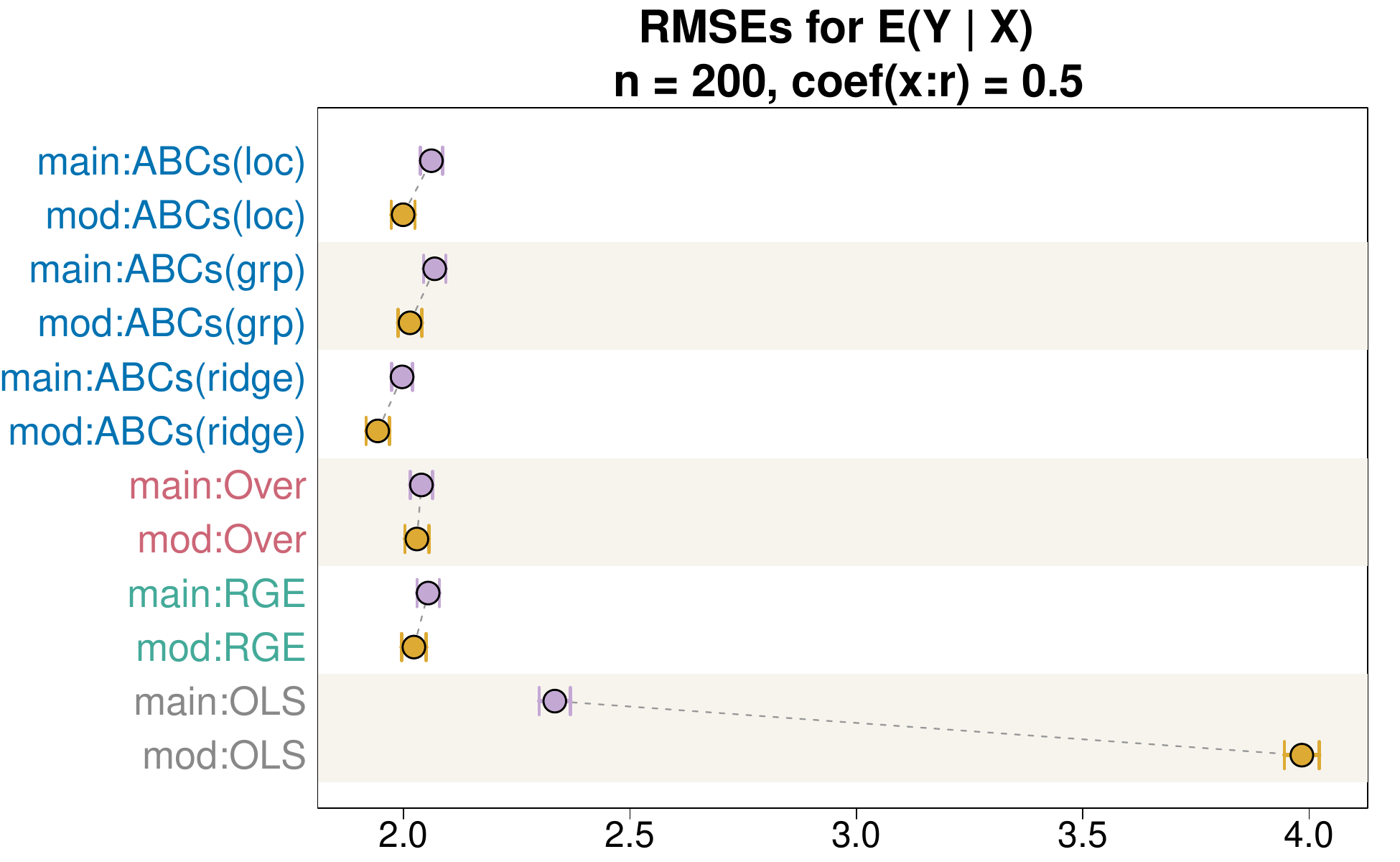}
\caption{\small RMSEs (mean $\pm 2$ SEs) for the regression function $\mu(\bm x)$ with $n=1000$ (top) and $n=200$ (bottom). Dashed lines connect the main-only and cat-modified versions of each method. When all categorical interactions are extraneous (left), BayesABCs methods keep the cat-modified version competitive with the main-only version. When some   are necessary (right), the cat-modified Bayesian models offer massive gains in accuracy over their main-only counterparts. 
\label{fig:sims-rmse-mu}}
\end{figure}

\begin{figure}[!h]
\centering
\includegraphics[width=.49\textwidth]{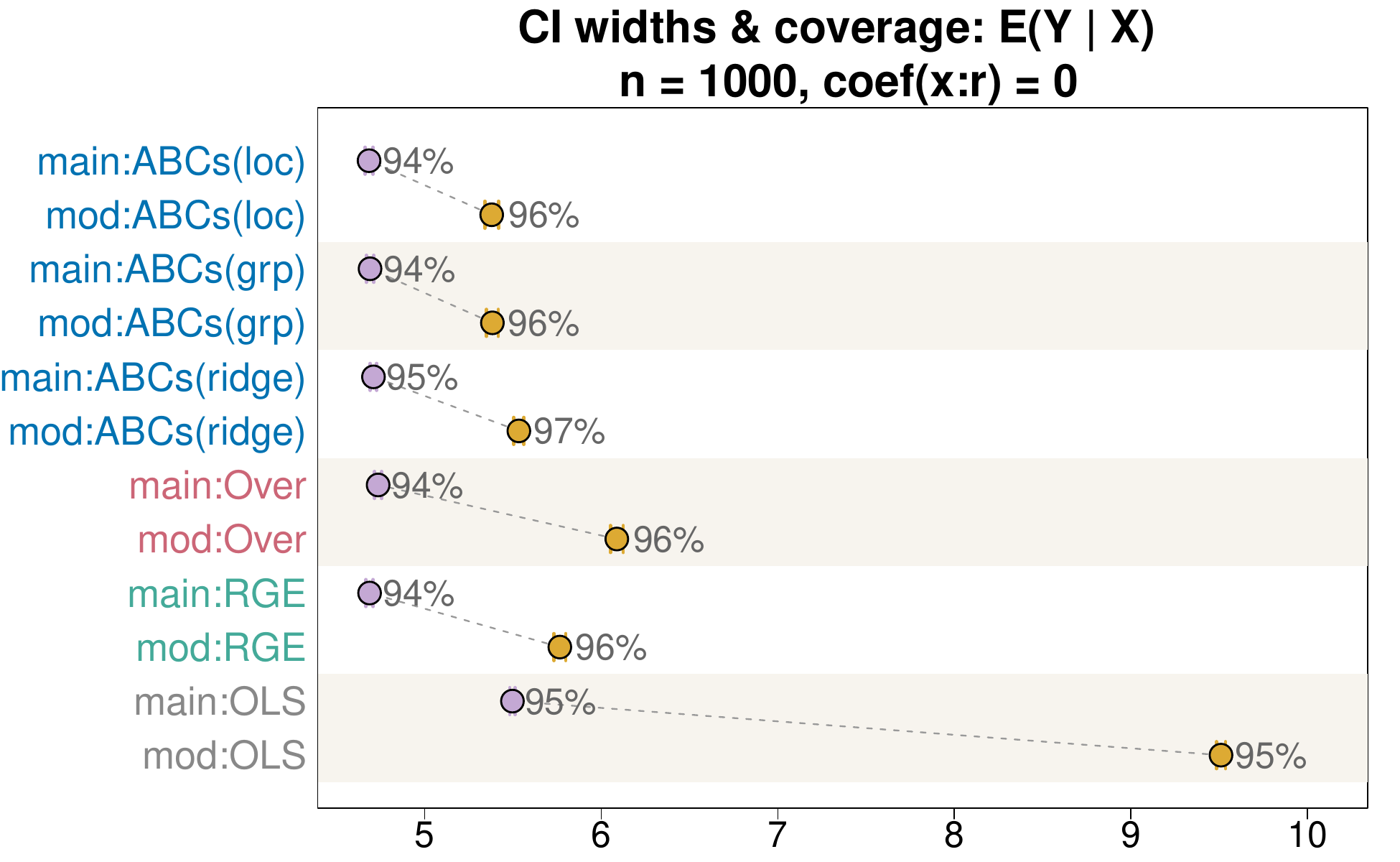}
\includegraphics[width=.49\textwidth]{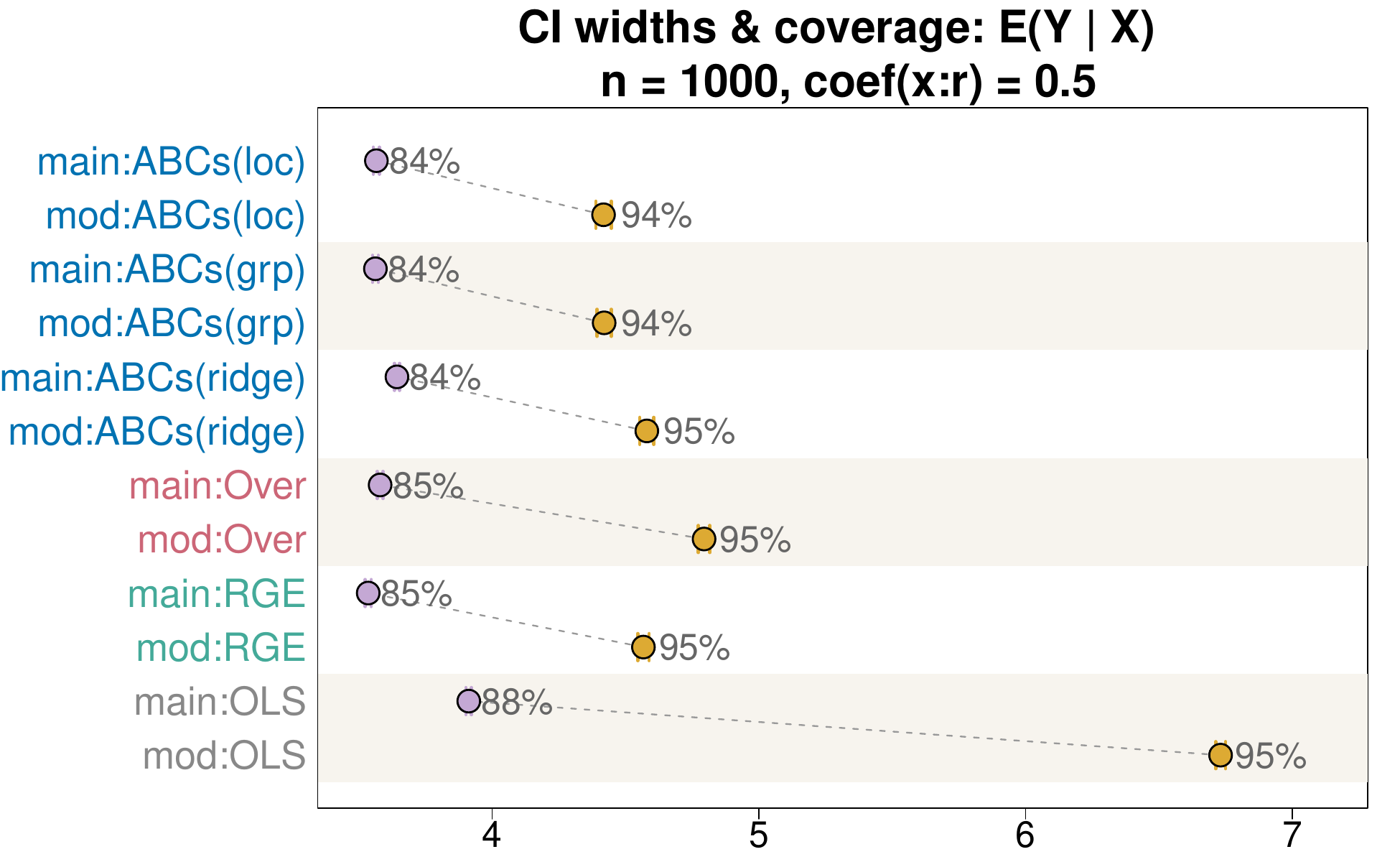}
\includegraphics[width=.49\textwidth]{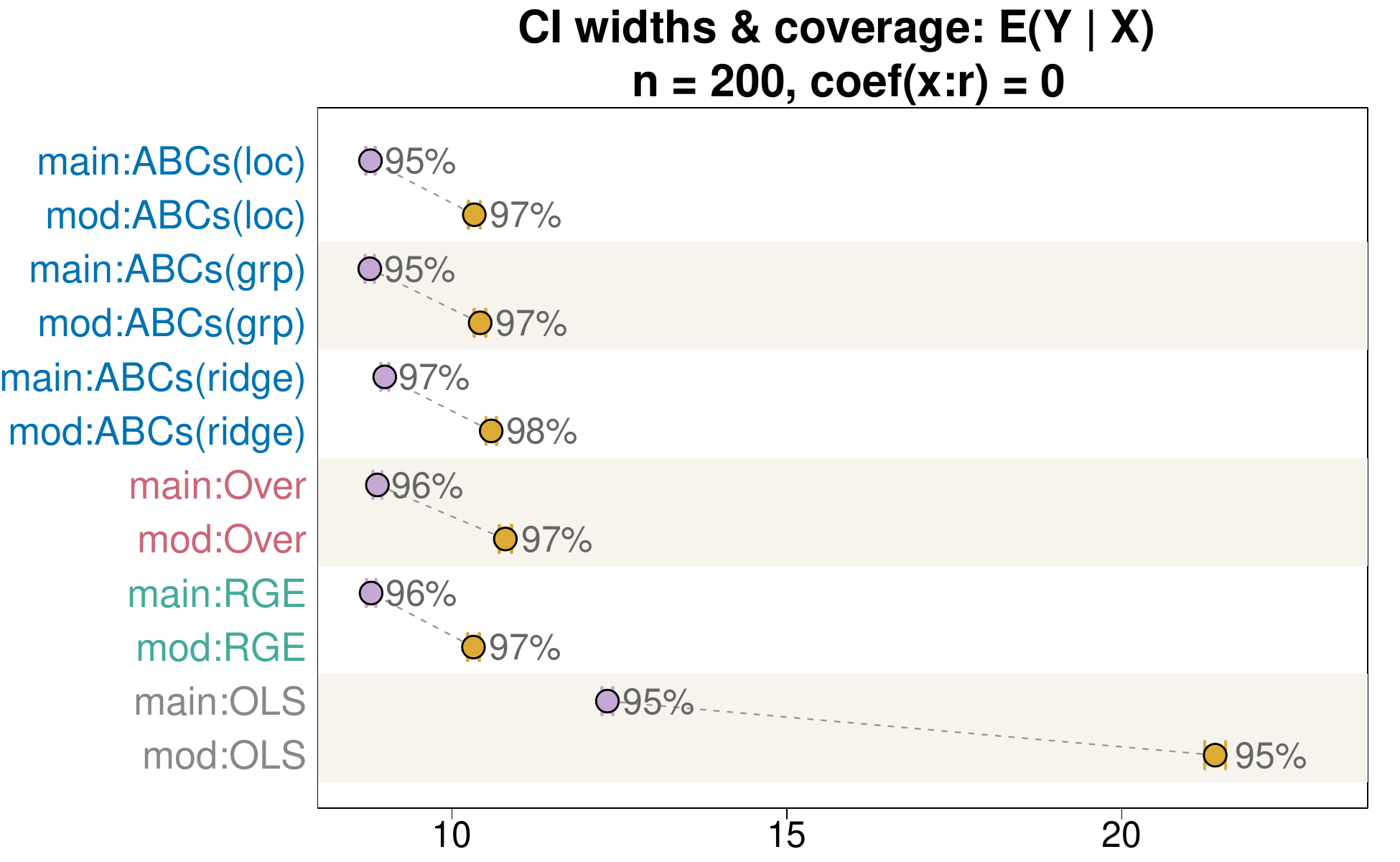}
\includegraphics[width=.49\textwidth]{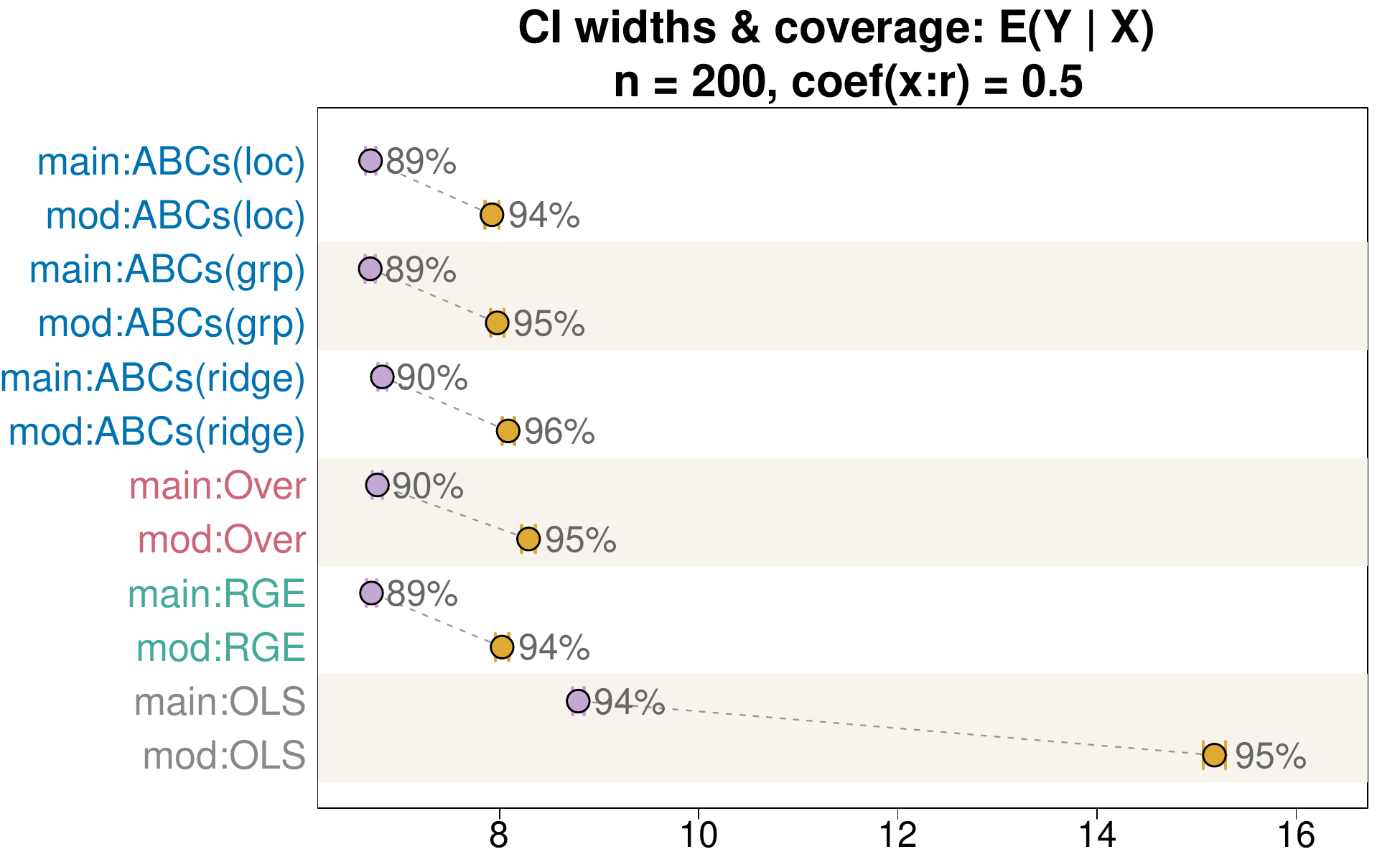}
\caption{\small Mean interval widths (mean $\pm 2$ SEs) and empirical coverage (annotations) for the regression function $\mu(\bm x)$ with $n=1000$ (top) and $n=200$ (bottom). Dashed lines connect the main-only and cat-modified versions of each method. When categorical interactions are extraneous (left), the cat-modified models are less precise than their main-only counterparts, although BayesABCs slightly narrow that gap. When some categorical interactions are necessary (right), the main-only models substantially undercover while \texttt{ABCs(grp)} and \texttt{ABCs(loc)} are most precise. 
\label{fig:sims-ci-mu}}
\end{figure}

Accompanying the results from Section~\ref{sec-sims-lm}, we include results for $n=200$ along with estimation and inference for $\mu(\bm x) = \mathbb{E}( y \mid \bm x)$. Point estimation accuracy and inference for the continuous main effect parameters $\bm \alpha$ with $n=200$ are in Figures~\ref{fig:sims-rmse-x-200} and \ref{fig:sims-ci-200}, respectively. These metrics for $\mu(\bm x)$ for $n \in\{200, 1000\}$ are in Figures~\ref{fig:sims-rmse-mu} and \ref{fig:sims-ci-mu}, respectively.

\section{Additional application results}\label{app-app}
\begin{figure}[!h]
\centering
\includegraphics[width=.8\textwidth]{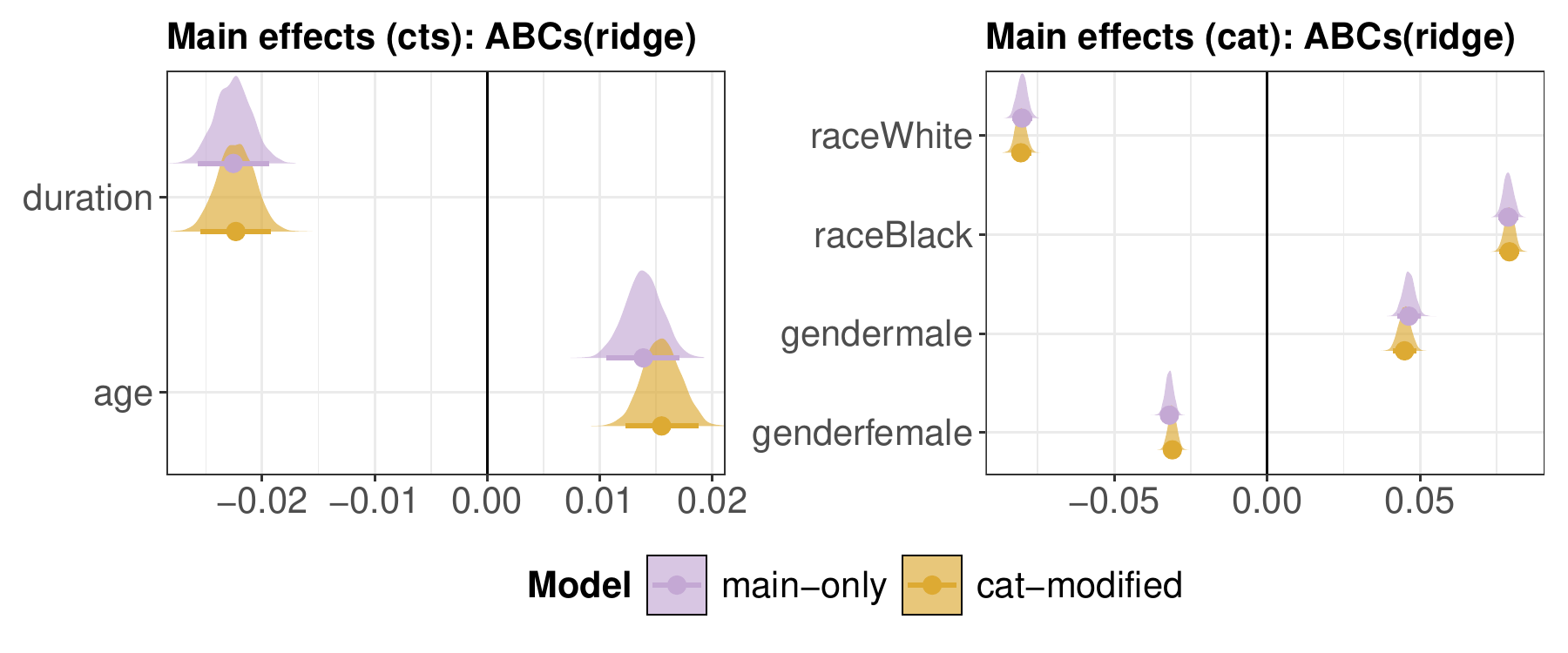}
\includegraphics[width=.8\textwidth]{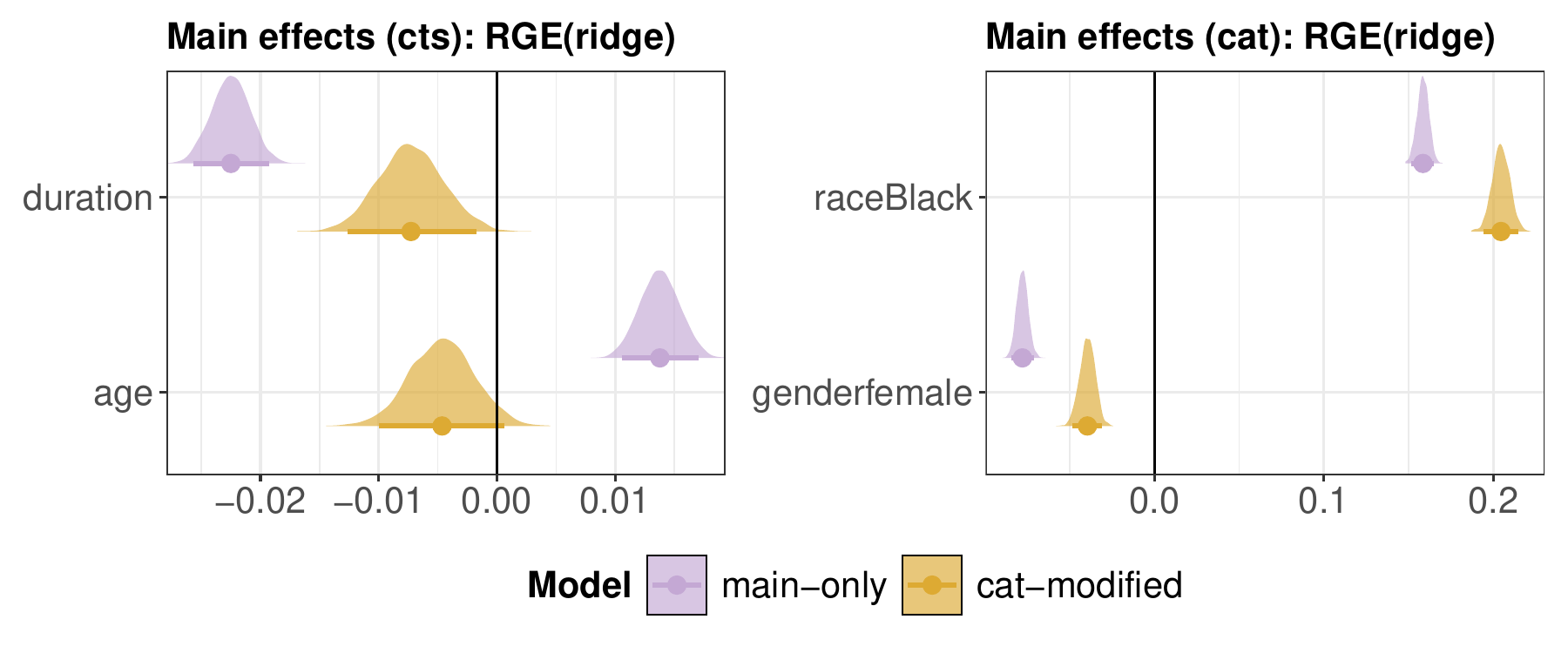}
\includegraphics[width=.8\textwidth]{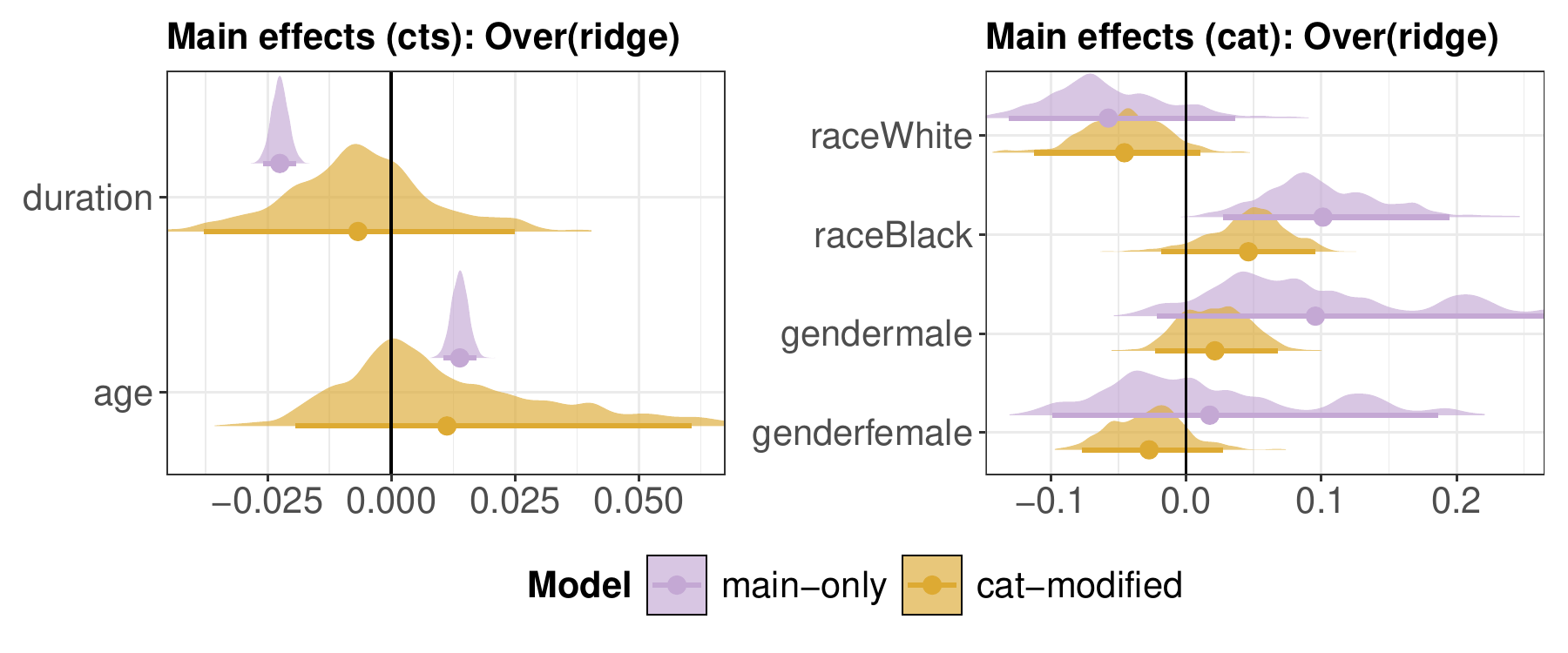}
\caption{\small Posteriors for continuous (left) and categorical (right) main effects under main-only (blue) and cat-modified (gold) models with BayesABCs (top), RGE (middle), and Over (bottom), now with ridge priors. %Points indicate posterior means and bars are 95\% equal-tailed intervals.
With BayesABCs, these posteriors are nearly identical for the main-only and cat-modified models. By comparison, RGE inference depends on the reference group and changes drastically from main-only to cat-modified models, whereas Over inference is unstable. 
\label{fig:panel-stack-ridge}}
\end{figure}

We include supporting results from the application to word error rates (WERs) in automatic speech recognition (ASR) systems (Section~\ref{sec-asr}). In Figure~\ref{fig:panel-stack-ridge}, we reproduce Figure~\ref{fig:panel-stack} but instead using ridge priors for each parameterization (ABCs, RGE, and Over). Even without the aggressive shrinkage priors, the results are similar. BayesABCs retains its posterior invariance and presents clear conclusions, while RGE suffers from the persistent issues of bias and reference group sensitivity. The posteriors under Over are different for the ridge prior compared to the horseshoe prior, but the exact same problems of nonidentifiability appear:  the posteriors are highly dispersed, multimodal, and erratic for both main-only and cat-modified models.

Next, we present the posterior analysis under the expanded model that includes an identifier for the speaker (\texttt{ID}) in each recording. Recall that we include 
\texttt{ID} in the main-only model \texttt{(1 | ID)} and \texttt{ID} interactions with age and duration in the cat-modified model \texttt{(1 + age + duration | ID)}. Now, the main-only and cat-modified models have 128 and 379 identified coefficients, respectively.

\begin{figure}[!h]
\centering
\includegraphics[width=.8\textwidth]{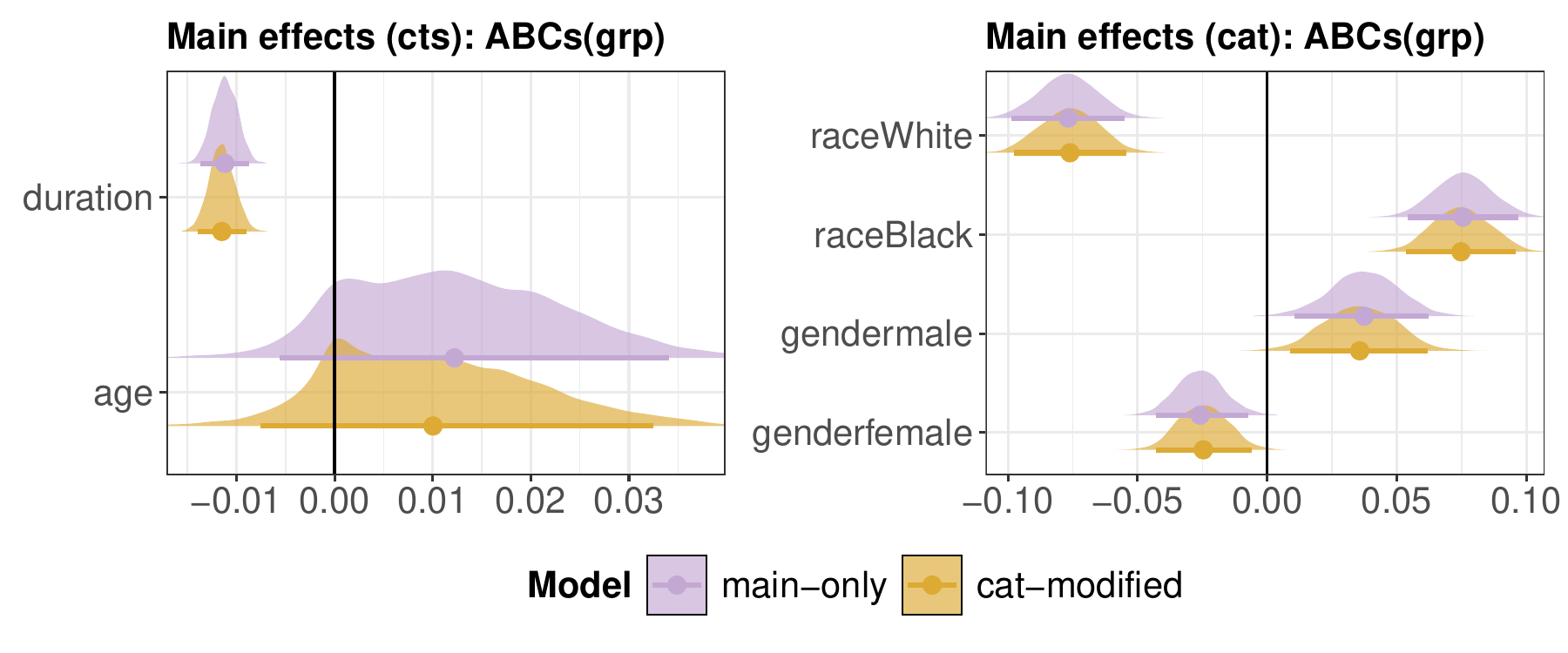}
\includegraphics[width=.8\textwidth]{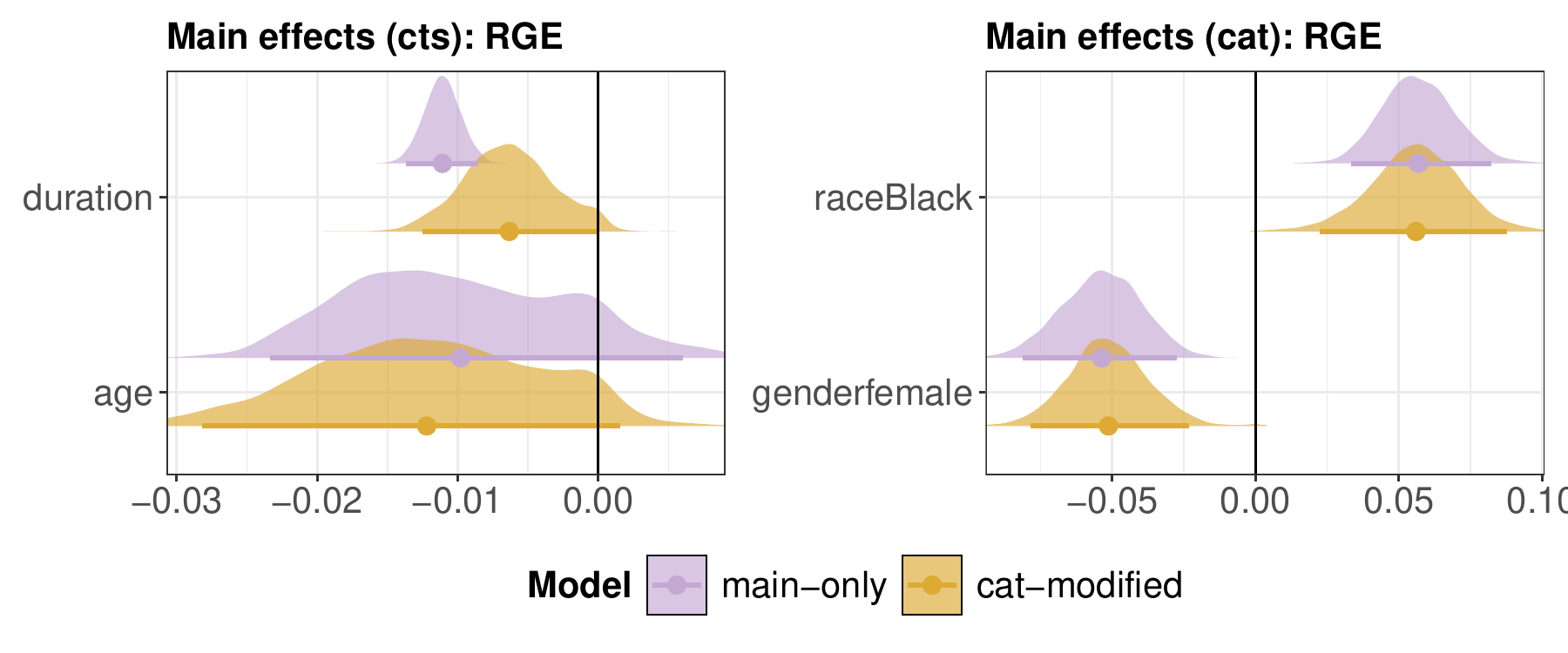}
\includegraphics[width=.8\textwidth]{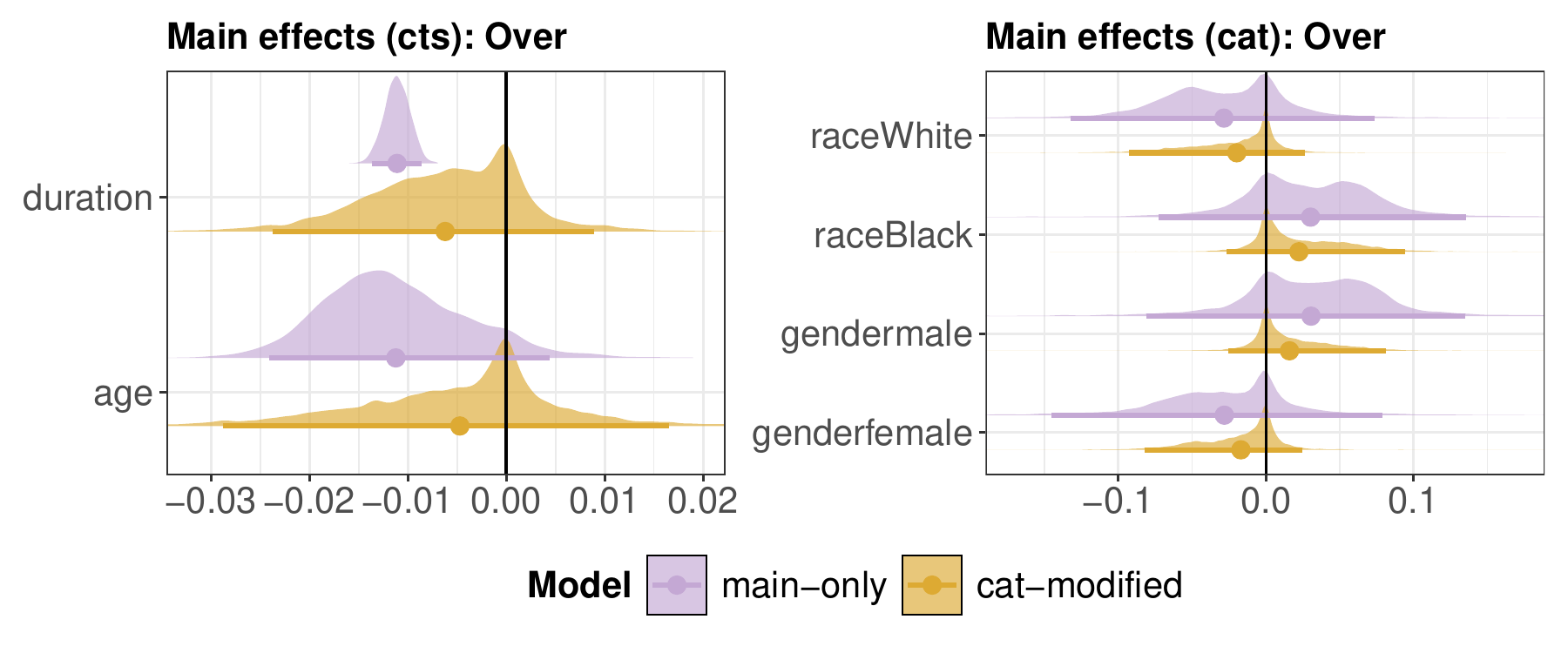}
\caption{\small Posteriors for continuous (left) and categorical (right) main effects under main-only (blue) and cat-modified (gold) models with BayesABCs (top), RGE (middle), and Over (bottom), now including a random intercept (main-only) and random slopes for age and duration (cat-modified) via speaker \texttt{ID}. Despite the additional high-dimensional covariate and its CX interactions, BayesABCs maintains posterior invariance. The problems with RGE and Over persist, and further the age effect changes sign due to their parameterization-prior combination. 
\label{fig:panel-stack-id}}
\end{figure}

\begin{figure}[!h]
\centering
\includegraphics[width=.8\textwidth]{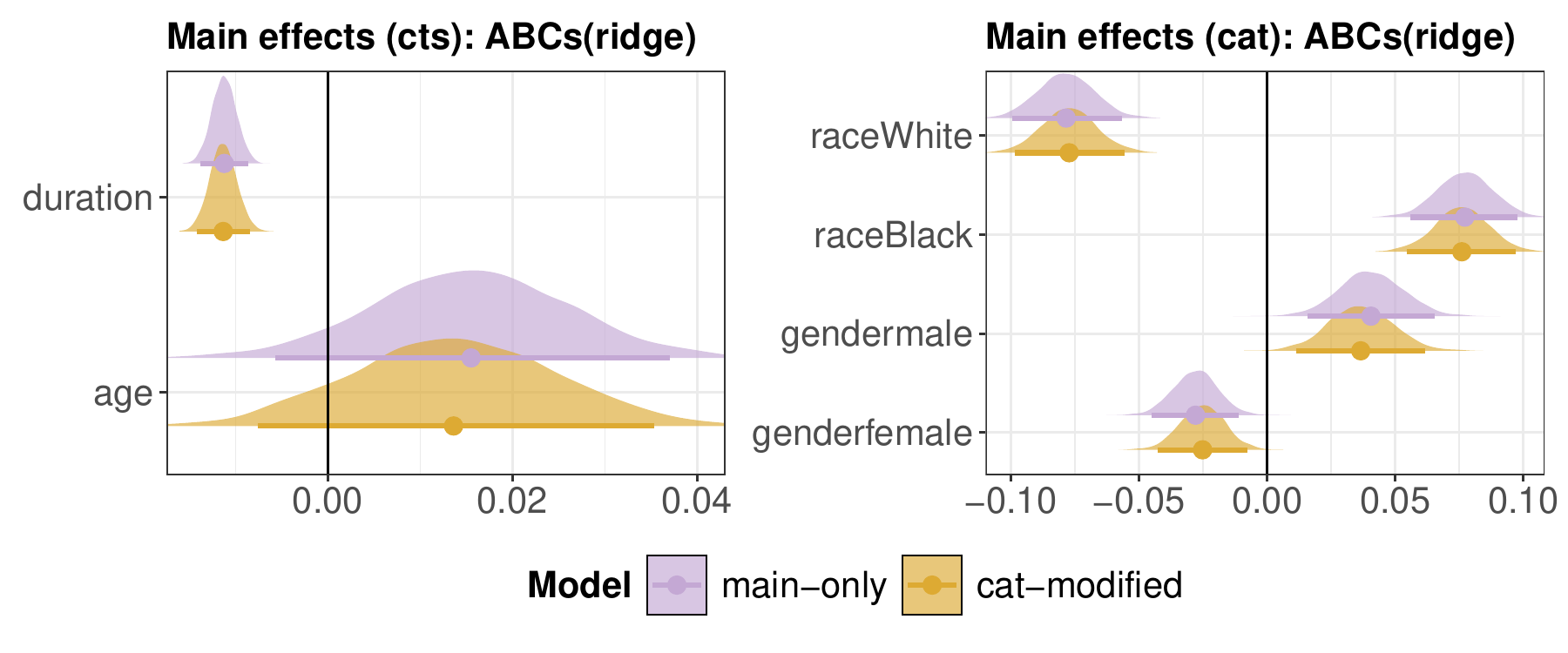}
\includegraphics[width=.8\textwidth]{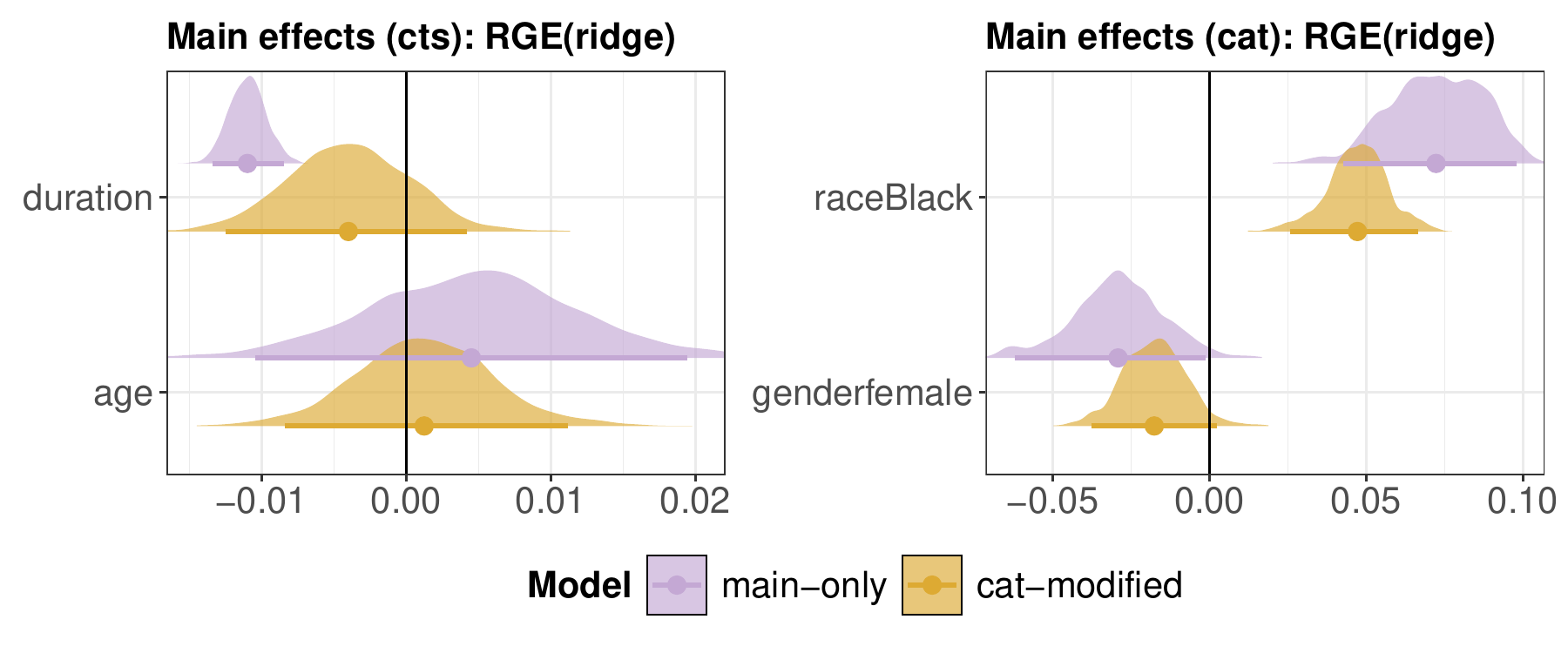}
\includegraphics[width=.8\textwidth]{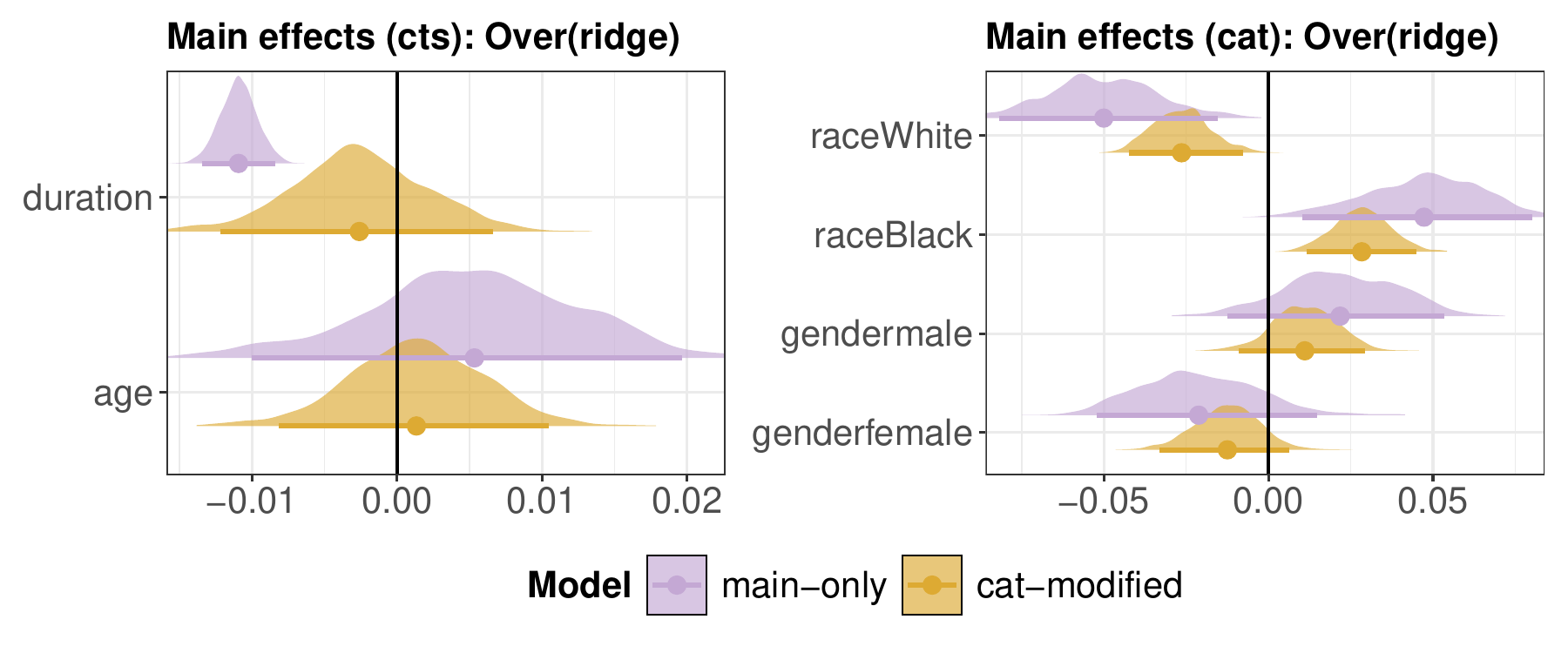}
\caption{\small Posteriors for continuous (left) and categorical (right) main effects under main-only (blue) and cat-modified (gold) models with BayesABCs (top), RGE (middle), and Over (bottom), now including a random intercept (main-only) and random slopes for age and duration (cat-modified) via speaker \texttt{ID} and using ridge priors for all models. Once again, BayesABCs uniquely maintains posterior invariance. RGE and Over posteriors avoid the age sign flip under this ridge prior, but the categorical effect posteriors remain erratic. 
\label{fig:panel-stack-id-ridge}}
\end{figure}

Figure~\ref{fig:panel-stack-id} presents the main effect posteriors for all models and parameterizations, using grouped shrinkage priors for BayesABCs (\texttt{ABCs(grp)}) and horseshoe priors for competing methods (\texttt{RGE} and \texttt{Over}); see Figure~\ref{fig:panel-stack-id-ridge} for a version with all ridge priors. Although these main-only and cat-modified models now include a high-dimensional categorical covariate and its interactions, BayesABCs once again maintains posterior invariance of main effects, leading to clear and robust conclusions across models. With \texttt{ID}, the age effect vanishes, but other conclusions are unchanged (higher WERs occur for Black speakers and male speakers and for shorter recordings). The limitations of RGE and Over reappear, regardless of whether horseshoe or ridge priors are used. Yet here, these parameterization-prior combinations create new difficulties: with horseshoe priors, RGE and Over estimate a \emph{negative} (but not significant) effect for age. This conflicts with results from the same model and parameterizations under ridge priors (Figure~\ref{fig:panel-stack-id-ridge}), all models without \texttt{ID} (Figures~\ref{fig:panel-stack} and \ref{fig:panel-stack-ridge}), and all BayesABCs models. The likely culprit is that the horseshoe prior for \texttt{ID} main effects and CX interactions with \texttt{ID} is a poor specification for random intercepts and slopes. Nonetheless, it is yet another impediment to clear conclusions---and one that does not occur for BayesABCs.

\begin{figure}[!h]
\centering
\includegraphics[width=.8\textwidth]{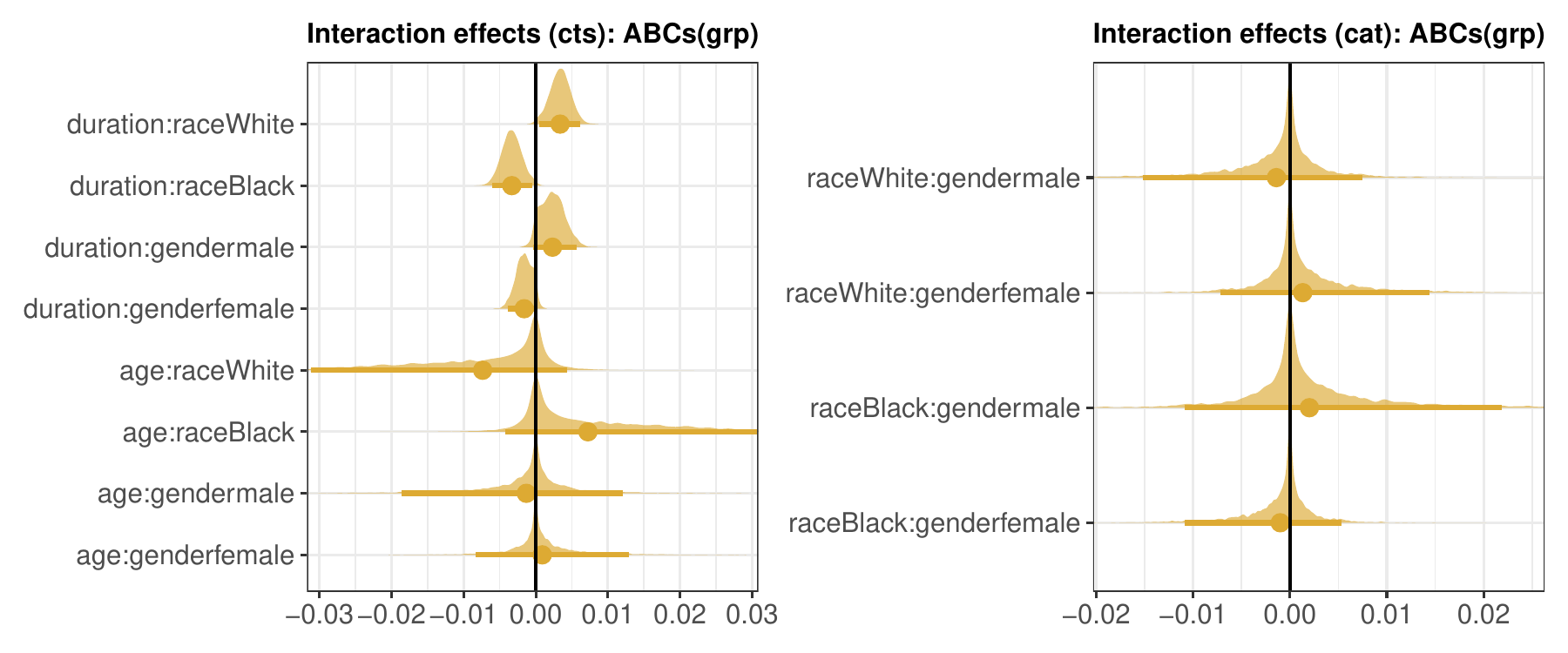}
\includegraphics[width=.8\textwidth]{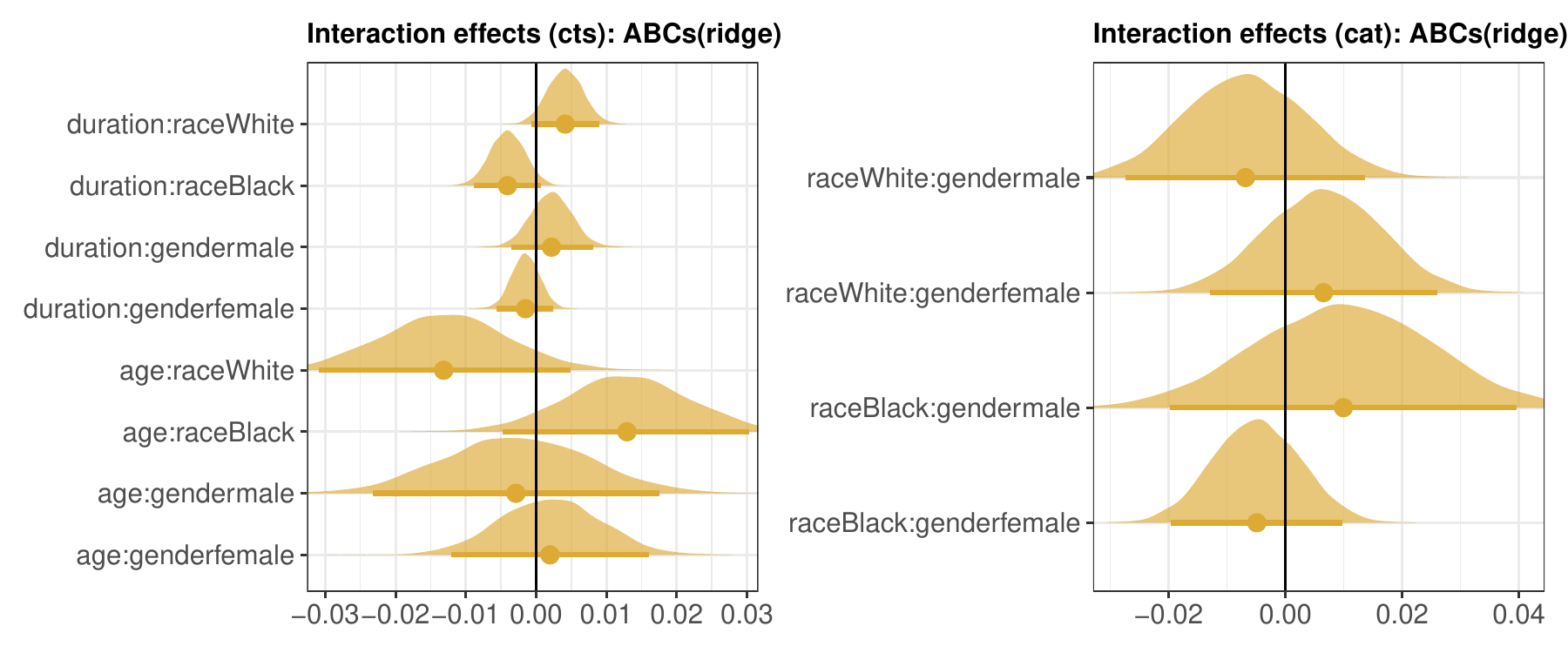}
\caption{\small BayesABCs posteriors for CX (left) and CC (right) interactions for ASR data, now including a random intercept and random slopes for age and duration via speaker \texttt{ID}. We present both grouped shrinkage priors (top) and ridge priors (bottom). 
\label{fig:interact-abcs-id}}
\end{figure}

Finally, we present the interaction effect posteriors under BayesABCs in Figure~\ref{fig:interact-abcs-id} for both grouped shrinkage priors (top) and ridge priors (bottom). After adjusting for \texttt{ID}, the age interactions with race and gender are attenuated, consistent with the age main effect. Similarly, the race-gender interaction vanishes. Yet some CX interaction effects persist: amplified duration effects still occur for Black speakers. These posteriors also show clear differences between the priors. The grouped shrinkage prior yields posteriors with tighter peaks, but also heavier tails in some cases (e.g., CX interactions with age).

% FIXME: remove if okay...
\clearpage
\section{Proofs}\label{app-proofs}

% Example for impossibility of orthogonality:
% For comparison, consider a more vanilla setting where covariates are added to a linear model. If these new covariates are \emph{orthogonal} to the original covariates, then OLS estimates of the original coefficients are unchanged, OLS standard errors are no larger, and inferences are either the same or more powerful.  However, this ideal scenario appears impossible for cat-modified models: such orthogonality generally cannot occur. To see this, let   $\bm x' = (x_1,\ldots,x_n)$ be a covariate and $\bm z_x = \mbox{diag}(\bm x) \bm z$  a categorical modifier, where $\bm z$ is a vector of indicator variables (this generalizes for a matrix that encodes all levels of a categorical variable). Orthogonality requires $\bm x'\bm z_x = \sum_{i=1}^n x_i^2 z_i = 0$. Since $z_i \in\{0,1\}$, we must have that $x_i = 0$ whenever $z_i = 1$. When $x_i$ is a continuous covariate, this is all but impossible; when $x_i \in\{0,1\}$ is an indicator for a categorical covariate, it is in fact hopeless. Orthogonality would require that $x_i$ and $z_i$ are never active simultaneously---which also means that this CC interaction is not estimable. 

\subsection{Establishing posterior invariance under linear constraints}
Because we apply the following results in different contexts (e.g., both for $\mathcal{M}_{01}$ and $\mathcal{M}_{012}$), we begin with general notation. Let $\bm y$ be $n\times 1$ and $\bm X$ be $n \times p$. The Gaussian linear model is 
\begin{equation}\label{gen-lm}
\bm y = \bm X \bm \theta + \bm \epsilon,  \quad [\bm \epsilon \mid \tau] \sim N_n(\bm 0, \tau^{-1} \bm I_n) 
\end{equation}
with $p+1$ unknown parameters ($\bm \theta,\tau$), for which we specify joint priors of the form 
\begin{equation} \label{gen-prior-app}
    p(\bm \theta, \tau) \propto \tau^{-1} p(\bm \theta \mid \tau)
\end{equation}
i.e., with a  Jeffreys prior on the precision. We will sequentially 1) marginalize over the precision, 2) marginalize over the precision and a block of regression coefficients, and 3) apply these results for models $\mathcal{M}_{01}$ and $\mathcal{M}_{012}$ under condition \eqref{cond} and the linear constraint $\bm A\bm\gamma = \bm 0$. Steps one and two are standard but we present them for completeness and to illustrate our general strategy. Step three requires some modifications, especially the treatment of the linear constraint.

\subsubsection{Marginalizing over the precision}
First, we compute the marginal (over $\tau$) posterior $p(\bm \theta \mid \bm y) = \int p(\bm \theta, \tau \mid \bm y) \ d \tau$ under two choices of priors for \eqref{gen-prior-app}. A flat prior on $\bm \theta$ yields the following:
\begin{lemma}\label{lem-flat-marg}
    Under the Gaussian linear model \eqref{gen-lm} and prior \eqref{gen-prior-app} with $p(\bm \theta \mid \tau) \propto 1$, the marginal posterior for the regression coefficients is 
    \[
    [\bm \theta \mid \bm y] \sim t_{n-p}(\bm{\hat\theta}, \hat S^2 \bm{\hat\Sigma}_\theta)
    \]
    where $\bm{\hat\theta} = (\bm X'\bm X)^{-1}\bm X'\bm y$ is the OLS estimator, $\hat S^2 = (n-p)^{-1}\Vert \bm e\Vert^2$ is the estimated OLS residual variance with $\bm e= \bm y - \bm X \bm{\hat\theta} $, and $\bm{\hat\Sigma}_\theta = (\bm X'\bm X)^{-1}$ is the (unscaled) OLS covariance. 
\end{lemma}
\begin{proof}[Proof (Lemma~\ref{lem-flat-marg})]
    The joint posterior is $p(\bm \theta, \tau \mid \bm y) \propto \tau^{n/2 - 1} \exp( -\tau \Vert \bm y - \bm X \bm \theta \Vert^2/2)$, which is proportional to the density of $\mbox{Gamma}(n/2, \Vert \bm y - \bm X \bm \theta \Vert^2/2)$. We adjust for the missing constant and apply the fact that the density integrates to one: 
    \begin{align*}
        p(\bm \theta \mid \bm y) &\propto \Gamma(n/2) (\Vert \bm y - \bm X \bm \theta \Vert^2/2)^{-n/2} \int \mbox{Gamma}(\tau; n/2, \Vert \bm y - \bm X \bm \theta \Vert^2/2)  \ d\tau \\
        &\propto (\Vert \bm y - \bm X \bm \theta \Vert^2)^{-n/2}
    \end{align*}
%which is proportional to . 
Applying the OLS fact that $\Vert \bm y - \bm X \bm \theta \Vert^2 = (n-p) \hat S^2 + (\bm \theta - \bm{\hat \theta})' \bm X'\bm X (\bm \theta - \bm{\hat\theta})$, %(e.g., Zellner, 1996, p.~65), 
we can  rewrite the marginal posterior 
\[
p(\bm \theta \mid \bm y) \propto \left\{1 + (n-p)^{-1} \hat S^{-2}(\bm \theta - \bm{\hat \theta})' \bm{\hat\Sigma}_\theta^{-1} (\bm \theta - \bm{\hat\theta})
\right\}^{-n/2}
\]
yielding the stated distribution. 
% $n-p$ degrees of freedom and (posterior) mean and covariance 
%\begin{align*}
    %E(\bm \theta \mid \bm y) &= \bm{\hat\theta}\\
    %\mbox{Cov}(\bm\theta \mid \bm y) &= \frac{n-p}{n-p-2} \hat S^{2} (\bm X'\bm X)^{-1}.
%\end{align*}
\end{proof}
Implicitly, this result requires that $\bm X$ is full rank. No such requirement is needed for a Gaussian prior on $\bm \theta$:
\begin{lemma}\label{lem-norm-marg}
    Under the Gaussian linear model \eqref{gen-lm} and prior \eqref{gen-prior-app} with $[\bm \theta \mid \tau] \sim N_p(\bm \theta_0, \tau^{-1}\bm \Sigma_{\theta_0})$, the marginal posterior for the regression coefficients is 
        \[
        [\bm \theta \mid \bm y] \sim t_{n}(\bm{\hat\theta}_{G}, \hat S_{G}^2 \bm{\hat\Sigma}_{\theta_{G}})
    \]
    where $\bm{\hat\Sigma}_{\theta_{G}} = (\bm X' \bm X +\bm \Sigma_{\theta_0}^{-1})^{-1}$,     $\bm{\hat\theta}_{G} = \bm{\hat\Sigma}_{\theta_{G}}(\bm X'\bm y + \bm \Sigma_{\theta_0}^{-1} \bm \theta_0),$ and 
        $\hat S_{G}^2 = n^{-1}(\Vert \bm y \Vert^2 + \Vert \bm\theta_0 \Vert_{\bm\Sigma_{\theta_0}^{-1}}^2 - \Vert \bm{\hat\theta}_{G} \Vert_{\bm{\hat\Sigma}_{\theta_{G}}^{-1}}^2)
        $ 
%    \begin{align*}
%        \bm{\hat\Sigma}_{\theta_{G}} &= (\bm X' \bm X +\bm \Sigma_{\theta_0}^{-1})^{-1}, \quad
%        \bm{\hat\theta}_{G} = \bm{\hat\Sigma}_{\theta_{G}}(\bm X'\bm y + \bm \Sigma_{\theta_0}^{-1} \bm \alpha_0) \\
%        \hat S_{G}^2 &= n^{-1}(\Vert \bm y \Vert^2 + \Vert \bm\theta_0 \Vert_{\bm\Sigma_{\theta_0}^{-1}}^2 - \Vert \bm{\hat\theta}_{G} \Vert_{\bm{\hat\Sigma}_{\theta_{G}}^{-1}}^2)
%    \end{align*}
    using the notation $\Vert \bm v \Vert_{\bm M}^2 = \bm v'\bm M\bm v$.
\end{lemma}
\begin{proof}[Proof (Lemma~\ref{lem-norm-marg})]
    Because the joint prior is $
p(\bm \theta, \tau) \propto \tau^{-1} \tau^{p/2} \exp\{-\tau (\bm\theta - \bm\theta_0)'\bm \Sigma_{\theta_0}^{-1}(\bm \theta - \bm\theta_0)/2\}$, the joint posterior is 
\[
p(\bm \theta, \tau \mid \bm y) \propto \tau^{(n+p)/2 - 1} \exp\left\{ -\frac{\tau}{2} (\Vert \bm y - \bm X \bm \theta\Vert^2 + \Vert\bm\theta - \bm\theta_0\Vert_{\bm\Sigma_{\theta_0}^{-1}}^2) \right\}.
\]
The same strategy works as in Lemma~\ref{lem-flat-marg}: we now use a $\mbox{Gamma}\{(n+p)/2, (\Vert \bm y - \bm X \bm \theta\Vert^2 + \Vert\bm\theta - \bm\theta_0\Vert_{\bm\Sigma_{\theta_0}^{-1}}^2)/2\}$ density, so the marginal posterior is 
\[
p(\bm \theta \mid \bm y)\propto (\Vert \bm y - \bm X \bm \theta\Vert^2 + \Vert\bm\theta - \bm\theta_0\Vert_{\bm\Sigma_{\theta_0}^{-1}}^2)^{-(n+p)/2}.
\]
Deriving the moments simply requires simplifying the interior term: $\Vert \bm y - \bm X \bm \theta\Vert^2 + \Vert\bm\theta - \bm\theta_0\Vert_{\bm\Sigma_{\theta_0}^{-1}}^2 = n\hat S_{G}^2  + \Vert \bm \theta - \bm{\hat\theta}_{G} \Vert_{\bm{\hat\Sigma}_{\theta_{G}}^{-1}}^2$, which proves the result. 
\end{proof}
Similar arguments show that marginally, $[\tau \mid \bm y] \sim \mbox{Gamma}(n/2, n\hat S_{G}^2/2)$, so this scale term $\hat S_{G}^{2} = 1/E(\tau \mid \bm y)$ is an estimator of the variance.

%\begin{align*}
%p(\bm \theta \mid \bm y) &\propto \left\{1 + \frac{1}{n} \hat S_{\mathcal{N}}^{-2}  (\bm \theta - \bm{\hat \mu}_\theta)'\bm Q_\theta (\bm \theta - \bm{\hat \mu}_\theta)\right\}^{-(n+p)/2} \\
%[\bm \theta \mid \bm y] &\sim t_{n}\left(\bm{\hat\mu}_\theta, \hat S_{\mathcal{N}}^2 \bm Q_\theta^{-1}\right)
%\end{align*}
%i.e., a $t$-distribution with $n$ degrees of freedom and (posterior) mean and covariance 
%\begin{align*}
%    E(\bm \theta \mid \bm y) &= \bm{\hat\mu}_\theta\\
%    \mbox{Cov}(\bm\theta \mid \bm y) &= \frac{n}{n-2}\hat S_{\mathcal{N}}^2  (\bm X' \bm X + \bm \Sigma_\theta^{-1})^{-1}
%\end{align*}
%Using the same arguments, we can show that marginally $[\tau \mid \bm y] \sim \mbox{Gamma}(n/2, n\hat S_{\mathcal{N}}^2/2)$. This shows that the recurring term $\hat S_{\mathcal{N}}^{2} = 1/E(\tau \mid \bm y)$ appears as an estimator of the variance $\tau^{-1}$. 

\subsubsection{Marginalizing over the precision and some regression coefficients}
Next, we partition the coefficients $\bm\theta = (\bm\beta' : \bm\alpha')'$ and covariates $\bm X = (\bm X_0 : \bm X_1)$ with respective dimensions (columns) $p_0$ and $p_1$. 
The goal is to derive the marginal posterior 
$p(\bm \alpha \mid \bm y) = \int \int p(\bm \beta, \bm \alpha, \tau \mid \bm y) \ d \bm \beta \ d \tau$. We mildly restrict \eqref{gen-prior-app} to use a flat prior for $\bm \beta$:
\begin{equation}\label{alt-prior}
    p(\bm \beta, \bm \alpha, \tau) \propto \tau^{-1} p(\bm \alpha \mid \tau).
\end{equation}
Our results are facilitated by the following simplification:
\begin{lemma}\label{lem-simple}
    Under the Gaussian linear model \eqref{gen-lm} and prior \eqref{alt-prior}, the marginal (over $\bm \beta$) posterior is
    \begin{equation}\label{gen-simp}
    p(\bm\alpha, \tau \mid \bm y) \propto \tau ^{(n - p_0)/2 - 1} \exp\Big[-\frac{\tau}{2} \Big\{ (\bm \alpha - \bm{\hat\alpha})'\bm{\hat \Sigma}_{\alpha}^{-1}(\bm \alpha - \bm{\hat\alpha})  + \Vert \bm e \Vert^2 \Big\}\Big] \ p(\bm \alpha\mid \tau),
\end{equation}
where $\bm{\hat \Sigma}_{\alpha} = (\bm E_{10}'\bm E_{10})^{-1}$  and $\bm{\hat\alpha} = \bm{\hat \Sigma}_{\alpha} \bm E_{10}'\bm e_0$, with  $\bm E_{10} = \bm X_1 - \bm H_0 \bm X_1$  and $\bm e_0 = \bm y - \bm H_0\bm y$ for $ \bm H_0 = \bm X_0(\bm X_0'\bm X_0)^{-1}\bm X_0$.
\end{lemma}
Lemma~\ref{lem-simple} accomplishes the first step of marginalizing over $\bm \beta$; then marginalizing over $\tau$ proceeds as in Lemmas~\ref{lem-flat-marg} and \ref{lem-norm-marg}. %In effect, we regress out $\bm X_0$ from both $bm y$ and $\bm X_1$. 
To prove Lemma~\ref{lem-simple}, we use an intermediate result:
\begin{lemma}\label{lem-inter}
    The squared error is $
     \Vert \bm y - \bm X \bm \theta\Vert^2  = (\bm \beta- \bm{\hat \Sigma}_{\beta} \bm \ell_{\beta})'\bm{\hat \Sigma}_{\beta}^{-1}(\bm \beta - \bm{\hat \Sigma}_{\beta} \bm \ell_{\beta}) + 
    (\bm \alpha - \bm{\hat\alpha})'\bm{\hat \Sigma}_{\alpha}^{-1}(\bm \alpha - \bm{\hat\alpha})   + \Vert \bm e \Vert^2$, where $\bm{\hat \Sigma}_{\beta} = (\bm X_0'\bm X_0)^{-1}$ and $\bm \ell_{\beta} = \bm X_0'(\bm y - \bm X_1\bm\alpha)$.  
\end{lemma}
We prove Lemma~\ref{lem-inter} below, but first apply it to prove  Lemma~\ref{lem-simple}. 
\begin{proof}[Proof (Lemma~\ref{lem-simple})]
    The marginal (over $\bm \beta$) posterior is 
    \[
         p(\bm\alpha, \tau \mid \bm y)  %\int p(\bm \beta, \bm\alpha, \tau \mid \bm y) \ d \bm\beta  
         \propto \tau ^{n/2} \int  \exp\left(-\frac{\tau}{2} \Vert \bm y - \bm X \bm \theta\Vert^2\right) \ d \bm \beta  \times p(  \bm \alpha, \tau) 
    \]
    Applying Lemma~\ref{lem-inter}, we simplify the integral: 
\begin{align*}
    \int  \exp\left(-\frac{\tau}{2} \Vert \bm y - \bm X \bm \theta\Vert^2\right) \ d \bm \beta &=  
    \int \exp\Big[-\frac{\tau}{2} \Big\{ 
    (\bm \beta - \bm{\hat \Sigma}_{\beta} \bm \ell_{\beta})'\bm{\hat \Sigma}_{\beta}^{-1}(\bm \beta - \bm{\hat \Sigma}_{\beta} \bm \ell_{\beta}) \\
    &\quad \quad\quad\quad + (\bm \alpha - \bm{\hat\alpha})'\bm{\hat \Sigma}_{\alpha}^{-1}(\bm \alpha - \bm{\hat\alpha})   + \Vert \bm e \Vert^2 \Big\}\Big] \ d \bm \beta \\
    &\propto \tau^{-p_0/2} \exp\Big[-\frac{\tau}{2} \Big\{(\bm \alpha - \bm{\hat\alpha})'\bm{\hat \Sigma}_{\alpha}^{-1}(\bm \alpha - \bm{\hat\alpha})   + \Vert \bm e \Vert^2 \Big\}\Big]
\end{align*}
since the top line is the only term involving $\bm \beta$ and is the density of $N(\bm{\hat \Sigma}_{\beta} \bm \ell_{\beta}, \tau^{-1} \bm{\hat \Sigma}_{\beta})$ times $\vert \tau^{-1} \bm{\hat \Sigma}_{\beta} \vert^{1/2} \propto \tau^{-p_0/2}$. Combining the exponents for $\tau$, including from the prior, proves the result.  
%Notably, this posterior is essentially that for standard Bayesian regression, but with two modifications: 1) $\bm X_1$ and $\bm y$ are replaced by the residuals after regressing out $\bm X_0$, i.e., $\bm E_{10}$ and $\bm e_0$, respectively, and 2) there is an adjustment to the exponent for the precision $\tau$. 
\end{proof}

\begin{proof}[Proof (Lemma~\ref{lem-inter})]
    Expanding $\Vert \bm y - \bm X \bm \theta\Vert^2 = \Vert \bm y \Vert^2 + \bm \theta ' (\bm X ' \bm X) \bm \theta - 2 \bm \theta'(\bm X'\bm y)$, we partition the quadratic term into 
\[
    \bm \theta ' (\bm X ' \bm X) \bm \theta = \bm \beta ' (\bm X_0 ' \bm X_0) \bm \beta + \bm \alpha ' (\bm X_1 ' \bm X_1) \bm \alpha + 2 \bm\beta'(\bm X_0'\bm X_1)\bm \alpha
\]
and the linear term into
\[
\bm \theta'(\bm X'\bm y) = \bm \beta'(\bm X_0'\bm y) + \bm \alpha'(\bm X_1'\bm y).
\]
Using the definitions of $\bm{\hat \Sigma}_{\beta}$ and $\bm \ell_{\beta}$, we ``complete the square" for $\bm \beta$:
\begin{align*}
    \Vert \bm y - \bm X \bm \theta\Vert^2 
    &= \Vert \bm y \Vert^2 + \bm \beta ' \bm{\hat \Sigma}_{\beta}^{-1}  \bm \beta -2  \bm \beta'\bm\ell_{\beta} + 
    \bm \alpha ' (\bm X_1 ' \bm X_1) \bm \alpha  - 2 \bm \alpha'(\bm X_1'\bm y)\\
    &= \Vert \bm y \Vert^2 + (\bm \beta -  \bm{\hat \Sigma}_{\beta}\bm\ell_{\beta})'\bm{\hat \Sigma}_{\beta}^{-1}(\bm \beta - \bm{\hat \Sigma}_{\beta}\bm\ell_{\beta}) - \bm \ell_{\beta}' \bm{\hat \Sigma}_{\beta} \bm \ell_{\beta} + \bm \alpha ' (\bm X_1 ' \bm X_1) \bm \alpha  - 2 \bm \alpha'(\bm X_1'\bm y).
\end{align*}
Next, we simplify the middle term:
\begin{align*}
    \bm \ell_{\beta}' \bm{\hat \Sigma}_{\beta} \bm \ell_{\beta} &= (\bm y - \bm X_1 \bm \alpha)' \bm X_0 (\bm X_0'\bm X_0)^{-1} \bm X_0' (\bm y - \bm X_1 \bm \alpha)\\
    &= (\bm y - \bm X_1 \bm \alpha)' \bm H_0 (\bm y - \bm X_1 \bm \alpha) \\
    &= \bm \alpha' \bm X_1' \bm H_0 \bm X_1 \bm \alpha - 2 \bm \alpha'\bm X_1' \bm H_0 \bm y + \Vert \bm H_0 \bm y \Vert^2
\end{align*}
so that the $\bm\alpha$ terms consolidate to
\begin{align*}
    - \bm \ell_{\beta}' \bm{\hat \Sigma}_{\beta} \bm \ell_{\beta} + \bm \alpha ' (\bm X_1 ' \bm X_1) \bm \alpha  - 2 \bm \alpha'(\bm X_1'\bm y) &= \bm \alpha' \bm{\hat \Sigma}_{\alpha}^{-1} \bm \alpha - 2 \bm \alpha'\bm \ell_{\alpha} - \Vert \bm H_0 \bm y\Vert^2
\end{align*}
where  $  \bm \ell_{\alpha}=\bm E_{10}'\bm e_0$ and using the fact that $\bm X_1' \bm X_1 - \bm X_1' \bm H_0 \bm X_1 = \bm E_{10}'\bm E_{10} =\bm{\hat \Sigma}_{\alpha}^{-1}$ and  $ \bm X_1'\bm y - \bm X_1'\bm H_0 \bm y = \bm E_{10}'\bm e_0 = \bm \ell_{\alpha}$. Now, ``complete the square" for $\bm\alpha$:
\begin{align*}
    \Vert \bm y - \bm X \bm \theta\Vert^2 &= \Vert \bm y \Vert^2 + (\bm \beta -  \bm{\hat \Sigma}_{\beta}\bm\ell_{\beta})' \bm{\hat \Sigma}_{\beta}^{-1}(\bm \beta -  \bm{\hat \Sigma}_{\beta}\bm\ell_{\beta})  \\
    & \quad \quad\quad + (\bm \alpha - \bm{\hat\alpha})'\bm{\hat \Sigma}_{\alpha}^{-1}(\bm \alpha - \bm{\hat\alpha})  - \bm \ell_{\alpha}' \bm{\hat \Sigma}_{\alpha}\bm \ell_{\alpha}  - \Vert \bm H_0 \bm y \Vert^2. 
\end{align*}
The last step is to show that $\Vert \bm e \Vert^2 = \Vert \bm y \Vert^2- \bm \ell_{\alpha}' \bm{\hat \Sigma}_{\alpha}\bm \ell_{\alpha}  - \Vert \bm H_0 \bm y \Vert^2$. We simplify the middle term, $\bm \ell_{\alpha}' \bm{\hat \Sigma}_{\alpha}\bm \ell_{\alpha} = \Vert \bm H_{10} \bm e_0 \Vert^2$ where $\bm H_{10} = \bm E_{10}(\bm E_{10}'\bm E_{10})^{-1}\bm E_{10}'$, and note that $\Vert \bm y \Vert^2 - \Vert \bm H_0 \bm y \Vert^2 = \Vert \bm e_0 \Vert^2$. Thus, the necessary result holds by the Frisch–Waugh–Lovell (FWL) theorem \citep{FW1933,Lovell1963}. 
\end{proof}

%\begin{align*}
%    \bm Q_0 &= \bm X_0'\bm X_0 \\
%    \bm \ell_0 &= \bm X_0'(\bm y - \bm X_1\bm\alpha)\\
%        \bm H_0 &= \bm X_0(\bm X_0'\bm X_0)^{-1}\bm X_0'\\
% \bm E_{10} &= \bm X_1 - \bm H_0 \bm X_1 \\
%    \bm e_0 &= \bm y - \bm H_0\bm y\\
%    \bm Q_1 &= \bm E_{10}'\bm E_{10} \\ 
%    \bm \ell_1 &= \bm E_{10}'\bm e_0
%    \end{align*}

With Lemma~\ref{lem-simple} in hand, we proceed to marginalize over $\tau$. For a flat prior on $\bm \alpha$, we have the following:
\begin{lemma}\label{lem-flat-marg-2}
Under the Gaussian linear model \eqref{gen-lm} and the prior \eqref{alt-prior} with $p(\bm \alpha \mid \tau) \propto 1$, the partial marginal posterior for the regression coefficients is 
    \[
    [\bm \alpha \mid \bm y] \sim t_{n-p}(\bm{\hat\alpha}, \hat S^2 \bm{\hat\Sigma}_{\alpha})
    \]
    where $\bm{\hat \Sigma}_{\alpha} = (\bm E_{10}'\bm E_{10})^{-1}$  and $\bm{\hat\alpha} = \bm{\hat \Sigma}_{\alpha} \bm E_{10}'\bm e_0$, with  $\bm E_{10} = \bm X_1 - \bm H_0 \bm X_1$  and $\bm e_0 = \bm y - \bm H_0\bm y$ for $ \bm H_0 = \bm X_0(\bm X_0'\bm X_0)^{-1}\bm X_0$.
\end{lemma}
\begin{proof}[Proof (Lemma~\ref{lem-flat-marg-2})]
    The result follows from Lemma~\ref{lem-simple} by noticing $(n-p)\hat S^2 = \Vert \bm e \Vert^2$ and recycling the arguments from Lemmas~\ref{lem-flat-marg} and \ref{lem-norm-marg} to obtain $p(\bm \alpha \mid \bm y) \propto \{(n-p)\hat S^2 + (\bm \alpha - \bm{\hat\alpha})'\bm{\hat \Sigma}_{\alpha}^{-1}(\bm \alpha - \bm{\hat\alpha})  \}^{-(n-p_0)/2}$. 
\end{proof}
By the FWL Theorem, the posterior mean and (unscaled) covariance for $\bm \alpha$ are identical to the OLS analogs. As expected, marginalizing over $\bm \beta$ under a flat prior has the same effect as regressing out $\bm X_0$. Next, we extend this result for a Gaussian prior: 
\begin{lemma}\label{lem-norm-marg-2}
    Under the Gaussian linear model \eqref{gen-lm} and the prior \eqref{alt-prior} with $[\bm \alpha \mid \tau] \sim N(\bm \alpha_0, \tau^{-1}\bm \Sigma_{\alpha_0})$,  the partial marginal posterior for the regression coefficients  is 
    \[
        [\bm \alpha \mid \bm y] \sim t_{n-p_0}(\bm{\hat \alpha}_{G}, \hat S_{\mathcal{M}_{01}(G)}^2 \bm{\hat \Sigma}_{\alpha_G} ) 
    \]
        where      
        $\bm{\hat \Sigma}_{\alpha_{G}} = (\bm E_{10}'\bm E_{10} + \bm\Sigma_{\alpha_0}^{-1})^{-1}$, $\bm{\hat \alpha}_{G} = (\bm E_{10}'\bm E_{10}+ \bm \Sigma_{\alpha_0}^{-1})^{-1} (\bm E_{10}'\bm e_0 + \bm \Sigma_{\alpha_0}^{-1} \bm \alpha_0)$, $\hat S_{\mathcal{M}_{01}(G)}^2  = \hat S_{0}^2 + (n-p_0)^{-1}(\Vert \bm\alpha_0 \Vert_{\bm\Sigma_{\alpha_0}^{-1}}^2  - \Vert \bm{\hat \alpha}_{G} \Vert_{\bm{\hat \Sigma}_{\alpha_{G}}^{-1}}^2)$, and  $\hat S_{0}^2 = (n-p_0)^{-1} \Vert \bm e_0 \Vert^2$.
    %\begin{align*}
     %   \bm{\hat \Sigma}_{\alpha_{\mathcal{N}}} &= (\bm E_{10}'\bm E_{10} + \bm\Sigma_{\alpha_0}^{-1})^{-1}, \quad
     %   \bm{\hat \alpha}_{\mathcal{N}} = (\bm E_{10}'\bm E_{10}+ \bm \Sigma_{\alpha_0}^{-1})^{-1} (\bm E_{10}'\bm e_0 + \bm \Sigma_{\alpha_0}^{-1} \bm \alpha_0) \\
      %  \hat S_{\mathcal{N},\mathcal{M}_{01}}^2   &= \hat S_{0}^2 + (n-p_0)^{-1}(\Vert \bm\alpha_0 \Vert_{\bm\Sigma_{\alpha_0}^{-1}}^2  - \Vert \bm{\hat \alpha}_{\mathcal{N}} \Vert_{\bm{\hat \Sigma}_{\alpha_{\mathcal{N}}}^{-1}}^2), \quad \hat S_{0}^2 = (n-p_0)^{-1} \Vert \bm e_0 \Vert^2.
   % \end{align*}
\end{lemma}
\begin{proof}[Proof (Lemma~\ref{lem-norm-marg-2})]
    We again apply Lemma~\ref{lem-simple} but now with  $
p(\bm \alpha, \tau) \propto \tau^{-1} \tau^{p_1/2} \exp\{-\tau (\bm\alpha - \bm\alpha_0)'\bm \Sigma_{\alpha_0}^{-1}(\bm \alpha - \bm\alpha_0)/2\}$. We can combine exponentials for $\tau$ and rearrange terms within the exponential for $\bm \alpha$: 
\[
p(\bm\alpha, \tau \mid \bm y) \propto \tau ^{(n - p_0 + p_1)/2 - 1} \exp\left[-\frac{\tau}{2} \left\{(\bm \alpha - \bm{\hat \alpha}_{G})'\bm{\hat \Sigma}_{\alpha_{G}}^{-1}(\bm \alpha - \bm{\hat \alpha}_{G})   +  (n-p_0)\hat S_{\mathcal{M}_{01}(G)}^2 \right\}\right] 
\]
where the scale term collects  $(n-p_0) \hat S_{\mathcal{M}_{01}(G)}^2 =  \Vert \bm e \Vert^2 +  \bm \ell_{\alpha}' \bm{\hat \Sigma}_{\alpha}\bm \ell_{\alpha}  + \Vert \bm\alpha_0 \Vert_{\bm\Sigma_{\alpha_0}^{-1}}^2  - \Vert \bm{\hat \alpha}_{G} \Vert_{\bm{\hat \Sigma}_{\alpha_{G}}^{-1}}^2 $. The final simplification occurs because   $\Vert \bm e \Vert^2 + \bm \ell_{\alpha}' \bm{\hat \Sigma}_{\alpha}\bm \ell_{\alpha} = \Vert \bm e \Vert^2 + \Vert \bm H_{10} \bm e_0 \Vert^2 = \Vert \bm e_0 \Vert^2$. Then the same steps as in Lemmas~\ref{lem-flat-marg} and \ref{lem-norm-marg} apply to marginalize over $\tau$.  % with $\bm H_{10} = \bm E_{10}(\bm E_{10}'\bm E_{10})^{-1}\bm E_{10}'$.
%where $\hat S_0^2 = (n-p_0)^{-1} \Vert \bm e_0 \Vert^2$, noting that $\Vert \bm e \Vert^2 + \bm \ell_1'\bm Q_1^{-1} \bm \ell_1 = \Vert \bm e \Vert^2 + \Vert \bm H_{10} \bm e_0 \Vert^2 = \Vert \bm e_0 \Vert^2$. 
%noting $\Vert \bm y \Vert^2 - \Vert \bm H_0 \bm y \Vert^2 = \Vert \bm y - \bm H_0 \bm y\Vert^2$. 
%Thus, the marginal posterior is 
%\[
%p(\bm \theta_1 \mid \bm y) \propto \Big\{(n-p_0) \hat S_{\mathcal{N},1}^2  + (\bm \theta_1 - \bm{\hat\mu}_1)'\bm Q_{\mathcal{N},1}(\bm \theta_1 - \bm{\hat\mu}_1)    \Big\}^{-(n-p_0 + p_1)/2}
%\]
%which is a $t$-distribution with $n-p_0$ degrees of freedom and (posterior) mean and covariance 
%\begin{align*}
%    E(\bm \theta_1 \mid \bm y) &= \bm{\hat\mu}_1 = (\bm E_{10}'\bm E_{10}+ \bm \Sigma_1^{-1})^{-1} (\bm E_{10}'\bm e_0 + \bm \Sigma_1^{-1} \bm \mu_1)\\
%    \mbox{Cov}(\bm\theta_1 \mid \bm y) &= \frac{n-p_0}{n-p_0-2} \hat S_{\mathcal{N},1}^2  (\bm E_{10}'\bm E_{10}+ \bm \Sigma_1^{-1})^{-1}. 
%\end{align*}
\end{proof}

These results prove the statements about $\mathcal{M}_{01}$ in Theorems~\ref{thm-flat} and \ref{thm-normal}. Next, we adapt and extend these results for  $\mathcal{M}_{012}$ with condition \eqref{cond} and subject to a linear constraint on the augmented parameters.

\subsubsection{Adaptations for  linearly-constrained $\mathcal{M}_{012}$ with condition \eqref{cond}}
Instead of marginalizing over $\bm \beta$ and then $\tau$, we insert an intermediate step that marginalizes over $\bm \gamma$. Like $\bm \beta$, $\bm \gamma$ contains regression coefficients; but unlike $\bm\beta$, $\bm\gamma$ features a constraint $\bm A \bm \gamma = \bm 0$ and different priors subject to this constraint. We leverage the reparameterization approach from Section~\ref{subsec-enforce} to obtain  $\bm X_{\mathcal{N}} = \bm X_2 \bm Q_{\mathcal{N}}$ and $\bm \gamma = \bm Q_{\mathcal{N}} \bm \gamma_{\mathcal{N}}$. Now, we  consider  the Gaussian linear model \eqref{gen-lm} with partitioned parameters $\bm\theta = (\bm\beta' : \bm\alpha' : \bm \gamma_{\mathcal{N}}')'$ and covariates   $\bm X = (\bm X_0 : \bm X_1 : \bm X_{\mathcal{N}})$ with respective dimensions (columns) $p_0$, $p_1$, $p_{\mathcal{N}}$, and update the  prior \eqref{alt-prior} to be
\begin{equation}\label{alt-prior-cm}
    p(\bm \beta, \bm \alpha, \bm \gamma_\mathcal{N}, \tau) \propto \tau^{-1} p(\bm \alpha \mid \tau)\  p(\bm \gamma_\mathcal{N} \mid \tau)
\end{equation}
with the ability to map between $\bm \gamma$ and $\bm \gamma_\mathcal{N}$ depending on convenience. 
For clarity, we refer to this model as $\mathcal{M}_{012}$ and use that designation for residuals, for example, $\bm e_{\mathcal{M}_{012}}$ are the OLS residuals from the model with $(\bm X_0 : \bm X_1 : \bm X_{\mathcal{N}})$, $\bm e_{\mathcal{M}_{01}}$ are the OLS residuals from the previous model with $(\bm X_0 : \bm X_1)$, and so forth.  

The first step is to marginalize over $\bm \beta$, akin to Lemma~\ref{lem-simple}: 
\begin{lemma}\label{lem-simple-cm}
    Under the augmented Gaussian linear model $\mathcal{M}_{012}$, the prior \eqref{alt-prior-cm} subject to $\bm A \bm \gamma = \bm 0$, and condition \eqref{cond}, the marginal (over $\bm \beta$) posterior is
    \begin{align*}%\label{gen-simp}
    p(\bm\alpha, \bm\gamma_{\mathcal{N}}, \tau \mid \bm y) \propto \tau ^{(n - p_0)/2 - 1} \exp\Big[-\frac{\tau}{2} \Big\{ &(\bm \alpha - \bm{\hat\alpha})'\bm{\hat \Sigma}_{\alpha}^{-1}(\bm \alpha - \bm{\hat\alpha}) \\
    &+ (\bm \gamma_{\mathcal{N}} - \bm{\hat\gamma}_{\mathcal{N}})' \bm{\hat\Sigma}_{\gamma_{\mathcal{N}}}^{-1} (\bm \gamma_{\mathcal{N}} - \bm{\hat\gamma}_{\mathcal{N}})  
       + \Vert \bm e_{\mathcal{M}_{012}} \Vert^2 \Big\}\Big] \\
       & \times  p(\bm \alpha \mid \tau) \ p(\bm\gamma_{\mathcal{N}} \mid \tau),
\end{align*}
%where $\bm{\hat \Sigma}_{\alpha} = (\bm E_{10}'\bm E_{10})^{-1}$  and $\bm{\hat\alpha} = \bm{\hat \Sigma}_{\alpha} \bm E_{10}'\bm e_0$, with  $\bm E_{10} = \bm X_1 - \bm H_0 \bm X_1$  and $\bm e_0 = \bm y - \bm H_0\bm y$ for $ \bm H_0 = \bm X_0(\bm X_0'\bm X_0)^{-1}\bm X_0$.
where   $\bm{\hat\gamma}_{\mathcal{N}} = \bm{\hat\Sigma}_{\gamma_{\mathcal{N}}}  \bm E_{Q0}'\bm e_0$, $\bm{\hat\Sigma}_{\gamma_{\mathcal{N}}} = (\bm E_{Q0}'\bm E_{Q0})^{-1}$, and $\bm E_{Q0} = \bm E_{20} \bm Q_{\mathcal{N}} = \bm X_{\mathcal{N}} - \bm H_0 \bm X_{\mathcal{N}}$.
\end{lemma}
\begin{proof}[Proof (Lemma~\ref{lem-simple-cm})]
The result follows by first applying Lemma~\ref{lem-simple} jointly to $(\bm \alpha': \bm \gamma_{\mathcal{N}}')'$. Note that this notation is overloaded: in terms of Lemma~\ref{lem-simple}, we submit  $(\bm \alpha': \bm \gamma_{\mathcal{N}}')'$ for $\bm \alpha$, $(\bm X_1: \bm X_{\mathcal{N}})$ for $\bm X_1$, and $(\bm E_{10}: \bm E_{Q0})$ for $\bm E_{10}$. Then returning to current notation, we apply condition \eqref{cond} to obtain the orthogonality $\bm E_{10}'\bm E_{Q0} = \bm E_{10}' \bm E_{20} \bm Q_{\mathcal{N}} = \bm 0$, which produces the simplification within the exponential term. 
\end{proof}

%\[(\bm \theta_1 - \bm Q_1^{-1}\bm\ell_1)'\bm Q_1(\bm \theta_1 - \bm Q_1^{-1}\bm\ell_1) = (\bm \alpha - \bm{\hat\alpha})' \bm{\hat\Sigma}_\alpha^{-1} (\bm \alpha - \bm{\hat\alpha}) + (\bm \gamma_{\mathcal{N}} - \bm{\hat\gamma}_{\mathcal{N}})' \bm{\hat\Sigma}_{\gamma_{\mathcal{N}}}^{-1} (\bm \gamma_{\mathcal{N}} - \bm{\hat\gamma}_{\mathcal{N}})\]

The second step is to marginalize over $\bm \gamma_{\mathcal{N}}$, which equivalently marginalizes over $\bm \gamma$ subject to the linear constraint. The key insight of Lemma~\ref{lem-simple-cm} is that the impact of $\bm\gamma$ will be felt \emph{only} in the terms that involve $\tau$---namely, the scaling terms and potentially the degrees of freedom.  

First, consider a  flat (constrained) prior $p(\bm \gamma \mid \tau) \propto \mathbb{I}\{\bm A \bm \gamma = \bm 0\}$, or equivalently, $p(\bm\gamma_{\mathcal{N}})\propto 1$. Marginalizing over $\bm \gamma_{\mathcal{N}}$ yields the following: 
\begin{lemma}\label{lem-simple-cm-2}
    Under the augmented Gaussian linear model $\mathcal{M}_{012}$, the prior \eqref{alt-prior-cm} with $p(\bm \gamma \mid \tau) \propto \mathbb{I}\{\bm A \bm \gamma = \bm 0\}$, and condition \eqref{cond}, the marginal (over $\bm \beta$ and $\bm \gamma$) posterior is
    \begin{align*}%\label{gen-simp}
    p(\bm\alpha, \tau \mid \bm y) \propto \tau ^{(n - p_0 - p_{\mathcal{N}})/2 - 1} \exp\Big[-\frac{\tau}{2} \Big\{ (\bm \alpha - \bm{\hat\alpha})'\bm{\hat \Sigma}_{\alpha}^{-1}(\bm \alpha - \bm{\hat\alpha})   + \Vert \bm e_{\mathcal{M}_{012}} \Vert^2 \Big\}\Big] \ p(\bm \alpha \mid \tau). 
\end{align*}
\end{lemma}
\begin{proof}[Proof (Lemma~\ref{lem-simple-cm-2})]
The result is derived similar to the proof of Lemma~\ref{lem-simple}: the terms in Lemma~\ref{lem-simple-cm} that involve $\bm \gamma_{\mathcal{N}}$ equal the density of $N(\bm{\hat\gamma}_{\mathcal{N}}, \tau^{-1} \bm{\hat\Sigma}_{\gamma_{\mathcal{N}}})$ times $\vert \tau^{-1} \bm{\hat\Sigma}_{\gamma_{\mathcal{N}}} \vert^{1/2} \propto \tau^{-p_{\mathcal{N}}/2}$. Thus, this term is all that contributes. 
\end{proof}

Lemma~\ref{lem-simple-cm-2} immediately proves the  second result in Theorem~\ref{thm-flat}, simply by applying the usual marginalization steps for $\tau$. This is essentially Lemma~\ref{lem-flat-marg-2}, but replacing $\hat S^2 = \hat S_{\mathcal{M}_{01}}^2$ with $\hat S_{\mathcal{M}_{012}}^2$ and $p_0$ with $p_0 + p_{\mathcal{N}}$. The second result in Theorem~\ref{thm-normal} similarly follows Lemma~\ref{lem-norm-marg-2}, but replacing $\hat S_{0}^2$ with $\hat S_{\mathcal{M}_{02}}^2$---i.e., the OLS residual variance using $(\bm X_0:\bm X_{\mathcal{N}})$ (instead of just $\bm X_0$)---and $p_0$ with $p_0 + p_{\mathcal{N}}$. For the third results in Theorems~\ref{thm-flat} and \ref{thm-normal}, we consider a Gaussian (constrained) prior $[\bm \gamma \mid \tau] \sim N(\bm \gamma_0, \tau^{-1}\bm \Sigma_{\gamma_0})\mathbb{I}\{\bm A \bm \gamma = \bm 0\}$. We apply Lemma~\ref{lem-simple-cm} to marginalize over $\bm \beta$, then marginalize over $\bm\gamma$, and lastly simplify the relevant scale terms.
%again marginalize over $\bm \beta$ and $\bm \gamma$: 
\begin{lemma}\label{lem-simple-cm-3}
    Under the augmented Gaussian linear model $\mathcal{M}_{012}$, the prior \eqref{alt-prior-cm} with $[\bm \gamma \mid \tau] \sim N(\bm \gamma_0, \tau^{-1}\bm \Sigma_{\gamma_0})\mathbb{I}\{\bm A \bm \gamma = \bm 0\}$, and condition \eqref{cond}, the marginal (over $\bm \beta$ and $\bm \gamma$) posterior is
    \begin{align*}%\label{gen-simp}
    p(\bm\alpha, \tau \mid \bm y) \propto \tau ^{(n - p_0)/2 - 1} \exp\Big[-\frac{\tau}{2} \Big\{ &(\bm \alpha - \bm{\hat\alpha})'\bm{\hat \Sigma}_{\alpha}^{-1}(\bm \alpha - \bm{\hat\alpha})    \\
    &+ 
   %(n - p_0 - p_1)\hat S_{\mathcal{N},1,\mathcal{M}_{012}}^2 
   \Vert \bm e_{\mathcal{M}_{01}}\Vert^2 + \Vert \bm \gamma_{0}\Vert_{\bm{\Sigma}_{\gamma_0}^{-1}}^2  - \Vert \bm{\hat\gamma} \Vert_{\bm{\hat\Sigma}_{\gamma}^{-}}^2
    \Big\}\Big] \ p(\bm \alpha \mid \tau). 
\end{align*}
where $\bm{\hat\gamma}  = \bm{\hat\Sigma}_{\gamma} (\bm E_{20}'\bm e_0 +    \bm{\Sigma}_{\gamma_0}^{-1} \bm \gamma_{0})$,    and $\bm{\hat\Sigma}_{\gamma} = \bm Q_{\mathcal{N}}\{\bm Q_{\mathcal{N}}'(\bm E_{20}'\bm E_{20} + \bm{\Sigma}_{\gamma_0}^{-1})\bm Q_{\mathcal{N}}\}^{-1}\bm Q_{\mathcal{N}}'$.
\end{lemma}
\begin{proof}[Proof (Lemma~\ref{lem-simple-cm-3})]
Lemma~\ref{lem-simple-cm} isolates the likelihood contributions from $\bm \gamma_{\mathcal{N}}$ as 
$\exp\{-\frac{\tau}{2} (\bm \gamma_{\mathcal{N}} - \bm{\hat\gamma}_{\mathcal{N}})' \bm{\hat\Sigma}_{\gamma_{\mathcal{N}}}^{-1} (\bm \gamma_{\mathcal{N}} - \bm{\hat\gamma}_{\mathcal{N}})\}$. Then, since the prior contributes $\tau^{p_{\mathcal{N}}/2}\exp\{-\frac{\tau}{2} (\bm Q_{\mathcal{N}}\bm \gamma_{\mathcal{N}} - \bm\gamma_0)'\bm\Sigma_{\gamma_0}^{-1}(\bm Q_{\mathcal{N}}\bm \gamma_{\mathcal{N}} - \bm\gamma_0)\}$, we aggregate all terms that involve $\bm\gamma_{\mathcal{N}}$ and regroup them appropriately: 
\begin{align*}
   % & \tau^{p_{\mathcal{N}}/2}  \exp\Big[-\frac{\tau}{2} \Big\{
    %    (\bm \gamma_{\mathcal{N}} - \bm{\hat\gamma}_{\mathcal{N}})' \bm{\hat\Sigma}_{\gamma_{\mathcal{N}}}^{-1} (\bm \gamma_{\mathcal{N}} - \bm{\hat\gamma}_{\mathcal{N}})
     %   + 
      %  (\bm Q_{\mathcal{N}}\bm \gamma_{\mathcal{N}} - \bm \gamma_{0})' \bm{\Sigma}_{\gamma_0}^{-1} (\bm Q_{\mathcal{N}}\bm \gamma_{\mathcal{N}} - \bm \gamma_{0})
    %\Big\}
    %\Big] \\
    %&= 
    \tau^{p_{\mathcal{N}}/2} \exp\Big[-\frac{\tau}{2} &\Big\{
        (\bm \gamma_{\mathcal{N}} - \bm{\hat\gamma}_{\mathcal{N}(G)})' \bm{\hat\Sigma}_{\gamma_{\mathcal{N}}(G)}^{-1} (\bm \gamma_{\mathcal{N}} - \bm{\hat\gamma}_{\mathcal{N}(G)})\Big\}\Big] \\
        & \quad \quad \times \exp\Big[-\frac{\tau}{2} \Big\{
            \Vert \bm H_{Q0} \bm e_0 \Vert^2 + \Vert \bm \gamma_{0}\Vert_{\bm{\Sigma}_{\gamma_0}^{-1}}^2  - \bm{\hat\gamma}_{\mathcal{N}(G)} ' \bm{\hat\Sigma}_{\gamma_{\mathcal{N}}(G)}^{-1} \bm{\hat\gamma}_{\mathcal{N}(G)}
        \Big\}\Big] 
\end{align*}
where  $\bm{\hat\Sigma}_{\gamma_{\mathcal{N}}(G)} = (\bm E_{Q0}'\bm E_{Q0} +  \bm Q_{\mathcal{N}}'\bm{\Sigma}_{\gamma_0}^{-1}\bm Q_{\mathcal{N}})^{-1} = \{\bm Q_{\mathcal{N}}'(\bm E_{20}'\bm E_{20} + \bm{\Sigma}_{\gamma_0}^{-1})\bm Q_{\mathcal{N}}\}^{-1} $ and $\bm{\hat\gamma}_{\mathcal{N}(G)} = \bm{\hat\Sigma}_{\gamma_{\mathcal{N}}(G)}(\bm E_{Q0}'\bm e_0 +  \bm Q_{\mathcal{N}}' \bm{\Sigma}_{\gamma_0}^{-1} \bm \gamma_{0}) = \bm{\hat\Sigma}_{\gamma_{\mathcal{N}}(G)} \bm Q_{\mathcal{N}}'(\bm E_{20}'\bm e_0 +    \bm{\Sigma}_{\gamma_0}^{-1} \bm \gamma_{0}) $.  The first term is proportional to a normal density in $\bm\gamma_{\mathcal{N}}$ and hence integrates to a constant (not involving $\tau$). The second term does not involve $\bm \gamma_{\mathcal{N}}$ and only contributes to the scale terms. 

In conjunction with Lemma~\ref{lem-simple-cm}, the aggregated scale terms are $\Vert \bm e_{\mathcal{M}_{012}} \Vert^2 + \Vert \bm H_{Q0} \bm e_0 \Vert^2 + \bm \gamma_{0}' \bm{\Sigma}_{\gamma_0}^{-1} \bm \gamma_{0} - \bm{\hat\gamma}_{\mathcal{N}(G)} ' \bm{\hat\Sigma}_{\gamma_{\mathcal{N}}(G)}^{-1} \bm{\hat\gamma}_{\mathcal{N}(G)}$. First, we simplify $\Vert \bm e_{\mathcal{M}_{012}} \Vert^2 + \Vert \bm H_{Q0} \bm e_0 \Vert^2 =  \Vert \bm e_0\Vert^2 - \Vert \bm H_{10} \bm e_0 \Vert^2 = \Vert \bm e_{\mathcal{M}_{01}} \Vert^2 = (n - p_0 - p_1)\hat S_{\mathcal{M}_{01}}^2$. This follows similarly to a previous argument that showed $\Vert \bm e_{\mathcal{M}_{01}} \Vert^2 = \Vert \bm e_0\Vert^2 - \Vert \bm H_{10} \bm e_0 \Vert^2$: here we apply the same reasoning jointly to $(\bm X_1:\bm X_{\mathcal{N}})$ and note that the hat matrix for $(\bm E_{10}:\bm E_{Q0}) $ is simply $\bm H_{10} + \bm H_{Q0}$ with $\bm H_{10}\bm H_{Q0}=\bm 0$, thus yielding  $\Vert \bm e_{\mathcal{M}_{012}} \Vert^2 = \Vert \bm e_0\Vert^2 - \Vert \bm H_{10} \bm e_0 \Vert^2- \Vert \bm H_{Q0} \bm e_0 \Vert^2$. Next, we show that $\Vert \bm{\hat\gamma}_{\mathcal{N}(G)} \Vert_{\bm{\hat\Sigma}_{\gamma_{\mathcal{N}}(G)}^{-1}}^2 = \Vert \bm{\hat\gamma} \Vert_{\bm{\hat\Sigma}_\gamma^{-}}^2$ by linking the posterior moments, $\bm{\hat\Sigma}_\gamma = \bm Q_{\mathcal{N}} \bm{\hat\Sigma}_{\gamma_{\mathcal{N}}(G)}\bm Q_{\mathcal{N}}'$ and $\bm{\hat\gamma}  = \bm Q_{\mathcal{N}} \bm{\hat\gamma}_{\mathcal{N}(G)}$, and then simplifying: 
    \begin{align*}
        \Vert \bm{\hat\gamma}\Vert_{\bm{\hat\Sigma}_\gamma^{-}}^2 &= \bm{\hat\gamma}' \bm{\hat\Sigma}_\gamma^{-} \bm{\hat\gamma} = (\bm E_{20}'\bm e_0 +    \bm{\Sigma}_{\gamma_0}^{-1} \bm \gamma_{0})' \bm{ \hat \Sigma}_\gamma (\bm E_{20}'\bm e_0 +    \bm{\Sigma}_{\gamma_0}^{-1} \bm \gamma_{0}) \\
        &= (\bm E_{20}'\bm e_0 +    \bm{\Sigma}_{\gamma_0}^{-1} \bm \gamma_{0})'
        \bm Q_{\mathcal{N}} \bm{\hat\Sigma}_{\gamma_{\mathcal{N}}(G)}\bm Q_{\mathcal{N}}'(\bm E_{20}'\bm e_0 +    \bm{\Sigma}_{\gamma_0}^{-1} \bm \gamma_{0})\\
        &= (\bm E_{20}'\bm e_0 +    \bm{\Sigma}_{\gamma_0}^{-1} \bm \gamma_{0})'
        \bm Q_{\mathcal{N}} \bm{\hat\Sigma}_{\gamma_{\mathcal{N}}(G)}\bm{\hat\Sigma}_{\gamma_{\mathcal{N}}(G)}^{-1} \bm{\hat\Sigma}_{\gamma_{\mathcal{N}}(G)}\bm Q_{\mathcal{N}}'(\bm E_{20}'\bm e_0 +    \bm{\Sigma}_{\gamma_0}^{-1} \bm \gamma_{0}) \\
        &= \Vert \bm{\hat\gamma}_{\mathcal{N}(G)} \Vert_{\bm{\hat\Sigma}_{\gamma_{\mathcal{N}}(G)}^{-1}}^2.
    \end{align*}
Combining these simplified scale terms, and noticing that there are no additions to the $\tau$ exponent (from Lemma~\ref{lem-simple-cm}),  yields the result. 
%Compared to the flat prior from before, there are no additions to the exponent of $\tau$ and thus no $p_{\mathcal{N}}$ adjustments to the degrees of freedom. 
\end{proof}
Applying the usual steps to marginalize over $\tau$, the  third results in Theorems~\ref{thm-flat} and \ref{thm-normal} follow immediately.

 \subsection{Corollaries demystifying the main condition}

\begin{proof}[Proof (Corollary~\ref{cor-anova-k})]
Since $\bm E_{10} = \bm 1$ and $\bm E_{20} = (\bm Z_1, \ldots, \bm Z_K)$,  $\bm E_{10}'\bm E_{20}$ is composed of  $\bm 1'\bm Z_k$, which records the totals for each group in categorical variable $k=1,\ldots,K$. These match the ABCs \eqref{abcs-gen-cat}, so the row space condition of $\bm A$ is satisfied. 
\end{proof}

\begin{proof}[Proof (Corollary~\ref{cor-anova})]
Without $\bm X_0$, we have $\bm E_{10} = \bm X_1$ and $\bm E_{20} = \bm X_2$. Now, $\bm X_1'\bm X_2$  is composed of blocks  $\bm 1'\bm Z_{1:2}$, $\bm Z_1'\bm Z_{1:2}$, and  $\bm Z_2 ' \bm Z_{1:2}$. The entries of 
$\bm 1'\bm Z_{1:2} $ are the totals for each ($\texttt{c}_1$, $\texttt{c}_2$) pairing. $\bm Z_1'\bm Z_{1:2}$ is $L_1 \times L_1L_2$ where the entries in each row $r$ are the total counts of $\texttt{c}_1 = r$ and  $\texttt{c}_2 = s$ for each $s=1,\ldots,L_2$, with zeros for $\texttt{c}_1 \ne  r$; and similar for $\bm Z_2'\bm Z_{1:2}$. These match the ABCs \eqref{abcs-cat-cat-gen}, so the rows of $\bm E_{10}'\bm E_{20} = \bm X_1'\bm X_2$ belong to $\mbox{row}(\bm A)$. 
\end{proof}

\begin{proof}[Proof (Corollary~\ref{cor-x})]
For the CX interaction in the ANCOVA model, it is sufficient to construct  conditions for which the rows of $\bm E_{01}'\bm X_2$ are proportional to  $\bm{\hat\pi}'$. Expanding terms, $\bm E_{01}'\bm X_2 = \bm X_1'\bm X_2 - \bm X_1' \bm H_0 \bm X_2 = \bm x'\mbox{diag}(\bm x)\bm Z_1 - \bm x' \bm H_0 \mbox{diag}(\bm x)\bm Z_1$. Direct simplification yields $\bm x'\mbox{diag}(\bm x)\bm Z_1 = (s_{x[1]}^2, \ldots, s_{x[L_1]}^2)$, where $s_{x[r]}^2 = \sum_{r_i = r} x_i^2$, and $\bm x'\bm H_0 \mbox{diag}(\bm x)\bm Z_1 = (n\hat\pi_1 \bar x_1^2,\ldots, n\hat\pi_{L_1} \bar x_{L_1}^2)$. It follows that $\bm E_{01}'\bm X_2 = (s_{x[1]}^2 - n\hat\pi_1 \bar x_1^2, \ldots, s_{x[L_1]}^2 - n\hat\pi_{L_1} \bar x_{L_1}^2) = n (\hat \pi_1 \hat\sigma_{x[1]}^2, \ldots, \hat \pi_{L_1} \hat\sigma_{x[L_1]}^2)$. Under the equal-scale condition $\hat \sigma_{x[1]}^2 = \cdots = \hat \sigma_{x[L_1]}^2$, it follows that $\bm E_{01}'\bm X_2 = n \hat \sigma_{x[1]}^2 \bm{\hat \pi}' \propto \bm{\hat \pi}'$. 
\end{proof}

\begin{proof}[Proof (Corollary~\ref{cor-cat-cat})]
    The central term $\bm E_{10}'\bm X_2$ has rows $\bm e_{10,j}'\bm X_2$ for $j=1,\ldots,p$. The entries of $\bm e_{10,j}'\bm X_2$ are given by $n\hat\pi_{rs} \bar e_{j[rs]}$, 
    where $\hat \pi_{rs} $ denotes the mutual group proportions of $\texttt{c}_1 =r$ and $\texttt{c}_2 =s$, i.e., the column means of $\bm Z_{1:2}$.  Now, if all $\bar e_{j[rs]} = \bar e_{j[11]}$ are invariant to $r,s$, it follows that $\bm e_{10,j}'\bm X_2 \propto \bm{\hat\pi}'$. When this occurs for all $j=1,\ldots,p$,   \eqref{cond} is satisfied. %This implication follows from a comment in the previous ABCs paper, i.e., that the joint cat-cat ABCs satisfy this mutual ABC, so $\bm{\hat\pi}'\bm Q_{\mathcal{N}} = \bm 0$. This condition $\bar e_{j[rs]} = \bar e_{j[11]}$ for all $(r,s)$ allows $\bm x_j$ to vary with \texttt{race} and \text{sex}, but only additively: the main effects in $\bm X_0$, which include \texttt{race} and \texttt{sex} main effects (among other potential variables), must not leave race:sex interactions in the residuals for $\bm x_j$. 
\end{proof}

\begin{proof}[Proof (Corollary~\ref{cor-cat-cts})]
    For ABCs with multiple CX interactions, the constraint matrix and thus $\bm Q_{\mathcal{N}}$  is block-diagonal. Here, the categorical modifier  is $\texttt{c}_1$ for all $j=1,\ldots,p$, so
    $\bm Q_{\mathcal{N}} = \mbox{bdiag}(\{\bm Q_{\bm{\hat\pi}}\}_{k=1}^p) = \bm I_p \otimes \bm Q_{\bm{\hat\pi}}$, where $\bm Q_{\bm{\hat\pi}}$ refers to the QR null space term computed from $\bm A = \bm{\hat\pi}'$ as in Section~\ref{subsec-enforce}. %Now, let $\bm Z_{x_k} = \mbox{diag}(\bm x_k)\bm Z_{1}$. 
    The condition \eqref{cond} occurs whenever %$\bm E_{01}'\bm Z_{x_k} \bm Q_{\bm{\hat\pi}} = \bm 0$ 
    %$\bm E_{01}'\mbox{diag}(\bm x_k)\bm Z_{1} \bm Q_{\bm{\hat\pi}} = \bm 0$ for  $k=1,\ldots,p$. Considering each row, it suffices to show that  
    $\bm e_{10,j}'\mbox{diag}(\bm x_k)\bm Z_{1} \propto \bm{\hat\pi}'$  for $j,k=1,\ldots,p$. A key point is that $\bm e_{10,j}'\bm Z_1 = \bm 0$ (i.e., the elements of $\bm e_{10,j}$ sum to zero within each group $\texttt{c}_1 = r$) because $\bm Z_1$ is included in $\bm X_0$. Thus, it is straightforward to show that $\bm e_{10,j}'\mbox{diag}(\bm x_k)\bm Z_{1}$ is $1 \times L_1$ with entries $\sum_{r_i=r} e_{10,ij} x_{ik} = n \hat\pi_r\widehat{\mbox{Cov}}_{[r]}(\bm e_{10,j}, \bm x_k)$. Under the equal-covariance condition, it follows that $\bm e_{10,j}'\mbox{diag}(\bm x_k)\bm Z_{1} = n \widehat{\mbox{Cov}}_{[1]}(\bm e_{10,j}, \bm x_k) \bm{\hat\pi}' \propto \bm{\hat\pi}'$.
\end{proof}

\subsection{Validity of Algorithm~\ref{alg:constr}}\label{app-proofs-repar}
In Section~\ref{subsec-enforce}, we showed how to incorporate ABCs into a Bayesian model by 1) conditioning on the linear constraint and 2) using the QR-based reparameterization \eqref{qr-con} to a lower-dimensional, unconstrained space. Here, we justify the latter approach. Recall that we consider the Gaussian linear model $[\bm y \mid \bm\theta] \sim N_n(\bm X \bm \theta, \tau^{-1} \bm I_n)$ with $p$-dimensional parameter $\bm \theta \in \mathbb{R}^p$ subject to the linear constraints $\bm A \bm \theta = \bm 0$, 
where $\bm A$ is $m \times p$ with rank $m$, so $\bm \theta$ contains $p_\mathcal{N} = p - m$ free parameters. The idea is to specify a preliminary prior, $\bm \theta \sim N_p(\bm 0, \bm \Sigma_\theta)$, and then condition on the linear constraint. Thus, we seek the constrained posterior 
$[\bm \theta \mid \bm y, \bm A \bm\theta = \bm 0]$ and aim to show that Algorithm~\ref{alg:constr} samples from it. 

First, we provide a direct computation of the constrained posterior. Beginning with the unconstrained posterior,
\[
[\bm \theta \mid \bm y] \sim N_p(\bm \mu_u, \tau^{-1}\bm \Sigma_u), \quad \bm \Sigma_u = (\bm X'\bm X + \bm\Sigma_\theta^{-1})^{-1}, \quad \bm \mu_u = \bm \Sigma_u \bm X'\bm y,
\]
and noticing that  ($\bm \theta, \bm A \bm \theta)$ is jointly Gaussian, it is straightforward to derive the conditional posterior distribution that enforces the constraints: 
\begin{align*}
[\bm \theta \mid \bm y, \bm A \bm\theta = \bm 0] &\sim N_p(\bm \mu_A, \tau^{-1}\bm \Sigma_A) \\
\bm \Sigma_A &= \bm \Sigma_u - \bm \Sigma_u \bm A'(\bm A\bm\Sigma_u\bm A')^{-1}\bm A \bm \Sigma_u, \\
\bm \mu_A &= \bm \mu_u - \bm \Sigma_u \bm A'(\bm A\bm\Sigma_u\bm A')^{-1}\bm A \bm \mu_u = \bm \Sigma_A \bm X'\bm y.
\end{align*}
However, this distribution is suboptimal for sampling: it features $p$-dimensional (or $p \times p$) quantities, even though the constrained parameter lies in a  $p_{\mathcal{N}}$-dimensional space with $p_{\mathcal{N}} = p-m$. Alternatively, it is possible to sample from the unconstrained posterior $[\bm \theta \mid \bm y]$ and apply a suitable adjustment to obtain a draw from the constrained posterior (e.g., \citealp{gelfand2010handbook}, Chapter~12.1.7.4), but this is also wasteful for repeatedly sampling.

Now, the goal is to show that Algorithm~\ref{alg:constr} produces a draw from this same distribution. For completeness, recall that $\bm\Sigma_{\mathcal{N}} = (\tau \bm X_{\mathcal{N}}'\bm X_{\mathcal{N}} + \bm Q_{\mathcal{N}}'\bm\Sigma_\theta^{-1} \bm Q_{\mathcal{N}})^{-1}$, $\bm \mu_{\mathcal{N}} = \tau \bm \Sigma_{\mathcal{N}} \bm X_{\mathcal{N}}' \bm y$, 
$\bm X_{\mathcal{N}} = \bm X \bm Q_{\mathcal{N}}$, 
and $\bm Q_{\mathcal{N}}$ derives from \eqref{qr-con}.  Since Algorithm~\ref{alg:constr} samples $[\bm \theta_{\mathcal{N}} \mid \bm y] \sim N_{p_\mathcal{N}}(\bm \mu_{\mathcal{N}}, \bm \Sigma_{\mathcal{N}})$ and sets $\bm \theta =\bm Q_{\mathcal{N}}  \bm \theta_{\mathcal{N}}$, it follows that the induced posterior is $[\bm Q_{\mathcal{N}}  \bm \theta_{\mathcal{N}} \mid \bm y] \sim  N_p(\bm \mu, \bm \Sigma)$ where  $\bm \Sigma = \bm Q_{\mathcal{N}} \bm \Sigma_{\mathcal{N}} \bm Q_{\mathcal{N}}'$  and  $\bm \mu = \bm Q_{\mathcal{N}} \bm \mu_{\mathcal{N}}$. Notice that $\bm \mu = \bm \Sigma \bm X' \bm y$ and $\bm \mu_A = \bm \Sigma_A \bm X'\bm y$. Thus, it is sufficient to show that $\bm \Sigma_A = \bm \Sigma$.

To do so, we re-derive the constrained posterior, but instead using the rotated parameters $\bm Q' \bm \theta = (\bm \theta_{\mathcal{R}}', \bm \theta_{\mathcal{N}}')'$ with  $\bm \theta =  \bm Q_{\mathcal{R}} \bm \theta_{\mathcal{R}} + \bm Q_{\mathcal{N}} \bm \theta_{\mathcal{N}}$, and then conditioning on  $\bm \theta_{\mathcal{R}} = \bm 0$. Building upon the unconstrained posterior, it follows that $[\bm Q' \bm \theta \mid \bm y] \sim N_p(\bm Q' \bm \mu_u, \bm Q' \bm\Sigma_u \bm Q)$.
 Define the blocked versions 
\[
\bm Q'\bm \mu_u = \begin{pmatrix} \bm m_{\mathcal{R}} \\  \bm m_{\mathcal{N}} \end{pmatrix}, \quad \bm Q'\bm\Sigma_u \bm Q = \begin{pmatrix} \bm S_{\mathcal{R}\mathcal{R}} & \bm S_{\mathcal{R}\mathcal{N}} \\ \bm S_{\mathcal{N}\mathcal{R}} & \bm S_{\mathcal{N}\mathcal{N}} \end{pmatrix}.
\]
Now, we obtain the constrained posterior distribution  $[\bm \theta \mid \bm y, \bm A \bm \theta = \bm 0]$ by first conditioning on $\bm \theta_{\mathcal{R}} = \bm 0$ and then pre-multiplying by $\bm Q_{\mathcal{N}}$, which we present in one step:
\[
[\bm \theta \mid \bm y, \bm A \bm \theta = \bm 0] \sim N_p\left\{\bm Q_{\mathcal{N}} (\bm m_{\mathcal{N}} - \bm S_{\mathcal{N}\mathcal{R}}\bm S_{\mathcal{R}\mathcal{R}}^{-1} \bm m_{\mathcal{R}}), \bm Q_{\mathcal{N}} (\bm S_{\mathcal{NN}} - \bm S_{\mathcal{NR}}\bm S_{\mathcal{RR}}^{-1} \bm S_{\mathcal{RN}}) \bm Q_{\mathcal{N}}'\right\}.
\]
Observing that $\bm \Sigma_{\mathcal{N}}^{-1} =  \bm Q_{\mathcal{N}}' \bm \Sigma_u^{-1} \bm Q_{\mathcal{N}}$ and $\bm \Sigma = \bm Q_{\mathcal{N}} \bm \Sigma_{\mathcal{N}} \bm Q_{\mathcal{N}}'$ , it suffices to show that $\bm Q_{\mathcal{N}}' \bm \Sigma_u^{-1} \bm Q_{\mathcal{N}} = (\bm S_{\mathcal{NN}} - \bm S_{\mathcal{NR}}\bm S_{\mathcal{RR}}^{-1} \bm S_{\mathcal{RN}})^{-1}$. Pertinent to the latter term, we simplify $\bm Q' \bm \Sigma_u^{-1} \bm Q = (\bm Q' \bm \Sigma_u \bm Q)^{-1}$ since $\bm Q$ is orthonormal. Relating it to the former term, the lower right block of $\bm Q' \bm \Sigma_u^{-1} \bm Q$ is exactly $\bm Q_{\mathcal{N}}' \bm \Sigma_u^{-1} \bm Q_{\mathcal{N}}$. Now recalling the defined block structure of $\bm Q'\bm\Sigma_u \bm Q$ and using block inversion, the equality follows from the Schur complement identity. Thus, Algorithm~\ref{alg:constr} indeed samples from the constrained posterior distribution $[\bm \theta \mid \bm y, \bm A \bm \theta = \bm 0]$. 
Similar arguments also prove Lemma~\ref{lem-prior-equiv}.

\section{Alternative mechanism to enforce identifiability}\label{app-centering}
Besides the conditioning strategy employed in Section~\ref{subsec-enforce}, it is also possible to enforce ABCs via centering. To illustrate this idea, recall the one-way ANOVA model from Example~\ref{ex-anova}. We may ensure $\bm{\hat \pi}'\bm \beta = 0$ by again specifying an unconstrained prior, but now centering instead of conditioning:
\begin{equation}\label{prior-cent}
    [b_c \mid \lambda_\beta] \stackrel{iid}{\sim} N(0, \lambda_\beta^2), \quad 
    \beta_c = b_c - \bm {\hat\pi}'\bm b,\quad c=1,\ldots,L.
\end{equation}
Like in Sections~\ref{subsec-enforce} and \ref{subsec-anova}, it is possible to analyze the induced prior on $\bm \beta$. Writing $\bm \beta = \bm M_{\hat\pi} \bm b$ with $\bm M_{\hat\pi} = \bm I_L - \bm 1 \bm{\hat\pi}'$, it follows that 
\[
[\bm \beta \mid \lambda_\beta] \sim N_L(\bm 0, \lambda_\beta^2 \bm M_{\hat\pi} \bm M_{\hat\pi}'). 
\]
Here, the prior variances are $\mbox{Var}(\beta_c) = \lambda_\beta^2(1 + \Vert \bm{\hat\pi} \Vert^2 - 2 \hat\pi_c)$ whereas the conditional version had 
$\mbox{Var}(\beta_c \mid \bm{\hat \pi}'\bm\beta = 0) = \lambda_\beta^2(1 - \hat \pi_c^2/\Vert \bm{\hat \pi} \Vert^2)$. Similar differences arise for the priors on the group means, $\mu_c = \alpha_0 + \beta_c$. Larger discrepancies emerge for the contrasts: here, $\delta_{c,c'} = \mu_c - \mu_{c'} = b_c - b_{c'}$ has variance $2\lambda_\beta^2$,  whereas for the conditional version, $\mbox{Var}(\delta_{c,c'}  \mid \bm{\hat\pi}'\bm\beta = 0) = \lambda_\beta^2\{ 2 - (\hat\pi_c - \hat\pi_{c'})^2/\Vert \bm{\hat\pi} \Vert^2\}$ (both are mean zero Gaussian). Thus, the prior for the contrasts under \eqref{prior-cent} are invariant to $\bm{\hat \pi}$, while for the conditional version the prior variances are smallest for imbalanced groups. We caution that the implications for posterior shrinkage cannot be learned only by inspecting the prior variance; see Section~\ref{subsec-anova}. 

%The drawback of the centered prior \eqref{prior-cent} is that it is difficult to construct shrinkage or sparsity priors for $\bm \beta$, which are the regression coefficients in the linear model. Even if \eqref{prior-cent} were replaced  with an aggressive shrinkage or sparsity prior on $b_c$, taking $b_c \to 0$ does not imply sparsity of the corresponding regression coefficient $\beta_c$, which instead would be determined by the aggregated parameters $\bm {\hat\pi}'\bm b$. For a numerical example, suppose $\bm{\hat \pi}' = (0.25, 0.25, 0.5)$ and $\bm b' = (2, 2, 0)$, so sparsity exists in $\bm b$. Under \eqref{prior-cent}, $\bm \beta' = (1, 1, -1)$ is not sparse. Thus, it is unclear what an appropriate prior would be for $\bm b$, especially if regularization of $\bm \beta$ is desired. 

The drawback of the centered prior \eqref{prior-cent} is that it is difficult to generalize (uniquely) for arbitrary linear constraints, including ABCs with CC interactions \eqref{abcs-cat-cat-gen}. Even so, it is possible to recast this prior within the framework of Section~\ref{subsec-enforce}, which enables both posterior computing (Algorithm~\ref{alg:constr}) and posterior invariance (Section~\ref{sec-theory}). To see this, notice that we can write $\bm \beta = \bm Q_{\mathcal{N}} \bm \beta_{\mathcal{N}}$, where  $\bm Q_{\mathcal{N}}$ is from the QR-decomposition of the linear constraints \eqref{qr-con} and $\bm \beta_{\mathcal{N}}$ is a lower-dimensional, unconstrained Gaussian vector with mean zero and covariance $ \lambda_\beta^2 \bm Q_{\mathcal{N}}' \bm M_{\hat\pi} \bm M_{\hat\pi}' \bm Q_{\mathcal{N}}$. This representation is central to our sampling algorithm and posterior invariance results, and both will apply for centered priors of the form \eqref{prior-cent}.

\end{document}